White Paper

# LARGE-SCALE AI IN TELECOM

## Charting the Roadmap for Innovation, Scalability, and Enhanced Digital Experiences

*GenAINet Emerging Technology Initiative*

# EDITORS

Ali Mokh
Antonio De Domenico
Athanasios Karapantelakis
Chongwen Huang
Christina Chaccour
Fathi Abdeldayem
Juan Deng
Keith Ball
Lina Bariah
Merouane Debbah
Navid Nikaein
Omar Hashash
Qiyang Zhao
Sihem Cherrared

# CONTRIBUTORS

**AALTO UNIVERSITY**
*Nassim Sehad*

**BEIJING INSTITUTE OF TECHNOLOGY**
*Jie Zeng*
*Yifan Yang*

**BRUNEL UNIVERSITY LONDON**
*Kehai Qiu*
*Kezhi Wang*

**BUBBLERAN**
*Navid Nikaein*

**CEA-LETI**
*Emilio Calvanese Strinati*

**CENTRAL SOUTH UNIVERSITY**
*Chao Zhang*

**CENTRALESUPÉLEC, UNIVERSITY OF PARIS SACLAY**
*Elissa Mhanna*
*Mohamad Assaad*
*Ryad Madi*
*Salah Eddine El Ayoubi*
*Zeinab Nehme*

**CENTRE TECNOLÒGIC DE TELECOMUNICACIONS DE CATALUNYA**
*Rasoul Nikbakht Silab*

**CHINA MOBILE COMMUNICATIONS CORPORATION**
*Jiajun Wu*
*Juan Deng*
*Liexiang Yue*
*Tianjiao Chen*
*Yanping Liang*
*Yingping Cui*

**CHINA TELECOM**
*Meiling Dai*
*Xiaoou Liu*
*Xingyu Shang*

**CHINA UNICOM**
*Jinan Li*
*Zhuang Zhou*

**EAST CHINA NORMAL UNIVERSITY**
*Kun Guo*

**EMIRATES INTEGRATED TELECOMMUNICATIONS COMPANY (DU)**
*Fathi Abdeldayem*
*Mohammad Al Refai*
*Najla Alkaabi*

**ERICSSON**
*Alexandros Nikou*
*Ali Mokh*
*Athanasios Karapantelakis*
*Christina Chaccour*
*Hakimeh Purmehdi*
*Johny Gemayel*
*Serveh Shalmashi*
*Timothy Murphy*

**EURECOM**
*Ilias Chatzistefanidis*
*Ioannis Pitsiorlas*
*Roberto Morabito*

**FENTECH**
*Julien Frison*
*Moussab Djerrab*

**GSMA**
*Louis Powell*

**HUAWEI**
*Antonio De Domenico*
*Chenghui Peng*
*Fei Wang*
*Yvan Pointurier*
*Zhe Liu*

**IMEC - GHENT UNIVERSITY**
*Adnan Shahid*
*Eli De Poorter*
*Jaron Fontaine*

**ITU**
*Vishnu Ram*

## KATIM

*Dheeraj Sharma*
*Dimitris Kalogiros*

## KHALIFA UNIVERSITY

*Lina Bariah*
*Merouane Debbah*
*Samson Lasaulce*
*Wassim Hamidouche*

## KING'S COLLEGE LONDON

*Na Yan*
*Nan Li*
*Sige Liu*
*Yang Su*
*Yansha Deng*

## KOREA UNIVERSITY

*Inkyu Lee*

## KTH ROYAL INSTITUTE OF TECHNOLOGY

*Amirreza Kazemi*
*Carlo Fischione*

## LIGHTON

*Iacopo Poli*
*Igor Carron*

## NANJING UNIVERSITY

*Bo Cheng*
*Haibo Zhou*
*Yu Sun*

## NANYANG TECHNOLOGICAL UNIVERSITY

*Chau Yuen*
*Dusit Niyato*

## NOKIA BELL LABS

*Gianluca Fontanesi*

## NORTHEASTERN UNIVERSITY

*Maxime Elkael*
*Michele Polese*
*Salvatore D'Oro*
*Tommaso Melodia*

## NORTHWESTERN POLYTECHNICAL UNIVERSITY

*Bo Yang*

## NVIDIA

*Keith Ball*
*Maria Amparo Canaveras Galdon*
*Mubeen Syed*
*Swastika Dutta*

## ORANGE

*Lecorvé Gwénolé*
*Mehdi Boudjelli*
*Sihem Cherrared*

## QUALCOMM

*Jinane Karam*

## RIMEDO LABS

*Adrian Kliks (Poznan University of Technology)*
*Marcin Dryjanski*
*Pawel Sroka (Poznan University of Technology)*

## SINGAPORE UNIVERSITY OF TECHNOLOGY AND DESIGN

*Tony Q.S. Quek*
*Zihan Chen*

## SUN YAT-SEN UNIVERSITY

*Xijun Wang*

## TECHNOLOGY INNOVATION INSTITUTE

*Faouzi Bader*
*Hang Zou*
*Qiyang Zhao*

## ULSAN NATIONAL INSTITUTE OF SCIENCE AND TECHNOLOGY

*Hoon Lee*

## UNIVERSITAT POMPEU FABRA

*Giovanni Geraci (Telefonica Research)*
*Mohamed Benzaghta*

## UNIVERSITY OF ELECTRONIC SCIENCE AND TECHNOLOGY OF CHINA

*Weilong Chen*

## UNIVERSITY OF GRANADA

*Marios Kountouris*

## UNIVERSITY OF HONG KONG

*Hongyang Du*
*Kaibin Huang*

## UNIVERSITY OF HOUSTON
*Zhu Han*

## UNIVERSITY OF LEEDS
*Maryam Hafeez*
*Muhammad Amir*
*Syed A. R. Zaidi*
*Zeinab Nezami*

## UNIVERSITY OF MICHIGAN
*Zitao Shuai*

## UNIVERSITY OF OULU
*Ahmed Elbakary*
*Chaouki Ben Issaid*
*Mehdi Bennis*

## UNIVERSITY OF YORK
*Ahmed Al-Tahmeesschi*
*Hamed Ahmadi*
*Swarna Bindu Chetty*

## VIRGINIA TECH
*Christo Kurisummoottil Thomas*
*Omar Hashash*
*Walid Saad*

## XIDIAN UNIVERSITY
*Xuelin Cao*

## YALE UNIVERSITY
*Ali Maatouk*
*Leandros Tassiulas*
*Rex Ying*

## ZHEJIANG UNIVERSITY
*Bohao Wang*
*Chongwen Huang*
*Fenghao Zhu*
*Howard H. Yang*
*Qianqian Yang*
*Rongpeng Li*
*Xiaoxue Yu*
*Xinquan Wang*
*Yuxuan Chen*
*Zhaohui Yang*
*Zhaoyang Zhang*
*Zhenyu Yang*
*Zirui Chen*

# EXECUTIVE SUMMARY

The rise of generative artificial intelligence (AI) as a novel frontier that uniquely merges advanced levels of intelligence with revolutionary user experiences is redefining the AI landscape for future cellular networks. In particular, the transition towards 6G systems has introduced a myriad of challenges inherent to their AI-native network design, requiring innovative solutions to enable real-time network orchestration, intelligent decision-making, and adaptive dynamic configurations. Meanwhile, the envisioned user experiences for 6G are growing increasingly complex, exceeding the capabilities offered by vintage wireless technologies and conventional AI solutions to satisfy their advanced demands.

With its disruptive impact evident across diverse fields, generative AI possesses immense potential to tackle these challenges, leveraging its exceptional capabilities to manage complex tasks, operate autonomously, and adapt seamlessly to scenarios beyond its training domain. Remarkably, generative AI provides a transformative opportunity for telecom and cellular networks to bridge this defined gap in 6G systems, thereby shifting towards a new era with cutting-edge AI innovations across the different system and user levels.

In essence, the introduction of generative AI into the telecom domain is primarily facilitated by a set of large-scale AI models denoted as **large telecom models (LTMs)**. These LTMs are specifically designed to tailor the abilities of large scale AI models to effectively meet the demands of the telecom ecosystem. The goal of this white paper is to shed light on the potential of LTMs to revolutionize the telecom functions and applications from the theoretical design, implementation, and deployment perspectives, while touching on the regulatory, standardization, and industrial frameworks that govern their realization in practice. To this end, this white paper provides an explanatory overview of LTMs and their distinctive role in the radio access network (RAN) and core network, while expanding the discussion to cover several key areas that include:

**Fundamentals of large-scale AI:** Reflecting on the generative architectures and models that compose large-scale AI, along with recent trends in handling multi-modal training data, pre-training and fine-tuning techniques, alignment techniques (e.g., reinforcement learning (RL) with human feedback), and deployment strategies on the network.
**From large-scale AI models to LTMs:** Moving beyond the state-of-the-art large scale AI models that can be vulnerable in the telecom domain, while highlighting the necessary modifications to the underlying theory of large-scale AI models to foresee the emergence of LTMs.
**LTMs for physical and MAC layer designs:** Addressing resource allocation, spectrum management, channel modeling, and mobility management, among others.
**LTMs for network management and optimization:** Spanning adaptive monitoring and control in emerging frameworks such as Open RAN networks (i.e., O-RAN), while highlighting the critical role of leveraging LTMs with RL to enable user-centric network optimization.

**Datasets for LTMs:** Supporting the deployment of LTMs with telecom-specific datasets and providing benchmarks with evaluation frameworks to assess the performance of LTMs.

**Hardware advancements and requirements for LTMs:** Focusing on the role of high computing platforms to accelerate the deployment of LTMs and how the convergence of the RAN with AI plays a role in enabling LTMs over future cellular networks.

**New use cases and applications of LTMs:** Encompassing distributed LTM frameworks over the edge, novel approaches for federated learning in LTMs, RL with LTMs interaction, intent-based management with LTMs, etc.

**Regulatory and ethical considerations for LTMs:** Emphasizing that data governance and accountability are crucial considerations to acquire trustworthy LTM operations.

**Industry insights into large-scale AI models and LTMs:** Including the current trends and ongoing projects in the industry that include large action models and on-device generative AI models, recent model breakthroughs such as TelecomGPT, and practical challenges that face LTMs such as the limited decoding rate and massive model sizes.

**Standardization activities and LTM roadmap:** Discussing the key efforts to bring forth LTMs through focus groups within regional bodies, while setting the roadmap for LTMs by defining their roles in network infrastructure, network management, business operations with the corresponding timeline for LTMs to reach their milestones.

Ultimately, this white paper serves as an inaugural roadmap for LTMs in networks and provides a basis for telecom experts and industry professionals to build on the state-of-art in LTMs to push the boundaries of large-scale AI models for next-generation wireless networks.

# Contents

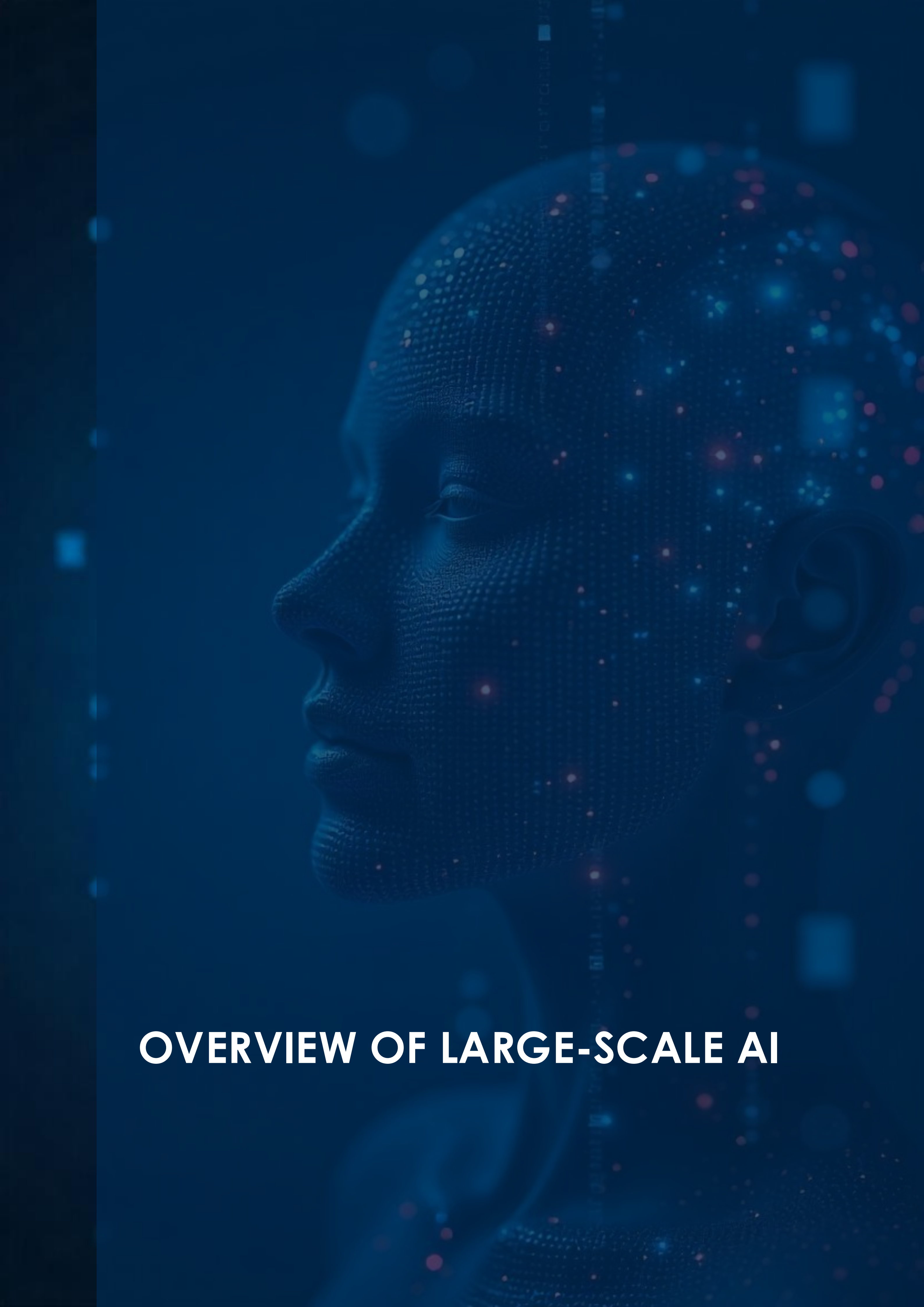

# OVERVIEW OF LARGE-SCALE AI

# 1 Overview of Large-Scale AI

## 1.1 Background and Definitions

Upcoming sixth generation of mobile networks (6G), are expected to provide more services than their predecessors, to an ever-growing number of users. The complexity of managing mobile networks is amplified by their distributed nature. To address this, Artificial Intelligence (AI) algorithms, including deep learning models and symbolic approaches must be deployed at scale across both the radio access network (RAN) and the core network. These algorithms serve as essential tools to automate network management and handle the growing complexity in a cost-efficient way.

While AI algorithms are already used in mobile networks, their functionality is typically confined to specific domains, resulting in siloed deployments that address only narrow aspects of network performance [1]. These AI systems struggle to generalize across diverse network environments. More complex challenges, such as those requiring human-like reasoning and planning, rely on symbolic techniques, which in turn depend on human-curated knowledge bases. However, these knowledge bases tend to be brittle and difficult to scale [2].

Generative AI (GenAI) algorithms have recently risen to prominence due to advancements in deep learning architectures such as transformers, which enable models to capture complex patterns and relationships within large datasets [3]. Techniques such as autoregressive modeling and diffusion processes allow these systems to generate high-fidelity outputs, ranging from text to images, by learning the underlying data distribution. This scalability, combined with the ability to prompt or fine-tune models using minimal supervision, has significantly reduced the reliance on human-annotated data, allowing GenAI models such as Large Language Models (LLMs) to outperform symbolic models in Question Answering (QA) tasks [4], but also deep-learning based Artificial Neural Networks (ANNs) such as Recurrent Neural Networks (RNNs), in tasks such as Automated Speech Recognition (ASR), Speech Translation (ST) and Text-to-Speech (TTS)[5]. Additionally, research shows that LLMs, when guided by well-structured prompts that instruct them to generate intermediate reasoning steps during the output process, can effectively solve complex arithmetic, symbolic and commonsense tasks involving advanced, System 2 reasoning [6]. Given this potential, it is promising to leverage GenAI in mobile networks at progressively larger scales and with greater decision authority, ultimately aiming for fully autonomous networks.

A recent survey highlighted multiple areas across all layers of mobile networks where research on the application of GenAI algorithms and models is actively underway [7]. In this paper, we focus on Large Telecom Models (LTMs). LTMs are Language Models (LMs) that require large amounts of resources, for example training data, but also compute, store and bandwidth for use in mobile networks.

## 1.2 Introduction to Scalable, Large Telecom Models

Figure 1 presents a simplified, layered architecture of a mobile network, highlighting various application areas for LTMs. The architecture includes a top-level exposure layer that enables third parties, such as enterprise customers, to interact with the mobile network. These interactions may involve specifying *intents*[1], such as requests for specific quality of service (QoS) parameters (e.g., limits on latency and packet drop rates, or guaranteed throughput). The management layer oversees operations of the mobile network, incorporating both automated processes (e.g., leveraging AI models) and human-driven activities, such as Field Service Operations (FSO) and

[1] In the context of autonomous networks, an intent refers to a high-level, declarative goal or objective that the system aims to fulfill without detailing the specific steps required.

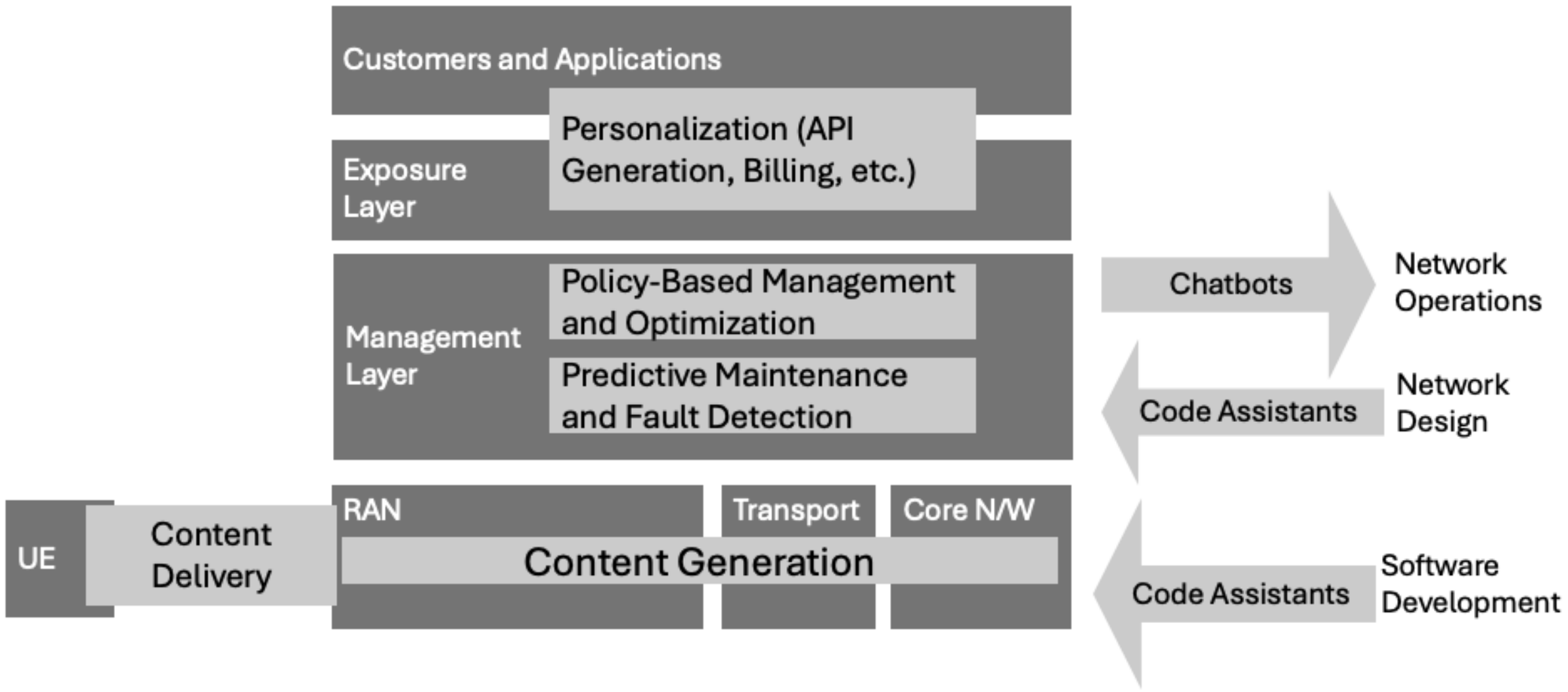

Figure 1: Overview of applications of LTMs in a mobile network.

monitoring through a Network Operations Centre (NOC). An infrastructure layer consists of various elements of a mobile network, such as RAN, core network, transport network between and through core and RAN, as well as the wireless interface between RAN and User Equipment (UE)[2]

Although LTMs are only a subset of AI models, they find broad applicability across multiple network layers. We generally categorize applications into two types: those that enhance mobile network capabilities by adding new functionalities or improving existing ones, and those that operate peripherally to the network's core functions, interacting directly with humans. Examples of such peripheral functions include network monitoring in a NOC, field maintenance operations, network design and deployment, as well as the development of network-related software components.

## 1.3 Large Telecom Models for Telecom Functions

From a top-down perspective, as illustrated in figure 1, LTMs play an important role in the exposure layer, serving as a link between the customer and network domains. Specifically, LTMs can translate high-level, abstract intents expressed in the customer's language—such as natural language in a Service-Level Agreement (SLA) document—into technical requirements that the management layer can interpret and configure in the network. Their ability to understand natural language and process technical documentation makes LTMs ideal for scaling this type of translation efficiently, thus thereby enhancing the personalization of the network experience for customers. Another facet of this personalization involves development of tailored subscription plans for customers, including billing structures and feature offerings that are specifically adapted to customer needs, rather than being generically designed by the mobile network operator. In another example of personalization,fa LTM agents can be used to abstract complexity of telecom-specific network exposure Application Program Interfaces (APIs) from the customer, such as those specified for Network Exposure Function (NEF) and Service Capability Exposure Function (SCEF) by 3GPP. In this case, a customer may use an abstract prompt, such as a natural language interface to request services from the network, such as monitoring of it's UE, and an LTM can translate this request to a sequence of API calls.

In the management layer, LTMs can facilitate the automation of network operations management. This encompasses two key aspects: first, the support LTMs can provide for field services, such as the installation, commission-ing, and maintenance of Radio Base Stations (RBSs). In this context, LTMs can serve as digital assistants, aiding field service personnel during on-site tasks. Often, RBSs are located in remote areas, making it costly to send personnel for fault repairs or scheduled maintenance. Additionally, staff must be proficient not only in the specific products on-site to diagnose issues and identify root causes but also must undergo safety training, especially since faults may occur on radio tower tops. LTM assistants can help mitigate some of these expenses by offering real-time assistance based on situational assessments to field service engineers. The second aspect involves automating trouble reporting in the NOC. Typically, identifying issues requires the aggregation and analysis of diverse data sets, including logs from various systems and visual data. LTMs, with their capacity to be effectively prompted for this information through approaches like Retrieval-Augmented Generation (RAG) [8], can process the data to generate a trouble report in natural language, as well as provide recommendations for fault resolution. LTMs can also be used for policy-based management, as they can dynamically interpret and implement complex network policies in real time.

[2]In Third Generation Partnership Project (3GPP) nomenclature, UE are the user terminals accessing the network (e.g., mobile phones).

In the infrastructure layer, we identify two separate uses for LTMs. The first is content generation, wherein LTMs, such as Generative Adversarial Networks (GANs), can be utilized to create synthetic data to augment real network observations. This is particularly useful in cases where real data is scarce or the network is overloaded with communication tasks, preventing it from observing and/or transporting actual data to where it is needed. For example, if data originates from UE, such as Channel State Information (CSI) reports, then the availability of this data depends on the presence of UE. However, there may be instances where UE are unavailable, such as when the network aims to generate a model of the wireless channel in a specific area or to create a coverage map. A generative model can be used in such cases to augment real-life network observations. The other aspect involves content delivery over the air interface, particularly for media-rich, high-resolution content like high-definition video. In 6G, adoption of Extended Reality (XR) applications is anticipated to rise, alongside the introduction of new devices such as virtual reality and augmented reality headsets. Technologies like semantic communication, which leverage encoder-decoder generative models such as Variational Autoencoders (VAEs) and transformers instead of transmitting raw data, can significantly reduce bandwidth requirements per XR session. This reduction would enable the network to accommodate a greater number of concurrent sessions.

## 1.4 Large Telecom Models for Peripheral Functions

In addition to being embedded in the mobile network itself, LTMs can also be used for functions that support the mobile network. One key area where LTMs can help is in network deployment and planning. They can analyze historical and real-time data to determine the best locations for new base stations. This analysis includes looking at factors like population density, user behavior, and existing network coverage. By understanding where users are and how they use the network, LTMs can suggest locations that will maximize coverage and minimize interference.

Additionally, LTMs can assist in designing network architectures by simulating different scenarios. For instance, they can model how changes in user demand or network traffic might affect performance. This helps engineers identify potential problems and make adjustments before actual deployment, saving time and resources.

In software development, LTMs can automate routine tasks such as coding, testing, and documentation. For example, they can take user requirements written in plain language and generate code snippets that fulfill those requirements. This can speed up the development process and reduce the chance of errors.

Moreover, LTMs can help create synthetic data for testing. When real data is limited or unavailable, this synthetic data can simulate real-world conditions, allowing developers to test their software thoroughly. They can also generate various test scenarios to ensure that network applications perform well under different conditions.

Another application of LTMs is as digital assistants, i.e. *chatbots* that assist users such as customers and mobile network operator personnel in navigating complex mobile standards, as well as product and network documentation. These chatbots can provide instant answers to queries about technical specifications, regulatory requirements, or best practices in mobile network operations. By offering easy access to this information, LTMs can help reduce the time engineers spend searching for documents or standards, allowing them to focus more on critical tasks.

Overall, using LTMs for network deployment, planning, and software development can lead to more efficient operations and better service for users. By analyzing data, simulating scenarios, automating tasks, and providing quick access to information, LTMs can support network engineers and developers in building and maintaining robust mobile networks.

## 1.5 Contribution

Mobile networks consist of a large number of interconnected nodes, which generate and transmit large amounts of data. This distributed architecture is challenging in context of large-scale LTM deployment. On a high level this paper aims to provide the reader with an understanding on the following challenges related to deployment of LTMs.

- Understanding the background context, including a review of state of the art (SoA) of LTMs algorithms and deployment architectures.
- Understanding the use-cases and mobile network infrastructure requirements for deployment of LTMs in mobile networks, both for training but also for inference.
- Understanding the current capabilities of network infrastructure and UE. to host and train LTMs.
- Describing metrics and datasets to train LTMs and evaluate the accuracy and credibility of their responses during inference-time.
- Understanding the regulatory framework, standardization activities and market trends for large-scale adoption of LTMs.

This paper functions as a multi-disciplinary guide for large-scale deployment aspects of LTMs. This is done in form of a deep-dive into SoA tools and methods for each of the scalability aspects highlighted above. The remainder of this section provides more information on each of these aspects and includes references to other sections of the paper, allowing readers to directly access the topics they are interested in.

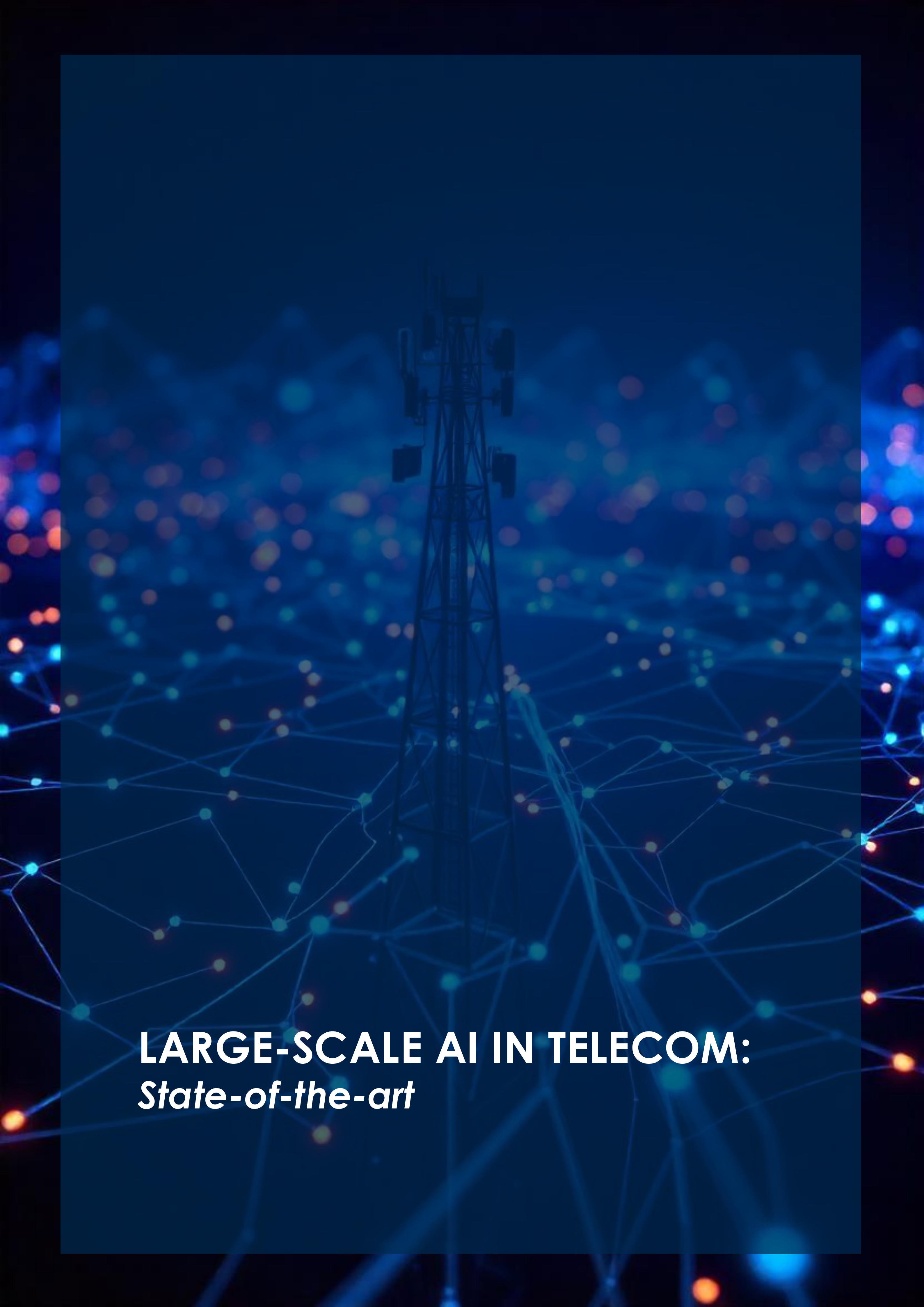

# LARGE-SCALE AI IN TELECOM:

*State-of-the-art*

# 2 Large-Scale AI in Telecom: State-of-the-art

## 2.1 Overview of Large-Scale AI in Telecom

6G network will be intelligent networks capable of responding in real-time to emerging demands and changing environments. It will support a wide range of applications and scenarios, such as the Internet of Things, smart grid, autonomous vehicles, intelligent agriculture, and so on. The design of Telecom network is transforming from pure communication oriented towards intent and goal oriented. It will not only deliver information to meet certain QoS requirements (throughput, latency, reliability), but also plan, configure, and optimize functionalities and protocols to environment conditions and user demands.

## 2.2 Large-Scale AI for Physical and MAC layer design

In this section, we discuss some fundamentals of the physical layer which would be enhanced by generative large models.

### 2.2.1 AI-Based MIMO Detection

MIMO detection refers to the process of recovering transmitted signals at the receiver end in a MIMO communication system, which is critical because, while MIMO allows for the simultaneous transmission of multiple data streams over the same frequency band, the signals from different antennas can interfere with each other due to multipath propagation and channel fading. The primary goal of MIMO detection is to separate these transmitted signals and recover the original data accurately. The main challenges in MIMO detection arise from interference between signals transmitted from different antennas, noise and other channel impairments, and the high-dimensional detection in systems with many antennas like massive MIMO. Depending on the number of antennas and the channel conditions, various detection algorithms are employed, each with trade-offs in terms of complexity and performance. The current methods for MIMO detection range from traditional linear approaches, like Zero Forcing and MMSE, to advanced non-linear (like Maximum Likelihood and Sphere Decoding), iterative, and machine learning-based techniques. Table 2 lists these methods.

Generative AI and large models (like deep learning) can significantly improve MIMO detection by offering advanced capabilities for handling complex and dynamic communication environments. These AI-driven methods are particularly useful in dealing with the high dimensionality, non-linearity, and real-time constraints typical of MIMO systems. Below are ways in which generative AI and large models can help improve MIMO detection:

*Learning-Based Detection*: Deep learning models, particularly large models like DNNs, can be trained to map received signals directly to transmitted symbols, bypassing traditional detection techniques. This is especially valuable in highly complex and dynamic MIMO systems such as massive MIMO, where traditional algorithms struggle with non-linearity and high interference levels. Generative models can learn complex non-linear relationships between transmitted and received signals, making them highly suitable for MIMO systems operating in complex environments, such as those with fading, interference, and mobility. Unlike traditional methods that require specific model assumptions, they also have easy real-time adaptation and deep learning models can generalize well to various channel conditions after training. As we observed, in large-scale MIMO systems, traditional methods can become computationally expensive. AI-based models can efficiently scale and handle these large systems, reducing computational load and improving detection speed.

*Generative Models for Channel Estimation*: Generative models such as GANs and VAEs can help enhance channel estimation and modeling in MIMO systems, which is crucial for accurate detection. These models can simulate realistic wireless channel environments and assist in generating more accurate channel matrices, improving de-tection performance. GANs can be used to generate synthetic, yet realistic, channel conditions based on a small amount of real-world data, allowing the MIMO detector to adapt to various channel conditions dynamically.

Table 2: Summary of the current ML detection methods

| Category | Method | Approach | Advantage | Disadvantage |
|---|---|---|---|---|
| **Linear Detection** | Zero Forcing (ZF) | Eliminates multi-antenna interference by inverting the channel matrix. | Simple and easy to implement. Good for high SNR environments. | Amplifies noise, poor performance in noisy or ill-conditioned channels. |
| | Minimum Mean Square Error (MMSE) | Balances between interference suppression and noise amplification. | Better than ZF in noisy channels. | Slightly more computationally complex than ZF. |
| | Matched Filter (MF) | Maximizes the received SNR for each transmitted stream. | Very simple and computationally efficient. | Suffers from interference between data streams. |
| **Non-Linear Detection** | Successive Interference Cancellation (SIC) | Detects the strongest signal first, subtracts its contribution from the received signal, and then detects the remaining signals iteratively. | Interference reduction between streams. | Error propagation can occur; the order in which the signals are detected affects performance. |
| | Maximum Likelihood (ML) Detection | Minimizes the Euclidean distance between the received signal and its estimated version based on the channel matrix. | Minimizing BER. | Computational complexity increases exponentially with the number of transmit antennas and modulation order. Impractical for large systems. |
| | Sphere Decoding (SD) | Reduces the complexity of ML detection by searching within a certain sphere around the received signal. | Near-ML performance with lower complexity. | Still computationally expensive for large MIMO systems, though much less so than ML. |
| **Iterative Detection** | Turbo Detection | Involves iterative feedback between the equalizer (or detector) and the decoder. | Close to optimal performance. | Requires multiple iterations, increasing detection delay and computational load. |
| | Belief Propagation (BP) | Uses a factor graph and performs probabilistic inference, passing ”messages” between nodes to refine probability estimates. | Efficient in structured MIMO systems, such as LDPC-based systems. | Complexity increases with the number of antennas; convergence issues can arise. |
| **Compressed Sensing-Based Detection** | | Exploits the sparsity of the signal in systems where the number of transmit antennas exceeds the number of receive antennas. | Suitable for underdetermined MIMO systems. | Works best when the transmitted signal has sparse characteristics. |
| **Hybrid Detection Techniques** | | Combines multiple detection methods to balance between complexity and performance. | Flexibility to different channel conditions. A good trade-off between performance and complexity. | Increased complexity. |

Moreover, this generated data could be used for supervised learning-based MIMO detectors, enhancing their robustness to different signal and channel scenarios, even when limited real-world data is available. By predicting channel states in real-time, they reduce the complexity of explicit channel estimation steps required for MIMO detection.

*AI-Enhanced Joint Detection and Decoding*: Deep learning-based architectures like DetNet (Detection Network) and RNNs can integrate MIMO detection and decoding processes. AI models can jointly perform both MIMO detection and channel decoding tasks, leveraging the correlation between these steps to enhance overall system performance. By leveraging the joint detection-decoding ability, these models reduce the overall complexity of processing while improving the detection performance. AI-driven architectures that mimic turbo decoding allow for efficient error correction and more accurate signal recovery in the presence of noise and interference.

*Adaptive Detection Using Reinforcement Learning (RL)*: this category of learning models can be employed to dynamically adapt the detection strategy based on real-time feedback from the environment. RL agents can learn optimal detection strategies over time, adjusting to changing channel conditions or system parameters to maintain performance. This is especially useful in fast-changing channel conditions, where static detection strategies may not perform well. RL-based models can optimize MIMO detection without requiring a complete model of the wireless environment, learning the best actions to minimize BER or maximize throughput. They approaches can gradually improve performance without requiring vast amounts of training data upfront, unlike supervised deep learning models.

### 2.2.2 Channel coding/decoding

Channel codecs (channel coder and decoder) are essential techniques in digital communications used to detect and correct errors that may occur during data transmission over noise, interference, and other channel impairments. The goal of channel coding is to enhance the reliability of data transmission by introducing redundancy into the transmitted information, allowing the receiver to detect and correct errors without the need for retransmission. It includes three parts: the channel encoding adds redundancy to data to enable error detection and correction. The channel decoder uses redundancy to detect and correct errors in the received data. Finally, error detection and correction which improves reliability by detecting and correcting errors in transmitted data.

Current channel codec methods are advanced techniques and they have evolved significantly to meet the demands of modern communication systems like 5G, satellite communications, and digital storage. The main methods include block codes, convolutional codes, turbo codes, and low-density parity-check (LDPC) codes. Each method has its own coding/decoding strategy based on the type of errors it aims to detect or correct. Below is a short explanation of the most common methods:

- Block Codes: Hamming, Reed-Solomon codes (used for bursting error correction, storage systems).
- Convolutional Codes: Used in real-time communication, Viterbi decoding is widely applied in satellite and mobile systems.
- Turbo Codes: Highly efficient for wireless systems, iterative decoding (4G, 5G).
- LDPC Codes: Widely used in modern wireless systems (Wi-Fi, 5G) and storage systems due to their near-capacity performance and efficient decoding.
- Polar Codes: The latest in channel coding, selected for 5G control channels, offering efficient decoding through successive cancellation.
- HARQ: Combines ARQ and FEC, retransmitting corrupted packets with additional redundancy for improved error correction.

Large and generative models can significantly enhance the efficiency and performance of channel coding/decoding by introducing data-driven techniques to improve error correction, decoding algorithms, and coding optimization. These improvements have been explored in recent research, with AI models showing promise in overcoming some of the traditional limitations of classical coding techniques. Here's how generative AI and large models can help:

*Improving Decoding Algorithms*: Generative AI and large deep learning models can be used to optimize and improve decoding algorithms, particularly in soft-decision decoding and iterative decoding (e.g., for Turbo codes, LDPC codes). Neural networks can be trained to learn the decoding process from received noisy signals and automatically perform error correction. Deep learning decoders can replace conventional methods like the Viterbi or belief propagation algorithms, resulting in more robust and faster decoding. RNNs, or even more advanced models like LSTMs, can learn sequential data and are particularly well-suited for decoding convolutional codes. These AI-based models can outperform traditional methods when trained on large datasets with varying noise levels. Finally, generative models can predict the likelihood of each received bit and perform soft decoding, which is particularly useful for LDPC and turbo codes. By leveraging AI, decoders can more efficiently handle noisy and corrupted signals and find the most likely transmitted sequence with higher accuracy.

*Designing New Codes*: Generative AI can be employed to design new coding schemes that outperform traditional codes like LDPC, Turbo, or Polar codes. By treating coding as a generative process, AI models can learn to create highly efficient codes optimized for specific communication environments. For instance, instead of relying on predefined codes like Reed-Solomon or Hamming, neural networks can generate codes tailored to specific channel characteristics or noise levels, offering better performance in specific applications. AI models like autoencoders can be trained to simultaneously learn encoding and decoding procedures, allowing the discovery of new codes that can handle errors more efficiently. Autoencoders can capture complex channel noise characteristics and adapt the code to the channel. AI models can also create adaptive error-correction codes that evolve based on the current communication environment. This adaptability is useful in scenarios where channel conditions change dynamically (e.g., 5G networks).

*Optimizing Hybrid Automatic Repeat Request (HARQ)*: Generative AI models can optimize HARQ systems by predicting when retransmissions are necessary and adjusting the coding redundancy dynamically to reduce the need for retransmissions, saving bandwidth and improving throughput. RL can optimize retransmission decisions in HARQ by learning from the network's feedback. AI models can predict the optimal amount of redundancy needed for each transmission, improving efficiency. Generative models can adaptively generate parity bits based on the channel conditions, instead of relying on fixed redundancy patterns. This adaptive approach allows HARQ systems to be more efficient, reducing retransmissions in challenging conditions. Moreover, GANs can create synthetic data that represent possible transmission errors, allowing the system to simulate different channel conditions and learn optimal HARQ strategies under various error rates.

*Enhancing Polar Codes for 5G*: Generative AI can enhance Polar Codes, which are used in 5G for control channels, by improving the successive cancellation decoding (SCD) process or designing more efficient decoding architectures. AI models can be applied to successive cancellation list decoding (SCLD), improving the decision-making process during decoding by better estimating the probability of each bit. This leads to better error correction performance, especially in short-block-length polar codes. In addition, polar codes rely on frozen bits to reduce complexity. Generative AI can dynamically optimize the selection of frozen bits based on the channel conditions, improving the overall code efficiency and error correction capability. AI-driven polar codes can be optimized for specific use cases and communication environments.

*Optimizing LDPC Codes*: LDPC (Low-Density Parity-Check) codes are widely used in wireless and storage systems. Generative AI can optimize LDPC decoders by improving belief propagation (BP) algorithms, or even developing entirely new graph-based decoding architectures based on learned data. Enhancing the belief propagation algorithm is possible by learning better message-passing strategies between variable and check nodes, improving convergence speed and decoding accuracy. Moreover, DNNs can be trained as decoders that simulate the message-passing process of LDPC decoding but with higher resilience to noise and interference.

### 2.2.3 Resource allocation

Resource allocation for the physical layer in wireless communication systems involves distributing and managing resources like power, time, frequency, and spatial domains to maximize data throughput, minimize interference, and improve the overall network quality of service (QoS). Efficient resource allocation is critical for ensuring that the physical layer can support higher data rates, better reliability, and lower latency, especially in complex networks like 5G and beyond. The main components of resource allocation at the physical layer are power allocation, subcarrier allocation, time and spatial resource allocations, adaptive modulation and coding, and interference management and coordination. The current methods focus on optimizing power, frequency, time, and spatial resources to enhance performance in terms of throughput, latency, and energy efficiency. The resource allocation methods at the physical layer are especially vital in complex systems like 5G, 6G, and massive IoT, where resource allocation must be adaptive and scalable. However, these methods continue to evolve with advances in AI and ML. Generative AI and large models can enhance resource allocation in the physical layer by improving real-time decision-making, predictive analytics, and adaptive optimization. These technologies provide a data-driven, adaptive, and proactive approach to managing resources, allowing for enhanced performance, efficiency, and resilience in complex and high-density networks such as 5G and future 6G systems.

*Dynamic Power Allocation*: Generative AI can learn and predict optimal power allocation strategies based on real-time and historical network conditions, resulting in more efficient use of energy and better interference management. RL can continuously optimize power allocation by learning from network feedback and prioritizing power efficiency.

*Frequency and Subcarrier Allocation*: AI models can learn and allocate frequency resources more effectively in systems like OFDMA by predicting user demand, channel quality, and minimizing interference among users. By leveraging historical channel data, generative models can predict channel quality for users, improving the efficiency of frequency allocation. AI models can also treat resource allocation as a multi-agent game, optimizing allocation strategies among users competing for spectrum.

*Time Slot Allocation and Scheduling*: Large AI models can improve scheduling strategies in time domain resource allocation, handling real-time variations in user demand and network conditions more adaptively. RL models can dynamically allocate time slots based on network feedback, optimizing for throughput, and reducing latency. Moreover, AI models can balance user demand and fairness, providing tailored time slot allocations based on priority and resource availability.

*Spatial Resource Allocation with Beamforming and MIMO*: Generative models and reinforcement learning can optimize beamforming and MIMO spatial resources, making it easier to manage multi-user interference and boost spectral efficiency. Generative models can predict optimal beam directions to maximize signal quality and reduce interference for each user. AI models group users with similar spatial characteristics, optimizing spatial resource allocation and minimizing interference.

*Adaptive Modulation and Coding (AMC)*: AI-based models can predict channel conditions and adapt modulation and coding schemes dynamically, optimizing throughput and error rate based on real-time conditions. For example, NNs predict channel states and automatically select the most appropriate modulation and coding scheme, maximizing data rates; or, RL models dynamically adapt AMC schemes based on channel feedback, ideal for rapidly changing wireless environments.

*Interference Management and Coordination*: Generative AI models can be used to simulate and predict interference patterns interference under various configurations, allowing for preemptive adjustments to resources and enhancing network performance under high user density. Also, ML can improve CoMP by predicting inter-cell interference, enabling collaborative resource allocation among base stations.

### 2.2.4 Reconfigurable intelligent surfaces (RIS)

Using higher frequency bands poses challenges with electromagnetic wave obstructions. RIS offer a solution by actively shaping and controlling electromagnetic waves to improve wireless network performance. Comprising passive elements or small antennas, RIS can adjust the phase, amplitude, and sometimes polarization of incoming signals, allowing them to steer, reflect, or scatter radio waves without needing active transmission power. RIS acts like a smart mirror, reflecting signals and steering them by adjusting the phase. It enhances signal strength at specific locations, focuses signals into narrow beams for minimal energy loss, and optimizes multiple signal paths in wireless communication, reinforcing desired paths and reducing interference. RIS can provide improved coverage and capacity, enhances the energy efficiency with relatively simple and inexpensive HW compared to traditional infrastructure. Nevertheless, controlling thousands of elements on RIS requires sophisticated optimization algorithms that can adapt in real-time to changes in the environment.

Current methods for RIS algorithms focus on how to control and manipulate the individual elements of the surface to achieve desired signal processing outcomes. These methods generally fall into categories based on how the elements are designed and how they are controlled. Here are the main methods and approaches currently used in RIS:

- Passive vs. Active RIS (energy consumption vs. signal amplification)
- Phase Shifting (discrete vs. continuous)
- Programmable Meta-materials (electrical control being the most common)
- Reflective vs. Transmissive RIS (based on signal interaction)
- Hybrid RIS (combining passive and active elements)
- Digital vs. Analog Control (accuracy vs. simplicity)
- AI-Driven Optimization (for dynamic adaptation)
- Wireless Power Transfer (for self-powered systems)
- High-Frequency RIS (mmWave and THz bands for next-gen communication)

Generative AI and large models like deep learning and reinforcement learning can significantly enhance the performance, optimization, and deployment of RIS in wireless communication. the complexity of optimizing RIS-assisted systems requires advanced solutions, and generative AI and large models can provide innovative ways to address these challenges. They improve real-time adaptive control through RL and deep generative models enables dynamic and efficient beamforming. Provide high quality channel estimation and prediction using GANs and large models allow RIS to anticipate environmental changes and optimize signal paths. Moreover, RIS deployment and energy efficiency optimization by leveraging AI-driven simulations and forecasting methods reduces operational costs and improves network performance. These models also can contribute to the following areas which either directly or indirectly can help RIS:

*Channel Estimation and Reflection Coefficient Optimization*: One of the key tasks in RIS-assisted systems is optimizing the reflection coefficients (phase shifts) of the RIS elements to maximize SNR at the receiver. This requires accurate CSI between the base station, RIS, and users, which can be challenging to obtain in practical scenarios, especially for large surfaces with many elements. Generative models can generate realistic CSI data, helping improve channel estimation accuracy in RIS systems. Furthermore, deep learning models can directly learn the mapping between the channel conditions and optimal reflection coefficients, leading to better and faster optimization compared to traditional iterative algorithms.

*End-to-End Learning for RIS-Assisted Communication*: Optimizing the RIS requires coordination between the transmitter, the RIS, and the receiver. Traditional methods rely on separate optimization stages, which may not be globally optimal. Moreover, RIS elements are passive and cannot actively adjust their behavior based on real-time feedback, making real-time adaptation challenging. End-to-end deep learning models can be used to jointly optimize the entire system, including the transmit beamforming, RIS phase shifts, and receiver processing. RL or Deep RL can further enable RIS to dynamically adjust its configuration based on real-time environmental feedback.

*Generative AI for Environment and Channel Modeling*: RIS systems heavily depend on the surrounding environment, including obstacles, reflectors, and scattering objects. Accurate environment modeling is essential to predict how RIS should be configured to enhance the wireless signal. However, real-world environments are complex and dynamic, making it difficult to model them accurately in real-time. Generative AI models, such as GANs and VAEs, can be used to create realistic 3D models of the wireless environment as well as the channel conditions, allowing RIS systems to be trained and tested in a wide variety of scenarios without requiring costly real-world measurements.

*Data-Driven Optimization of RIS Hardware Parameters*: RIS devices are typically designed with fixed hardware characteristics, such as the number of reflecting elements and the phase-shifting capabilities. However, the hardware design may not always be optimal for every deployment scenario, especially in dynamic environments with varying interference, mobility, or user density. Data-driven optimization of hardware parameters, based on generative large models, allows RIS devices to be more adaptable and efficient, ensuring that they are designed for maximum flexibility and performance in real-world scenarios. By training AI models on large datasets that reflect different deployment scenarios, the hardware design can be fine-tuned to maximize the performance across a wide range of environments.

*AI for Real-Time Control of RIS in Mobile Environments*: In mobile environments, where users and objects are constantly moving, the wireless channel can change rapidly, and the optimal RIS configuration may need to be updated in real-time. Traditional optimization algorithms are often too slow to keep up with these rapid changes. RL-based approaches enable RIS to adapt quickly and efficiently to real-time changes in the environment, such as user mobility, changes in interference, or obstacles. This improves the robustness and performance of RIS in dynamic scenarios.

*Joint RIS and Base Station Beamforming Using AI*: The coordination between the RIS and the base station is critical to optimizing the overall system performance. Beamforming at the base station and phase shifting at the RIS need to be jointly optimized, which is computationally intensive and challenging in practice. Deep learning models can be used to jointly optimize both the beamforming at the base station and the phase shifts at the RIS, considering the wireless channel characteristics and user positions. This approach allows the system to maximize the received signal power at the user while minimizing interference to other users, reducing the BER, improving throughput, and enhancing coverage in challenging environments.

*Generative Models for RIS-Assisted Beam Prediction*: In systems like mmWave and massive MIMO, beam alignment between the transmitter and receiver is crucial, and RIS can assist by reflecting signals in optimal directions. Generative models, such as GANs, can be used to predict the optimal beams and RIS configurations based on partial channel information or user mobility patterns. These models can generate likely channel realizations, allowing the system to make more accurate beam predictions, improve the accuracy of beam alignment in RIS-assisted systems, reducing the time needed for beam training and improving throughput in dynamic environments.

*Federated Learning for Distributed RIS Control*: In large-scale deployments with multiple RISs, centralized control may become infeasible due to communication overhead, privacy concerns, and scalability issues. Distributed learning mechanisms such as federated learning can be applied to enable distributed RIS control. Each RIS can independently learn from local data, while periodically sharing model updates with a central server. This allows the global model to improve without requiring raw data exchange. This will improve scalability and privacy. It also reduces the need for constant communication between the RISs and the central controller.

### 2.2.5 MIMO-IM (Index modulation)

MIMO-IM is an advanced communication technique that combines MIMO technology with Index Modulation (IM), with the goal of improving spectral and energy efficiency by modulating both data symbols and antenna indices for information transmission. Traditional MIMO system employs all the antennas while MIMO-IM activates only a subset of antennas – the indices of the active antennas carry additional information. Although MIMO-IM increases the spectral efficiency without requiring additional power or bandwidth, there are major performance challenges. Jointly detection of both the data symbols and the active antenna indices requires more sophisticated detection algorithms which can increase computational complexity of the detection, particularly for large-scale MIMO systems. MIMO-IM highly depends on accurate CSI estimation for the correct detection, which becomes even more challenging due to the dynamically changing of the active antennas. Finally, MIMO-IM is sensitive to interference, especially when deployed in dense networks. Managing interference while detecting both antenna indices and symbols requires advanced signal processing techniques.

Current methods for MIMO-IM systems aim to improve spectral efficiency, energy efficiency, and detection performance by leveraging the unique features of IM while addressing the inherent challenges of detection complexity, interference, and channel estimation. Table 3 summarizes these methods, their advantages, and limitations. The information in this table indicates although deterministic AI is helpful, their training and accessing to the right volume of the data are main problems. In contrast, Generative AI and large models can significantly enhance MIMO-IM systems by optimizing various aspects such as channel estimation, signal detection, resource allocation, and performance under complex scenarios. The optimization of MIMO-IM systems is highly complex, and generative AI can bring transformative benefits by addressing these challenges.

Gen AI and the large models can help MIMO-IM in the following areas: *Improving Channel Estimation in MIMO-IM Systems*: MIMO-IM systems rely on CSI for efficient detection and performance. Traditional channel estimation methods, especially in high-mobility or fading environments, may not provide the necessary accuracy, leading to sub-optimal performance. Generative models, such as VAEs or GANs, can be used to generate synthetic CSI data to enhance the training and accuracy of deep learning models for channel estimation. Furthermore, deep learning models can be trained to estimate the channel more efficiently in a data-driven manner, bypassing the need for traditional, complex channel estimation algorithms. By leveraging Gen AI to model the channel, MIMO-IM systems can benefit from more accurate and robust channel estimation, leading to improved signal detection and overall system performance.

*Enhancing Signal Detection with Deep Learning*: MIMO-IM systems introduce an additional layer of complexity in the signal detection process because data is transmitted not only through signal modulation but also through the indices of the activated antennas. Traditional detection techniques (e.g., Maximum Likelihood detection) may suffer from high computational complexity, especially for large MIMO-IM configurations. Deep learning models, such as CNNs or RNNs, can be trained to perform signal detection in MIMO-IM systems. These models can learn the complex relationships between the transmitted signals, the antenna indices, and the received signal, resulting in efficient and accurate detection. AI-based detection methods can significantly reduce the computational complexity while improving detection accuracy, particularly in scenarios with high interference or complex channel conditions.

*Joint Optimization of Antenna Selection and Modulation*: In MIMO-IM, selecting the active antennas and modulating the signal simultaneously can be a complex task, as it involves a large combinatorial search space. Traditional algorithms may not efficiently find the optimal antenna selection and modulation strategy, especially in large-scale MIMO systems. RL or Deep RL can be applied to jointly optimize antenna selection and modulation in MIMO-IM systems. By learning the optimal policy through interaction with the environment, RL agents can adapt to changing channel conditions and interference patterns, ensuring optimal resource usage and performance. RL-based optimization methods allow MIMO-IM systems to dynamically adapt to varying conditions, leading to improved spectral efficiency, energy efficiency, and overall performance.

*End-to-End Learning of MIMO-IM System Components*: MIMO-IM systems consist of multiple components, such as transmitter design, channel estimation, signal detection, and decoding. Optimizing each component separately may not lead to globally optimal performance, as the interactions between components are complex and non-linear. End-to-end learning can be applied to jointly optimize the entire MIMO-IM system, from transmission to detection and decoding. By training a neural network on the full system, the model can learn the optimal configurations for each component, considering their interactions. This learning provides a unified approach that optimizes the entire MIMO-IM system, leading to better performance compared to isolated optimization of individual components.

Table 3: Summary of current MIMO-IM methods

| **Method** | **Description** | **Advantages** | **Limitations** |
|---|---|---|---|
| Spatial Modulation | • Only one antenna is activated at a time to transmit the data symbol<br>• The index of the active antenna carries additional information | • Low complexity<br>• Reduced interference<br>• Reduced power consumption | • Suffers from limited spectral efficiency |
| Generalized Spatial Modulation | • A subset of antennas is selected and can be activated simultaneously | • Improved spectral efficiency | • Increased detection complexity |
| Quadrature Spatial Modulation | • Information is encoded in both spatial (antenna indices) and the signal phase | • Better spectral efficiency without increasing complexity | • Requires precise phase synchronization between transmitter and receiver (impractical) |
| Enhanced Spatial Modulation | • Combines Spatial Modulation with traditional modulation techniques<br>• Both antenna indices and transmitted symbols are modulated using higher-order modulation schemes | • Increased data rate | • Higher power requirements<br>• More complex detection algorithms at the receiver |
| Dual-Mode Index Modulation | • In addition to antenna indices, antenna modes are also selected<br>• Antenna modes are ON, OFF, or transmitting with different power levels | • More transmitted bits by utilizing different transmission modes | • Increased complexity of transmitter and receiver design |
| Compressed Sensing-Based MIMO-IM Detection | • Compressed sensing techniques exploit the sparse nature of the transmitted signal | • Computationally efficient, particularly for large-scale systems | • High sensitivity to sparsity of signal and quality of channel estimation |
| Hybrid MIMO-IM Techniques | • Combination of advanced techniques such as OFDM, NOMA, or Massive MIMO | • Enables MIMO-IM to be applied in broader scenarios, such as in 5G networks | • Increased system complexity<br>• Requires more advanced hardware or signal processing capabilities |
| Iterative Detection and Decoding | • Receiver iteratively refines its estimates of transmitted symbols and active antenna indices | • Useful in low SNR or high interference scenarios<br>• Lower bit error rates | • Increased processing time<br>• Not suitable for low-latency applications |
| Error Control Coding with MIMO-IM | • Error control coding, such as LDPC or Turbo Codes, integrated with MIMO-IM | • Enhanced robustness against errors | • Added complexity of encoding and decoding<br>• Increased computational load |
| Deep Learning-Based MIMO-IM Detection | • CNN and RNN models learn relationships between received signals, active antenna indices, and transmitted symbols for efficient detection in challenging environments | • Highly adaptable to different channel conditions, interference patterns, and system configurations | • Requires large amounts of data for training<br>• High computational complexity, especially for large-scale systems |

*Generative AI for Channel and Interference Simulation*: Designing and testing MIMO-IM systems in realistic environments can be challenging due to the wide variety of channel conditions, interference patterns, and mobility scenarios. Acquiring large amounts of training data for AI models in such environments is expensive and time-consuming. Generative AI models, such as GANs, can be used to simulate realistic channel conditions and interference patterns, allowing MIMO-IM systems to be trained and tested in a wide range of scenarios without needing extensive real-world measurements. These synthetic datasets can help train AI models that perform channel estimation, signal detection, and interference management. Using generative AI for data augmentation allows MIMO-IM systems to be more robust to real-world conditions, leading to better performance when deployed in diverse environments.

*AI for Low-Latency Detection and Resource Allocation*: MIMO-IM systems, particularly in real-time applications such as 5G and beyond, require low-latency detection and resource allocation to meet the stringent performance requirements. Traditional algorithms may introduce significant delays, especially as the size of the system grows. Lightweight AI models, such as pruned neural networks or quantized models, can be designed for low-latency signal detection and resource allocation. These models can quickly process the received signals and allocate resources (such as power or antennas) with minimal computational overhead. AI-based models can significantly reduce the detection and resource allocation time in MIMO-IM systems, making them suitable for real-time applications without sacrificing performance.

*Reinforcement Learning for Adaptive Antenna Configuration*: The optimal configuration of antennas in MIMO-IM systems may vary depending on the channel conditions, user mobility, and interference. Static or pre-defined antenna configurations may not always lead to optimal performance in dynamic environments. RL can be used to adaptively configure the antennas in real-time, based on feedback from the environment. An RL agent can learn which antenna configurations maximize system performance under different conditions, dynamically adjusting the active antenna indices to optimize throughput or minimize error rates. RL-based adaptive antenna configuration improves the flexibility and adaptability of MIMO-IM systems, enabling them to maintain high performance in rapidly changing environments.

*AI-Assisted Error Correction*: MIMO-IM systems can be prone to errors due to imperfect detection of both the transmitted symbols and the active antenna indices. Traditional error correction methods, such as LDPC or Turbo codes, may not be sufficient to handle the unique challenges posed by MIMO-IM. Deep learning-based error correction models, such as neural decoders, can be trained to correct both symbol errors and index detection errors in MIMO-IM systems. These models can learn from large datasets of transmitted and received signals to improve the error correction process. AI-based error correction improves the reliability and robustness of MIMO-IM systems, particularly in challenging channel conditions where traditional error correction may fail.

The above-mentioned methods are compared in Table 4 in terms of the AI approach as well as the model size. Overall, generative AI and large models provide powerful tools to enhance the performance, efficiency, and adaptability of MIMO-IM systems. By leveraging AI for channel estimation, signal detection, resource allocation, and error correction, MIMO-IM systems can achieve better spectral efficiency, lower latency, and higher reliability. AI-driven optimization of antenna selection and modulation strategies further improves the system's ability to adapt to dynamic environments, making it a key enabler for future communication technologies such as 5G and beyond.

### 2.2.6 Joint Approaches

Sometimes it is important to combine some physical functions together, for example joint symbol detection and channel estimation, or joint equalization and decoding. By leveraging predictive modeling, data synthesis, and adaptive optimization, generative models provide better noise resilience, faster processing times, and improved accuracy, even in complex environments like massive MIMO systems and high-mobility scenarios. Here's how generative AI and large models can help in these areas:

*Joint Symbol Detection and Channel Estimation*: In wireless systems, symbol detection and channel estimation are interdependent processes. Traditional methods often handle these independently, which can lead to suboptimal performance in noisy or high-interference environments. Generative AI models can simulate channel conditions in advance, allowing for better symbol detection under varying noise levels and interference. They can synthesize data to reduce the need for extensive pilot signals, allowing for better channel estimation with less overhead. Some example methods could be GANs or VAEs. Large neural networks (e.g., RNNs) can jointly learn channel characteristics and symbols, enhancing detection reliability, especially in fading channels. Overall, generative AI and large models can jointly optimize both, improving accuracy and efficiency.

Table 4: Comparison of Deterministic vs. Generative Models

| Application | Deterministic AI | Generative AI | Training Model Size | Inference Model Size |
|---|---|---|---|---|
| Improving channel estimation | x | x | Medium to Large | Medium to Large |
| Enhancing signal detection | x | | Medium to Large | Small to Medium |
| Joint optimization of antenna selection and modulation | x | | Medium to Large | Small to Medium |
| End-to-End Learning of MIMO-IM System Components | x | x | Large | Large |
| Generative AI for Channel and Interference Simulation | | x | Large | Medium to Large |
| AI for Low-Latency Detection and Resource Allocation | x | x | Medium to Large | Small |
| Reinforcement Learning for Adaptive Antenna Configuration | x | x | Medium to Large | Medium |
| AI-Assisted Error Correction | x | x | Medium to Large | Medium to Large |

*Joint Equalization and Decoding*: Joint equalization and decoding involve handling inter-symbol interference (ISI) and correcting channel-induced distortions, particularly challenging in high-mobility and high-interference environments. Generative models can simulate interference scenarios, allowing adaptive equalization techniques that dynamically adjust to changes in interference. Deep learning models, such as CNNs, can jointly perform equalization and decoding tasks, learning the structure of both interference and noise to improve decoding reliability. Finally, RL models can optimize equalization strategies in real-time, adapting to dynamic environments and improving decoding performance. In general, generative AI and large models offer predictive analytics and adaptive equalization techniques that improve decoding accuracy and speed.

*End-to-End Learning for Integrated Channel Estimation, Detection, and Decoding*: End-to-end learning models use neural networks to combine channel estimation, symbol detection, equalization, and decoding, creating an optimized, single-model pipeline that adapts to complex and fast-changing environments. By jointly learning all steps, generative models reduce processing time and computational complexity, ideal for real-time applications. This kind of models are adaptive, training across various signal, interference, and noise conditions, making them resilient to channel impairments and improving error rates. GANs are used to synthesize realistic channel effects, training models to better adapt to real-world interference and noise scenarios.

### 2.2.7 Wireless Spectrum Sensing

Wireless standards continuously evolve, leading to new ways to connect devices and a massive increase in connected devices. This rapid expansion presents a significant challenge regarding RF spectrum availability. Spectrum sensing enables the detection of unused spectrum bands, known as spectrum holes. This allows secondary users without dedicated licenses to access the spectrum not currently used by primary, licensed users. This opportunistic spectrum access enables more efficient spectrum usage by continuously monitoring the spectrum and identifying spectrum holes. This technique is known as Dynamic Spectrum Access (DSA).

Wireless Spectrum Sensing can be categorized into three main types: traditional methods such as energy detection, matched filtering, cyclo-stationary feature detection, and receiver metrics-based approaches; machine learning and deep learning-based methods; and, more recently, large-scale AI approaches. Traditional spectrum sensing, particularly energy detection, involves comparing the computed signal energy against a threshold to determine the presence or absence of a signal. However, this method is challenged by noise, interference, incomplete data, and environmental variability. These challenges make it difficult to set an optimal threshold for detection. While promising, machine learning and deep learning techniques struggle with generalization—models that perform well on a specific data set fail to achieve similar accuracy on unseen data.

Shao et al. [9] propose a framework for adapting and enhancing LLMs for wireless communication systems and suggests using few-shot learning with LLMs for spectrum sensing. This technique is important when deep learning models require collecting large amounts of labeled data, which is difficult. With just a few examples, LLMs can effectively learn the task and perform comparably to optimal detectors, such as energy detectors, especially in scenarios with varying signal-to-noise ratios.

Traditional RF sensing methods face noise, interference, and incomplete data challenges. Wang et al. [10] propose an RF sensing framework based on GenAI for IoT systems. GenAI enhances multi-modal data fusion, which is essential for IoT systems that depend on diverse sensor inputs, such as RF signals, images, and audio. By combining these different data types, GenAI-driven systems create more intelligent and comprehensive sensing solutions. GenAI techniques, like GANs, VAEs, and Diffusion Models (DMs), can generate high-quality synthetic data, de-noise signals, and fill in the missing information. These abilities significantly strengthen the reliability of RF sensing systems.

Automatic Modulation Classification (AMC) can be used for wireless spectrum sensing by identifying the modulation schemes of primary users (PUs). By classifying modulation scheme used by PUs, AMC enables wireless radios to intelligently detect whether a spectrum band is occupied or available for secondary use, reducing interference with licensed users. This integration of AMC in spectrum sensing system improves detection accuracy, especially in low Signal-to-Noise Ratio (SNR) environments, compared to traditional methods like energy detection. Olaloye et al. [11] proposed the use of machine learning models, such as Multi-Layer Perceptron (MLP) and have demonstrated high accuracy in classifying modulation types. Therefore, validated the use of AMC in real-time DSA.

## 2.3 Large-Scale AI for Network Management and Optimization

### 2.3.1 Large-Scale AI in User-centric Network Optimization

User-centric Network Optimization has become a focal point in next-generation network optimization. This approach is crucial because it addresses the diverse needs and preferences of individual users, leading to improved overall QoE. Traditional uniform service delivery often results in varying levels of user satisfaction. User-centric optimization can be applied in various scenarios, such as personalized content delivery, adaptive video streaming, and dynamic resource allocation in mobile networks. Despite the emergence of several User-centric Network Optimization methods, accurately assessing user requirements, particularly subjective experiences, remains a challenge. Some studies have incorporated psychological laws to approximate users' subjective QoE [12]. However, these approaches often fail to capture the complexity of real-world applications. An alternative solution involves using Reinforcement Learning with Human Feedback (RLHF) paradigms to train management models. This method requires ongoing QoE feedback from experts, which is expensive, raises ethical concerns, and is difficult to implement in real time. These limitations lead to our first research question:

Large-scale AI presents significant potential in User-centric Network Optimization due to its ability to process vast amounts of user data during training, enabling it to simulate user QoE effectively. LLMs-empowered generative agents can process and understand complex instructions in natural language, serving as a universal interface for various tasks, including evaluation [13, 14]. Research in [13] demonstrates ChatGPT's capacity to evaluate textual content across human-aware criteria such as quality, tone, and coherence. These evaluations lay the groundwork for extending LLM functionality to other domains. Further studies in [15] assess the potential of LLMs like ChatGPT in Computational Social Science, examining their performance in classification and generative tasks in a zero-shot manner. Results indicate that LLMs show fair agreement with humans and can enhance the annotation process. Recent work in [16] reveals that modern role-playing LLMs can effectively mimic specific personality traits, achieving an 82.8% alignment with human perceptions. In the context of Large-Scale AI in User-centric Network Optimization, AI can serve two primary roles:

- *Active Solution Generation:* Large-scale AI can actively generate network optimization solutions. Scalable model architectures suitable for decision-making include transformer-based models with attention mechanisms, graph neural networks for network topology understanding, and hierarchical reinforcement learning models for multi-level decision processes. For example, the authors in [17] propose an innovative LLM-enabled Mixture of Experts (MoE) approach for network optimization. This method leverages the LLM's advanced reasoning capabilities to analyze user objectives and constraints, select specialized DRL experts, and determine the decision weights for each participating expert. The LLM acts as a dynamic gate network, managing the selection and integration of expert models to address new and complex optimization tasks. This approach demonstrates the potential of large-scale AI in adapting to diverse user requirements and generating effective solutions for network optimization problems without the need to train new models for each specific task.

- *Passive Optimization Support:* Large-scale AI can function as a component of optimization algorithms, providing subjective QoE assessments. LLM-empowered generative agents offer a powerful mechanism to provide human-aware subjective QoE feedback for generated content. A QoE feedback scheme using these agents can simulate diverse user personalities. By utilizing prompts and assigning one agent per user, generative agents can mimic users with varied subjective preferences, delivering evaluations of received services. For example, the authors in [18] propose a Reinforcement Learning with LLMs Interaction (RLLI) framework for distributed GenAI services. This approach leverages LLM-empowered generative agents to simulate user feedback. The framework uses the Big Five personality model as a basis for configuring generative agents, aligning with research showing that LLMs can effectively simulate these personality traits. By designing prompts that include specific Big Five trait scores, the system enables generative agents to mimic diverse user personalities. These agents then evaluate generated content, providing subjective QoE feedback that reflects individual preferences. This method offers a scalable and efficient alternative to human feedback, demonstrating improved performance in maximizing sum QoE compared to conventional methods.

## 2.4 Large-Scale AI for Un-crewed Aerial Vehicles (UAVs)

Un-crewed aerial vehicles (UAVs) have recently gained significant attention due to their exceptional autonomy, mobility, and adaptability. These attributes have expanded their use across a broad spectrum of applications, including surveillance, search and rescue missions, healthcare, and maritime communications [19]. The convergence of advancements in UAV technology and AI has yielded significant benefits across a wide range of applications. For instance, AI-enabled UAVs utilize facial recognition to enhance security applications, and real-time video analysis enables monitoring remote areas. In agriculture, UAVs equipped with AI models assess crop health, enabling precision farming that increases revenues. Additionally, AI-driven UAVs optimize logistics by enhancing route planning and inventory management, thereby streamlining warehouse operations and increasing delivery efficiency [20, 21]. Among these advancements, large-scale AI models have recently attracted considerable attention in the UAV sector [22]. The capabilities of these models in real-time data processing, natural language understand-ing and generation, content recommendation, sentiment analysis, automated response, language translation, and content summarizing have paved the way for new opportunities within the UAV domain.

Recent literature [23, 24, 25, 26] has investigated the integration of large-scale AI models into UAV communication systems to enhance interaction between human operators and UAVs, as well as among the UAVs themselves. Traditionally, UAVs have relied on pre-programmed commands, offering limited dynamic interaction capabilities. However, the incorporation of such large-scale AI models; i.e., LLMs, introduces support for natural and intuitive communication methods. For instance, LLMs can interpret and respond to commands in natural language, making UAV control more straightforward and enabling the management of complex, real-time mission adjustments. This evolution transforms UAVs into more adaptable and practical tools across a wide range of applications. For example, in [27] the authors provided a framework that utilizes GPT-3 to enhance the intuition of human-UAV interactions. The framework leverages NLP techniques to allow users to control UAVs using simple language commands, eliminating the need for complex programming knowledge. By translating user instructions into executable code, such a framework enables UAVs to carry out tasks and provide feedback in natural language, significantly simplifying the control process. Another application is provided in [28], in which the authors presented a framework that integrates OpenAI's GPT-3.5-Turbo model with an UAV simulation systems (i.e, PX4/Gazebo simulator), to develop a natural language-based drone control system. The system's architecture is designed to facilitate seamless interaction between the user and the UAV simulator through a chatbot interface, enabled by a Python-based middleware. This middleware processes natural language inputs from the user, relays them to the ChatGPT model using the OpenAI API, retrieves the generated responses, and translates them into commands that the simulator can interpret, thereby enhancing the interactivity and accessibility of the UAV simulation system.

Large-scale AI models enable UAVs to react instantly to dynamic environmental changes and communication demands. The adaptive learning capabilities of such models enable continuous improvement in operational strategies by leveraging incoming data, thereby enhancing decision-making processes. In [27], the authors introduced a vision-based autonomous planning system for UAVs designed to enhance safety. The system predicts the trajectories of dynamic obstacles and generates safer flight paths by utilizing NanoDet for precise obstacle detection and Kalman Filtering for accurate motion estimation. In another work [29], the authors integrated GPT models and computer vision technologies into autonomous inspection UAVs to enhance their functionality in indoor environments. The proposed system enables UAVs to analyze images captured during flight to generate detailed object dictionaries. These dictionaries enable the UAVs to recognize and understand various elements within their environment, allowing them to dynamically adapt their behavior in response to both anticipated and unforeseen conditions.

Additionally, large-scale AI models can enhance UAVs' autonomous decision-making by leveraging communication context or environmental data [30]. For instance, during a search and rescue operation, live video feeds and text reports from multiple UAVs can be analyzed and synthesized using multi-model LLMs to recommend areas of focus or adjust search patterns accordingly [23]. UAVs can also operate in ad-hoc and mesh configurations to form dynamic networks without the need for pre-existing infrastructure. This capability is especially valuable in situations where establishing permanent network infrastructure is impractical, such as in disaster response. Such self made networks continuously discover new neighbors and can dynamically adjust routes based on the network's topology and traffic conditions, thereby improving scalability and flexibility [31].

Large-scale AI also contribute to simulating and modeling the behavior of networks under different scenarios, aiding in the planning and decision-making processes for UAV deployments. The GPT series can simulate various communication scenarios for UAV training by generating realistic mission scenarios and responses. This allows operators to undergo comprehensive training, equipping them to handle different situations more effectively and enhancing their preparedness for real-world operations [32]. Large-scale AI models can also assist UAVs in understanding network traffic patterns, enabling them to recommend adaptive protocols that reduce latency and increase throughput, especially under the varying user conditions often encountered in these networks.

Large-scale AI models can be also utilized to analyze data from the UAVs themselves, including operational logs and flight data, to predict possible failures, maintenance needs, and potential malicious attacks, before they happen [33]. This predictive capability can significantly enhance the reliability and lifespan of UAVs, thereby reducing downtime and maintenance costs. In [34], the authors developed enhanced security and forensic analysis protocols for UAVs to support the growing use of drones across various sectors, including those at risk of criminal misuse. They introduced a named entity recognition system to extract information from drone flight logs. This system employs fine-tuned BERT and DistilBERT models with annotated data, significantly improving the identification of relevant entities essential for forensic investigations of drone-related incidents.

## 2.5 Large-Scale AI for Telecom Use Cases

### 2.5.1 ”Qiming” Network Large Model Case Study

China Telecom’s ”Qiming” Network Large Model is designed to optimize and automate network operations through AI-driven processes. As modern telecommunications networks become increasingly complex, the need for advanced tools capable of real-time decision-making has grown. The ”Qiming” model leverages vast amounts of data and professional network knowledge to assist in various tasks such as network planning, maintenance, monitoring, troubleshooting, and performance optimization. The model’s generative capabilities, combined with knowledge retrieval and intent recognition, aim to enhance network autonomy, reduce manual intervention, and improve operational efficiency.

The ”Qiming” Network Large Model employs innovative large model architectures, including incremental training and feedback optimization algorithms. These ensure that the model evolves over time to adapt to new network challenges and requirements. It is also capable of handling vast amounts of network data, significantly improving operational efficiency and reducing the need for manual input. Despite these advantages, the model faces challenges such as the high computational cost associated with processing large datasets and training the model. Moreover, the model may struggle with generalization when encountering entirely new or unforeseen network conditions, which requires ongoing optimization and updates. The ”Qiming” Network Large Model exemplifies how large-scale AI can be leveraged to tackle complex network management challenges. It serves as a vital tool for China Telecom, driving the company’s network automation efforts and enhancing the efficiency of its operations.

### 2.5.2 ”Qiming” Network Large Model Operational Workflow

The workflow of the ”Qiming” model, as illustrated in the accompanying diagram, outlines a multi-step process:

1.User Intent and Input: Network operation and maintenance staff initiate the process by providing a specific user intent, such as a request for network optimization or troubleshooting.

2.Querying Network Knowledge: The Network Large Model interacts with the knowledge base by querying for relevant network knowledge. This step includes retrieving professional knowledge that assists in decision-making processes such as network optimization or answering user queries.

3.Querying Network Data: The model queries network data from databases and other sources, distinguishing between real-time and non-real-time data. This data could include statistics and network status, which is crucial for diagnosing issues and providing accurate recommendations. The interaction with the BSS (Business Support System), OSS (Operational Support System), and MSS (Management Support System) allows for the extraction of operational data and relevant metrics for a comprehensive analysis.

4.Decision-Making: Based on the gathered knowledge and data, the model makes decisions that pertain to the network operation. This could involve generating answers, recommending solutions, or optimizing network functions. Network Operation Response: The Network Large Model then dispatches the decision back to the staff or directly interacts with the network components. The network operates based on the model's decisions, and an acknowledgment (ACK) is sent back, confirming the successful implementation of the operation.

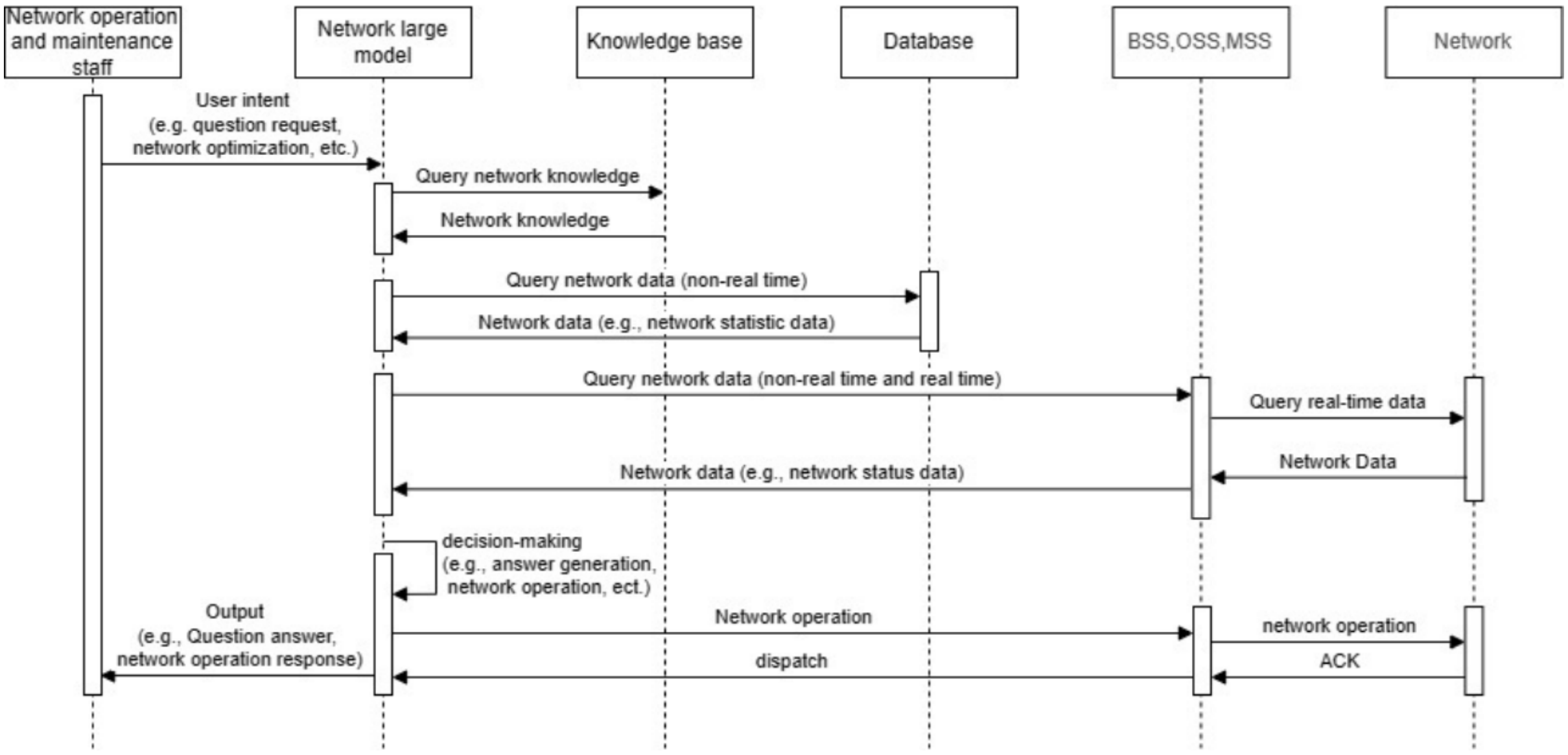

Figure 2: "Qiming" Network Large Model Operational Workflow

#### 2.5.3 "Qiming" Network Large Model Application Scenarios

**Intelligent Network Operations:** The model enables the automation of network tasks across the entire lifecycle, from planning and construction to maintenance and optimization. By using advanced algorithms, it ensures efficient, real-time responses to network issues.

**Fault Diagnosis and Prevention:** The model's ability to process both historical and real-time data allows it to predict potential network failures and provide preventive measures. This helps reduce downtime and ensures a smooth network experience for users.

**Task Decomposition and Orchestration:** The model can break down complex tasks into manageable steps, providing intelligent task orchestration. This capability improves the speed and accuracy of network maintenance and troubleshooting.

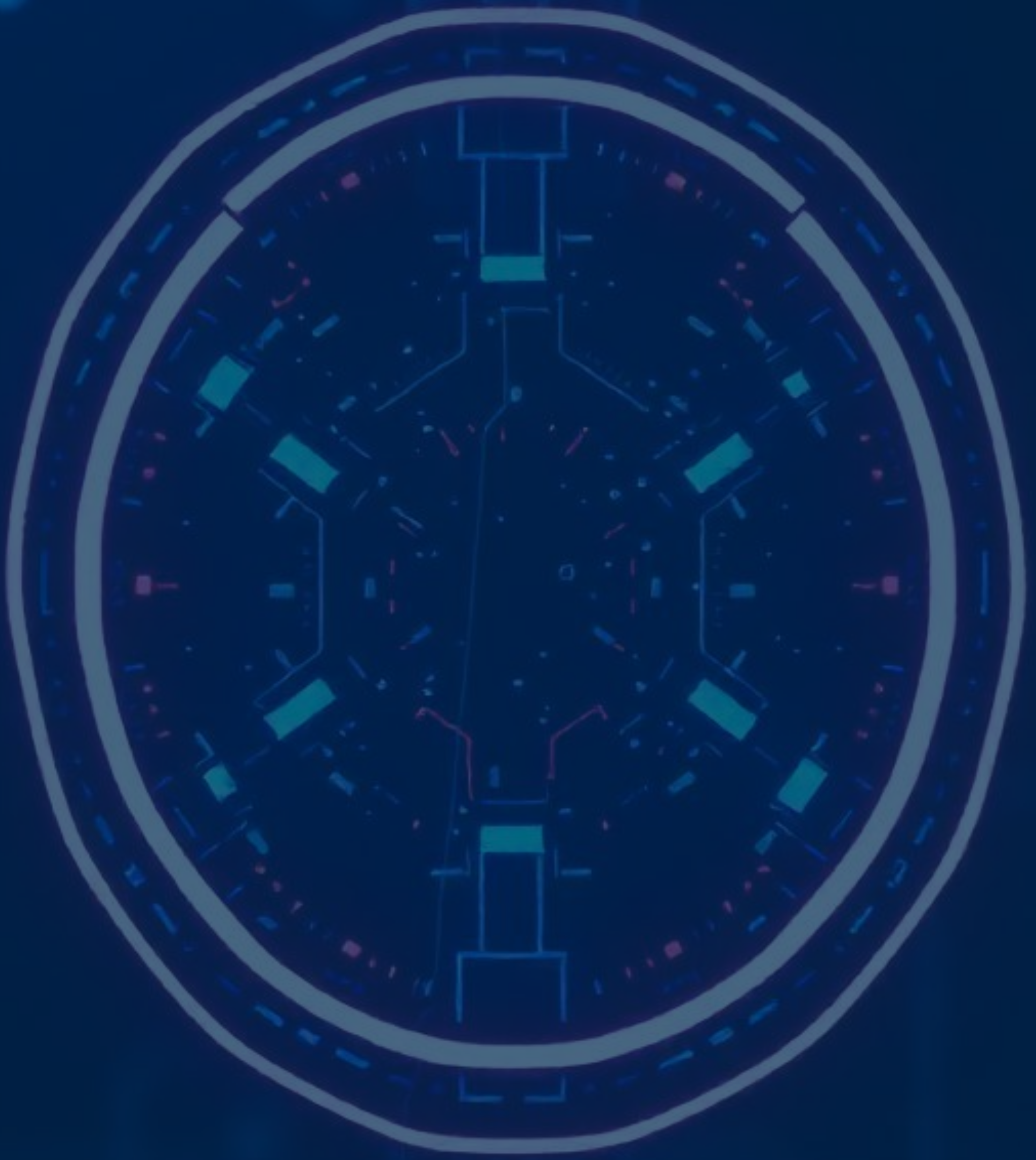

# AI THEORY OF LARGE TELECOM MODELS

# 3 AI Theory of Large Telecom Models

## 3.1 From Language to Telecom Models: Challenges and Necessary Modifications

While state-of-the-art LLMs excel in versatile NLP tasks like question answering and sentence completion, ensuring similar performance upon integrating them into telecoms requires further modifications to their underlying theory and mechanisms. For instance, simply relying on a GPT architecture in dealing with network parameters and KPIs such as SNR, QoS, and channel gains can lead to erroneous mistakes. This is mainly due to the limitations inherited from text-based models which are transferred to telecom models, leading to the following drawbacks:

- **Limited abstract telecom knowledge:** The attention mechanism that perfectly captures the sophisticated correlations between tokens (or generally words) falls short in capturing the other relations (i.e., causal, mathematical, etc.) that govern such *telecom tokens*. In fact, LLMs build their own knowledge that may not necessarily reflect the real-world phenomenon. For instance, an LLM may not properly understand the causal relation between increasing the transmission frequency and the elevated propagation losses encountered by a wireless signal. Clearly, one of the most prominent implications of this limited knowledge is the tendency of large models to hallucinate when generating their repose.

- **Lack of mathematical foundations:** In general, LLMs build on their captured patterns to define the mathematical operations that govern the different telecom tokens. Hence, they lack the proper mathematical foundations that enable them to freely manipulate and verify the tokens. For instance, LLMs may struggle to prove how the Rx antenna measurements (e.g., Reference Signal Received Power (RSRP)) mathematically flow from the underlying theorems and equations of wireless signal propagation (e.g., pathloss). Accordingly, LLMs cannot calculate how the captured parameters can be proven based on the channel information and transmit signals. This can potentially hinder the validity of the LLM results in different situations or limit the applications of LLMs in situations that require reasoning and planning.A simple example of this can be in the form of a network design problem. Although an LLM can elaborate on the design questions that relate power at the Tx antenna and the pathloss, it it may not be able to capture that doubling the propagation distance would actually decrease the power at the Rx by a factor of 4. Thus, state-of-art LLMs cannot fully apprehend the telecom formulations behind these telecom tokens.

- **Static performance:** As LLMs are trained on massive datasets up to a certain point in time, they show a static performance that may become non-relevant when it comes to dynamic and non-stationary settings such as those introduced in the RAN. Unlike text and generally language that is mostly static, a dynamic environment such as the RAN demands telecom models to admit evolving knowledge paradigms.

- **Absence of guardrails:** Unlike most LLMs that can be used to boost productivity and enhance performance, the role of LLMs in telecoms may demand autonomously taking critical decisions that drive the network operations. To this end, these actions must adhere to specific rules and abide by the guidelines set by regulatory bodies (e.g., FCC, ETSI, etc.). Nevertheless, the state-of-art LLMs do not impose any guardrails in their design. For instance, an LLM may not set the transmit power of a base station above a predetermined threshold as it may threaten to harm individuals.

To address these drawbacks, large telecom models must consider necessary modifications into their foundational language architectures before being implemented in the telecoms sector. Next, we shed light on emerging AI approaches, such as causality and neurosymbolic AI, that can potentially fill in the aforementioned gaps in the AI theory behind large telecom models.

#### 3.1.1 Grounding via causality

First, large telecom models must enable the *grounding* of their telecom tokens so that they harness true meaning and acquire full understanding abilities about their data. Here, grounding is the process of anchoring the generated responses of these models into real-world knowledge. In the telecom terms, it is the ability of large telecom models to root the telecom tokens (or embedded representations) into the physical world and wireless phenomenon. This ensures that large telecom models maintain coherence to the real-world context and true physical phenomenon. Effectively, this takes place by integrating the absent logical mechanisms to complement their knowledge gauge. As denoted earlier, these models may lack to capture the causal dimensions between the tokens. Hence, one method to enable grounding can be through the framework of causal reasoning. This can facilitate a level of causal understanding that refers to identifying cause-and-effect relationships among various features within wireless tokens [35]. For instance, a vector of channel measurements in a wireless environment can be interpreted using a causal graph that identifies the relationships among scattering objects and multipath characteristics such as angle of arrival, delay, and path gains [36]. In particular, *causal discovery* methods [36] can be leveraged to identify the cause and effect relations among the network variables that can further be arranged as structural causal models. As it may be challenging to extract the causal variables and identities when dealing with high-dimensional observations (e.g., Rx antenna measurements), causal representation learning presents an effective solution to map these observations into low-dimensional representations that capture only the relevant causal variables. In fact, the captured representations could be further clustered on the basis of their similarity into general representations. Consequently, this can reduce the embeddings space while still ensuring that distinct representations remain differentiable from each other. Henceforth, causal discovery and causal representation learning enable can filter the high-dimensional wireless observations to a minimally compact and sufficient embedding space suitable for large scale telecom models with evolving knowledge. Furthermore, telecom models could additionally bolster their performance through identifying the relevant causal variables from a telecom specific dataset. In particular, leveraging RAG can equip large telecom models with the necessary wireless and telecom knowledge. As shown in Fig. 3, adopting RAG from a telecom specific source dataset can enable bolster the performance of telecom models.

#### 3.1.2 Alignment via RAG

Alignment ensures that the outputs of large telecom models align with the guidelines of MNO. In particular, these guidelines should comply with government regulations and adhere to the goals of system designers. Hence, telecom models must ensure that they achieve these guidelines without contradicting their initial premise and refine their available settings to abide by the guardrails. For instance, while various complex modulation schemes and beamforming principles can be discussed in telecom literature, supporting such approaches might not necessarily align with the limited set of transmission schemes approved by 3GPP standards. Therefore, it is essential to ensure that telecom models adhere to the standards and regulations. Furthermore, 3GPP standards are updated periodically. Hence, it is crucial that telecom models remain consistent with these updates. One notable approach to keep large telecom models up to date with standards is to leverage RAG with a dynamically evolving database to extract relevant contextual information for the wireless tasks enabled by the foundation model. Moreover, this alignment requires an adopting an *RL with wireless feedback* mechanism that ensures that the responses from the telecom model maximizes the average reward. Here, the reward can be defined as proportional to the QoE of network agents. Notably, establishing this dynamically evolving database requires concerted efforts from industry, academia, and standardization bodies.

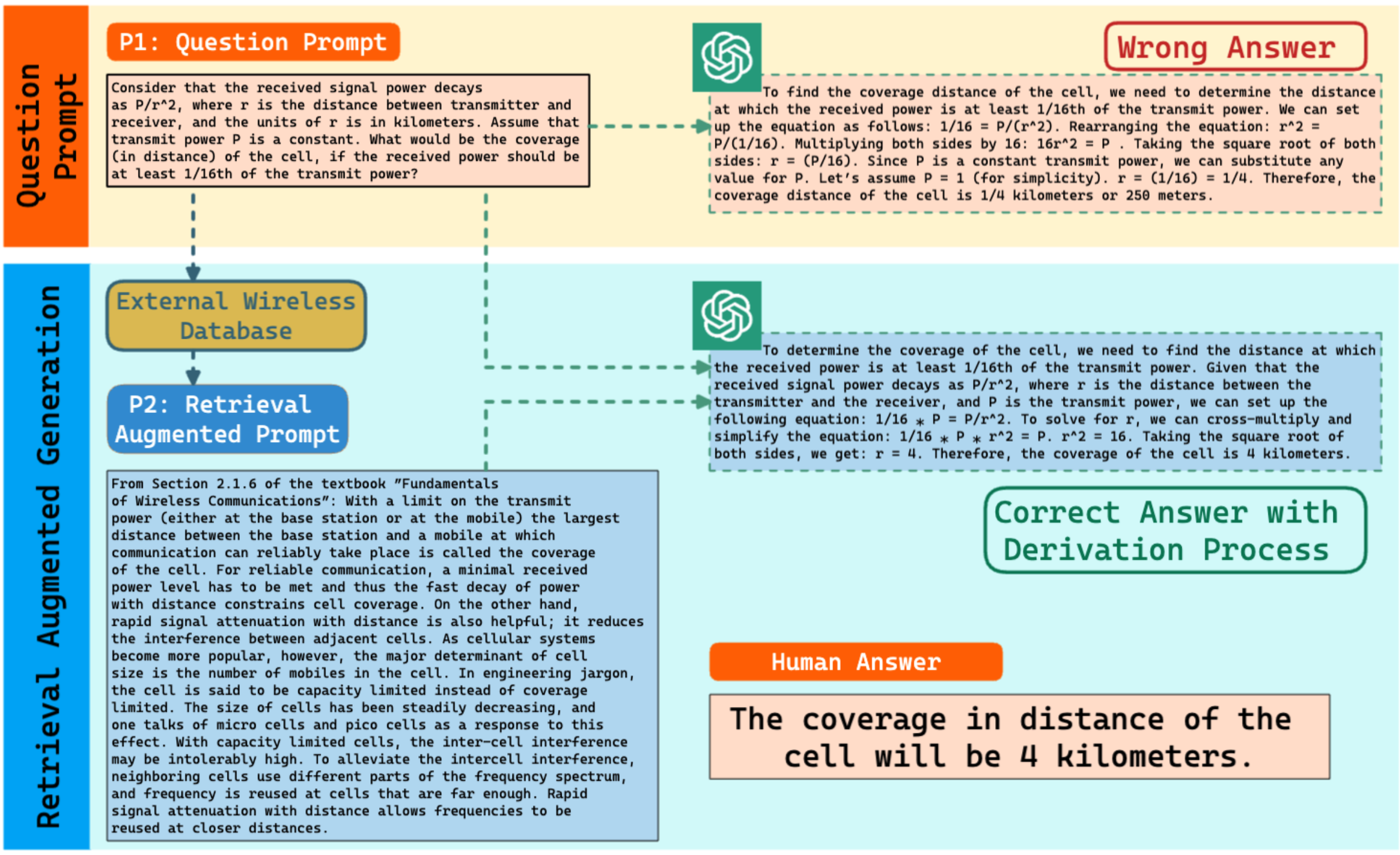

Figure 3: A sample mathematical Q/A pair from the wireless specific dataset [37].

#### 3.1.3 Dynamic performance via instructibility

Additionally, LLMs struggle to operate in real-time environments and lack adaptability to changing wireless conditions and tasks. To address this, large telecom models must incorporate *instructibility*, enabling them to adjust their parameters and behavior in response to evolving environments and tasks.

#### 3.1.4 Neuro-symbolic AI as a cornerstone for mathematical reasoning

Recent advances in LLMs focus on scaling AI models to enhance generalization capabilities. While the human brain requires only a few symbolic rules and experiences to generalize behavior to unseen scenarios, LLMs need trillions of parameters to acquire knowledge for generalization. Despite this massive scale, they often fail to perform deductive reasoning, making them vulnerable to extreme or uncommon scenarios. Inspired by human intelligence, a promising approach is to build a hybrid system that combines the best of both worlds: a symbolic component that represents rule-based logic and background knowledge, enabling logical reasoning, and a neural component that allows for generalization of their behavior under epistemic uncertainty. Such hybrid models are called neurosymbolic AI. Neurosymbolic AI models allow to build sample efficient telecom models. Moreover, they help to build instructible wireless sytems, wherein their parameters can be dynamically adapted in response to the environment or user feedback.

As highlighted in [38], next-generation AI models for telecom must exhibit long-term planning capabilities. Here, *planning* refers to the capability of network components to propose a sequence of actions—encompassing both network configurations and actions of connected autonomous agents—by predicting future environmental states. Such planning must ensure that connected agents maintain a high quality of experience while satisfying network intent without interruption. These capabilities enable networks to configure actions such as beamforming and power allocation, alongside autonomous agent control policies, ensuring that the quality of experience for network agents (involves both autonomous agents part of cyber-physical-systems and the network infrastructure) is maximized while satisfying the network intent without interruption. Here, intent refers to specific goals that the network must achieve, such as maximizing the network's sustainability. Herein, traditional AI methods such as deep reinforcement learning is not sufficient to perform real-time control and network actions, due to the overhead in retraining. Moreover, such data-driven AI models are not trustworthy, and hence their decisions cannot be trusted as we move towards building autonomous networks. Herein, causal inference enables performing interventions and counterfactuals [36] on the learned causal world model, that describes the interactions of network with the autonomous agents. Interventions and counterfactuals [36] enable analyzing the impact of network actions on quality of experience of network agents. Using such effect analysis, network can compute optimal actions that remediate any deviations from expected quality of experience for the network agents.

One promising approach to instill mathematical reasoning is through invertible symbolic regressions [39] that learns underlying equations that describes the physical processes from the data. Such symbolic equation learners can be used to learn non-linear mathematical equations using symbolic expressions that cannot be described using model-based systems. Such symbolic expressions enhance the explainability of the AI models compared to using black-box models, that lack interpretability.

## 3.2 On the Interplay between Data Compression and LLMs

The connections between large language models and data compression operates in two complementary directions. On the one hand, the principles of compression are inherently relevant to the design and operation of language models neural networks. However, deploying LLMs in resource-constrained scenarios requires applying compression techniques to meet performance and infrastructure requirements.

LLMs and data compression share a fundamental goal: reducing redundancy while retaining meaningful information. As shown in recent work [40], optimizing the conditional probability of the next token in language models is similar to the principle of arithmetic (source) coding to minimize the average coding length. This equivalence highlights that language models, through their next-token prediction objective, inherently perform a form of compression.

This natural alignment suggests that the internal mechanisms of LLMs can be fine-tuned to improve both their predictive accuracy and efficiency. Techniques such as quantization and pruning, which traditionally belong to the field of data compression, can be applied to LLMs without compromising their performance. The interplay between data compression and LLMs not only improves inference efficiency but also creates opportunities for deploying lightweight models in dynamical telecom environments.

The deployment of LLMs in wireless networks where latency, memory, and energy constraints are critical requires signification model compression. Without compression, the size and computational demands of LLMs pose significant challenges for real-time operations at the edge.

Quantization reduces the bit-width used to represent weights and biases from standard 32-bit floating-point to lower-precision formats, such as 8-bit or 4-bit. It comes in two primary forms: Quantization-Aware Training (QAT) and Post-Training Quantization (PTQ). QAT integrates quantization into the training process, allowing the model to adjust its parameters to mitigate precision loss. PTQ applies quantization after training, making it more practical for pre-trained models. For instance, BitNet [41] proposes a highly efficient quantization scheme by reducing weights to ternary values (-1, 0, or 1). This achieves a quantization rate of 1.58 bits per parameter, resulting in a 20x reduction in size compared to 32-bit floating-point models, while maintaining comparable accuracy and significantly improving memory, latency, and energy efficiency.

Beyond scalar quantization, advanced methods like vector quantization and variable-length coding offer additional opportunities for model compression. Vector quantization leverages correlations among weights by simultane-ously compressing groups of parameters, reducing redundancy more effectively. This technique connects with pruning techniques, where unimportant weights are set to zero and excluded from computations, further enhancing efficiency. Indeed, vector quantization can be seen as a way of jointly discarded irrelevant NN parameters and approximating the selected ones. Additionally, recent work on lossless entropy coding [42] demonstrates that even the exponent bits in floating-point numbers (FP16/32) can be compressed without compromising accuracy. For instance, lossless coding has achieved over 50% reductions in model size, providing substantial savings in network and storage costs. Whereas users making use of very large language models such as GPT might expect it to accomplish and more and more diverse tasks, the deployment of LLMs in wireless networks is expected to be more targeted. By delineating the set of goals the LLM has to accomplish, it is possible to further compress the model. Over the past years, goal-oriented compression techniques have emerged. In [43][44] for instance, it is shown how to tailor quantization to the goal function. This approach can be reused in model compression by prioritizing layers or parameters based on their influence on task accuracy.

In summary, the interplay between LLMs and compression spans two crucial aspects. First, LLMs inherently align with the principles of compression through their predictive objectives, as highlighted by their mathematical equivalence to arithmetic coding. Second, deploying LLMs in telecom environments necessitates applying advanced compression techniques to meet the stringent requirements for latency, memory, and energy efficiency. Techniques like quantization, pruning, and low rank factorization are critical not only for efficient deployment but also for enabling real-time AI capabilities in future 6G networks. This dual perspective on LLMs and compression highlights the importance of continued research at the intersection of AI and information theory, ensuring that LLMs can operate efficiently in diverse and resource-constrained environments.

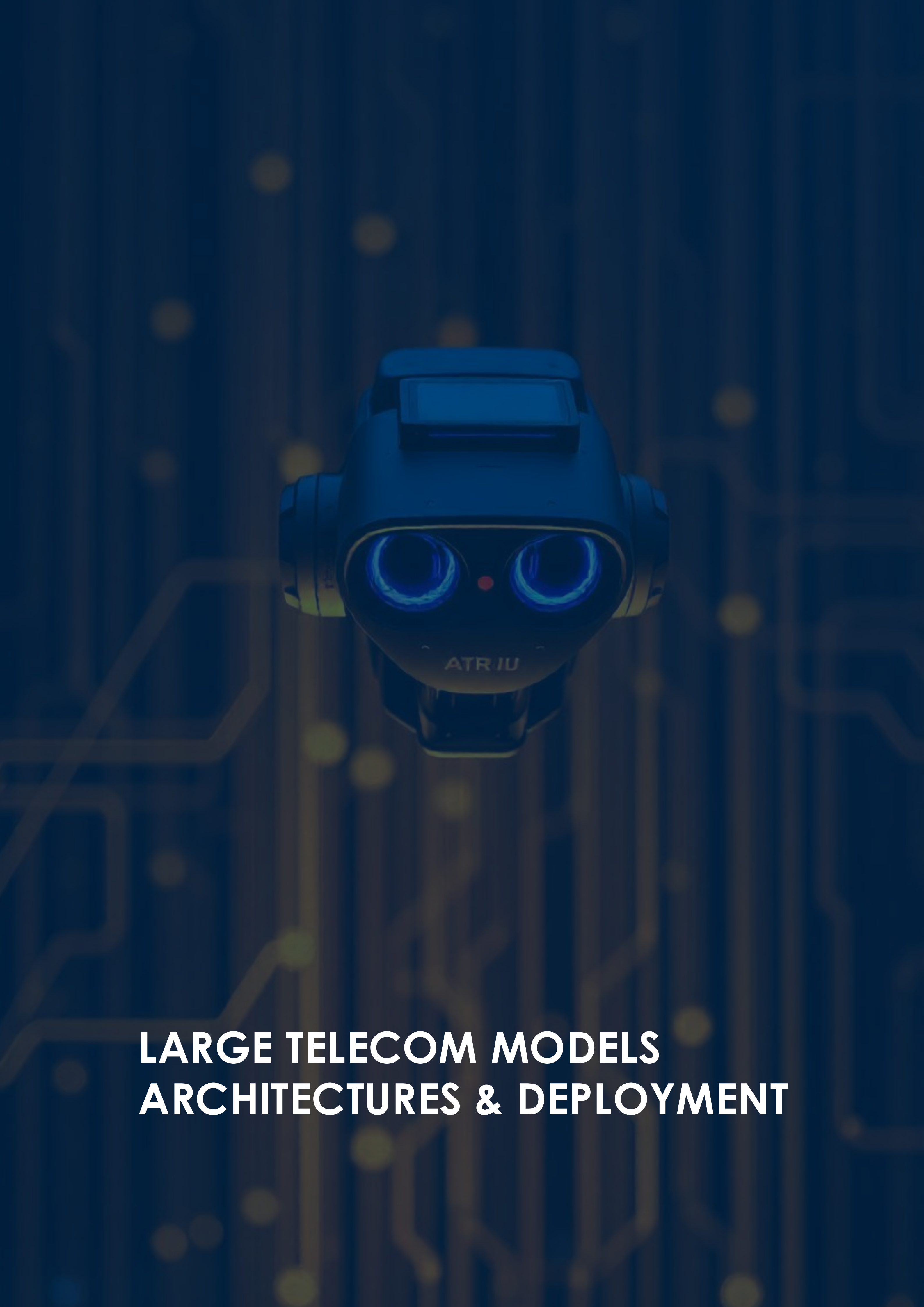

# LARGE TELECOM MODELS ARCHITECTURES & DEPLOYMENT

# 4 Large Telecom Models Architectures & Deployment

## 4.1 Neural Network Architecture

This section provides a comprehensive overview of foundational and advanced architectures critical for developing Large AI systems in telecommunications. It begins with an exploration of Long Short-Term Memory (LSTM) networks, highlighting their role in sequence modeling and temporal data analysis. The discussion then transitions to the transformative impact of Transformer architecture, detailing its evolution into encoder-only, decoder-only, and encoder-decoder configurations for tasks like classification, generation, and translation. Cutting-edge techniques such as Mixture of Experts (MoE) are introduced, showcasing their efficiency in scaling models. The section concludes by contrasting uni-modal architectures, designed for single data modalities, with multimodal architectures that integrate diverse data types, emphasizing their potential to enhance AI capabilities in complex, data-rich environments.

### 4.1.1 Pre-Transformer Architectures

Before the advent of transformers, neural network architectures like Recurrent Neural Networks (RNNs), Long Short-Term Memory (LSTM) networks, and Gated Recurrent Units (GRUs) were the cornerstone of sequence modeling. These architectures were designed to process sequential data, such as text, time series, and speech, by maintaining contextual information through recurrent connections.

*Recurrent Neural Networks (RNNs)*

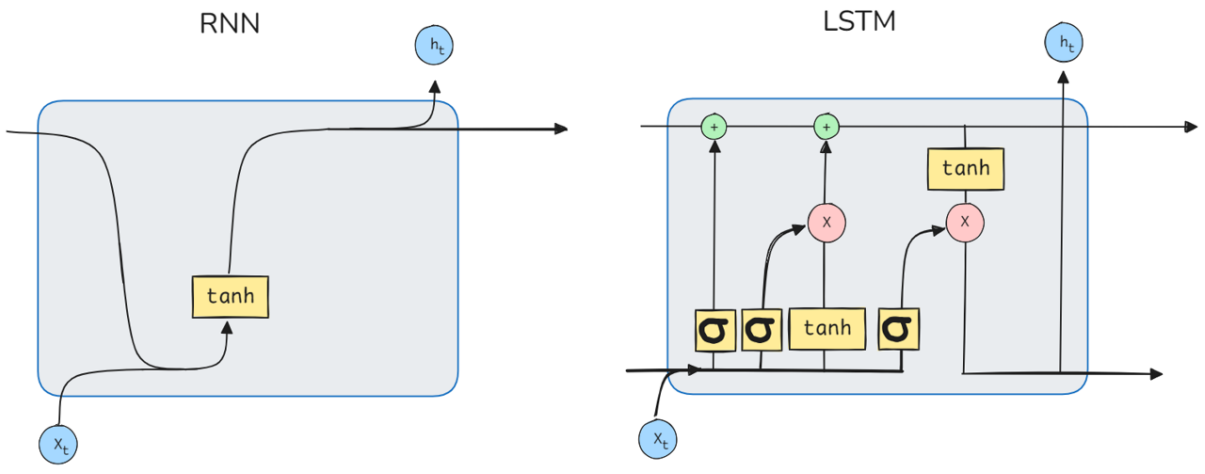

Figure 4: RNN and LSTM cells Architecture

RNNs were among the first neural network architectures designed to handle sequential data, making them particularly suitable for NLP tasks. They process input sequences one element at a time, maintaining a hidden state that captures information from previous time steps. This ability to consider context made RNNs valuable for tasks such as language modeling and machine translation. However, traditional RNNs faced challenges in capturing long-term dependencies due to the vanishing gradient problem, limiting their effectiveness in processing longer sequences [45].

*Long Short-Term Memory (LSTM) Networks*

To address the limitations of traditional RNNs, Hochreiter and Schmidhuber introduced LSTM networks in 1997, though they gained widespread adoption around 2014. LSTMs introduced memory cells and gating mechanisms that allow models to retain important information over long sequences, significantly improving their ability to handle long-range dependencies in text, as shown in Figure4 [46]. Key features of LSTMs include memory cells for storing information over extended periods, as well as Input, forget, and output gates to control information flow. LSTMs were groundbreaking in the early evolution of NLP. Their ability to process sequential data and understand long-term dependencies made them indispensable for a variety of NLP tasks, including sentiment analysis, machine translation, and text generation. For years, LSTMs were the dominant architecture in sequential data modeling and played a pivotal role in advancing the field.

Despite these notable strengths, LSTM faced several challenges in the modern AI landscape [47, 46]:

- Scalability: LSTMs struggle to scale effectively to the large parameter counts achieved by transformer models.
- Computational Efficiency: Training LSTMs is computationally intensive, particularly for large-scale models.
- Parallel Processing: The sequential nature of LSTM computation hinders efficient parallelization, a critical factor for modern high-performance architectures.

While LSTMs dominated sequence modeling for years, their computational inefficiency and challenges with scalability paved the way for alternative architectures.

### 4.1.2 Transformers

Transformers have revolutionized natural language processing tasks by outperforming previous architectures such as RNN and LSTM models, particularly in handling long-range dependencies and parallelizing computations [48]. A key innovation in transformers is the introduction of *self-attention* mechanisms and *positional encoding* (Figure 5). *Self-attention* enable the model to weigh the relevance of different input tokens dynamically, allowing for efficient modeling of dependencies without the need for sequential processing as required by RNNs or LSTMs. *Positional encoding* compensates for the lack of inherent order understanding in transformers by embedding position information into input tokens [48]. These advancements, along with transformers' ability to scale to massive datasets, are major factors behind their widespread success across diverse tasks such as machine translation, text generation, and more [49, 50].

Before applying attention mechanisms, input sequences such as words or tokens are first transformed into dense vector representations through an **embedding layer**. In high-dimensional space, similarity between embeddings can be measured using the **dot product similarity** or the **cosine similarity**. The dot product of two vectors $\mathbf{v}_1$ and $\mathbf{v}_2$ is computed as:

$$\mathbf{v}_1 \cdot \mathbf{v}_2 = \sum_{i=1}^{d} v_{1i} v_{2i}, \tag{1}$$

where a larger dot product indicates higher similarity. Alternatively, **cosine similarity** measures the angle between two vectors, normalized by their magnitudes:

$$\text{cosine}(\mathbf{v}_1, \mathbf{v}_2) = \frac{\mathbf{v}_1 \cdot \mathbf{v}_2}{||\mathbf{v}_1|| ||\mathbf{v}_2||}, \tag{2}$$

which provides a scale-invariant measure of similarity, with values ranging between -1 (completely dissimilar) and 1 (identical).

The attention mechanism leverages dot product similarity to compute the relevance between embeddings in a sequence. Given an input sequence $X \in R^{n \times d}$, where $n$ is the sequence length and $d$ is the embedding dimension, the model projects the sequence into **queries** $Q$, **keys** $K$, and **values** $V$ as follows:

$$Q = XW_Q, \quad K = XW_K, \quad V = XW_V, \tag{3}$$

where $W_Q, W_K, W_V \in R^{d \times d_k}$ are learnable projection matrices. The attention scores are computed by taking the dot product of the query and key matrices, scaled by $\frac{1}{\sqrt{d_k}}$ to mitigate large values, and applying softmax to get the

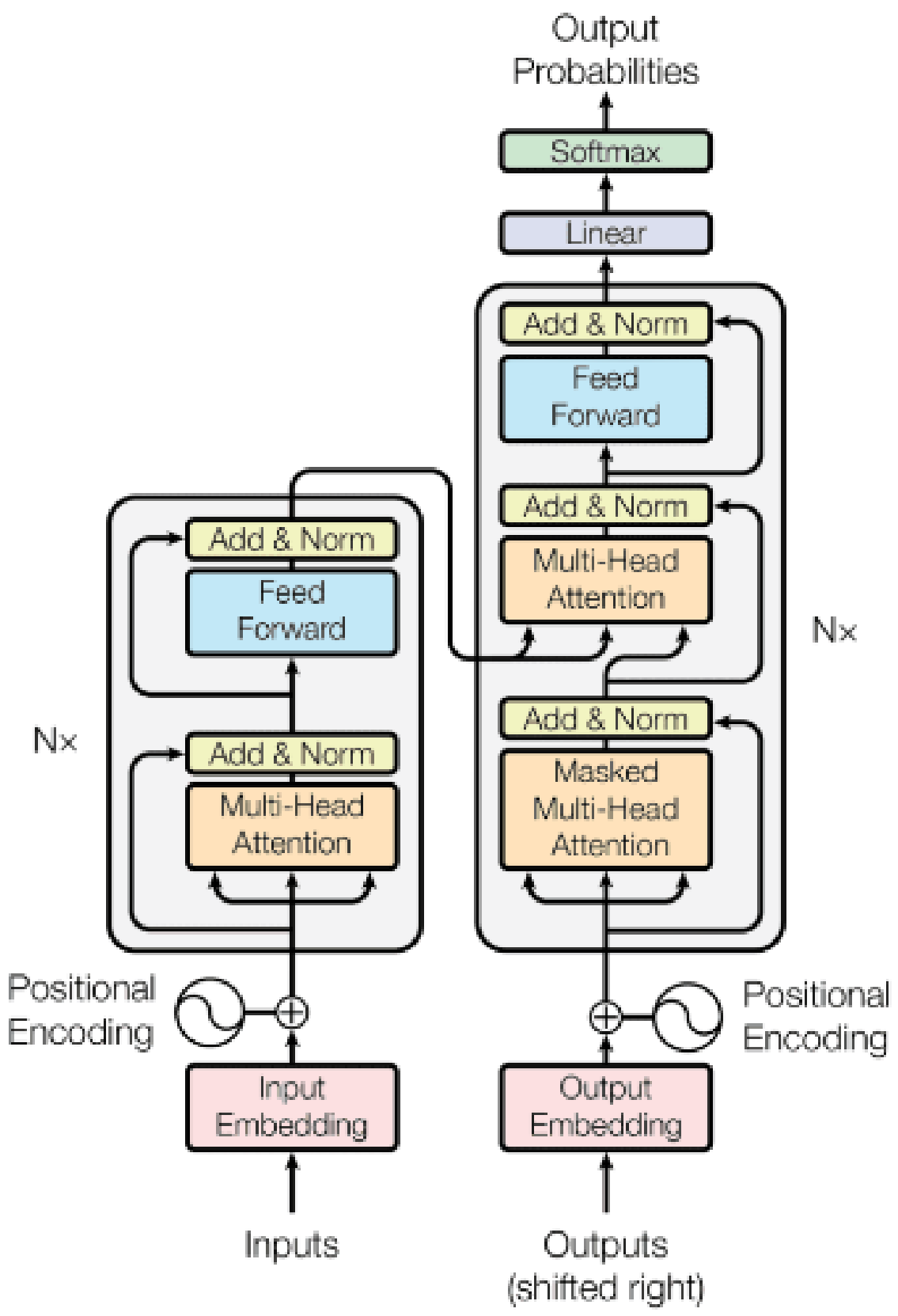

Figure 5: Transformers Architecture

attention weights:

$$\text{Attention}(Q,K,V) = \text{softmax}\left(\frac{QK^T}{\sqrt{d_k}}\right)V. \tag{4}$$

These attention weights are then used to compute a weighted sum of the value vectors, allowing the model to attend more to relevant parts of the input sequence.

To capture different types of relationships between tokens, the Transformer uses **multi-head attention**. Multiple attention heads operate in parallel, with each head computing its own set of attention scores and output. The outputs are concatenated and projected back to the original embedding dimension $d$:

$$\text{MultiHead}(Q,K,V) = \text{Concat}(\text{head}_1,\ldots,\text{head}_h)W_O, \tag{5}$$

where $W_O \in R^{hd_k \times d}$ is the learned output projection, and $h$ represents the number of heads.

In the encoder, attention is applied through **self-attention**, where each token in the input sequence attends to all other tokens. The queries, keys, and values are derived from the same input sequence $X$:

$$Q = XW_Q, \quad K = XW_K, \quad V = XW_V, \tag{6}$$

In the decoder, self-attention is modified by applying **causal masking** to prevent tokens from attending to future tokens during training, enforcing an autoregressive structure:

$$\text{MaskedAttention}(Q,K,V) = \text{softmax}\left(\frac{QK^T}{\sqrt{d_k}} + M\right)V, \tag{7}$$

where $M$ is the mask matrix blocking future positions.

In addition to self-attention, the decoder includes **cross-attention**, where queries $Q$ come from the decoder hidden states, and keys $K$ and values $V$ are derived from the encoder output $H_{\text{enc}}$:

$$Q = YW_Q^{\text{dec}}, \quad K = H_{\text{enc}}W_K^{\text{enc}}, \quad V = H_{\text{enc}}W_V^{\text{enc}}. \tag{8}$$

*Encoder only, decoder only, encoder-decoder*

Transformers can be specialized by utilizing either the encoder or decoder based on the specific task requirements. Encoder-only models, such as BERT and RoBERTa [49, 51], are particularly effective for tasks that necessitate semantic understanding, such as classification and language processing. In contrast, decoder-only models like GPT-3 [50] are tailored for generative tasks, where the objective is to predict tokens sequentially to generate coherent text. The autoregressive structure allows these models to maintain contextual information across sequences, resulting in high-quality language generation. Finally, Encoder-decoder models like T5 [52], are beneficial for sequence-to-sequence tasks, including translation, summarization, and captioning. In this setup, the encoder creates a representation of the input, while the decoder generates the corresponding output. Cross-attention mechanisms are employed to align the input and output sequences, facilitating effective information transfer between the two components [53]. Table 5 presents some Architectural Modifications of Transformer Models for Various Tasks.

Table 5: Variations in Transformer Model Architectures for Various Tasks

| Task | Model | Input-Output | Modifications | Use Case |
|---|---|---|---|---|
| **Sequence Classification** | BERT (AE) | Input: $\mathbf{x}_{1:N}$, Output: $\mathbf{y} \in C$ | Fully connected layer added, Softmax layer for classification. | Detecting anomalies in telecom networks [54] |
| **Question Answering (QA)** | BERT (AE), GPT-3 (AR), T5 (AE-AR) | Input: $\mathbf{x}_{1:N}, \mathbf{c}$, Output: BERT: Start and End token positions. T5, GPT3: $\mathbf{a}_{1:M}$ | BERT: Span prediction head for start and end tokens. GPT-3: x T5: x | BERT: Extractive QA. GPT-3: Generative QA. T5: Abstractive and extractive QA. [54, 52] |
| **Text Summarization** | BART, Pegasus, T5 (AE-AR) | Input: $\mathbf{x}_{1:N}$, Output: $\mathbf{s}_{1:M}$ | x | Summarizing technical documents [55] |
| **Machine Translation** | MarianMT (AE-AR) | Input: $\mathbf{x}_{1:N}$, Output: $\mathbf{t}_{1:M}$ | x | Translating regulatory documents [56] |
| **Named Entity Recognition (NER)** | RoBERTa (AE) | Input: $\mathbf{x}_{1:N}$, Output: $\mathbf{e}_{1:N} \in E$ | Conditional Random Field (CRF) layer for structured output. | Extracting named entities from customer conversations [57] |
| **Dialogue Systems** | GPT-3 (AR) | Input: $\mathbf{x}_{1:N}$, Output: $\mathbf{r}_{1:M}$ | x | Automating real-time customer support [58] |
| **Time Series Prediction** | Transformer (AE) | Input: $\mathbf{x}_{1:N}$, Output: $\mathbf{y}_{1:M}$ | Temporal self-attention layer added to encoder. | Predicting network traffic and demand forecasting [57] |
| **Speech Recognition** | Wav2Vec 2.0 (AE) | Input: Speech signal $\mathbf{x}$, Output: $\mathbf{y}$ (text) | CNN layers for feature extraction from raw waveforms. | Converting customer calls to text for analysis [59] |

### 4.1.3 Mixture of Experts (MoE)

The Mixture of Experts (MoE) is a machine learning framework designed to handle complex tasks by dividing them into smaller sub-tasks that are distributed across specialized models, known as "experts" (Figure 6). The core idea is to utilize a gating network that dynamically selects the most relevant experts for each input, thereby allowing only a small subset of experts to be activated, which significantly reduces computational complexity. This selective activation not only optimizes resource usage but also enhances the scalability of the model [60, 61]. At a high level, the MoE architecture consists of several key components:

**Experts:** These are deep neural networks, often feedforward networks, where each expert is trained to specialize in solving a particular part of the problem space. The experts are responsible for processing the input and providing output relevant to their specific domain. While the input to all experts may be the same, their learning objectives differ based on the part of the task they are specialized in.

**Gating Network:** The gating network takes the input and outputs a set of scores that indicate which experts are most suited for handling the input. These scores are typically produced using a softmax function, creating a probability distribution over the experts. The gating network determines which experts will be activated based on the input data, ensuring only the most relevant experts are involved.

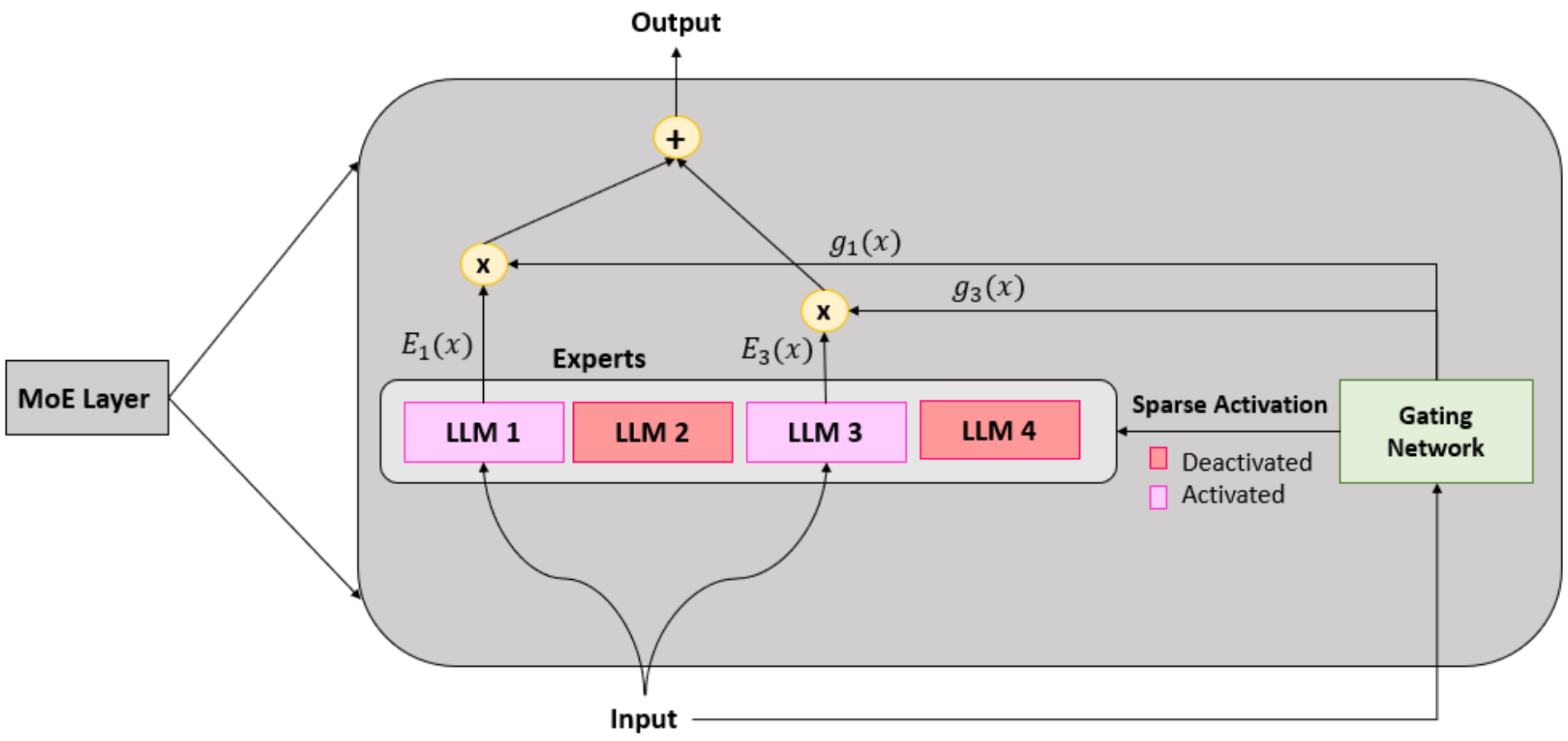

Figure 6: Mixture of Experts (MoE) Layer

**Sparse Activation:** MoE operates with sparse activation, meaning only a small number of experts are activated at any given time. This drastically reduces the computational load and makes MoE models highly efficient for large-scale tasks. The sparse selection of experts is guided by the gating network, which ensures that only the best-suited experts contribute to the model's output for a particular input.

**Output Combination:** The outputs of the selected experts are combined, typically using a weighted sum where the weights come from the gating network. The combined output is then used as the final prediction or decision. Mathematically, if we have $\mathbf{x}$ as the input, the output $y(\mathbf{x})$ of the MoE model can be expressed as:

$$y(\mathbf{x}) = \sum_{i=1}^{N} g_i(\mathbf{x}) \cdot E_i(\mathbf{x}) \tag{9}$$

where:

- $N$ is the total number of experts.
- $g_i(\mathbf{x})$ represents the gating weight for expert $i$, obtained through the softmax layer of the gating network.

- $E_i(\mathbf{x})$ denotes the output of the $i$-th expert.

**Training and Backpropagation:** During training, only the experts selected by the gating network are involved in processing the input and receiving gradient updates. The model is trained end-to-end via backpropagation, and gradients flow through both the experts and the gating network. To ensure balanced usage of experts, a load-balancing term is often included in the loss function, encouraging the gating network to distribute tasks evenly across experts. The overall loss function is given by:

$$L = L_{\text{task}} + \lambda \cdot L_{\text{balance}} \tag{10}$$

where $L_{\text{task}}$ is the primary task loss (e.g., cross-entropy), $L_{\text{balance}}$ is the regularization term for load balancing, and $\lambda$ is a regularization coefficient.

## 4.2 Advancements in Cross-Modality Translation: From Unimodal Processing to Multimodal Generative Architectures

### 4.2.1 Uni-modal Architectures

In previous work, researchers have focused extensively on translating content from one modality to another. For instance, the fields of image-to-text (captioning) and text-to-image generation have garnered significant research interest, with many works exploring approaches to model these tasks. Image captioning predominantly uses encoder-decoder architectures, where a Convolutional Neural Network (CNN) encodes images and generates captions via a Recurrent Neural Network (RNN) or Transformer decoder, as explored in multiple studies [62, 63, 64, 65] (see Figures 7 and 8). Similarly, text-to-image generation remains a challenging task, where variational autoencoders and Generative Adversarial Networks (GANs) have emerged as the dominant approaches. Text-to-image generation using VAEs or GANs generally involves conditioning the model on an input, which is a text embedding. This conditioning allows the model to generate images that are semantically aligned with the provided text. In this approach, low-resolution images are initially created and then iteratively refined into high-resolution outputs [66, 67, 68, 69] (see Figure 9).

A breakthrough in Generative Models was the use of the transformer architecture - considered the de-facto standard for natural language processing tasks - to other modalities. For instance, Alexey et al. introduced in [70] the Vision Transformer (ViT) architecture which represents a paradigm shift in image recognition, applying the transformer model directly to sequences of image patches rather than relying on convolutional networks (CNNs) (see Figure 10). In the ViT framework, an image is divided into fixed-size patches, typically 16x16 pixels, which are flattened and linearly embedded to produce a sequence of vectors. Each vector in this sequence, akin to a token in text-based transformers, represents a distinct patch of the image. These embeddings are augmented with position embeddings to retain spatial information and then passed through a standard transformer encoder comprising multi-head self-attention and feedforward layers. ViT's reliance on self-attention layers instead of convolutions enables it to model long-range dependencies across patches globally from the earliest layers, contrasting CNNs where local receptive fields incrementally grow with depth. This architecture inspired advancements in unified multimodal architectures, where Transformers are leveraged to process both text and images simultaneously.

In addition to cross-modal text-visual generation, some studies have shown great results in audio processing and generation alongside texts and images. For example, a study on cross-modal audio-visual generation [71] introduces a conditional GAN-based architecture designed to generate one modality (either audio or visual) based on the input from the other. Their architecture comprises two networks: a Sound-to-Image (S2I) network for generating images from sound and an Image-to-Sound (I2S) network for generating audio spectrograms from images.

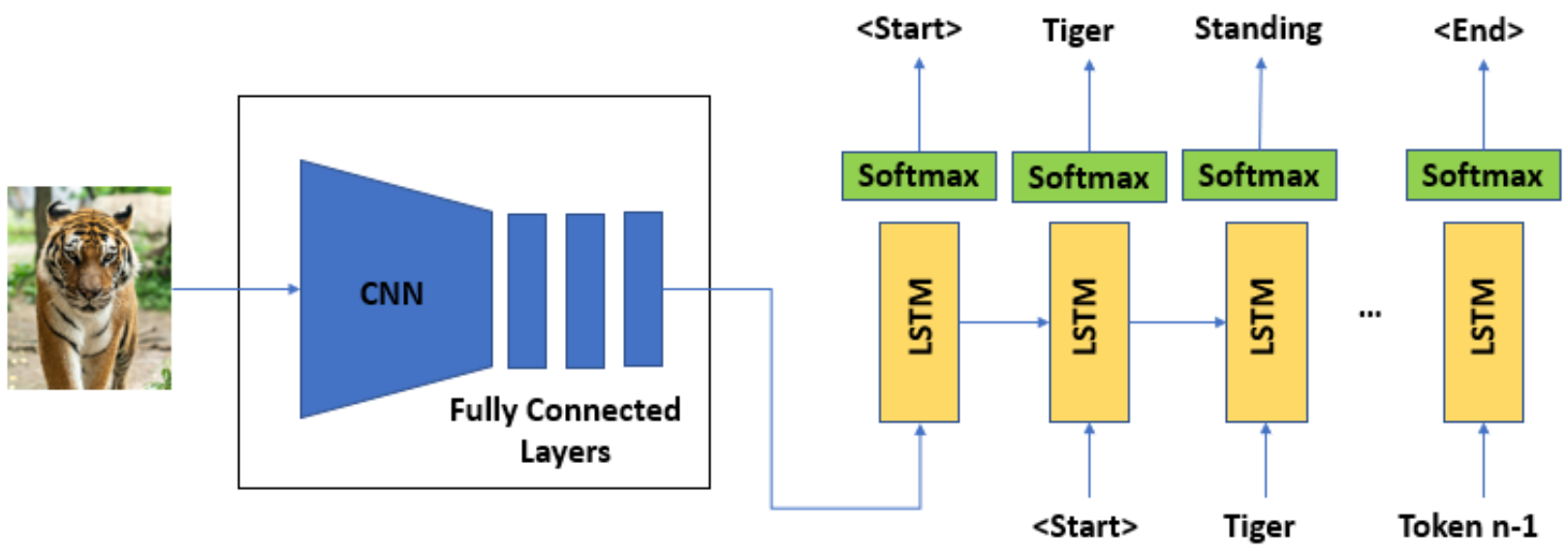

Figure 7: Image-to-Text Generation with CNN-LSTM Architecture

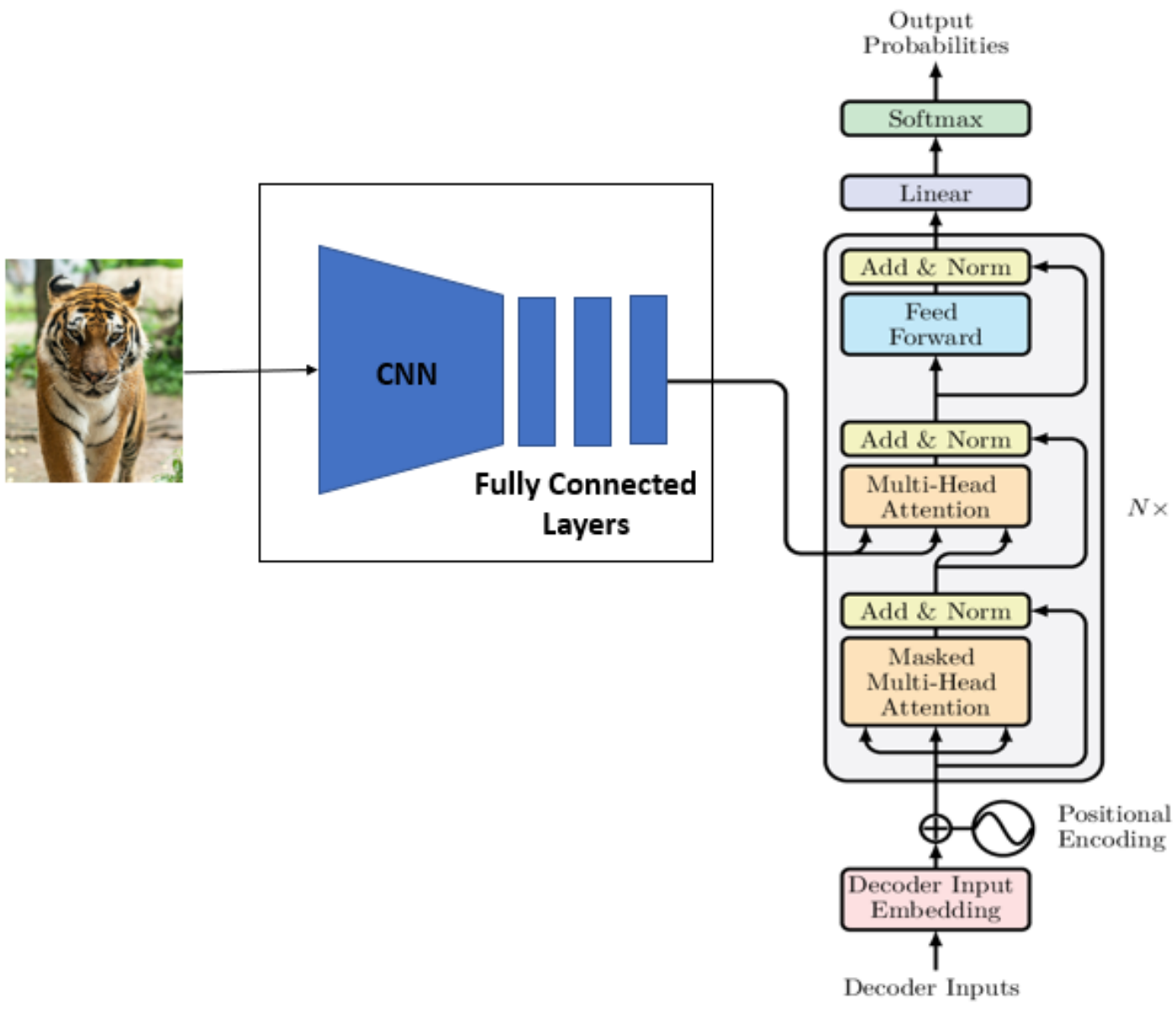

Figure 8: Image-to-Text Generation with CNN-Transformer Architecture

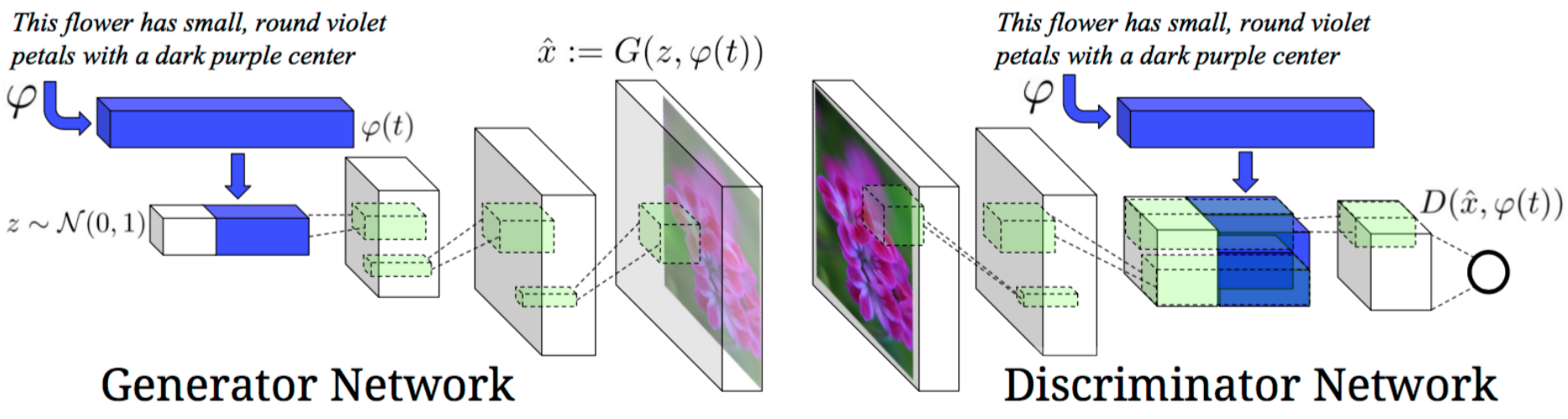

Figure 9: Text-to-Image Generation with text-conditional convolutional GAN Architecture as described in [66]

Each network has an encoder, generator, and discriminator. For the S2I network, sound input is transformed into a log-mel spectrogram representation (LMS), which is then processed by a CNN-based encoder. This encoding is combined with a noise vector and fed into the generator, which produces an image that visually represents the sound (such as a musical instrument being played). The I2S network operates similarly, using CNNs to encode images and generate a spectrogram corresponding to the expected audio output for that image (see Figure 11).

#### 4.2.2 Multi-modal Architectures

Multimodality, by contrast to Unimodality which involves working with a single type of data, involves processing and integrating information from multiple types of data, such as text, images, audio, and video, within a single

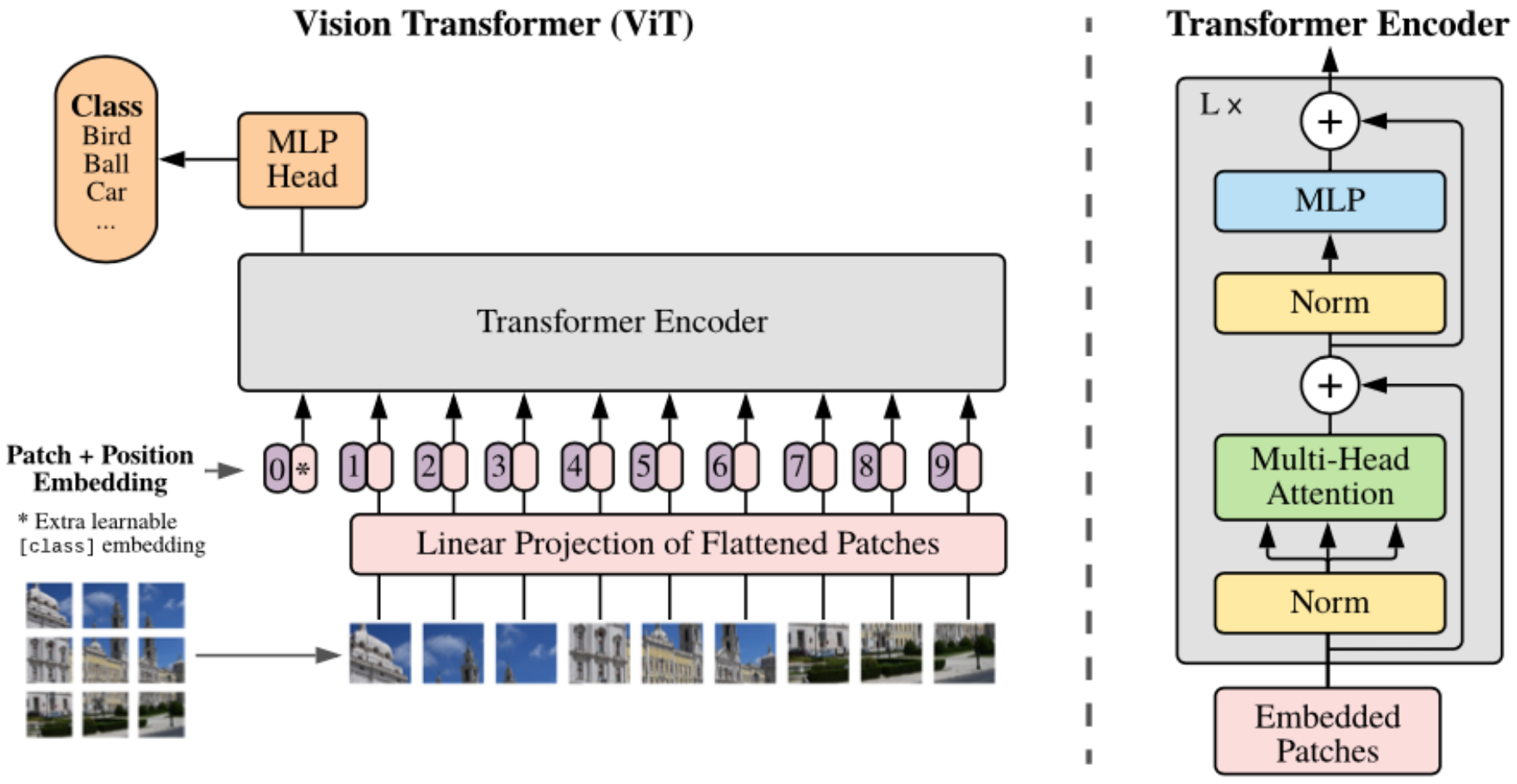

Figure 10: Vision Transform (ViT) architecture as described in [70]

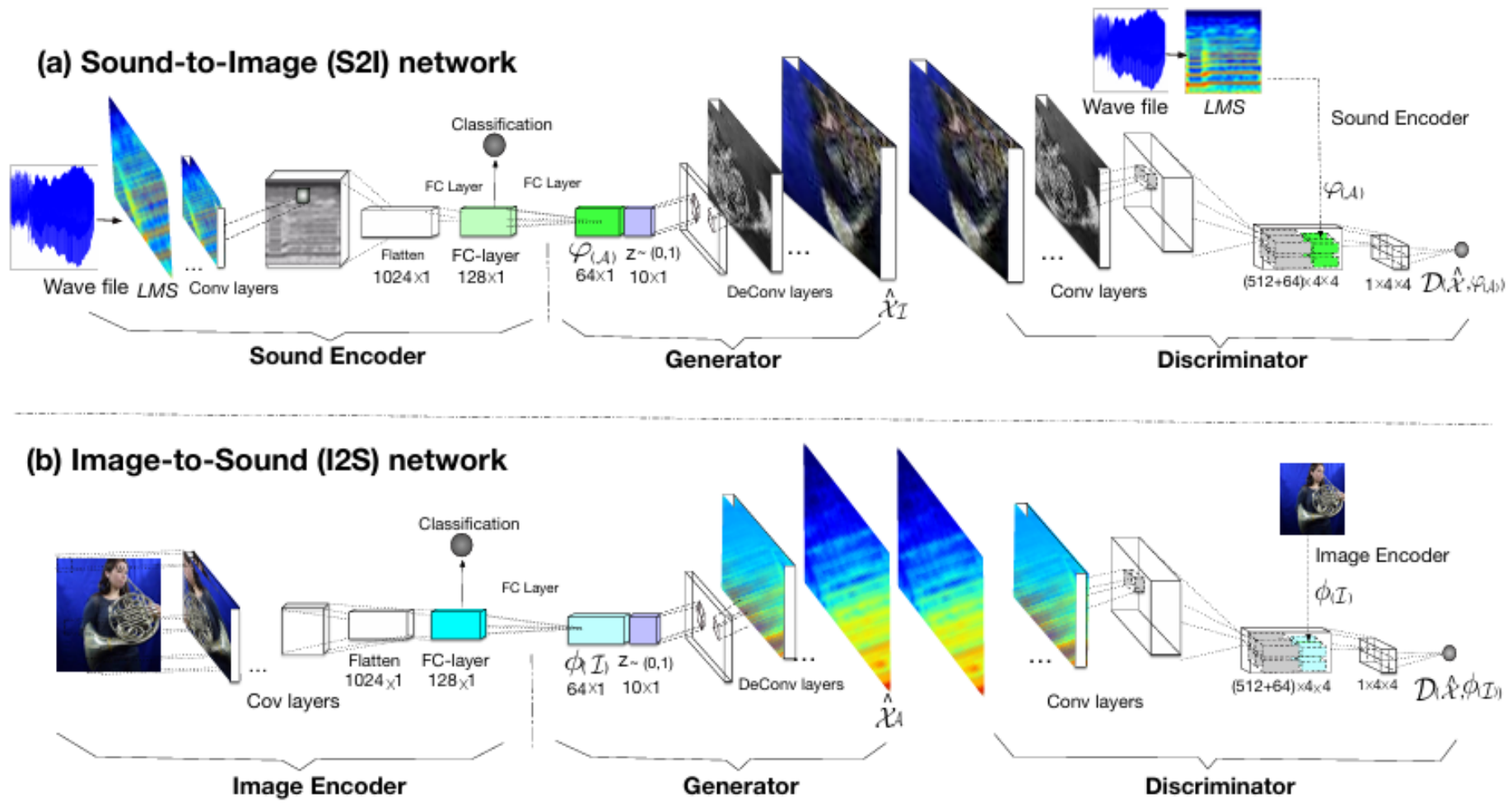

Figure 11: Cross-modal text-visual generation consisting of two networks: (a) an S2I GAN network and (b) an I2S GAN network. The network is described in [71]

model. Multimodal models aim to learn richer and more comprehensive representations by capturing relationships across these diverse data types. This ability is crucial in advancing AI towards more human-like understanding and responses, as it enables models to utilize complementary information across different sources. Generative AI models are built to produce new data that closely resembles a given training set. Multimodal generative AI extends this by integrating multiple data types. For example, a multimodal model trained on both images and text can generate an image based on a textual description or, conversely, create a description of an image. Similarly, a multimodal model trained with text and audio data can convert text into realistic speech or transcribe spoken words into text. This enhanced versatility across modalities allows these models to generate more contextually rich and adaptable output.

The architecture of a multimodal large language model (LLM) is fundamentally built on a transformer backbone, adapted to process and integrate different types of input data [72]. While, there is no unified description for multimodal models architecture, a simple way to look at Multimodal models is a sequential call to unimodal models. For instance, speech inputs can be handled by Automatic Speech Recognition (ASR) systems like Whisper, converting audio to text. Images can be processed using vision-language models (e.g., CLIP, ViT, CNN) for text extraction. These intermediate outputs, such as text from speech or images, can be then passed to GPT for reason-ing, generation, or integration with user-provided text. For output, models like DALL-E can handle text-to-image conversion, while Text-to-Speech systems synthesize speech from generated text. In contrast, advanced Multimodal architectures incorporate cross-modal attention layers with separate encoders and decoders tailored to each modality to enable inter-modal interactions or a shared encoder-decoder architecture, where modality-specific embeddings are used to preprocess inputs according to their type, ensuring compatibility across modalities [73] (see Figure 12).

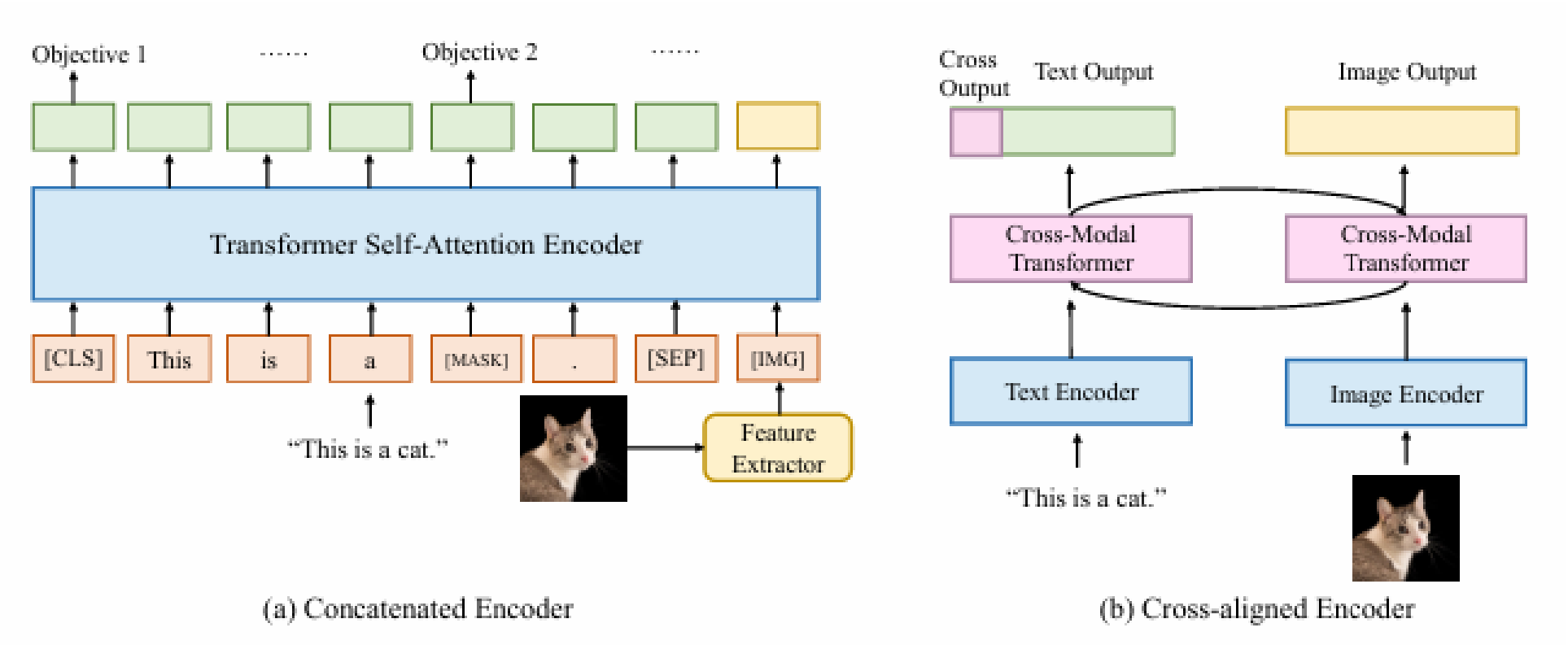

Figure 12: Vision language encoders: concatenated encoders and cross-aligned encoders as described in [72]

#### 4.2.3 Variations in Self-Attention Mechanism for Multimodal Learning

Self-attention mechanisms form the backbone of Transformer models. For multimodal scenarios, the interaction between different modalities is achieved through specialized self-attention designs. These designs can be categorized based on their mode of interaction, computational complexity, and sequence handling. In this section, we describe the main self-attention variants for multimodal Transformers, illustrating their principles, strengths, and limitations (see Figure 13).

Let $X_A$ and $X_B$ denote the inputs from two arbitrary modalities $A$ and $B$, and $Z_A$ and $Z_B$ represent their respective token embeddings. The output $Z$ denote the token embedding produced by the multimodal interactions. $T_f()$ refer to the processing of the Transformer layers. The formulations discussed are modality-agnostic and can be generalized to multiple modalities.

**Early Summation:** Early summation is a straightforward approach where token embeddings from different modalities are combined through a weighted element-wise sum at each token position. The resultant embeddings are then processed by the Transformer layers.

$$Z = T_f(\alpha Z_A \oplus \beta Z_B) \tag{11}$$

$$Q_{AB} = (\alpha Z_A \oplus \beta Z_B) W_Q^{AB}, \quad K_{AB} = (\alpha Z_A \oplus \beta Z_B) W_K^{AB}, \quad V_{AB} = (\alpha Z_A \oplus \beta Z_B) W_V^{AB} \tag{12}$$

Here, $Q_{AB}$, $K_{AB}$, and $V_{AB}$ represent the query, key, and value matrices after summation. Early summation is computationally efficient, but relies on manually tuned weights, which may limit its flexibility in capturing complex interactions.

**Early Concatenation:** In early concatenation, token embeddings from multiple modalities are concatenated along the sequence dimension and processed as a unified input sequence.

$$Z = T_f(\mathrm{C}(Z_A, Z_B)) \tag{13}$$

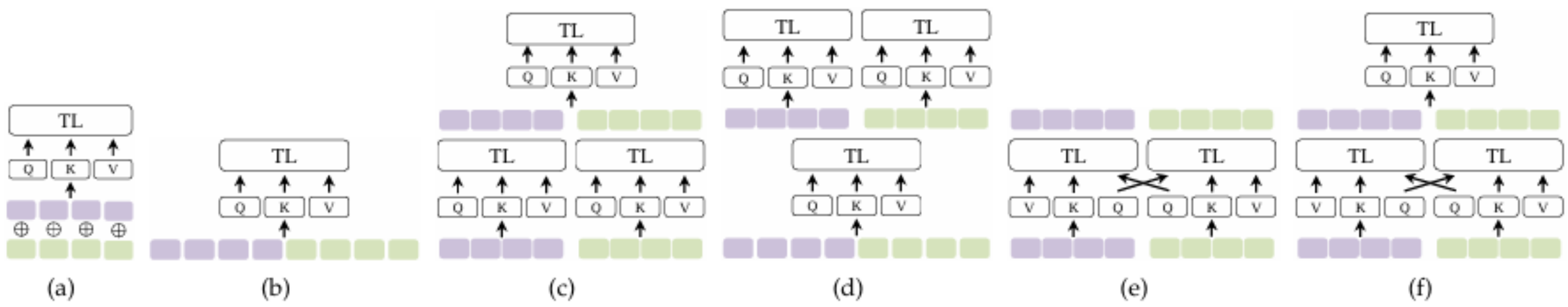

Figure 13: Transformer-based cross-modal interactions as described by Xu, Peng and Zhu, Xiatian and Clifton, David A in "Multimodal learning with transformers: A survey" [74]: (a) Early Summation, (b) Early Concatenation, (c) Hierarchical Attention (multi-stream to one-stream), (d) Hierarchical Attention (one-stream to multi-stream), (e) Cross-Attention, and (f) Cross-Attention to Concatenation. Q: Query embedding; K: Key embedding; V: Value embedding; TL: Transformer Layer.

This method enables each modality to condition on the context of the others, providing richer multimodal interactions. However, it increases the sequence length, leading to higher computational costs.

**Hierarchical Attention (Multi-Stream to One-Stream):** Hierarchical attention employs independent Transformer streams for each modality, whose outputs are concatenated and fused by another Transformer.

$$Z = T_{f3}\left(\mathrm{C}\left(T_{f1}(Z_A), T_{f2}(Z_B)\right)\right) \tag{14}$$

This approach represents a late interaction/fusion strategy and can be viewed as a specialized form of early concatenation.

**Hierarchical Attention (One-Stream to Multi-Stream):** In this variant, concatenated multimodal inputs are encoded by a shared Transformer, followed by separate streams for each modality.

$$C(Z_A, Z_B) = T_{f1}\left(\mathrm{C}(Z_A, Z_B)\right) \tag{15}$$

$$Z_A = T_{f2}\left(Z_A\right), \quad Z_B = T_{f3}\left(Z_B\right) \tag{16}$$

This method balances cross-modal interactions while preserving modality-specific features.

**Cross-Attention:** Cross-attention enables interaction by exchanging query embeddings between modalities. Each modality attends to the other's context.

$$Z_A = \mathrm{MHSA}(Q_B, K_A, V_A) \tag{17}$$

$$Z_B = \mathrm{MHSA}(Q_A, K_B, V_B) \tag{18}$$

While efficient, cross-attention may fail to capture global context due to its limited self-context representation within each modality.

**Cross-Attention to Concatenation:** This method combines cross-attention with concatenation for global context modeling. Cross-attention streams are concatenated and processed by another Transformer layer.

$$Z_A = \mathrm{MHSA}(Q_B, K_A, V_A) \tag{19}$$

$$Z_B = \mathrm{MHSA}(Q_A, K_B, V_B) \tag{20}$$

$$Z = T_f\left(\mathrm{C}(Z_A, Z_B)\right) \tag{21}$$

This approach mitigates the limitations of standalone cross-attention by integrating global context.

Finally, each variant has trade-offs in terms of computational complexity, sequence length, and interaction richness. For instance, early concatenation provides comprehensive context but is computationally expensive, while cross-attention is efficient but context-limited.

**Some Telecom Use Cases for Multimodal Models**

In telecom, multimodal generative models represent a promising frontier. As the telecom landscape evolves, the ability of multimodal generative models to synthesize data, predict outcomes, and offer actionable insights positions them as indispensable tools for future networks. For instance, the paper [75] presents a multimodal deep learning model for mobile network traffic prediction, combining CNNs and GNNs to leverage grid and graph data representations for capturing spatiotemporal dependencies (see Figure 14).

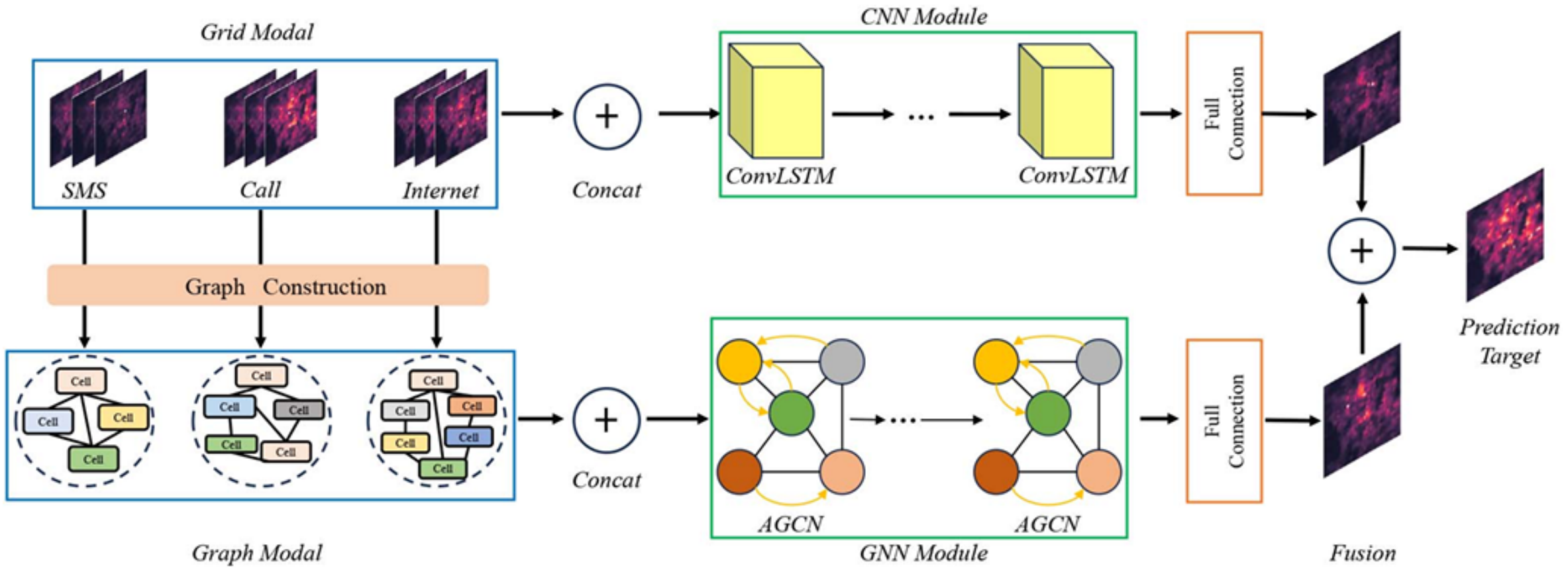

Figure 14: The proposed framework for mobile traffic prediction as described in [75]

Grid data is modeled using ConvLSTM layers, which combine convolutional operations with temporal modeling to capture localized patterns. For graph data, an Adaptive Graph Convolutional Network (AGCN) employs self-adaptive adjacency matrices to dynamically learn spatial relationships beyond static proximity. The grid and graph outputs are fused using a parametric fusion layer with learnable weights, optimally balancing their contributions. Experiments show this hybrid CNN-GNN framework outperforms baselines across metrics, particularly for SMS and call traffic, though Internet traffic remains challenging due to variability. Ablation studies validate the importance of integrating both modalities and the effectiveness of ConvLSTM and AGCN. This work highlights the benefits of fusing grid and graph data for accurate mobile traffic prediction, with applications in resource allocation, congestion management, and energy efficiency.

Another paper [76] introduces a framework for intent-based network management in 6G systems, leveraging multimodal large language models (LLMs) to translate high-level user intents into machine-readable configurations for network orchestrators. The framework consists of three main components: a Dialogue Block, an LLM, and a Translator. (see Figure 15)

The Dialogue Block serves as the interface for receiving user inputs, such as service descriptions and deployment details. It extracts key information and forwards it for processing. The LLM processes these intents using few-shot learning to specialize in networking domains like Enhanced Mobile Broadband (eMBB), Ultra-Reliable Low Latency Communications (URLLC), and Massive Machine Type Communications (mMTC). It generates NEST JSON outputs, which describe network slices while incorporating traditional KPIs and modern KVIs, such as security and energy efficiency. The Translator converts the NEST JSON into TMF-compliant Service Order APIs to ensure compatibility with platforms like OpenSlice. This step maintains adherence to industry standards, such as GST templates, without altering the original NEST JSON.

The framework was validated through a proof of concept where high-level user intents were successfully transformed into Operate API calls for network slice creation. The Dialogue and Translator blocks were implemented as microservices on Kubernetes, while OpenAI's GPT-3.5-turbo was employed for intent processing. The outputs were integrated into OpenSlice for execution following feasibility checks. The LLM was fine-tuned for classifying intents and dynamically generating NEST JSON templates tailored to specific use cases. During inference, user inputs were processed into TMF Service Orders for execution by OpenSlice, demonstrating the framework's capability for seamless Network-as-a-Service (NaaS) deployment.

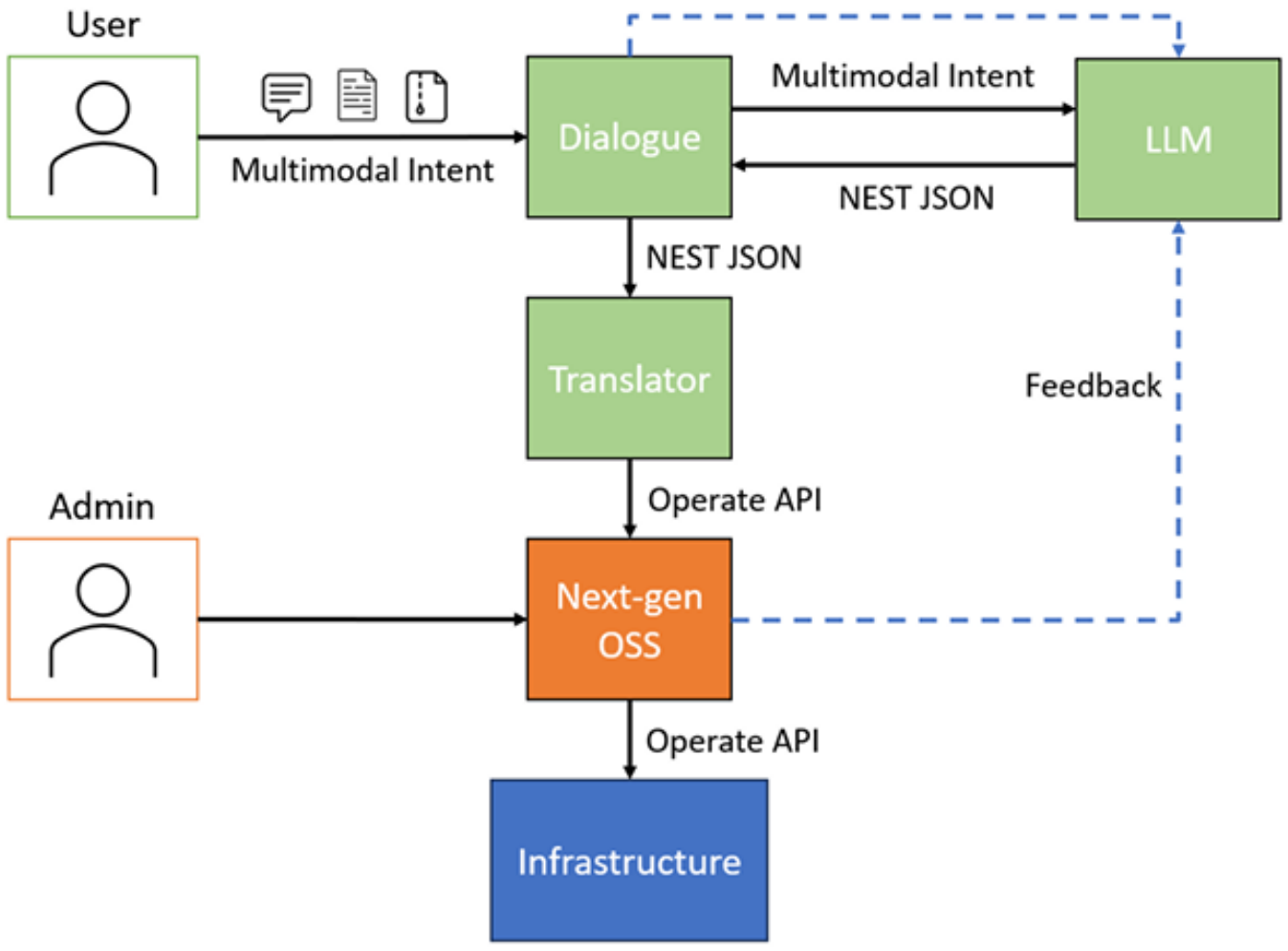

Figure 15: The proposed framework's architecture for intent-based network management in 6G systems as described in [76]

Finally, multimodal generative models unveil transformative opportunities in telecommunications, driving advancements in network operations, service delivery, and user experience. Generative AI will undoubtedly serve as the cornerstone of next-generation mobile communication systems, where large language models (LLMs) proficiently interpret multimodal intents and deliver precise, actionable responses. This integration promises to redefine efficiency, adaptability, and intelligence in the telecom industry, laying the foundation for a more connected and seamless digital future.

## 4.3 Diffusion model

### 4.3.1 Principle of diffusion model

**Forward process.** The forward process (i.e., training process) is performed by iteratively adding Gaussian noise to the initial distribution $z_0 \sim p(z)$ over $T$ time steps, gradually approaching an isotropic Gaussian distribution $z_T \sim \mathcal{N}(0,\mathbf{I})$ [77]. This process can be viewed as a discrete-time Markov chain, at time step $t \in [0,T]$, the forward process of $z_t$ is expressed as

$$z_t = \sqrt{1-\beta_t} z_{t-1} + \sqrt{\beta_t}\varepsilon, \tag{22}$$

where $\beta_t \in (0,1)$ is the noise scheduling function, typically modeled as a monotonically increasing linear function of $t$, and $\varepsilon \sim \mathcal{N}(0,\mathbf{I})$.

From the score-based perspective (i.e., the gradient of the log probability density with respect to the data $z_t$ at each noise scale $\beta_t$), the forward stochastic differentiable equation (SDE) can be expressed as

$$\begin{aligned} \mathrm{d}^{(f)} z_t &= -\frac{\beta_t}{2} z_t \,\mathrm{d}t + \sqrt{\beta_t}\,\mathrm{d}w_t \\ &= f(z_t,t)\,\mathrm{d}t + g(t)\,\mathrm{d}w_t, \end{aligned} \tag{23}$$

where $f(z_t,t) = -\frac{\beta_t}{2} z_t$ is the drift term, $g(t) = \sqrt{\beta_t}$ is the diffusion coefficient, $\mathrm{d}w_t$ is the standard Wiener process.

**Reverse process.** The reverse process (i.e., inference process) aims to recover the original input data $z_0$ from the noisy sample $z_T \sim \mathcal{N}(0,\mathbf{I})$. In DDPM, the reverse-time SDE becomes

$$\begin{aligned}\mathrm{d}^{(r)} z_t &= \left[f(z_t,t) - g(t)^2 \nabla_{z_t} \log p(z_t)\right] \mathrm{d}t + g(t)\, \mathrm{d}\overline{w}_t \\ &= \left[-\frac{\beta_t}{2} z_t - \beta_t \nabla_{z_t} \log p(z_t)\right] \mathrm{d}t + \sqrt{\beta_t}\, \mathrm{d}\overline{w}_t,\end{aligned} \tag{24}$$

where $s(z_t,t) = \nabla_{z_t} \log p(z_t)$ is the score function, which is intractable and needs to be approximated using a neural network $s_\theta(z_t,t)$.

Since $\nabla_{z_t} \log p(z_t) = \nabla_{z_t} \log p(z_t \mid z_0)$, we can approximate $\nabla_{z_t} \log p(z_t) \simeq s_\theta(z_t,t)$ for the reverse process in (24) by solving the following minimization problem during the training in the forward process [78]:

$$\theta^* = \underset{\theta}{\operatorname{argmin}} E_{z_t,z_0} \left[ \| s_\theta(z_t,t) - \nabla_{z_t} \log p(z_t|z_0) \|_2^2 \right], \tag{25}$$

where the trained score network $s_\theta(z_t,t)$ can be denoted by using Tweedie's identity as

$$s_\theta(z_t,t) = \nabla_{z_t} \log p(z_t) = -\frac{1}{\sqrt{1-\bar{\alpha}_t}} \varepsilon_\theta(z_t,t), \tag{26}$$

where $\alpha_t = 1 - \beta_t$ and $\bar{\alpha}_t = \prod_{i=1}^{t} (1-\alpha_i)$, and the parameter $\varepsilon_\theta(z_t,t)$ is the learned noise estimator at time step $t$.

### 4.3.2 Stable Diffusion model for wireless communication

The received signal can be denoted as

$$z' = hz + n \tag{27}$$

where $h$ is the channel gain and $n$ is the noise subject to $h \sim \mathcal{N}(0,\sigma^2)$.

**Known $h$.** Leveraging the diffusion model as the prior, it is straightforward to modify (24) to derive the reverse process of stable diffusion from the posterior distribution

$$\begin{aligned}\mathrm{d}^{(r)} z_t &= [-\frac{\beta_t}{2} z_t - \beta_t \nabla_{z_t} \log p(z_t \mid z')] \mathrm{d}t + \sqrt{\beta_t} \mathrm{d}\overline{w}_t \\ &= [-\frac{\beta_t}{2} z_t - \beta_t (\nabla_{z_t} \log p(z_t) + \nabla_{z_t} \log p(z'|z_t,h))] \mathrm{d}t + \sqrt{\beta_t} \mathrm{d}\overline{w}_t,\end{aligned} \tag{28}$$

By discretizing the reverse process in (28), we have

$$z_{t-1} = \frac{1}{\sqrt{\alpha_t}} (z_t + \beta_t [s_\theta(z_t,t) + \nabla_{z_t} \log p(z'|z_t,h)]) + \sqrt{\beta_t} \mathcal{N}(0,\mathbf{I}). \tag{29}$$

To solve the reverse process in (29), the main challenge lies in the posterior distribution $p(z'|z_t,h)$. Although the relationship between the received noisy SI $z'$ and the transmitted SI $z_0$ is known, the relationship between the intermediate data $z_t$ in the $t$th step of the forward process and $z'$ remains unknown. To tackle this issue, we express $p(z' \mid z_t)$ as

$$p(z' \mid z_t) = \int p(z' \mid z_0)\, p(z_0 \mid z_t)\, \mathrm{d}z_0, \tag{30}$$

where the mean of $p(z_0 \mid z_t)$ can be approximated by a delta function as

$$p(z_0 \mid z_t) \simeq \delta_{\mathbf{E}[z_0|z_t]}(z_0). \tag{31}$$

To estimate the $E[z_0|z_t]$, we can use the well-trained noise estimator $\varepsilon_\theta(z_t,t)$ in the forward process (25) to obtain the estimation $E[z_0|z_t] = \hat{z}_t$ as

$$\hat{z}_t = \frac{1}{\sqrt{\alpha_t}}(z_t - \sqrt{1-\bar{\alpha}_{t-1}}\varepsilon_\theta(z_t,t)). \tag{32}$$

Using (32), the approximation $p(z_0|z_t)$ leads to the following formula for the gradient of the log-likelihood:

$$\nabla_{z_t} \log p(z'|z_t,h) = -\frac{1-\bar{\alpha}_t}{\beta_t(1-\bar{\alpha}_{t-1})}\nabla_{z_t}||z' - hz_t||^2. \tag{33}$$

**Unknown *h*.** Notably, the above is only applicable when the instantaneous channel gain $h$ is known, and hence cannot be directly used for the scenarios with imperfect estimation of $h$ (e.g., Massive MIMO communications systems [79]). To solve this issue, a parallel SD (PSD) denoiser can be used to jointly estimate the channel gain and remove the noise.

- Forward Process: Since $z$ and $h$ are independent, the posterior probability is given by

  $$p(z_0,h|z') = p(z'|z_0,h)p(z_0)p(h). \tag{34}$$

  Therefore, we can train two separate forward processes for $z_o$ and $h$ as similar to (26) and (32), respectively.

- Reverse Process: Similar to the reverse process in (28) and (29), the reverse process of $h_t$ can be expressed as

  $$\mathrm{d}^{(r)}h_t = \left[-\frac{\beta_t}{2}h_t - \beta_t(\nabla_{h_t}\log p(h_t) + \nabla_{h_t}\log p(z'|z_t,h_t))\right]\mathrm{d}t + \sqrt{\beta_t}\mathrm{d}\overline{w}_t. \tag{35}$$

  and

  $$h_{t-1} = \frac{1}{\sqrt{\alpha_t}}\left(h_t + \beta_t\left[\nabla_{h_t}\log p(z'|z_t,h_t) + s_\vartheta(h_t,t)\right]\right) + \sqrt{\beta_t}\mathcal{N}(0,\mathbf{I}).$$

  Due to the sparse structure of wireless channel, we use $\ell_1$ regularization to sparse the channel gain by augmenting the diffusion prior thereby better stabilizing the reconstruction. The estimated channel gain is then updated as

  $$h_{t-1} = h_{t-1} - \alpha(||z' - h_{t-1} * z_{t-1}||_2 + \phi||h_{t-1}||), \tag{36}$$

  where $\phi$ is the regularization strength.

### 4.3.3 Deployment

#### Deployment Strategies

- Network edge deployment: In large-scale AI deployment within telecom, network edge deployment plays a pivotal role by bringing computation closer to users, improving the real-time processing of data and enabling more dynamic services. The edge layer in a network architecture leverages local computing resources, such as base stations or regional servers, to execute tasks that traditionally required a centralized cloud infrastructure. This distributed deployment reduces latency, optimizes bandwidth usage, and offers enhanced personalization capabilities.

  The NetGPT framework [80] exemplifies this by implementing a collaborative architecture between cloud and edge resources, allowing smaller versions of LLMs to be deployed at the network edge while more resource-intensive computations are relegated to the cloud. For example, an edge LLM can handle location-based tasks such as completing prompts with local context without the need to query the cloud for every request. This reduces network traffic and improves response times.

  Moreover, edge deployment offers substantial advantages in resource management and cost-efficiency. By splitting the workload between edge and cloud, network providers can avoid overburdening any single part of the infrastructure. Edge nodes process localized and less demanding tasks, while the cloud handles more complex tasks, such as multi-modal content generation or deeper inference processing. This division not only reduces the load on cloud servers but also optimizes the use of edge resources, enhancing both performance and cost-effectiveness.

In addition to latency reduction, edge computing significantly contributes to data privacy and security. Since sensitive user data can be processed locally at the edge rather than transmitted over long distances to centralized servers, there is less exposure to potential cyber threats. For telecom applications involving confidential or location-sensitive data (e.g., healthcare or financial services), edge deployment mitigates privacy concerns by reducing the number of data transfers and points of potential interception.

Furthermore, network edge deployment of LLMs facilitates contextual awareness in real-time decision-making. Edge LLMs can utilize real-time data streams from local devices to tailor responses based on environmental factors like network conditions or user behavior. This real-time, contextually aware capability improves the quality of service (QoS) by dynamically adjusting service provisioning, such as adjusting the level of content complexity depending on network congestion or local device performance.

In summary, deploying LLMs at the network edge enables telecom operators to offer low-latency, personalized services while maintaining a high degree of data security. The architectural flexibility of cloud-edge collaboration also ensures that services can scale efficiently to meet the demands of growing user bases without the need for massive cloud infrastructure expansion.

- Mobile (on-device) deployment: Mobile (on-device) deployment represents a paradigm shift in how large-scale AI models, particularly LLMs, are utilized in modern telecom environments. The rise of powerful mobile processors and advances in model compression techniques have made it feasible to deploy AI models directly on user equipment (UE), such as smartphones, IoT devices, and wearables. This enables real-time AI-driven services to operate locally on devices without relying on constant network connectivity to cloud or edge servers.

  However, deploying LLMs on mobile devices is not without challenges. LLMs, by their nature, are computationally intensive. For instance, the latest LLMs, like GPT-3 and LLaMA-7B, contain billions of parameters, making them difficult to fully deploy on devices with limited computing power and memory capacity. Consequently, various strategies have emerged to address this issue, including split inference and adaptive layer splitting.

  In split inference, the LLM is partitioned into layers, with the initial layers processed locally on the device and the intermediate outputs transmitted to edge servers or the cloud for further processing. This strategy significantly reduces the computational burden on the device while allowing it to contribute to the overall inference process. For example, user input can be preprocessed locally to generate intermediate features that are then sent to the edge or cloud for final prediction or response generation. This approach enables a hybrid model where both device and edge/cloud resources are utilized in tandem.

  To further enhance mobile deployment, techniques such as quantization, model pruning, and knowledge distillation are employed to shrink model size and reduce the computation required without sacrificing significant performance. These methods allow even resource-constrained devices to execute scaled-down versions of LLMs while maintaining high levels of accuracy. Knowledge distillation, for example, trains a smaller “student” model to replicate the behavior of a larger “teacher” model, enabling the student model to operate .

An important advantage of mobile deployment is its potential for offline functionality. By deploying LLMs directly on devices, telecom providers can enable services that do not require continuous network access. This is crucial in areas with poor connectivity or high data costs. For instance, an LLM could be used for voice recognition, language translation, or contextual assistance locally, enabling uninterrupted service even in network-constrained environments.

Mobile deployment also allows for greater personalization. Since the AI model operates locally, it can directly access user-specific data stored on the device, such as personal preferences, browsing history, and location data, to provide more accurate and tailored responses. This localized computation also ensures better privacy protection, as sensitive user data need not be transmitted across networks, reducing the risk of unauthorized access or data breaches.

Finally, on-device deployment reduces the dependency on centralized infrastructure, which can result in reduced latency and a more resilient service. In scenarios where network reliability is critical, such as emergency services or navigation assistance, the ability for mobile devices to process AI tasks locally ensures a more consistent user experience.

**Deployment Challenges**

- **Memory and Storage Challenges:** To meet the deployment requirements of LLMs/FMs, substantial increases in storage and memory are essential to manage the streaming of collected or generated data during inference and the updating of model parameters during training or fine-tuning. Conventional network architectures and hardware may not adequately support these elevated storage and memory demands. The scaling law of LLMs and FMs indicates that increasing the parameter size enhances accuracy and enables emergent abilities. For instance, achieving multi-task comprehension capabilities necessitates a significantly larger model size (i.e., number of parameters) compared to the size required for demonstrating specific arithmetic reasoning abilities in LLMs and FMs [81]. Consequently, deploying LLMs and FMs within current network architectures and edge devices presents significant challenges, particularly in heterogeneous networks. The models are often too large to fit into the memory of these architectures and devices, leading to a "*memory wall*" that obstructs their deployment. Even when deployed, these challenges would also impede LLMs from achieving real-time responsiveness, especially in on-device deployments, which is critical for network operations and time-sensitive applications. In light of this, accelerated and lightweight LLM deployment strategies can be implemented, and collaborative FM inference schemes can be explored in a task-oriented manner.

- **Latency:** The 5G-and-Beyond network imposes strict latency requirements. In FM-integrated networks, the non-negligible latency introduced by the integration of FMs is heavily task-dependent, making it uncertain how such integration would meet these stringent latency demands.

## 4.4 Large Language Model Training

Training is how LLMs gain the ability to understand language and follow instructions [50],[82]. In this section, we will discuss the three main types of modeling in LLMs: Pre-training, where the model learns to predict the next token given a set of tokens [83]; instruction training, which uses supervised learning techniques to teach a pre-trained model how to follow instructions; and alignment, a supervised methodology to train the model to align with human preferences [82].

### 4.4.1 Pretraining

Pre-training is the foundational process through which Large Language Models (LLMs) acquire their language understanding and generation capabilities. This process relies on large-scale corpora and extensive model architectures. In this section, we will first explore the methodologies employed for pre-training, followed by a description of the essential steps to prepare data for the training process. Next, we will discuss fundamental hyperparameter optimization techniques to facilitate convergence during training. Finally, we will outline specific strategies to address the substantial computational demands of pre-training, with a focus on reducing training time.

**Pretraining Methodologies**

In this section, the different techniques used for pre-training are described. The key difference lies in the training objective: autoregressive models focus on sequential token prediction, masked language models reconstruct missing tokens to leverage bidirectional context, denoising methods recover corrupted inputs, and contrastive techniques emphasize learning semantic distinctions.

*Autoregressive Training*

This is the most common training used by the GPT and Llama model families. The model is trained to predict the next token in a sequence given the previous tokens (left-to-right or causal modeling). The objective is to learn a unidirectional representation of text. Based on [50] the training objective is to predict the next token based on previous tokens. That means

$$x = (x_1, x_2, \ldots, x_T)$$

a sequence of tokens representing a text input and the parameter of an LLM, the causal language modeling task is to minimize the negative log-likelihood loss expressed as:

$$\mathscr{L}(x, \theta) = -\sum_{t=1}^{T} \log P(x_t \mid x_{<t}),$$

where $x < t$ denotes the token sequence before token $x_t$.

Autoregressive pretraining serves as the foundation for creating models capable of handling various text generation tasks, including question answering, story writing, summarization, and translation. By predicting the next token in a sequence based on preceding tokens, this method enables models to generate coherent and contextually relevant text [84].

*Masked Language Modeling (MLM)*

In this type of training, the model is trained to predict masked tokens within a sequence. A percentage (e.g., 15%) of tokens is randomly masked, and the model learns to predict them using their surrounding tokens. MLM enables the model to consider the context of both, preceding and following words in a sequence [49]. BERT (Bidirectional Encoder Representations from Transformers) and its derivatives are the most well-known models that use this technique. The formula for predicting a masked token in Masked Language Modeling (MLM) can be expressed as:

$$\mathscr{L} = -\sum_{t \in \text{masked}} \log P(x_t \mid x_{\text{context}})$$

where $x_t$ is the masked token, and $x_{\text{context}}$ is the context around $x_t$.

Because bidirectional models leverage information from both directions in a sequence, they are valuable for tasks such as text classification, sentiment analysis, and named entity recognition (NER), where understanding the full context is key [49].

*Denoising Autoencoding*

With this technique, the model is trained to reconstruct the original input from a corrupted version. Corruptions can include masking, shuffling, or adding noise. The model is trained to predict these masked or corrupted parts based on the remaining context [85]. The model formula can be represented as follows:

$$\mathscr{L} = -\sum_{t=1}^{T} \log P(x_t \mid \text{corrupted}(x))$$

Two examples of models trained in this technique are BART and T5. In the case of BART, noise (token deletion or permutation) is applied, and the model is trained to reconstruct the original input [86]. In the case of T5 the noise is applied to a group of consecutive span tokens [87]. The models trained with De-noising excel at reconstructing corrupted inputs by learning robust, context-aware representations, enabling them to effectively handle noisy or incomplete data. This strength makes them particularly good for text summarization (capturing long-range dependencies), question answering (generating coherent answers from noisy contexts), and text generation [86].

*Contrastive Pretraining*

Contrastive involves learning representations by contrasting positive and negative pairs of examples. The model is trained to distinguish between similar and dissimilar data points in a multidimensional space where semantically similar inputs are close together and dissimilar data points are far apart [88]. The training formula can be represented as follows:

$$\mathscr{L} = -\log \frac{\exp(\text{sim}(h_i, h_j))}{\sum_k \exp(\text{sim}(h_i, h_k))}$$

Where $\text{sim}(h_i, h_j)$ measures the similarity between representations.

The NLP models trained with contrastive learning can produce high-quality sentence embeddings that can be used for tasks such as semantic similarity, information retrieval, and text clustering [89].

**Pretraining Data Preprocessing**

Large Language Models (LLMs) are trained on datasets containing hundreds of billions to trillions of tokens sourced from diverse domains to ensure comprehensive language understanding. The table below shows a summary of the usual type of data and data sources commonly used for model pretraining [90]

Data obtained from the sources listed in Table 6 cannot be directly utilized for training in its raw form. To ensure the effectiveness of the training process, it is essential to apply several preprocessing steps. These steps encompass widely employed techniques such as filtering to remove low-quality or irrelevant content, de-duplication to eliminate redundant data, and privacy protection to safeguard sensitive information. Additionally, the data must be converted into numerical representations that can be processed by the model, a step known as tokenization. The following section provides a detailed overview of these preprocessing techniques and their importance in preparing data for pre-training.

*Pretraining Filtering*

Filtering is a critical step in preparing the pre-training corpus for Large Language Models (LLMs), as unfiltered data can introduce biases and reduce linguistic diversity [91]. Effective filtering strategies range from simple heuristics that remove low-quality text [92], to training models specifically designed to identify and exclude undesirable content. For example, in [84], a classifier was trained using high-quality data to filter the corpus, while in [87], filtering was achieved by training a model to flag data with abnormally high perplexity during pretraining for T5.

Table 6: Summary of Data Sources for LLM Pretraining [90]

| Type of Source | Description | Typical Sources |
|---|---|---|
| Webpages | Provide diverse linguistic knowledge by crawling data from the web; includes both high-quality (e.g., Wikipedia) and low-quality text. | CommonCrawl, filtered web data |
| Conversation Text | Enhances conversational competence and performance on Q&A tasks; processed into tree structures for multi-party data. | PushShift.io Reddit corpus, online social media conversations |
| Books | Offers formal, long texts beneficial for linguistic knowledge, long-term dependency modeling, and coherent text generation. | Books3, BookCorpus2, Pile dataset |
| Multilingual Text | Enhances language understanding and generation in multiple languages, useful for translation and multilingual tasks. | BLOOM (46 languages), PaLM (122 languages), FLM (Chinese and English corpora) |
| Scientific Text | Improves understanding of scientific knowledge and reasoning tasks; includes complex data requiring specific preprocessing. | arXiv papers, scientific textbooks, math webpages, scientific publications |
| Code | Enhances program synthesis, reasoning abilities, and accurate execution logic; supports solving programming tasks. | Stack Exchange (programming Q&A), GitHub (public software repositories, code with comments/docstrings) |

Filtering strategies can also be tailored to specific domains, enabling the creation of domain-specialized language models. Notable examples include BloombergGPT, optimized for financial applications [93], and BioBERT, designed to support tasks in the biomedical field [94]. These customized approaches ensure that domain-relevant knowledge is prioritized while maintaining high data quality.

*De-duplication*

Duplication in training data adversely affects model performance, particularly by damaging the internal structures responsible for generalization [95]. Consequently, de-duplication of training data is a critical preprocessing step prior to pretraining. This process must be conducted at multiple levels, addressing repeated words, duplicate sentences within the same document, and redundant content across different documents [96]. De-duplication techniques for pretraining datasets can be categorized into three main approaches: hashing, similarity-based methods, and filtering, each addressing duplication at various levels of granularity. Hashing techniques are efficient for detecting duplicates by creating unique hash values for data. Examples include ”Exact Match Hashing” [97], which identifies duplicates by comparing hash values of entire texts or portions, and MinHash [98] or Locality Sensitive Hashing (LSH), which detects near duplicates by analyzing similarities in hashed token subsets. Similarity-based methods focus on identifying duplicate or semantically similar content by comparing text features. Examples include Shingling and Jaccard Similarity [99], which break text into overlapping chunks and compute similarity scores, and advanced approaches like TF-IDF Vector Similarity [100] and Embedding-Based Similarity [101] using models like BERT or Sentence-BERT. Multilingual embeddings extend these techniques to identify duplicates between languages [102]. Filtering techniques offer scalable solutions for large datasets. For example, Bloom Filters efficiently identify duplicates without storing raw data [103], and Entropy-Based Filters flag repetitive texts with low token distribution entropy [104].

*Privacy protection*

Privacy protection in LLM training requires removing personally identifiable information (PII) from the pre-training corpus. A straightforward and effective approach is to use rule-based methods, such as keyword spotting, to detect and eliminate PII, such as names, addresses, and phone numbers [105].

**Tokenization**

Tokenization is the process of segmenting raw text into individual tokens, which are the inputs for LLMs. Three main types of tokenizers are commonly used: Byte-Pair Encoding (BPE) [106], which iteratively merges frequent token pairs and is popular in models like GPT-2 and LLaMA; WordPiece, originally developed for Google's voice search, which merges token pairs based on the likelihood increase it provides for training data and is used in BERT [107]; and Unigram Tokenization [108], which starts with a large set of substrings and removes the least likely tokens iteratively, as seen in models like T5. While existing tokenizers can be used, custom tokenizers tailored to the pre-training corpus can improve model performance, especially for specific domains [109].

**Hyper-parameter Optimization** Understanding hyperparameters is a critical aspect of training Large Language Models (LLMs), as they significantly impact the learning process, convergence speed, and overall model performance. Proper tuning of hyperparameters, such as learning rate, batch size, and regularization parameters, is essential for efficient training.

**Batch Size** Batch size, the number of training samples processed before updating model parameters, significantly affects LLM training and accuracy. Small batch sizes provide frequent updates and may enhance generalization but can introduce noisy gradients, potentially destabilizing training. Large batch sizes improve hardware utilization and gradient stability but require careful learning rate scaling to maintain convergence and avoid overfitting. Balancing batch size is critical for efficient and effective training of LLMs [110].

**Learning Rate** Learning rate, a critical hyperparameter in LLM training, determines the magnitude of weight updates during optimization. In NLP models, an incorrectly tuned learning rate can lead to slow convergence or model divergence, affecting both training efficiency and final accuracy. Adaptive learning rate schedules, such as warm-up and decay strategies, are often used in LLM frameworks to stabilize training and enhance performance [111]. By optimizing these settings, models can achieve improved accuracy and better generalization to unseen data. The following section provides a detailed discussion of the hyperparameter optimization techniques referenced above [111].

**Regularization** Regularization techniques, such as dropout, weight decay, and label smoothing, are critical to mitigating overfitting in LLM training by controlling model complexity and improving generalization. Dropout randomly deactivates neurons to reduce reliance on specific features, weight decay penalizes large weights to simplify models, and label smoothing prevents overconfidence in predictions. These methods improve the robustness and accuracy of the model in unseen data [112]. Gradient clipping, while not a regularization method in the traditional sense of penalizing model complexity, acts as a form of training stabilization that mitigates instability and overfitting caused by large, unbounded updates [45].

**Optimizer** An optimizer in LLM training is an algorithm that adjusts model parameters to minimize the loss function, directly affecting convergence speed and final accuracy. Popular choices include Adam [113], which combines momentum and adaptive learning rates for stability and speed; AdaGrad, effective for sparse data; and AdamW (Loshchilov  Hutter, 2019), which improves generalization by decoupling weight decay regularization from gradient updates.

**Training Parallelism** Training large language models (LLMs) requires distributing the computational load due to the immense scale of datasets and model architectures, which exceed the capacity of a single device. To address these challenges, various parallelism techniques and optimization strategies, such as data parallelism, pipeline parallelism, tensor parallelism, and mixed precision training, are employed. These methods enable scalable and efficient pretraining by reducing computational bottlenecks and optimizing resource utilization. This section provides an in-depth discussion of these techniques and their roles in modern LLM workflows.

*Data parallelism*

Data Parallelism is a fundamental approach to improving training throughput by replicating model parameters and optimizer states across multiple GPUs. The training corpus is distributed among GPUs, allowing each GPU to process its assigned data and perform forward and backward propagation independently to compute gradients. These gradients are then aggregated to update the model across all GPUs, making the approach highly scalable by increasing the number of GPUs [114].

Traditional Single GPU training processes all computations on a single device, offering simplicity but limited scalability [115]. The Single Parameter Server architecture centralizes parameter updates, enabling multiple workers to share gradients, though it can become a bottleneck [116]. To address this, Distributed Parameter Server systems partition parameters across multiple servers, reducing update latency [117] . Alternatively, the work in [118] employs a ring-allreduce algorithm for distributed gradient updates, improving communication efficiency among GPUs. More advanced approaches, such as the distributed parameter server with Apex Mixed precision [119], leverage reduced precision arithmetic (FP16) to enhance memory efficiency and training speed. Similarly, Apex Mixed-Precision is applied in [120] to achieve robust multi-node scalability. A comparative study of these techniques concluded that the Distributed Parameter Server with Apex is optimal for single-node scenarios, while Horovod with Apex stands out as the most robust solution for multi-node environments, offering a balance of speed, memory efficiency, and scalability [121].

*Pipeline Parallelism*

If Data parallelism focuses on splitting the data across devices, pipeline parallelism focuses on splitting the model across devices. The model is divided into sequential stages, with each stage assigned to a different device. Data flows through these stages in a pipeline fashion, enabling concurrent computation across different parts of the model. Consecutive layers assigned to the same GPU minimize communication overhead of transmitting hidden states or gradients. Although this technique can reduce the training time a straightforward implementation can lead to lower GPU utilization due to idle waiting (”bubbles”) between GPUs. To address this, GPipe [122] uses micro-batch processing to maximize hardware utilization, and employs gradient checkpointing to reduce memory overhead and PipeDream [123] introduces overlapping computation and communication to minimize idle time and weight stashing to eliminate the problem of weight inconsistencies across micro-batches and ensure training convergence.

*Tensor Parallelism*

Tensor Parallelism [114] is a technique for decomposing large language models (LLMs) across multiple GPUs by splitting their parameter matrices. For instance, in the matrix multiplication operation $Y = XA$, the parameter matrix $A$ can be divided into submatrices $A_1$ and $A_2$ by columns, expressed as $Y = [XA_1, XA_2]$. By assigning $A_1$ and $A_2$ to different GPUs, the operation can be performed in parallel, with the final result obtained through inter-GPU communication. To overcome the known issues with Tensor Parallelism, like communication overhead, memory constraints, and underutilization of computational resources during the pretraining, the study in [124] introduces an interleaved pipeline parallelism schedule that improves throughput by over 10%, while maintaining a scalable memory footprint.

*Mixed Precision Training*

While 32-bit has been common in older models like BERT, recent studies [125] have adopted 16-bit floating-point numbers (FP16) to reduce memory usage and communication overhead. FP16 also leverages the architecture of GPUs like NVIDIA A100, which have twice as many FP16 computation units as FP32, enhancing computational efficiency. However, FP16 can result in reduced computational accuracy [125], impacting model performance. To mitigate this, Brain Floating Point (BF16), which allocates more exponent bits and fewer significant bits than FP16, was introduced [126].

**Instruction Tuning** As opposed to pretraining, which involves exposing large language models (LLMs) to vast unstructured datasets using unsupervised or self-supervised learning to develop general language understanding [49], instruction tuning focuses on fine-tuning these models by training them on specific instruction-response pairs to better align with user intent [127]. While pretraining equips the model with a broad foundation of linguistic knowledge, instruction tuning refines this foundation by leveraging curated datasets to optimize the model's performance on specific downstream tasks.

**Supervised Fine-Tuning**

Supervised Fine-Tuning (SFT) involves training the model on labeled datasets consisting of input-instruction-output triplets. The objective is to adapt the pre-trained model to predict outputs aligned with human-provided instructions. The objective function minimizes the negative log-likelihood of the target sequence $y = \{y_1, y_2, \ldots, y_T\}$, conditioned on the input $x$:

$$\mathscr{L}_{\text{SFT}} = -\sum_{t=1}^{T} \log P(y_t \mid y_{<t}, x; \theta),$$

where: - $\theta$: Model parameters. - $y_{<t}$: Tokens up to $t-1$. - $P(y_t \mid y_{<t}, x; \theta)$: Model's predicted probability for the next token.

**Parameter-Efficient Fine-Tuning Techniques**

These techniques involve modifying several parameters to fine-tune the model, making them computationally efficient for instruction training.

*Adapters*

Adapters [128] introduce small bottleneck layers into the model architecture. During fine-tuning, only these adapter layers are updated, while the rest of the model remains frozen. Adapters add a bottleneck structure, where the output is defined as:

$$h_{\text{adapter}}(x) = W_{\text{up}} \cdot \text{ReLU}(W_{\text{down}} \cdot h(x)),$$

where: - $h(x)$: Output from the frozen pre-trained model layer. - $W_{\text{down}}$: Down-projection matrix. - $W_{\text{up}}$: Up-projection matrix.

*Prefix Tuning*

Prefix Tuning prepends a trainable sequence (prefix) to the input of each transformer layer, allowing task adaptation without altering the model's main parameters [129].

The loss function for Prefix Tuning is:

$$\mathcal{L}_{\text{Prefix}} = -\sum_{t=1}^{T} \log P(y_t \mid y_{<t}, p, x; \theta),$$

where: - $p$: Trainable prefix added to the input. - $\theta$: Pre-trained model parameters (frozen).

The advantages are that it keeps the model intact.

*Low-Rank Adaptation*

LoRA introduces low-rank matrices into the attention layers of the model. These matrices are fine-tuned while the rest of the model remains frozen [130].

The weight update is represented as:

$$\Delta W = W_{\text{down}} \cdot W_{\text{up}},$$

where: - $W_{\text{down}}$: Low-rank down-projection matrix. - $W_{\text{up}}$: Low-rank up-projection matrix.

This technique is memory-efficient and enables fine-tuning large models on limited hardware

*QLoRA (Quantized LoRA)*

QLoRA extends LoRA by using quantized weights to further reduce memory usage while maintaining performance. QLoRA uses quantized weights for LoRA updates:

$$\mathcal{L}_{\text{QLoRA}} = \mathcal{L}_{\text{LoRA}} \quad \text{with quantized weights.}$$

Advantages: Further reduces computational requirements. Suitable for fine-tuning on consumer-grade GPUs.

### 4.4.2 Alignment

Alignment is a training methodology to ensure that a model's behavior aligns with human preferences. The goal is to train models that produce outputs which are not only accurate but also safe, ethical, and useful according to human-defined criteria [131]. In alignment, the focus is on adapting a model's predictions or behaviors to meet specific expectations, even in ambiguous or high-stakes scenarios. This is particularly important for large language models (LLMs), where unintended outputs (e.g., biased or harmful content) could lead to significant consequences. There are mainly three techniques that fall into the alignment training methodologies.

#### Reinforcement Learning with Human Feedback

RLHF is a training method that aligns large language models (LLMs) with human preferences by incorporating human feedback into the training process. Unlike traditional instruction methodologies that rely only on supervised fine-tuning with predefined datasets, RLHF involves an interactive approach where human evaluators assess and rank model outputs. These human preferences are then used to fine-tune the model through reinforcement learning, leading it to produce outputs that better align with human expectations [82]. In this work the authors suggest applying Reinforcement learning by training a reward function to capture human preferences, then Proximal Policy is used to generate outputs that maximize the reward model's scores and align the model's behavior with the human preference.

*Reward Model*

The reward model is trained to predict human preferences using a pairwise comparison loss that ensures that the reward model learns to assign higher scores to the outputs preferred by human evaluators. The reward model is trained to minimize the following loss function:

$$\mathscr{L}(\phi) = -\frac{1}{\binom{K}{2}} E_{(x,y_w,y_l)\sim\mathscr{D}} \left[\log\left(\sigma\left(r_\phi(x,y_w) - r_\phi(x,y_l)\right)\right)\right]$$

Where:

- $\mathscr{L}(\phi)$: Loss function of the reward model with parameters $\phi$.
- $r_\phi(x,y)$: Reward score predicted by the reward model for input $x$ and response $y$.
- $y_w$: The response labeled as "preferred" by human evaluators.
- $y_l$: The response labeled as "less preferred" by human evaluators.
- $\sigma$: Sigmoid function, which transforms the difference in scores into a probability value.
- $\mathscr{D}$: Dataset of prompts $x$ and pairs of responses $(y_w, y_l)$ used for training.
- $\binom{K}{2}$: Number of possible pairwise comparisons in a set of $K$ responses.
- $E$: Expectation over all samples $(x, y_w, y_l)$ from the dataset $\mathscr{D}$.

*Proximal Policy Optimization (PPO)*

The reinforcement learning algorithm applied is Proximal Policy Optimization (PPO), which fine-tunes the supervised fine-tuned (SFT) model by optimizing responses to customer prompts based on rewards from a reward model while incorporating a per-token KL divergence penalty to prevent over-optimization

$$\text{objective}(\phi) = E_{(x,y)\sim\mathscr{D}_{\text{RL}}}\left[r_\phi(x,y) - \beta\log\left(\frac{\pi_\phi^{\text{RL}}(y\mid x)}{\pi_\phi^{\text{SFT}}(y\mid x)}\right)\right] + \gamma E_{x\sim\mathscr{D}_{\text{pretrain}}}\left[\log\pi_\phi^{\text{RL}}(x)\right]$$

Where:

- $\phi$: Parameters of the learned RL policy.
- $\pi_\phi^{\text{RL}}(y \mid x)$: Probability of generating response $y$ given prompt $x$ under the RL-trained policy.
- $\pi_\phi^{\text{SFT}}(y \mid x)$: Probability of generating response $y$ given prompt $x$ under the supervised fine-tuned (SFT) policy.
- $\mathscr{D}_{\text{RL}}$: Dataset used during RL fine-tuning, which includes prompts and corresponding rewards.
- $r_\phi(x,y)$: Reward score determined by the reward model for the prompt-response pair $(x,y)$.
- $\mathscr{D}_{\text{pretrain}}$: Pretraining distribution used for additional gradient updates.
- $\beta$: KL reward coefficient, controlling the strength of the KL divergence penalty.
- $\gamma$: Pretraining loss coefficient, controlling the weight of pretraining gradients in the PPO update. For "PPO" models, $\gamma$ is set to 0.

- $E$: Expectation over the respective datasets.
- log: Logarithm function.

*Direct Policy Optimization*

In [132], DPO (Direct Preference Optimization) is proposed as an alternative to PPO for aligning LLMs. DPO simplifies the alignment process by directly optimizing language models based on human preference data, eliminating the need for intermediate reward modeling and complex reinforcement learning steps inherent in methods like PPO. The core of DPO involves adjusting the model's policy $\pi_\theta$ to maximize the likelihood of preferred responses over less preferred ones. This is achieved through a binary cross-entropy loss function:

$$\mathscr{L}_{\text{DPO}}(\pi_\theta; \pi_{\text{ref}}) = -E_{(x,y_w,y_l)\sim\mathscr{D}}\left[\log\sigma\left(\beta\log\frac{\pi_\theta(y_w \mid x)}{\pi_{\text{ref}}(y_w \mid x)} - \beta\log\frac{\pi_\theta(y_l \mid x)}{\pi_{\text{ref}}(y_l \mid x)}\right)\right]$$

**Where:**

- $\mathscr{L}_{\text{DPO}}$: Loss function for Direct Preference Optimization.
- $\pi_\theta$: Policy being optimized, parameterized by $\theta$.
- $\pi_{\text{ref}}$: Reference policy used as a baseline for comparison.
- $(x, y_w, y_l)$: A triplet from the dataset $\mathscr{D}$, where $y_w$ is the preferred response and $y_l$ is the less preferred response for input $x$.
- $\sigma$: Sigmoid function, which maps the input to a probability between 0 and 1.
- $\beta$: Scaling factor that controls the weight of the log-probability differences.
- $\mathscr{D}$: Dataset of preference triplets used for training.
- $E$: Expectation over all sampled triplets from the dataset $\mathscr{D}$.
- log: Natural logarithm.

### 4.4.3 Distributed Fine-tuning within Telecom

In modern telecommunications, LLMs often require fine-tuning to meet the specific needs of different business applications. Network-based fine-tuning presents a new paradigm, as 6G networks are increasingly equipped to support AI-as-a-Service (AIaaS) and Compute-as-a-Service (CaaS) [133], providing a robust infrastructure for model training [134].

With these advancements, the network itself can serve as a distributed computing platform capable of handling the computational demands of fine-tuning large-scale models [134]. This shift enables more efficient utilization of the network's resources, allowing for the dynamic offloading of training tasks and model updates across the network. Such a network-driven fine-tuning approach eliminates the need for central servers and reduces the computational burden on individual devices, opening the door to new possibilities for optimizing model performance across diverse applications in real-time.

### 4.4.4 Existing distributed learning frameworks:

Federated Learning (FL) [135] enables distributed devices to train models locally using their own data and then share only the model updates with a central server, ensuring privacy by avoiding the transmission of raw data. A common aggregation strategy used in FL is FedAvg, where each client's model update is weighted based on the size of its local data. While FL enhances privacy, it suffers from several limitations, particularly its reliance on a single central server, which makes it vulnerable to server failure and reduces scalability. Furthermore, training large models on resource-constrained devices remains challenging due to limited computational resources such as processing power, memory, and bandwidth. The "straggler problem" is another significant issue, where slower devices or network connections can disrupt the synchronization process, delaying updates from the entire network and hindering the efficiency of the learning process.

Split Learning (SL) [136] provides a solution for resource-constrained environments by offloading the majority of deep neural network (DNN) computations to a central server, with clients only processing the initial layers of the model using their local data. Clients then transmit intermediate activations (or "smashed data") to the server, where further computations are performed. While SL reduces local computation and memory demands on clients, it introduces substantial communication overhead, as large volumes of data must be transmitted to the central server. Additionally, relay-based SL approaches, such as Random Walking Snakes (RWS) [137], further reduce server dependence by segmenting the model and activating a sequence of clients in a "snake-like" pattern, where model segments are passed between clients. While RWS offers some relief from the central server bottleneck, it still suffers from high communication costs and increased vulnerability to disruptions, particularly if any client fails to participate or experiences network issues. Furthermore, the heterogeneous nature of the data across different clients (Non-IID data) introduces the risk of overfitting or catastrophic forgetting, as clients may not have sufficiently representative or consistent data distributions.

Besides, by adopting an incremental update strategy where model parameters are updated gradually, Snake Learning allows for a much more relaxed synchronization requirement is allowed. Each node in the network updates and shares a segment of the model in a sequential manner, resembling the way the snake eats items and grows, but in this case, it grows by consuming model parameters in a structured way. The updates are done in a serpentine manner, meaning that nodes incrementally propagate updates through each layer of the model, ensuring that each part of the model is progressively refined through collaboration across the network. This reduces the need for real-time and frequent global synchronization, a major bottleneck in traditional distributed learning systems.

Hybrid methods that combine FL and SL are also being explored to leverage the strengths of both frameworks. One such approach, SplitFed Learning [138], integrates FL with SL by using two servers: one handles the server-side model computations, while the other manages client-side synchronization and model updates via FedAvg. This parallel processing improves communication efficiency but also increases the complexity of the system. Another hybrid approach, Accelerated FL [139], reduces the need for frequent communication by adopting a local-loss-based training method, where separate local loss functions are used for the client-side model and the server-side model, allowing updates to be made independently without requiring gradients from the central server. This reduces the real-time synchronization requirements but still faces challenges related to the communication overhead between the federal server and clients. Despite these efforts to combine FL and SL, these hybrid models still suffer from persistent issues like communication bottlenecks, server reliance, and maintaining model consistency across decentralized nodes.

**Key challenges:**

While existing distributed learning frameworks such as Federated Learning, Split Learning, and their hybrid variants have made significant strides in enabling distributed model training across edge devices, they still encounter several key challenges.

- Communication Synchronization Challenges: Unstable wireless connections, especially in mobile environments, disrupt the synchronization needed for distributed learning, resulting in slower or inconsistent updates. Asynchronous methods can exacerbate this by introducing issues like ”model staleness” and training instability.

- Heterogeneous, Dynamic Resource Availability: Unlike cloud computing, 6G resources are shared across services, creating dynamic shifts in the availability of computation power, which complicates the efficient processing of learning tasks. Nodes with heterogeneous hardware capabilities further complicate the deployment of effective distributed learning frameworks.

- Data Heterogeneity: Non-IID data distributions across nodes lead to biases and affect model convergence and stability, diminishing generalization abilities. Advanced strategies beyond simple data-level interventions (such as augmentation) are needed to address these challenges.

- Model Consistency and Stability: Frameworks that rely on asynchronous updates or decentralized models often face challenges in maintaining model consistency. Disruptions caused by slow nodes or network failures can lead to ”model staleness” or inconsistent updates, potentially hindering the stability and convergence of the model training process.

# DATASETS

# 5 Datasets

As telecommunications evolve into the era of 6G, integrating large models into the domain has become a cornerstone for innovation. These models hold immense potential for automating complex tasks, such as interpreting technical standards, optimizing networks, and analyzing channel behaviors. However, their successful application in telecommunications faces unique challenges: the specialized and highly technical nature of the field necessitates curated, domain-specific datasets and novel adaptation techniques. To address these challenges, researchers have developed a suite of datasets and tools designed to benchmark, pre-train, and fine-tune LLMs for telecom applications. From benchmarking tools like TeleQnA to domain-specific pre-training datasets such as TSpec-LLM and OpenTelecom, and advanced ray tracing datasets supporting 6G research, these resources empower large models to navigate the complexities of telecom systems. Together, they unlock new possibilities for AI-driven automation, resource optimization, and decision-making in next-generation communication networks, transforming LLMs into indispensable allies in the telecommunications domain.

## 5.1 Benchmarking and Testing Datasets

Benchmarking and testing datasets play a critical role in evaluating and enhancing the capabilities of LLMs in specialized domains like telecommunications. These datasets serve two primary purposes: assessing the performance of LLMs in understanding domain-specific knowledge and providing targeted resources for fine-tuning models to excel in specialized tasks. For instance, TeleQnA acts as a benchmarking tool by offering a diverse set of multiple-choice questions designed to evaluate an LLM's telecom knowledge across various subdomains. This dataset highlights the strengths and limitations of general-purpose and fine-tuned models, facilitating focused improvements. Similarly, TSpec-LLM provides a comprehensive dataset of 3rd Generation Partnership Project (3GPP) standards, enabling LLMs to better grasp the complexities of telecom technical documents. Together, these datasets address key challenges in applying LLMs to telecom, ensuring their ability to interpret, process, and respond to intricate technical requirements accurately. A summary of the relevant telecom datasets is listed in Table 7.

### 5.1.1 TeleQnA

TeleQnA is a benchmark dataset to assess the telecom knowledge of LLMs [140]. This dataset is composed by 10000 multiple-choice questions and answers related to different resource publications, research overview, standard specifications, standard overview, and telecom lexicon. The fraction of multiple-choice questions on the different topics are shown in Figure 16.

To construct a comprehensive dataset covering the multifaceted domain of telecommunications, a substantial number of questions is required. This task is further complicated by the specialized nature of telecom knowledge, demanding expertise to craft pertinent questions, answers, and explanations. Moreover, the telecom documents we collected often contain highly intricate information, making it infeasible for a team of human experts to generate questions and answers that comprehensively cover the diverse range of telecom subdomains. To address these challenges, TeleQnA uses two LLMs to generate and validate the questions/answers and integrates human-in-the-loop to verify the grammar of the questions and filter our duplicated and/or degenerated questions. TeleQnA can be used to test the telecom knowledge of both general-purpose LLMs as well as fined-tuned models. An overview of the entire process can be found in Figure 17

### 5.1.2 TSpec-LLM

Understanding telecom standards requires navigating a wide range of technical documents, including those produced by the 3GPP. This process can be both time-consuming and labor-intensive. While LLMs offer potential assistance in managing the extensive 3GPP knowledge base, the effectiveness of these models heavily depends on the availability of an inclusive dataset for proper pre-training and fine-tuning. In this section, we present *TSpec-LLM*, an open-source, comprehensive dataset that encompasses all 3GPP documents from Release 8 to Release 19 (spanning the years 1999 to 2023).

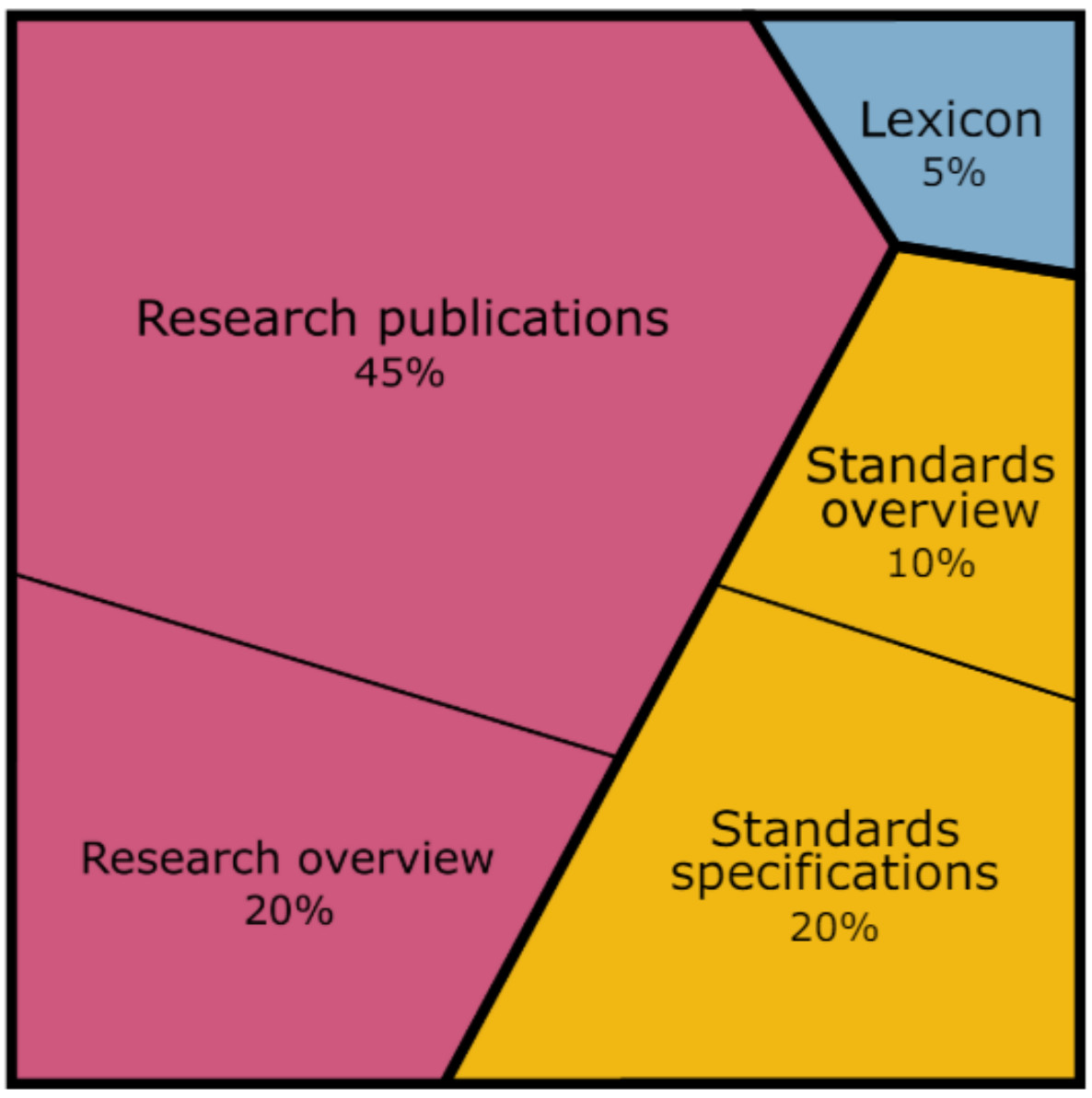

Figure 16: Distribution of the TeleQnA dataset among the categories of the collected source materials [140].

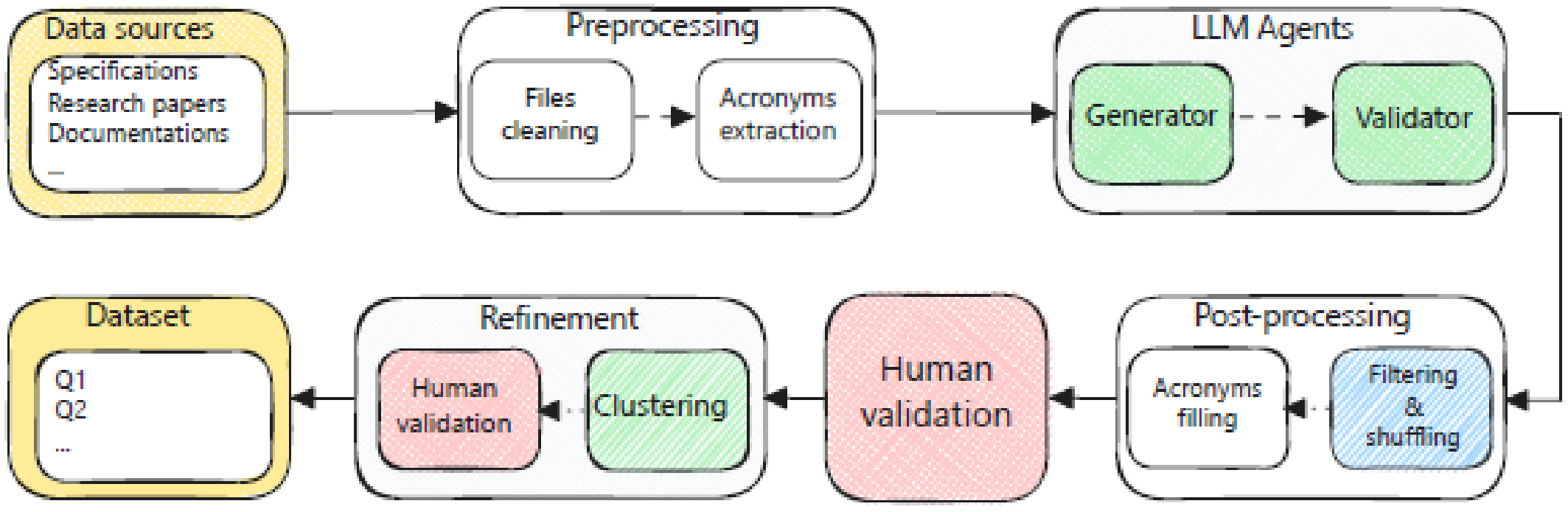

Figure 17: A high-level overview of the TeleQnA generation process [140].

Studies have shown that base LLM models are proficient in addressing general telecom-related queries; however, they struggle with more complex questions concerning standards. Despite being trained on extensive web data, these models find it challenging to process technical specifications related to wireless communication technologies [54, 140, 141]. Although such specifications are publicly accessible, their intricate nature—featuring tables, formulas, and figures—poses significant difficulties for LLMs. This complexity hinders the models' ability to extract pertinent information and deliver accurate responses to user queries, thereby limiting the effectiveness of state-of-the-art LLMs in generating precise answers.

An inclusive dataset is essential for the pre-training and fine-tuning of LLMs to ensure they effectively grasp the complexities of the telecom domain. With such a dataset, LLMs can serve as a valuable tool for engineers and researchers, offering an assistant model that can autonomously comprehend and organize 3GPP technical documents, reducing the need for human intervention. The *TSpec-LLM* dataset offers a well-organized, comprehensive, open-source dataset designed for research involving LLMs. It maintains the integrity of the original content from the tables and formulas found within the 3GPP specifications. Additionally, *TSpec-LLM* encompasses a complete collection of all 3GPP documents from Release 8 to Release 19, covering the period from 1999 to 2023. This extensive dataset totals 13.5 GB, containing 30,137 documents and 535 million words. Importantly, each document within a specific release retains its original structure, with the contents neither sampled nor consolidated into a single file.

*TSpec-LLM Dataset Creation:* The dataset comprises processed documentation files from the 3GPP standards, which have been converted to markdown (.md) format to enhance compatibility with natural language processing applications. This dataset is specifically designed for engineers and researchers working with LLMs in the telecommunications sector. The documents were sourced from the 3GPP website [142] using the open-source tool *download3gpp 0.7.0* [3], which efficiently retrieves all documents from all releases and series into a designated directory. The dataset was subsequently processed using a custom-designed Python script, which is available open-source [4]. This script employs the command-line version of LibreOffice to process files in parallel, significantly accelerating the conversion process. The headless conversion approach is optimized for server-side operations and batch-processing tasks, resulting in a well-structured and versatile dataset ideally suited for natural language processing tasks involving LLMs in the telecom domain.

We utilized a Python script to analyze the file sizes of markdown documents within the `3GPP-clean` directory. This analysis spanned all releases and versions of the 3GPP documentation, focusing on the total size of the .md files. The findings were organized by version and release and compiled into a report, which was saved in JSON format. Fig. 18 presents the total size in megabytes for each release, alongside the total word counts in millions.

[3] `https://pypi.org/project/download_3gpp`

[4] `https://huggingface.co/datasets/rasoul-nikbakht/TSpec-LLM/blob/main/process-3GPP.ipynb`

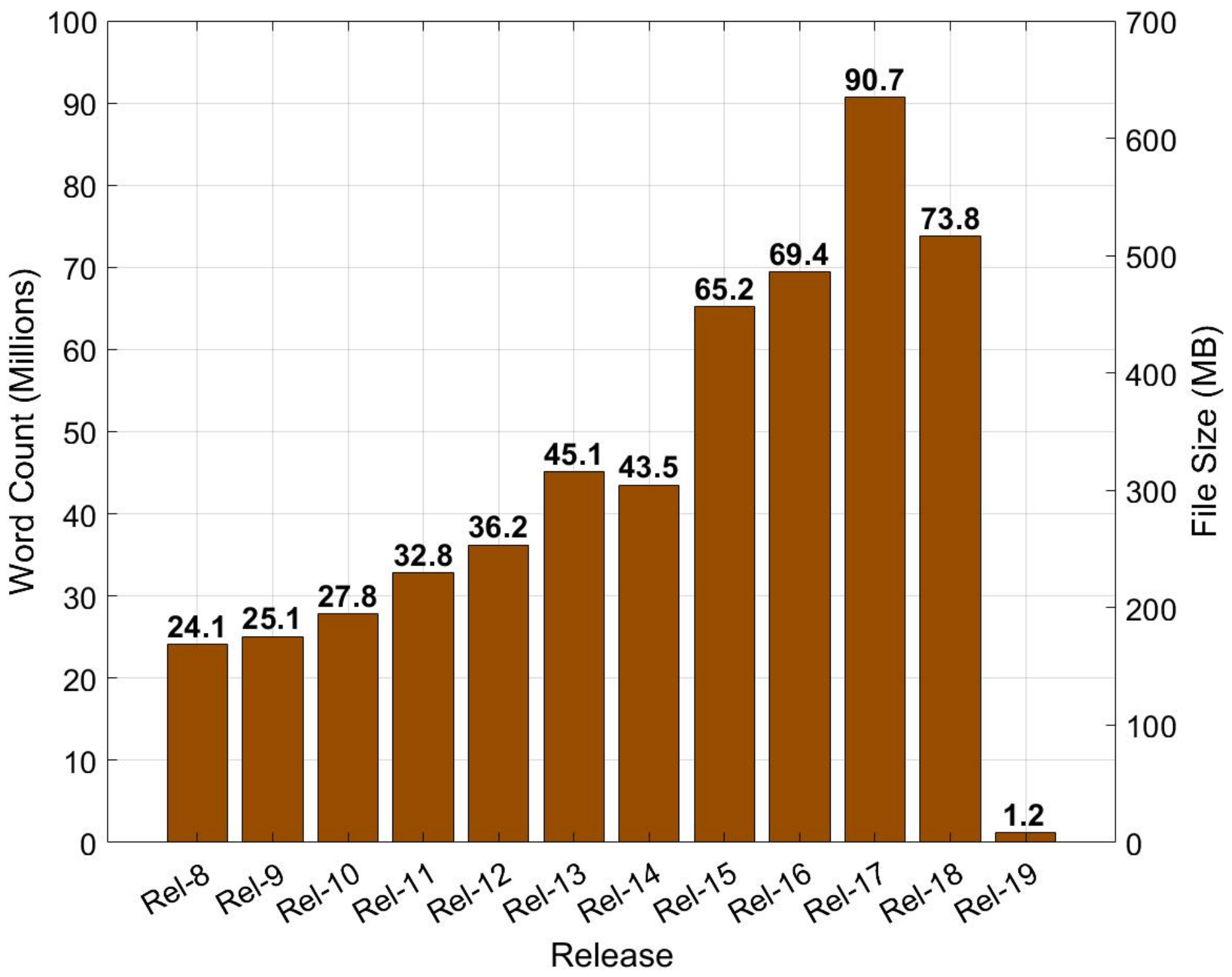

Figure 18: Word counts and file sizes for the TSpec-LLM dataset across various 3GPP releases. Data cut-off is December 2023.

Fig.19 presents an example from the *TSpec-LLM* dataset, illustrating how the content of Table 7.8-2 from [143] is preserved within the markdown files. This type of content is crucial as it includes system parameters, configurations, and other technical details essential for understanding and utilizing telecom standards. Additionally, Fig. 20 provides a comparison between an equation in its original .docx format and its corresponding representation in *TSpec-LLM* using LaTeX. This LaTeX format enhances the processing capabilities of state-of-the-art LLMs, making it easier to handle complex mathematical expressions.

```
Table 7.8-2: Simulation assumptions for full calibration

+-----------------------------+-------------------------------------+
| Parameter                   | Values                              |
+=============================+=====================================+
| Scenarios                   | UMa, UMi-street Canyon, Indoor-office |
|                             | (open office)                       |
+-----------------------------+-------------------------------------+
| Carrier Frequency           | 6 GHz, 30 GHz, 60GHz, 70GHz         |
+-----------------------------+-------------------------------------+
| Bandwidth                   | 20MHz for 6GHz, and 100MHz for 30GHz, |
|                             | 60 GHz and 70 GHz                   |
+-----------------------------+-------------------------------------+
| BS Tx power                 | 44 dBm for UMi-Street Canyon, 49 for |
|                             | UMa at 6GHz                         |
|                             |                                     |
|                             | 35 dBm at 30GHz, 60 GHz and 70 GHz  |
|                             | for UMa and UMi-Street canyon       |
|                             |                                     |
|                             | 24 dBm for Indoor for all carrier   |
|                             | frequencies                         |
+-----------------------------+-------------------------------------+
```

Figure 19: Illustration of the TSpec-LLM dataset content, highlighting the simulation parameters as detailed in [143, Table 7.8-2].

Table 7: Existing datasets for training and testing llm in the telecommunication domain.

| Dataset | Year | Training | Testing | Size | Description | Open source |
|---|---|---|---|---|---|---|
| TeleQnA [140] | 2023 | ✗ | ✓ | 10,000 samples | mcq from multiple Telecom sources. | HuggingFace |
| Tele-Eval [145] | 2024 | ✗ | ✓ | 750,000 pairs | Open-ended qna from the Tele-Data dataset. | HuggingFace |
| ORAN-Bench-13K [146] | 2024 | ✗ | ✓ | 13,952 samples | mcq from 116 O-RAN specification documents. | GitHub |
| Telco-DPR [147] | 2024 | ✗ | ✓ | 1,741 samples | Synthetic qna from four 3gpp technical documents. | HuggingFace |
| SPEC5G [148] | 2023 | ✓ | ✓ | 134M words | Cellular network specification documents and 13 online websites. | HuggingFace |
| TSpec-LLM [144] | 2024 | ✓ | ✗ | 535M words | 3gpp documents from Release 8 (1919) to Release 19 (2023). | HuggingFace |
| Tele-Data [145] | 2024 | ✓ | ✗ | 2.5B tokens | Telecom documents from different sources: ArXiv, 3gpp, Wikipedia. | HuggingFace |
| Open Telecom [149] | 2024 | ✓ | ✗ | 1.7B tokens | 3gpp documents & filtering RedPajama-1T with telecom keywords. | ✗ |

Cluster delay is updated as:

$$\tilde{\tau}_n(t_k) = \begin{cases} \tilde{\tau}_n(t_{k-1}) - \frac{\hat{r}_{\text{rx},n}(t_{k-1})^T \bar{v}_{\text{rx}}(t_{k-1}) + \hat{r}_{\text{tx},n}(t_{k-1})^T \bar{v}_{\text{tx}}(t_{k-1})}{c}\Delta t & \text{for } k > 0 \\ \tau_n(t_0) + \tau_\Delta(t_0) + \frac{d_{\text{3D}}(t_0)}{c} & \text{for } k = 0 \end{cases} \qquad (7.6\text{-}9)$$

(a) Original Equation 7.6-9 from [143] (.docx).

```
[Cluster delay]{.underline} is updated as:

${\widetilde{\tau}}_{n}\left( t_{k} \right) = \left\{ \begin{matrix}
{\widetilde{\tau}}_{n}\left( t_{k - 1} \right) -
\frac{{\widehat{r}}_{rx,n}\left( t_{k - 1} \right)^{T}
{\overline{v}}_{\text{rx}}\left( t_{k - 1} \right) +
{\widehat{r}}_{tx,n}\left( t_{k - 1} \right)^{T}
{\overline{v}}_{\text{tx}}\left( t_{k - 1} \right)}
{c}\mathrm{\Delta}t & \text{for\ }k > 0 \\
\tau_{n}\left( t_{0} \right) +
\tau_{\mathrm{\Delta}}\left( t_{0} \right)
+ \frac{d_{3D}(t_{0})}{c} & \text{for\ }k = 0 \\
\end{matrix} \right.\ $ (7.6-9)
```

(b) Equation representation in TSpec-LLM (.md).

Figure 20: Example of the TSpec-LLM dataset content, displaying: (a) an equation in the original document (.docx) [143, Eqn. 7.6-9] and (b) its corresponding representation in TSpec-LLM (.md).

The *TSpec-LLM* dataset is hosted on Hugging Face [5] and it can be downloaded by following the installation guide provided. To evaluate the efficacy of the dataset, we selected a representative sample of 3GPP documents, generating corresponding technical questions, and assessing the baseline performance of various state-of-the-art LLMs. Subsequently, we implemented a RAG framework to enhance the LLMs' capabilities by retrieving relevant context from the *TSpec-LLM* dataset. For detailed information on the evaluation performance, readers are referred to [144].

## 5.2 Pre-Training and Instruction Tuning Datasets

Pretraining and instruction tuning datasets are essential for adapting LLMs to the specialized needs of the telecommunications domain. Pretraining datasets, such as those used by Orange TelcoLM [150] and OpenTelecom, provide a foundation by exposing LLMs to large volumes of telecom-specific content, including technical standards, research papers, and curated web data. These datasets ensure the models develop a robust understanding of telecom terminology, concepts, and contextual nuances. Instruction tuning datasets, like the Telecom Instructions Dataset, further refine this knowledge by teaching models to follow domain-specific instructions and complete specialized tasks, such as protocol generation, mathematical modeling, and technical classification. Together, these datasets enable LLMs to perform more accurately and effectively across a wide range of telecom applications, bridging the gap between general-purpose AI and domain-specific expertise.

[5] https://huggingface.co/datasets/rasoul-nikbakht/TSpec-LLM

### 5.2.1 Training Datasets Used by Orange TelcoLM

Domain adaptation of language models requires carefully curated datasets that capture the specific terminology, concepts, and knowledge of the target domain. In this section, we describe the datasets assembled by Orange to create TelcoLM, a language model specifically adapted to the telecommunications domain. This collection effort represents an example of how domain-specific data can be gathered and structured for language model adaptation. Two types of data are considred: raw texts and instructions. Raw texts can be used for continued pretraining of a generic LM or to generate instructions. Instructions are there to perform instructing-tuning. They can also be used to build test sets.

#### Raw Text Collection

The pretraining corpus for TelcoLM contains 803M tokens, gathered from various telecommunications related sources. The data collection process followed a three-pronged approach:

**Technical Documentation (42% of tokens):** This includes technical specifications and white papers from standards organizations, accounting for 28% of the tokens. The sources are 3GPP, ETSI (European Telecommunications Standards Institute), and RFC (Request for Comments) documents. Additionally, research papers from arXiv's Networking and Internet Architecture category contribute 13% of the tokens.

**Web Content (11% of tokens):** This portion is derived from Common Crawl archives filtered for telecommunications-related domains, including industry organizations like 5GAcia, 5GAmericas, and GSMA, equipment manufacturers such as Nokia and Huawei, and technical communities like IPv6 and Juniper. Furthermore, Stack Exchange archives from relevant categories such as Network Engineering, Signal Processing, Security, and Quantum Computing are included.

**Selected Content through Importance Sampling (47% of tokens):** Using the Data Selection via Importance Resampling (DSIR) method [151], additional content was selected from various sources. This includes Stack Overflow (9%) with questions and answers related to telecommunications, Wikipedia (12%) with articles about telecommunications concepts and technologies, Open Web Mathematical Texts (10%) relevant to telecommunications engineering, and ArXiv Papers from RedPajama (15%) selected based on their relevance to telecommunications.

#### Instruction Dataset

The instruction dataset comprises 80,000 instruction-output pairs, created through three main approaches:

**Technical Document Transformation:** The source materials include standards and specifications from 3GPP, ATIS, and ETSI. The process involves dividing documents into paragraphs and prompting GPT-4 to generate instruction-output pairs. Post-processing is performed to remove references to implicit information such as figures, tables, and equations. Additionally, secondary prompting is used to expand short answers.

**Q&A Pair Transformation:** This approach utilizes Stack Exchange discussions, specifically focusing on the Network Engineering category, Digital Signal Processing topics, and security-related discussions. The TeleQnA dataset, a subsample of 4,000 pairs, is also used. Multiple-choice questions from this dataset are converted to an open-ended format, and chain-of-thought prompting is added to generate detailed explanations.

**Quality Control Measures:** Several quality control measures are implemented, including length filtering to re-move instructions with outputs shorter than 200 characters, reference cleaning to eliminate instructions referring to external elements, and relevance verification using the Zephyr-7b model to classify telecommunications relevance.

The dataset preparation involved several technical challenges. For PDF documents, different extraction methods were employed depending on the content type: the Nougat visual transformer was used for technical and mathematics-heavy documents, while PDFMiner was applied to documents from Common Crawl. Web content processing followed a pipeline inspired by recent work on large-scale web datasets [152, 153, 92], including language detection, document-wise and line-wise filtering, exact and fuzzy deduplication, and content quality assessment.

This dataset collection effort demonstrates the complexity of assembling domain-specific training data for language models. The combination of authoritative sources (technical standards), community knowledge (Q&A sites), and filtered web content provides a broad coverage of telecommunications concepts, from formal specifications to practical applications.

### 5.2.2 OpenTelecom (Pre-Training) Dataset

Mainstream general-purpose LLMs lack knowledge and know how in telecom domain. For instance, SotA LLMs such as GPT-4 fails almost half part of the specification-related problem in TeleQnA. Therefore, it is of great interest to enhance or adapt general-purpose LLMs into telecom-specific LLMs with telecom-relevant knowledge. Howver, despite various amount of textual documents or RF signal in telecom domain, there are few curated dataset available for LLMs for both pre-training and fine-tuning. One possible direction is to filter or select the telecom-relevant contents from existing general pre-training dataset. OpenTelecom in [telecomgpt] represents how such approach can be implemented easily.

To select Telecom-specific data from a general-purpose dataset, a non-exhaustive 700 keywords list from telecom textbook, section of definitions and abbreviations from open 3GPP technical specifications and etc is built. 6 criteria are set to improve the quality and the effectiveness of these keywords in help identifying telecom-relevant content while minimizing the amount of irrelevant contents introduced by polysemy of vocabularies: domain specficity, frequency in telecom discourse, distinctiveness with telecom, authority and standards, timeliness and emerging technologies, clarity and avoidance of ambiguity.

Then for each data sample, two quantities to evaluate its relevance to telecom domain are used: number of telecom keywords (same keyword will only be counted once) and telecom keyword density. The density for a text of $N$ words with a total number of $M$ telecom keywords matches is $M/\log(N+1)$. The reason for using the logarithm of the number of total words rather itself is to compensate those long documents. RedPajama-1T, an open pre-training dataset consisting 7 subsets: Commoncrawl, C4, Github, Books, Arxiv, Wikipedia and StackExchange, with 1.2 billion text tokens is used to showcase the approach. In addition to filtering from general-purpose text, 3GPP and IEEE standards are extracted following the methodology in [understand telecom language, Lina]. The final dataset of OpenTelecom contains approximately 1.7B tokens telcom-relevant texts. The details of the category is summarized in 8.

### 5.2.3 Telecom Instructions Dataset

However, applying LLMs effectively to the highly specialized and technical field of telecommunications necessitates fine-tuning on task-specific datasets. Instruct tuning, where models are trained to follow domain-specific instructions, is a promising approach to address this challenge. By building and leveraging instruct datasets specifically curated for telecom applications, such as resource allocation, interference management, and spectrum sharing, LLMs can be adapted to understand and execute telecom-relevant instructions efficiently. These datasets can include real-world scenarios, annotated technical standards (e.g., 3GPP documents), and expert-designed tasks, enabling the LLMs to generate actionable insights tailored to the needs of wireless networks. Such instruct tuning empowers telecom-specific LLMs to support decision-making in network planning, optimize resource utilization, and even provide automated troubleshooting, ultimately driving innovation and operational efficiency in next-generation wireless systems.

Table 8: OpenTelecom dataset for Continual Pre-training (in million tokens)

| Category | Training | Validation | Percentage (%) |
|---|---|---|---|
| 3GPP Standard | 193 | 1.9 | 11.49 |
| IEEE Standard | 7.5 | 0.07 | 0.45 |
| Paper (arxiv) | 893 | 9 | 53.17 |
| Books | 1.9 | 0.02 | 0.11 |
| Patent (C4) | 253.2 | 2.6 | 15.08 |
| StackExchange | 51.9 | 0.5 | 3.09 |
| Wikipedia | 18.9 | 0.2 | 1.13 |
| Code (Github) | 260.1 | 2.6 | 15.49 |
| Total | 1679.5 | 16.89 | — |

Our instruction dataset, referred as **Telecom Instruct** in what follows consists of different critical tasks in telecom domain:

- **MCQ Answering**: Select all correct answers from a MCQ.
- **Open-ended Question Answering**: Answer telecom-relevant question from standards, research papers or patents in an open-ended manner.
- **Technical documents classification**: Classify text from different Tdocs into the related working group as documented in the library of SDO.
- **Math Modeling**: Generate accurate math equation such as channel models for given text description of system model and problem formulation in the context of telecom with predefined notations.
- **Code Generation**: Generate script/function for a given task/functionality in telecom domain such as sending signal indicator and extracting MAC address from a frame.
- **Code Infilling**: Infill incomplete script based on the context and the targeted functionality. This task is also known as FIM cite[bavarian2022efficient] task which is beneficial for developers or researchers to improve efficiency when generating telecom-relevant scripts.
- **Code Summary**: Summarize the core functionality of a given script, including identifying if the script is telecom-relevant or not.
- **Code Analysis**: Detail the operational logic behind the function, emphasizing the knowledge and principles in telecom domain.
- **General Instruction**: Explain concepts, describe specifications, identify problems, propose solutions, sum-

marize content in Telecom standards, patents, and papers.

- **Protocol Instruction**: Generate the protocol workflows in Telecom standard following a human prompt.

It has been that such telecom-relevant instruction dataset can be used to adapt LLMs to telecom-specific LLMs and outperforms the base models for tasks included in the dataset.

## 5.3 Advanced Dataset Design Datasets

Advanced dataset design focuses on creating specialized resources tailored to solving complex, task-specific challenges in telecommunications. Unlike pretraining and instruction tuning datasets, which primarily build a foundational understanding of telecom knowledge and enhance a model's general adaptability, advanced datasets are crafted to address specific applications such as network optimization or system-level problem-solving. These datasets emphasize structured, task-oriented outputs, making them indispensable for practical implementations. For example, Tele-Data provides a comprehensive resource for pretraining and continual learning by combining diverse telecom content from standards and research papers, enabling models to handle complex domain-specific tasks. Similarly, the Dataset for Network Optimization offers structured data for tasks like computation offloading and scheduling, directly supporting real-world efficiency improvements. Together, these datasets bridge the gap between foundational model capabilities and applied telecom innovations, enabling AI to drive smarter and more efficient systems.

### 5.3.1 Tele-Data

Adapting LLMs to the telecommunication domain requires continual pretraining on telecom-specific datasets. One relevant dataset for this process in the literature is Tele-Data [154], the first open-source collection of telecommunications materials composed of four primary sources: (1) scientific papers from arXiv, (2) 3GPP standards, (3) Wikipedia articles relevant to telecommunications, and (4) telecommunications-related websites sourced from Common Crawl dumps. This variety of sources provides comprehensive coverage of telecommunications knowledge, facilitating the transfer of expertise across different aspects of the domain. The dataset is available on HuggingFace[6] and has been used for the continual pretraining of Tele-LLMs, the first open-source series of specialized LLMs for the telecom industry [154]. These models are also accessible on HuggingFace[7].

**ArXiv**

**Curation**. One of the largest repositories of open-access research on telecommunications consists of preprints submitted to arXiv. As of March 2024, the combined snapshot for the computer science and electrical engineering categories includes approximately 610k papers. However, since these categories overlap and cover topics beyond telecommunications, targeted filtering is required to extract relevant material. To address this, the curation process of Tele-Data employs a language model-based filtering approach. Specifically, the Mixtral 8x-7B-Instruct[8] model was used, providing it with each paper's abstract to assess its relevance to the telecommunications and networking domain. The model was prompted to give a binary Yes or No response on whether the paper is relevant. The logits of the 'Yes' and No' tokens were then utilized to classify the paper accordingly.

**Cleaning**. After curation, a thorough cleaning process was applied to the papers, involving: (1) removal of comments, (2) flattening of LaTeX sources, (3) substitution of user-defined macros with standard LaTeX commands, (4) elimination of LaTeX native commands, and (5) standardization of citation formats and removal of markup changes. Figures and tables were excluded to focus on inline text and equations. Further details on this process, along with an evaluation of the cleaning methods, are available in [154].

[6]https://huggingface.co/datasets/AliMaatouk/Tele-Data
[7]https://huggingface.co/collections/AliMaatouk/tele-llms-66de81a0c1e1f09e2a6c78ce
[8]`https://huggingface.co/mistralai/Mixtral-8x7B-Instruct-v0.1`

#### Standards

**Curation**. Standards are essential in telecommunications, ensuring interoperability across technologies from multiple vendors. These standards are developed and maintained by recognized organizations such as 3GPP, IEEE, and ITU. Given its open-access nature, Tele-Data focuses on incorporating 3GPP documents. The latest specifications for each standard in every series were retrieved through the 3GPP FTP portal[9], resulting in a collection of approximately 2.8k documents.

**Cleaning**. After curation, the standards documents underwent a detailed cleaning process. Non-essential sections, such as related works and appendices, were removed, and figures and tables were excluded to prioritize inline text and equations. One notable challenge is that equations in .doc files are encoded in XML, unlike the LaTeX format used in arXiv papers. To address this, all .doc files were converted to .docx format, followed by the use of docx2tex[10] to convert them into LaTeX format. This conversion ensures consistent formatting of equations across document types, streamlining the training process. Finally, the same cleaning pipeline applied to arXiv papers was used on the converted standards LaTeX files, ensuring uniform cleanliness and coherence throughout the dataset.

#### Wikipedia

Wikipedia serves as another valuable source of telecommunications material, offering articles that cover both technical and domain-specific content. To curate relevant content, the English subset of the Wikipedia dataset[11], containing 6.4 million samples, was utilized. Given the computational cost of applying pure LLM-based filtering to such a large corpus, a two-step process was employed:

1. **Keyword Filtering:** A list of 100 telecom-related keywords, such as telecommunications, base station, Wi-Fi, and 5G, was defined. Articles containing any of these keywords were flagged for further evaluation. This step reduced the dataset from 6.4 million to approximately 70k articles.

2. **LLM-based Content Evaluation:** In the second step, the flagged articles were evaluated using the Mixtral 8x-7B-Instruct model. The first 10,000 characters of each article were provided to the model, which was prompted to return a Yes or No response based on the article's relevance and technical content. This filtering ensures the exclusion of non-technical content, such as articles focusing on the history of telecom operators.

Following this process, 19.5k articles with technically relevant telecommunications content were curated from Wikipedia.

#### Websites

The final source of telecommunications material is the Common Crawl dataset, a web archive containing data from across the internet. To avoid issues with duplicates, non-English content, and offensive material in the raw dumps, the refined web dataset[155] was used. This curated version of Common Crawl includes approximately 1 billion rows across 2.8 terabytes of data.

To further refine the dataset, Wikipedia articles were filtered out to avoid redundancy. The same two-step process used for Wikipedia articles was applied to extract telecommunications-related content from the refined web dataset. Additionally, content from well-known telecommunications blogs, such as ShareTechNote, was incorpo-rated to enhance relevance. The final collection contains content from 740k website links, providing extensive coverage of telecommunications information available on the web. Examples of Tele-Data are provided below. The [...] symbol is inserted below to reduce the size of the strings.

[9] `https://www.3gpp.org/ftp/`
[10] `https://github.com/transpect/docx2tex`
[11] `https://huggingface.co/datasets/wikimedia/wikipedia`

**ID:** arxiv_14326
**Category:** **arxiv**
**Content:** Flexible-Position MIMO for Wireless Communications: Fundamentals, Challenges, and Future Directions\n \n Abstract\n \n The flexible-position multiple-input multiple-output (FLP-MIMO), such as fluid antennas and movable antennas, is a promising technology for future wireless communications [...]
**Metadata:**
**Arxiv_id:** 2308.14578
**Title:** Flexible-Position MIMO for Wireless Communications: Fundamentals, Challenges, and Future Directions
**Abstract:** The flexible-position multiple-input multiple-output (FLP-MIMO), such as [...]

**ID:** standard_2413
**Category:** **standard**
**Content:** 3rd Generation Partnership Project; \n Technical Specification Group Core Network and Terminals;\n Interworking between the Public Land Mobile Network (PLMN)\n supporting packet based services with\n Wireless Local Area Network (WLAN) Access and\n Packet Data Networks (PDN)\n (Release 12)\n Foreword\n This Technical Specification (TS) has been produced [...]
**Metadata:**
**Series:** 29
**Release:** 12
**File_name:** 29161-c00

**ID:** wiki_5438
**Category:** **wiki**
**Content:**A backbone or core network is a part of a computer network which interconnects networks, providing a path for the exchange of information between different LANs or subnetworks. A backbone can tie together diverse networks [...]
**Metadata:**
**Title:** Backbone network
**Url:** https://en.wikipedia.org/wiki/Backbone%20network

**ID:** web_71187
**Category:** **web**
**Content:**1. Field of the Invention\n The present invention relates generally to methods of addressing data packets destined to a host in a communications network, and particularly to a method of defining an address for a mobile terminal/host [...]
**Metadata:**
**Url:** http://www.google.com/patents/US6147986

### 5.3.2 Dataset for Network Optimization

Data, models, and computing power form the three pillars of the big model era. In the telecom field, domain-specific data is especially crucial. However, due to the specialized and niche nature of research in this area, telecom studies rarely release their data. Whether dealing with professional numerical datasets or domain-specific knowledge bases, there is significant room for improvement in the efficiency and availability of data collection and preparation within the telecom community.

In the context of large models, [149] introduced the first foundation model in the telecom field, [156, 157] pioneered real-world deployment in mobile networks to support downstream tasks, and [140, 144] were the first to create and release standard natural language datasets. The momentum generated by these efforts in model implementation and data construction is worth further promotion.

For task-specific models and applications, the diversity and complexity of datasets have increased significantly. However, if the telecom community aims to keep pace with advancements in data mining and AI, it is crucial to accumulate data for specialized tasks publicly. To contribute to this effort, in [158] we open-sourced the MSMU dataset and its accompanying tools for the multi-server multi-user computation offloading problem [159]. This dataset includes solved instances in the form of graph data structures, where the input consists of edge features representing offloading cost information from users to their associated servers, and node features representing device types. The output comprises edge selections, indicating offloading decisions, and edge weights, representing the allocation ratio of computing resources. The instance graphs in this dataset range from 9 to 88 nodes, offering tens of thousands of examples.

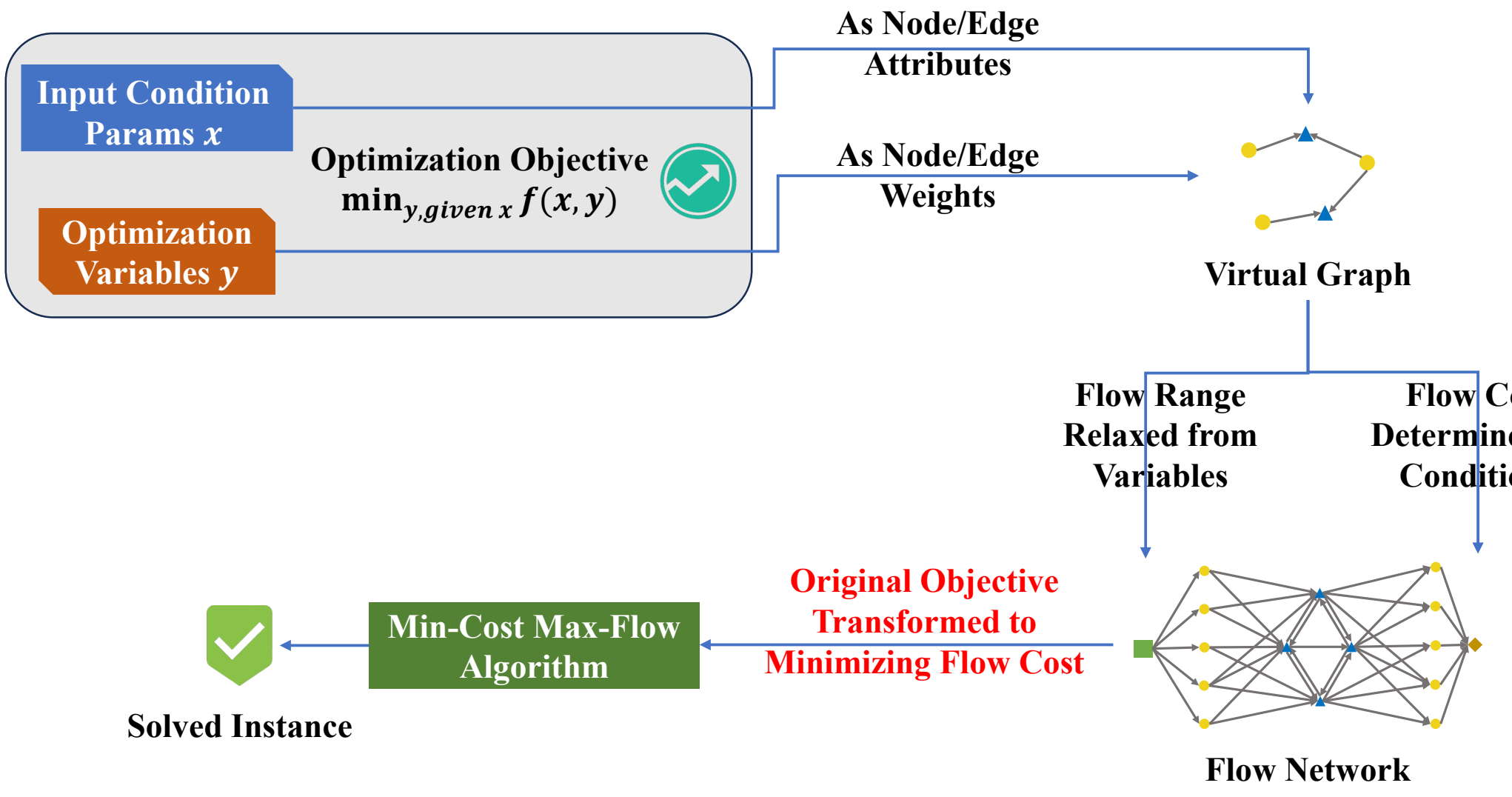

Figure 21: Network optimization problem data production tool based on MCMF.

As shown in Fig. 21, our tool can generate data with different objective functions or scales as needed. We have open-sourced the data generation tool, which is based on the minimum-cost maximum-flow (MCMF) algorithm. The approach involves using the network optimization inputs as attributes and variables on a virtual graph, thereby constructing a flow network that transforms the original optimization objective into a minimum flow cost problem. This allows for the direct application of various network flow algorithms, such as MCMF. This tool efficiently provides optimal solutions for single-objective problems and sub-optimal solutions for multi-objective problems. With sufficient time, it can also achieve optimal solutions for multi-objective problems. The algorithm has polynomial time and space complexity. For graph optimization problems with up to 1,000 nodes, an i7-13700F can compute a solution in under 1 second, with larger scales still maintaining acceptable time performance. By switching between different basic network flow algorithms, the tool can be adapted to label most graph optimization problems, which encompass a wide range of network optimization tasks.

This dataset and toolset will serve as a valuable resource for advancing the development of task-specific datasets in telecommunications. By providing an open-source solution, we aim to encourage further research and collaboration, accelerating progress in network optimization and related areas.

## 5.4 Ray Tracing and 6G-Specific Datasets

Ray tracing and 6G-specific datasets are critical for bridging the gap between theoretical research and real-world implementation in next-generation wireless communication systems. Unlike pretraining or task-specific datasets, which focus on telecom knowledge or structured problem-solving, these datasets emphasize high-fidelity simulations and real-world scenarios to capture the complexities of 6G environments, including advanced propagation phenomena and network dynamics. Their integration with LLMs enables AI systems to leverage precise data for tasks like channel modeling, beam tracking, and intelligent network planning. For instance, datasets such as Sionna, DeepMIMO, and WAIR-D provide detailed ray tracing outputs that simulate electromagnetic wave propagation in diverse environments, supporting applications like CSI prediction and RIS optimization. Additionally, the Dataset for Network AI introduces high-resolution, real-world data designed for resource scheduling and dynamic system management in 6G networks. Together, these datasets serve as essential resources for training AI models to achieve robust, adaptive, and intelligent performance in next-generation communication networks.

### 5.4.1 Raytracing Dataset

Ray tracing datasets are a focal point in both academia and industry, aiming to simulate the propagation of electromagnetic waves in specific scenarios to obtain electromagnetic properties and channel characteristics at every point in space as close to the real world as possible. These datasets enable data analysis, model training, algorithm validation, and performance prediction.

Current channel models can be broadly categorized into three types: statistical models, geometry-based stochastic models, and deterministic models. Ray tracing datasets employ deterministic models, which are built based on electromagnetic wave propagation theory and actual environmental measurements. For a given scenario—whether indoor or outdoor—along with the base station location and antenna configurations, deterministic models leverage ray tracing techniques to simulate channel characteristics. This involves accounting for the reflection, refraction, and scattering of each ray path to acquire precise information, such as angle, delay, and polarization, at specific locations. Because deterministic models can accurately simulate the propagation characteristics of electromagnetic waves in a given environment, they enable precise predictions of channel behavior. As a result, ray tracing-based data generation has become the mainstream approach for creating intelligent communication datasets today.

Ray tracing datasets play a crucial role in the development and deployment of large models for communication systems. Large models, such as generative models and deep neural networks, require extensive and high-quality data for effective training. Ray tracing datasets, with their deterministic modeling capabilities, provide precise electromagnetic and channel characteristics, making them ideal for powering large models in communication tasks.

These datasets enable AI models to learn complex channel behaviors, such as path loss, multipath effects, and beamforming patterns, with high accuracy. By incorporating ray tracing data, these models can achieve enhanced performance in tasks like channel state information (CSI) prediction, beam tracking [160], and intelligent reflecting surface (RIS) optimization [161, 162]. Furthermore, ray tracing datasets allow large models to generalize across various scenarios, including indoor and outdoor environments, different antenna configurations, and dynamic mobility patterns, ensuring robust and adaptive communication strategies.

The synergy between ray tracing datasets and large models not only enhances the precision of communication simulations but also accelerates innovations in next-generation wireless networks, such as 6G. This connection empowers communication systems to evolve toward real-time, data-driven intelligence, with improved signal quality, resource management, and energy efficiency.

**Sionna**

The Sionna dataset generation approach combines classical scene reconstruction and ray tracing.

**5.4.1.1 1. Scene Reconstruction** The construction of Sionna scenarios is carried out in Blender. In the scene reconstruction phase, we first obtain a Google Maps API key and select a specific area in OpenStreetMap(OSM) as shown in Fig. 22. This generates the basic LoD1 geometry of the buildings in that region as shown in Fig. 23. Then, in Blender, various materials are assigned to the building surfaces, characterized by predefined dielectric constants and magnetic permeabilities. The scene is exported to an XML file using the Mitsuba plugin. Using Python and Mitsuba, the OptiX ray tracing core on the GPU is utilized, enabling GPU-accelerated parallel ray tracing.

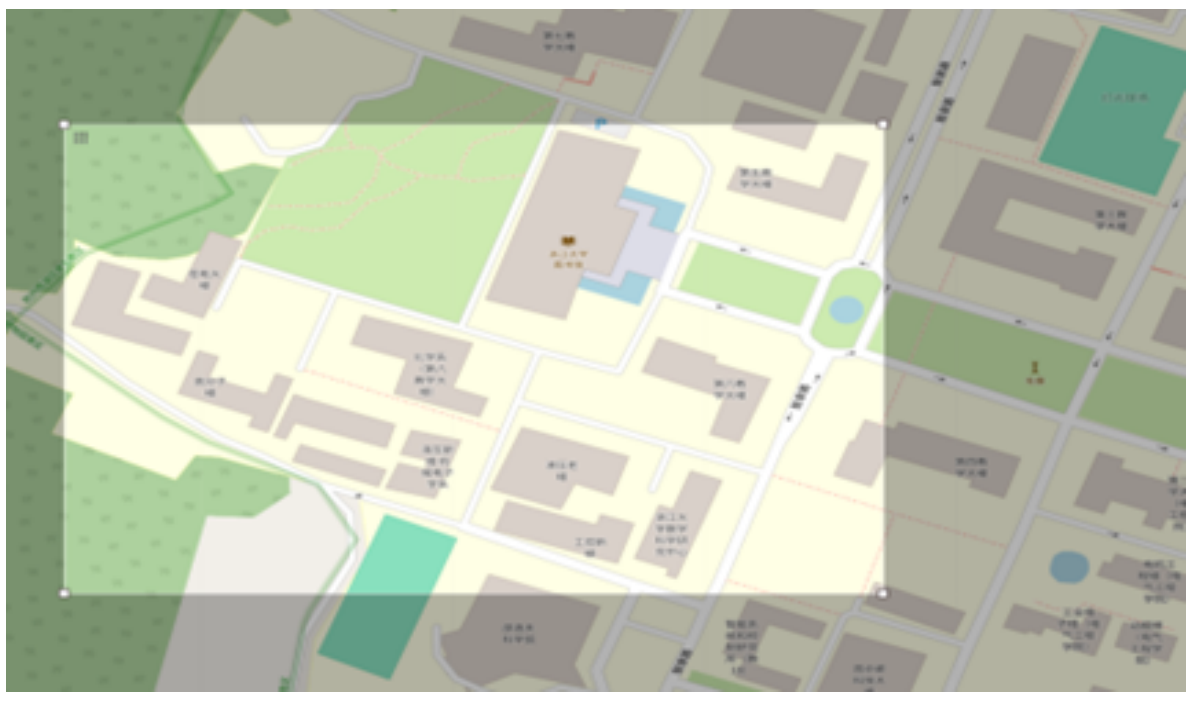

Figure 22: OSM selected area

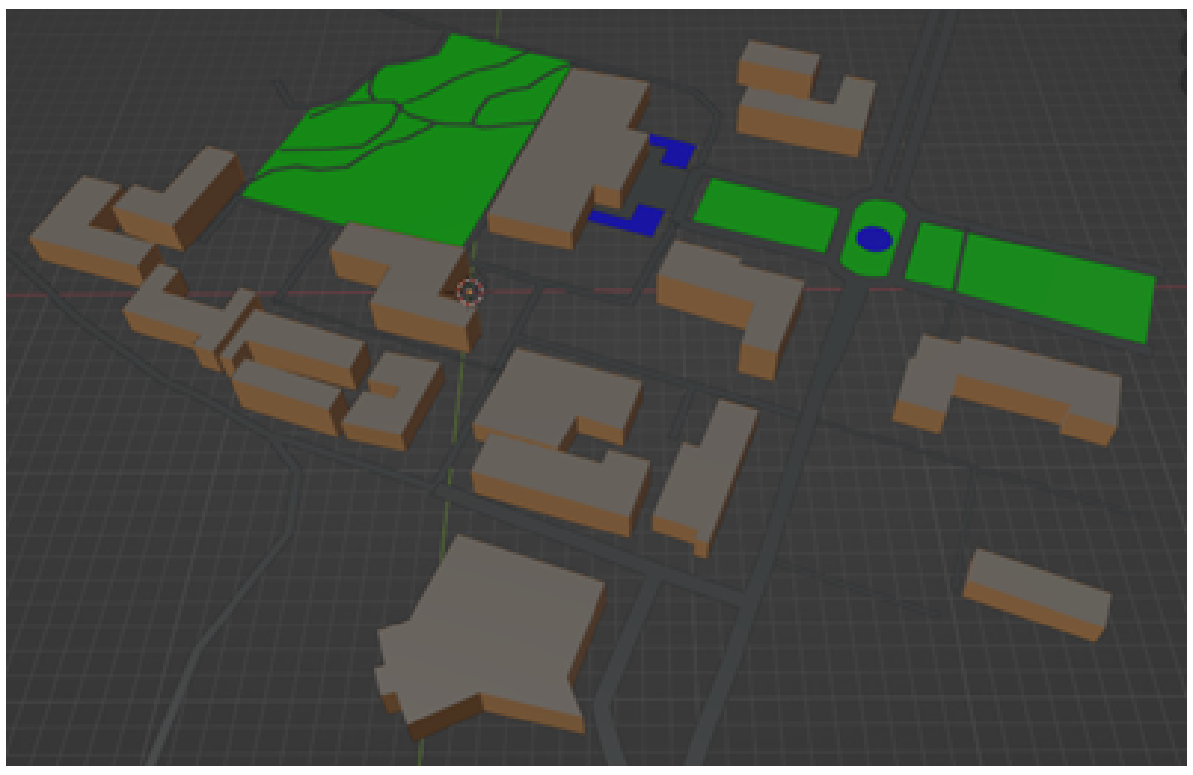

Figure 23: LoD1 scene construction example

**5.4.1.2 2. Ray Tracing** For ray tracing, the reflection, refraction, scattering, and diffraction effects of electromagnetic waves in the air are determined using geometric optics (GO) and uniform theory of diffraction (UTD). These effects are correlated with the material properties of the surfaces in contact with electromagnetic waves, and the electromagnetic field intensity at each point in the air is computed.

**2.1 Geometric Optics** Electromagnetic waves emitted from the transmitter are spherical waves radiating in all directions from the source point. However, when the wavelength is relatively short, the wavefront of the electromagnetic waves can be approximated locally as a plane wave. This allows the use of geometric optics to describe the field strength at the receiver. Ray tracing using geometric optics involves the following steps:

- **Finding possible propagation rays:** According to Snell's law of reflection and refraction, all potential ray paths between the transmitter and receiver are identified. This step is theoretically straightforward but computationally intensive, necessitating efficient ray tracing methods. To reduce computational effort and improve efficiency, only the strongest rays (with the least number of reflections and refractions) are typically considered.

- **Calculating Snell reflection and transmission coefficients:** Assuming the electromagnetic waves are plane waves, the reflection and transmission coefficients are computed for each reflection and refraction point. A key constraint is that the wavelength of the emitted electromagnetic wave must be significantly smaller than the distance between the first reflection point and the transmitter.

- **Calculating wavefront curvature:** For each ray, the amplitude of the electromagnetic wave is corrected based on the curvature of the wavefront at the boundary. The curvature of the boundary must be of the same order of magnitude as the wavelength of the electromagnetic waves.

- **Vector summation:** The amplitudes and phases of all ray paths are summed vectorially.

According to geometric optics, the field strength at the receiver is expressed as:

$$E = E_0 A_0 e^{-jk_0 r_0} + \sum_{i=1}^{N_r} \mathrm{RE}_i A_i e^{-jk_i r_i} + \sum_{j=1}^{N_t} \mathrm{TE}_j A_j e^{-jk_j r_j}$$

where $N_r$ and $N_t$ denote the total number of reflected and refracted rays, respectively; $r_n$ is the propagation distance of the $n$-th ray; $k_i$ represents the wave number of the $i$-th ray in the medium; $A_i$ is the propagation factor of the $i$-th ray, determined by the material properties of the reflecting and refracting media; and $E_{i,j}$ are the unit field strength vectors at reflection and refraction points, respectively.

Although geometric optics has certain limitations, it effectively addresses numerous challenges in high-frequency communication, such as antenna and wave propagation, accurately describing the field strength at the receiver.

**2.2 Uniform Theory of Diffraction** Geometric optics only considers direct, reflected, and refracted electromagnetic waves, failing to explain diffraction into shadow regions. When rays encounter rough surfaces, edges, or curved surfaces, they cannot penetrate shadow zones, and geometric optics predicts zero field strength in these areas, which is inconsistent with real-world observations. To resolve the discontinuity of the electromagnetic field in geometric optics, the uniform theory of diffraction (UTD) is introduced to correct the field strength in shadow regions.

Diffraction is categorized based on the geometric characteristics of the diffracting surface:

**a. Edge Diffraction**

Edge diffraction occurs when rays encounter the edge of an object in a homogeneous medium. Unlike reflection, where a single incident ray produces a single reflected ray, an infinite number of diffracted rays are generated from a single incident ray at the edge. These rays lie on a cone with the diffraction point as the apex. The cone's semi-apex angle equals the angle between the incident ray and the edge $\theta_o$. When the incident ray is perpendicular to the edge ($\theta_o = \pi/2$), the cone degenerates into a planar disk perpendicular to the edge.

The diffracted field at a diffraction point is given by:

$$E_d = DE_i A_d$$

where $D$ is the diffraction coefficient matrix containing vertical and horizontal information; $E_i$ represents the incident field matrix; and $A_d$ is the spreading factor, dependent on the distance between the source and the field point and the curvature of the edge.

**b. Apex Diffraction**

Apex diffraction occurs when rays encounter the apex of an object. Diffracted rays emanate in all directions from the apex, forming spherical wavefronts. These fields attenuate faster than edge diffraction fields, and their general asymptotic representation is often difficult. In most cases, apex diffraction fields are neglected.

**DeepMIMO**

Remcom's Wireless Insite is an RF propagation modeling software that integrates satellite imagery and CAD processing to deliver advanced radio wave propagation models. As a commercial product, it offers 3D ray tracing, ray-based fast methods, and empirical models to analyze radio wave propagation and wireless communication systems in specific locations. With comprehensive modeling, simulation, and post-processing functionalities, Wireless Insite efficiently and accurately predicts the propagation characteristics of electromagnetic waves and communication channels in complex environments, including urban, indoor, rural, and mixed-path scenarios, enabling AI models to be trained and tested on more realistic scenarios [163].

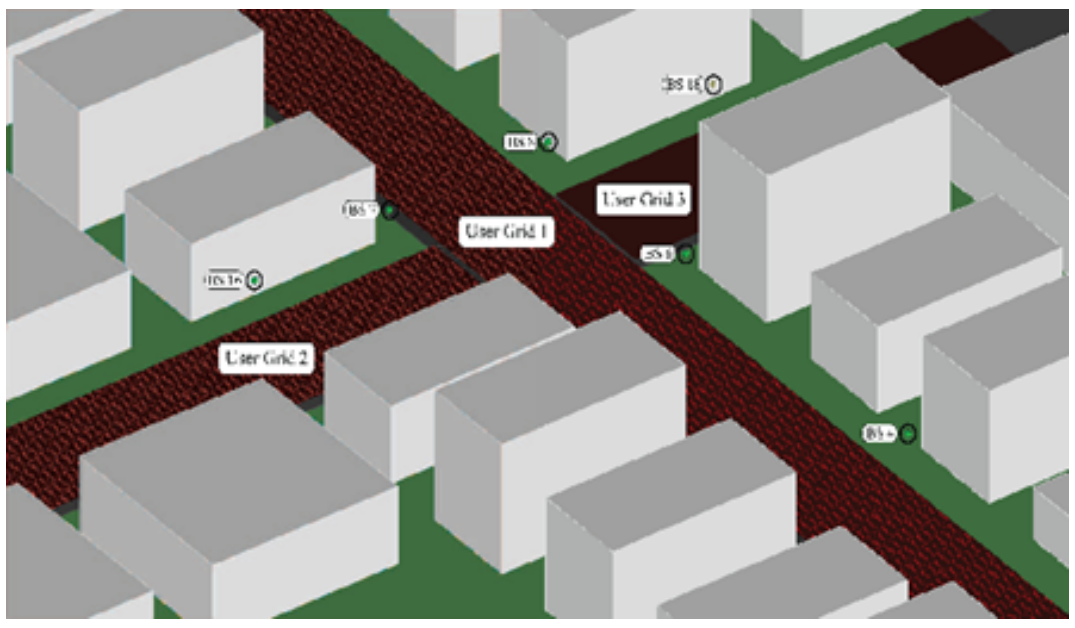

Figure 24: Example of DeepMIMO Outdoor Scenario

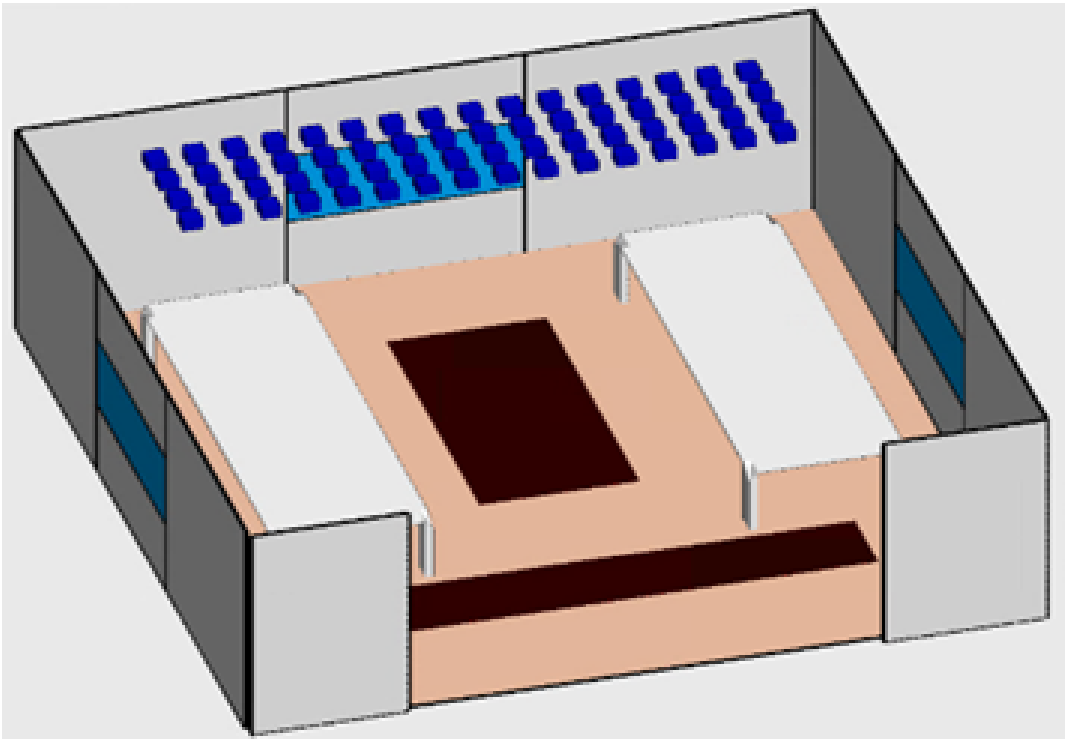

Figure 25: Example of DeepMIMO Indoor Scenario

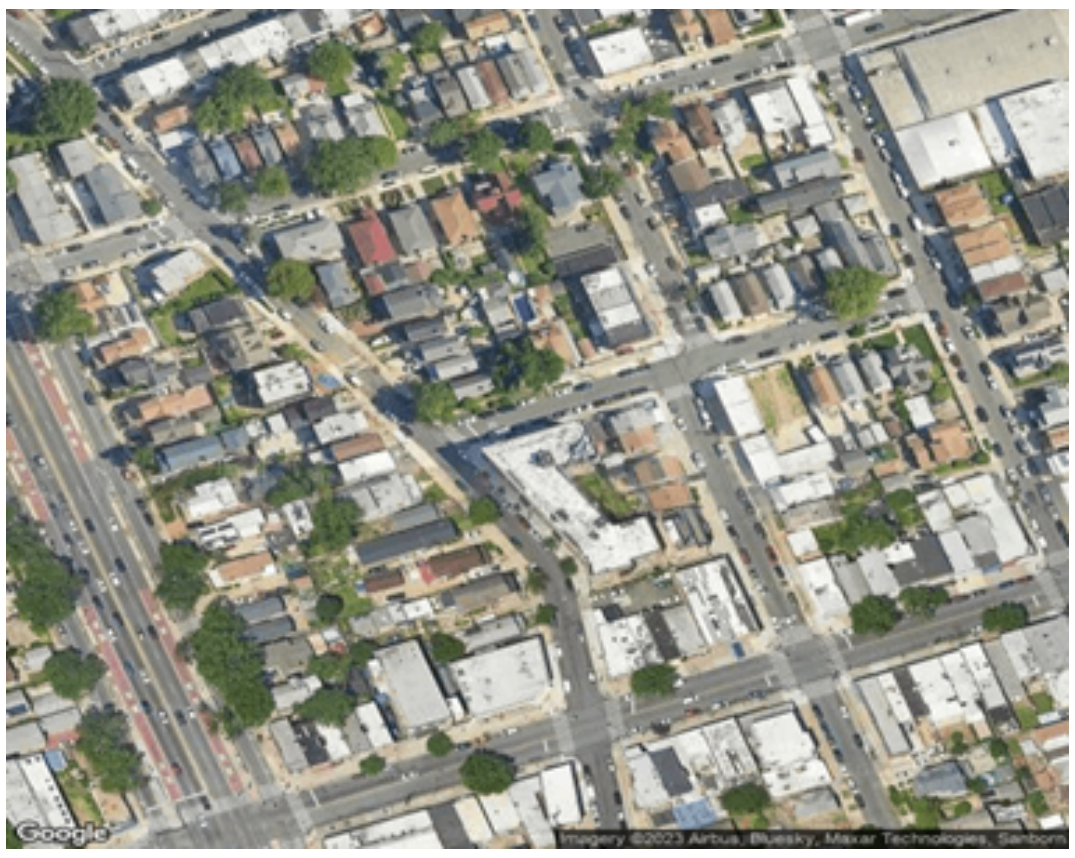

Figure 26: Example of a DeepMIMO City Overview

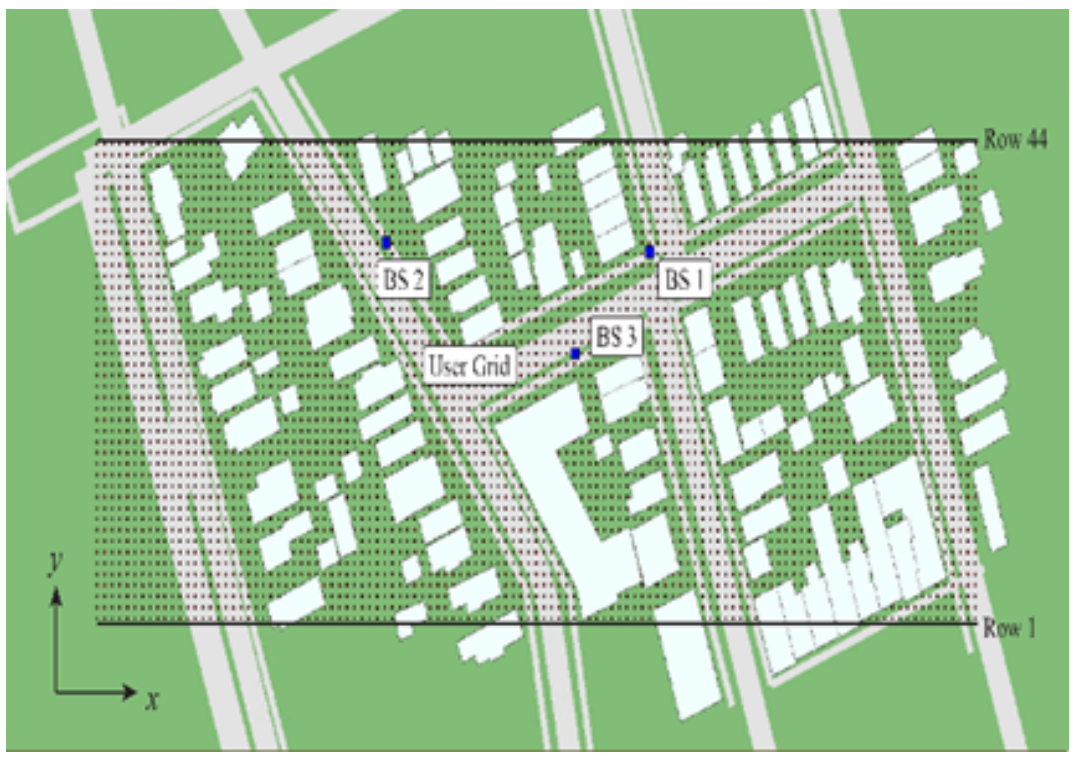

Figure 27: Example of a 3D model of a city in DeepMIMO

The DeepMIMO dataset [164] leverages Wireless Insite's capabilities in map generation by creating scenes in the following formats. Fig. 24 and Fig. 25 illustrate the outdoor and indoor scenarios supported by DeepMIMO, respectively. Meanwhile, Fig. 26 and Fig. 27 show New York's remote sensing imagery and the corresponding generated 3D scene map.

**WAIR-D**

The Wireless AI Research Dataset (WAIR-D) [165] is a versatile and user-friendly dataset that simulates realistic environments for a wide range of wireless AI applications [166, 167, 168, 169]. It encompasses a variety of tasks, including sensing tasks such as device localization and environment reconstruction, MIMO tasks such as reflection system modeling and beamforming, as well as PHY tasks like CSI feedback and channel estimation. The features are as follows.

- Realistic: The dataset is built on a foundation of realism, with 10,000 scenarios randomly selected from actual maps of over 40 major cities worldwide. The building layouts are carefully designed to mirror real-world environments, providing a more accurate representation of the complexities of urban landscapes.
- Flexible: To cater to diverse research needs, the dataset offers a range of flexible features, including:

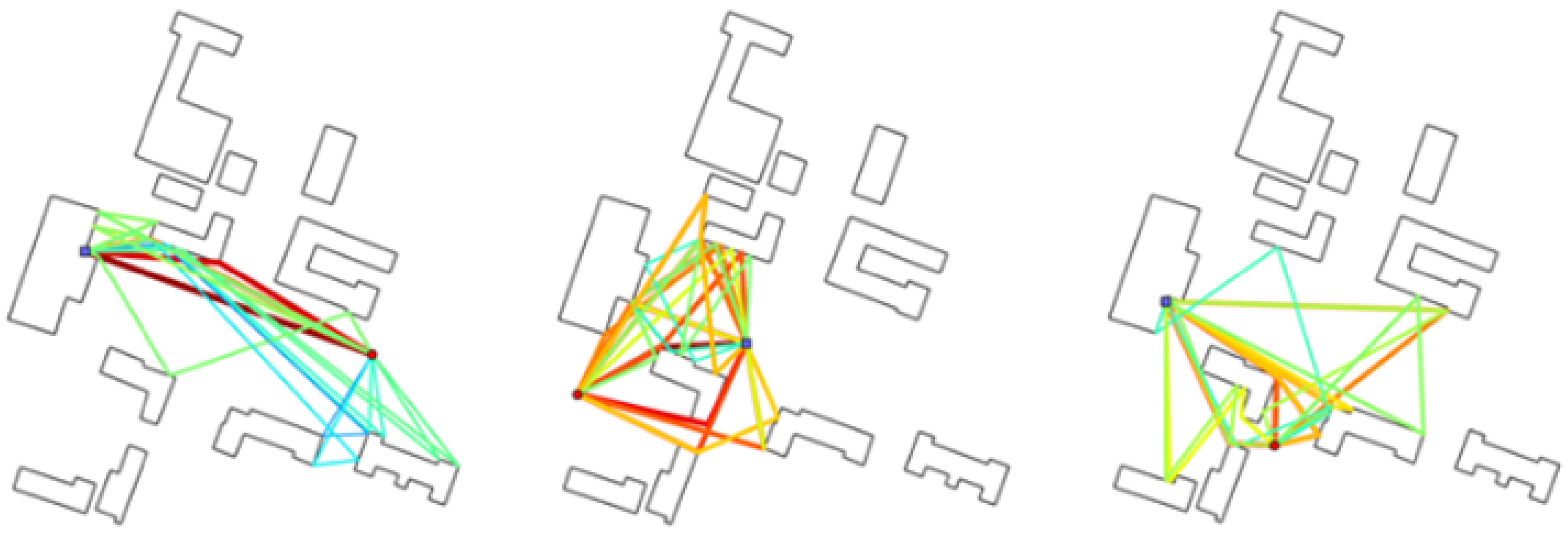

Figure 28: Ray tracing example for the WAIR-D dataset

  - Ray-tracing generated rays, allowing users to simulate complex propagation environments
  - Customizable communication parameters, enabling users to tailor the data to their specific research requirements
  - Data generation capabilities, providing users with the flexibility to create data tailored to their specific wireless AI tasks

- Easy-to-use: The dataset prioritizes usability and provides a comprehensive set of tools to get users started quickly:
  - Data generation code, allowing users to easily generate data for their research needs
  - Data preprocessing code, streamlining the data preparation process
  - Sample task training code, providing a starting point for users to develop and test their wireless AI models

The ray tracing example for the WAIR-D dataset is shown in Fig. 28. Given the positions of the base station and the user, the total number of rays can be determined. Rays of different colors represent different propagation paths.

#### 5.4.2 Dataset for Network AI

With the advent of the 6G era, the deep integration of communication and AI technologies has become a key driving force for the development of wireless communication systems. In this process, data, as a core element, is critical for enhancing AI models' capabilities and expanding their application potential. However, the lack of high-quality public datasets has limited further research and development of AI models in the communication field.

To address this challenge, we leverage the intelligent network open innovation platform to develop a collection of AI+6G datasets. These include AI air interface channel series data, intelligent planning of antennas and reconfigurable intelligent surface (RIS) for smart seaports, channel state information (CSI) compression feedback, real-world CSI measurements, computational resource scheduling for network AI, and wireless resource scheduling in cell-free scenarios. These datasets aim to assist global researchers in analyzing, exploring, and addressing challenges related to Network AI, thereby advancing and innovating communication technologies. The rest of this subsection is a detailed introduction to these datasets.

### AI air interface channel dataset

This dataset is designed for 6G AI air interface channel simulations, supporting advanced features such as large-scale MIMO near-field communication and high-speed mobility. It offers flexible configurations across multiple regions, antennas, frequency bands, and communication links, catering to diverse research requirements.

Currently, the dataset spans the 3.5 GHz, 28 GHz, and 60 GHz frequency bands, facilitating investigations into scenarios such as blockage, beamforming, and uplink/downlink channel prediction. The datasets for these frequency bands can be accessed via the following URLs:

- 3.5 GHz Dataset
- 28 GHz Dataset
- 60 GHz Dataset

The data is generated using high-precision ray-tracing simulations of outdoor urban street environments, specifically non-line-of-sight (NLOS) scenarios with multiple user regions. Within each region, user points are positioned at intervals of 0.25 or 0.5 meters, covering 12 user regions and encompassing over 180,000 user points. Base stations (BSs) are strategically deployed along the streets, with eight BSs configured differently to ensure full scene coverage. The distribution of BSs and user regions, along with the detailed parameter configurations of BSs and user terminals, are illustrated in Fig.29 and Table 9, respectively.

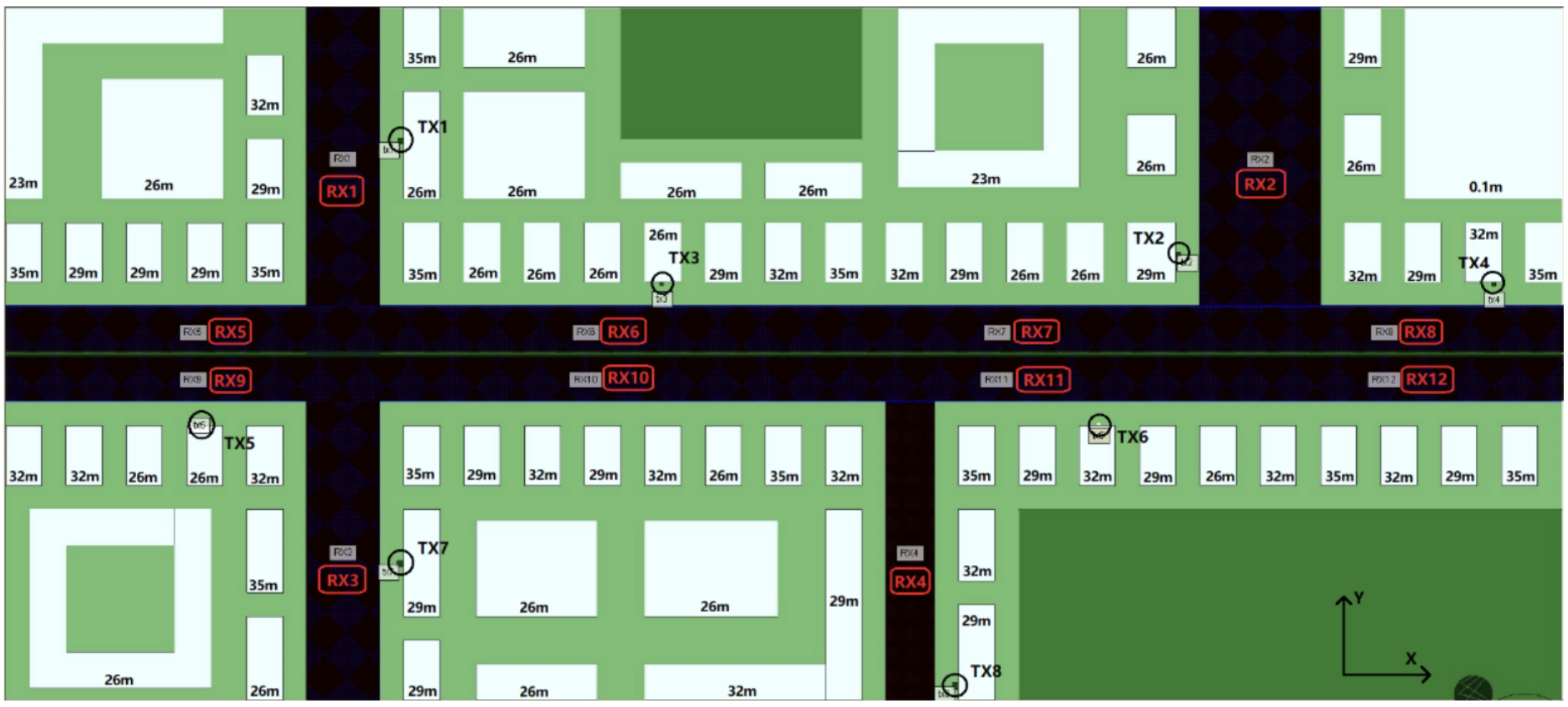

Figure 29: The distribution of BSs and user regions.

These datasets include two primary types of data generated through high-precision ray-tracing simulation platforms. The data contains critical parameters such as delay, horizontal and vertical departure angles, horizontal and vertical arrival angles, phase, power, and path loss, totaling billions of samples. The simulation scenarios are modeled on outdoor street environments, incorporating common scatterers like buildings and vegetation to closely approximate real-world conditions. Multiple BSs with varying antenna array configurations and user areas with a high density of user points ensure the dataset is diverse and comprehensive, capable of supporting a wide range of research efforts.

### Antennas and RIS intelligent planning dataset for smart seaports

This dataset is specifically designed to address the challenges of wireless network planning in seaport environments. It is based on real-world seaport scenarios and utilizes digital twin technology to construct a high-precision

Table 9: Detailed parameter configurations of BS and user terminals.

(a) BS parameters.

| BS ID | Coordinate | Antenna |
|---|---|---|
| TX1 | (162.5, 233.72, 20.04) | $4 \times 2 \times 2$ |
| TX2 | (486.01, 186.55, 10.11) | Single antenna |
| TX3 | (272.04, 173.58, 20.06) | $8 \times 4 \times 4$ |
| TX4 | (617.15, 173.3, 20.03) | $4 \times 4 \times 2$ |
| TX5 | (79.72, 116.2, 20.08) | Single antenna |
| TX6 | (452.99, 116.05, 9.94) | $2 \times 2 \times 2$ |
| TX7 | (163.18, 55.8, 20.06) | $4 \times 4 \times 4$ |
| TX8 | (393.72, 4.81, 19.92) | $8 \times 4 \times 4$ |

(b) User terminal parameters.

| Grid | Height | Column | Row | Spacing | Points | Antenna |
|---|---|---|---|---|---|---|
| RX1 | 1.5 | 61 | 251 | 0.5 | 15311 | $2 \times 2 \times 1$ |
| RX2 | 1.5 | 101 | 248 | 0.5 | 25048 | $2 \times 1 \times 2$ |
| RX3 | 1.5 | 61 | 250 | 0.5 | 15250 | $2 \times 2 \times 1$ |
| RX4 | 2 | 81 | 500 | 0.25 | 40500 | Single antenna |
| RX5 | 1.5 | 311 | 39 | 0.5 | 12129 | Single antenna |
| RX6 | 1.5 | 341 | 39 | 0.5 | 13299 | Single antenna |
| RX7 | 1.5 | 341 | 39 | 0.5 | 13299 | Single antenna |
| RX8 | 1.5 | 301 | 39 | 0.5 | 11739 | Single antenna |
| RX9 | 1.5 | 311 | 39 | 0.5 | 12129 | Single antenna |
| RX10 | 1.5 | 341 | 39 | 0.5 | 13299 | Single antenna |
| RX11 | 1.5 | 341 | 39 | 0.5 | 13299 | Single antenna |
| RX12 | 1.5 | 301 | 39 | 0.5 | 11739 | Single antenna |

model. The model encompasses twin modeling of base stations, wireless channels, terminals, and services, while also incorporating RIS twin capabilities. It supports network planning and performance validation in dynamically evolving seaport scenarios. The dataset includes cell and RIS planning results, coverage, and network performance metrics, offering billions of samples for research and development purposes. For detailed information about the dataset, please visit: antennas and RIS intelligent planning dataset for smart seaports.

Leveraging the multi-scenario information of antennas and RIS in seaport environments, the dataset enables AI-driven predictive analysis of various future scenarios. This facilitates the preemptive selection of wireless network element locations and parameter optimization, effectively reducing resource consumption while enhancing key network performance metrics.

In seaport scenarios, advanced wireless communication technologies enable remote intelligent operation of unmanned container trucks, a critical feature for building smart seaports. These unmanned trucks have stringent requirements for network quality along their movement trajectories. Ensuring that the planned site locations meet the demands of complex seaport environments—characterized by dynamic changes in containers, gantry cranes, and unmanned trucks—remains a significant challenge and focus of network planning.

This task constructs a high-precision digital twin capability for seaport environments based on real-world seaport scenarios. The digital twin encompasses base stations, wireless channels, terminals, and services, and incorporates RIS twin capabilities. These features support dynamic network planning and performance validation for seaport scenarios. The task not only provides simulation capabilities for seaport communications and validation of AI models but also makes a wealth of multidimensional datasets available for researchers.

**CSI compression feedback dataset**

In 5G technology, MIMO enhances spectrum and energy efficiency by deploying large-scale antenna arrays at base stations. To fully realize the potential performance gains of large-scale MIMO, base stations require accurate downlink CSI for channel-adaptive transmission optimizations, such as precoding. Traditional compressed sensing (CS) methods rely heavily on prior assumptions about channel structures. However, the CSI matrix is only approximately sparse in the angle-delay domain, and variations between adjacent elements are often correlated. These methods necessitate complex prior assumptions and struggle to guarantee reliable recovery performance.

To overcome these limitations, academia and industry have turned to deep learning, leveraging its powerful optimization and fitting capabilities. By training neural networks to learn channel structure information, deep learning provides superior reconstruction performance and enables rapid and accurate CSI recovery from low-compression feedback data. The CSI compression feedback dataset offers 600,000 simulation samples, featuring channel models for both LOS and NLOS scenarios. For more information, please visit: CSI compression feedback dataset.

This dataset combines simulated and real-world measurements to generate channel feature matrix data. It presents researchers with the challenge of compressing user-side channel feature information, transmitting it through the channel, and recovering it at the receiver side. The ultimate goal is to restore channel state information as accurately as possible, minimizing losses. Exploring this dataset provides researchers with valuable tools for optimizing channel-adaptive transmission techniques.

The primary objective of this dataset is to utilize AI-driven feature extraction and information compression feedback mechanisms. AI models are trained to compress channel information from the user side. The compressed data is transmitted through the channel and reconstructed at the receiver, aiming to restore channel state information with minimal loss under predefined compression bit constraints. A smaller number of compressed bits reduces transmission resource requirements but may degrade recovery accuracy and feedback performance. To address this, the dataset requires model designs for two scenarios: feedback vectors of 36 bits (low-bit feedback) and 128 bits (high-bit feedback). The final performance evaluation is based on a weighted average of the scores across both scenarios.

#### Real-world CSI measurements dataset

The dataset was meticulously collected from real terminal devices in commercial network scenarios. The collection process focused primarily on outdoor open areas, encompassing two modes: free-space scenarios and call scenarios. This approach ensures that the dataset captures diverse communication characteristics under varying conditions. To enhance comprehensiveness and diversity, multiple test points were established, with rotation tests conducted at four angles (0°, 90°, 180°, 270°) at each point. This not only increases the dataset's practical utility but also aligns it closely with real-world communication environments.

Specifically, data collection was conducted in outdoor open-area (NLOS) single-cell fixed-point tests. Test points were set at 10-meter intervals, resulting in 52 primary points. Additionally, nine extended points within a 1-meter radius of each primary point were included. Parameter configurations reflect the actual commercial network settings and were obtained through terminal signaling collection, as detailed in Table 10. The tests comprised two scenarios: free-space scenarios (with no uplink or downlink data transmission) and call scenarios (where terminals conducted voice calls), using two different models of real terminal devices.

The dataset comprises over 7,488 samples, including more than 3,700 entries from both free-space and call scenarios, forming a CSI measurement dataset. This dataset holds significant value for the communications field, bridging the gap between theoretical models and real-world scenarios that simulation tools often cannot cover. It offers accurate and comprehensive data to support research in wireless communication, signal processing, and communication system optimization.

The rigorous collection process and diverse characteristics of the dataset provide a solid foundation for its application in the communications domain. It serves as a concrete reference for exploring and optimizing communication technologies. For more details, visit the dataset at For more information, please visit: real-world CSI measurements dataset. The release of this dataset is poised to drive technological progress and innovation in the communications field, paving the way for the development of future communication systems.

#### Computational resource scheduling for network AI dataset

In 6G mobile infrastructure, the network evolves beyond providing connectivity to offer distributed computing services tailored to the demands of AI applications. Through flexible and dynamic computational resource scheduling, the system ensures the efficient allocation of resources, delivering ubiquitous, high-quality computing services. The dataset for computational resource scheduling in network AI comprises 9 data types, with over 7,000,000 samples in total. For access to the dataset, please visit: computational resource scheduling for network AI dataset.

#### Wireless resource scheduling dataset for cell-free scenarios

With the rapid advancement of cell-free and coherent transmission technologies, the amount of scheduling resources in wireless communication systems has significantly increased. Traditional greedy scheduling algorithms face challenges in terms of implementation and computational complexity when dealing with a large number of transmission points (TRPs) and user combinations. AI technologies have demonstrated immense potential in decision-making and management tasks by learning rules to output optimal strategies, aligning closely with the functional requirements of schedulers.

Table 10: Actual network parameter configuration.

| Parameter | | Configuration |
|---|---|---|
| Duplex, waveform | | TDD, OFDM |
| Scrnario | | Actual scenario |
| Frequency range | | FR1, 2.6GHz |
| Antenna port at UE | | 4 |
| BS antenna height | | 30m |
| Numerology | Slot/Non-slot | 14 OFDM symbol slot |
| | SCS | 30 kHz |
| CSI-RS | Frequency domain allocation | other: 001111 |
| | Nrof ports | 8 |
| | First OFDM symbol in time domain | 13 |
| | Cdm-type | fd-CDM2 |
| | Density | 1 |
| | Starting RB | 0 |
| | Nrof RBs | 272 |
| | Periodicity | slots 40 |
| CSI-report config | Periodiciy | slots 160 |
| | Codebook type | type 1 |
| | Subtype | typel-singlepanel |
| | Nrof antenna ports | more than 2 |
| | n1-n2 | 4-1-typel-singlepanel-restriction |
| | 4-1-typel-singlepanel-restriction | 1111111111111111 |
| | Typel-singlepanel-ri-restriction | 00001111 |

The wireless resource scheduling dataset for cell-free scenarios includes 16,000 samples. Each sample contains the following features: the current channel characteristics of 120 users, including PMI and channel gain values; the predicted channel characteristics (PMI and channel gain values) at the scheduling time (5 ms later); and historical scheduling rates. For access to the dataset, please visit: wireless resource scheduling dataset for cell-free scenarios.

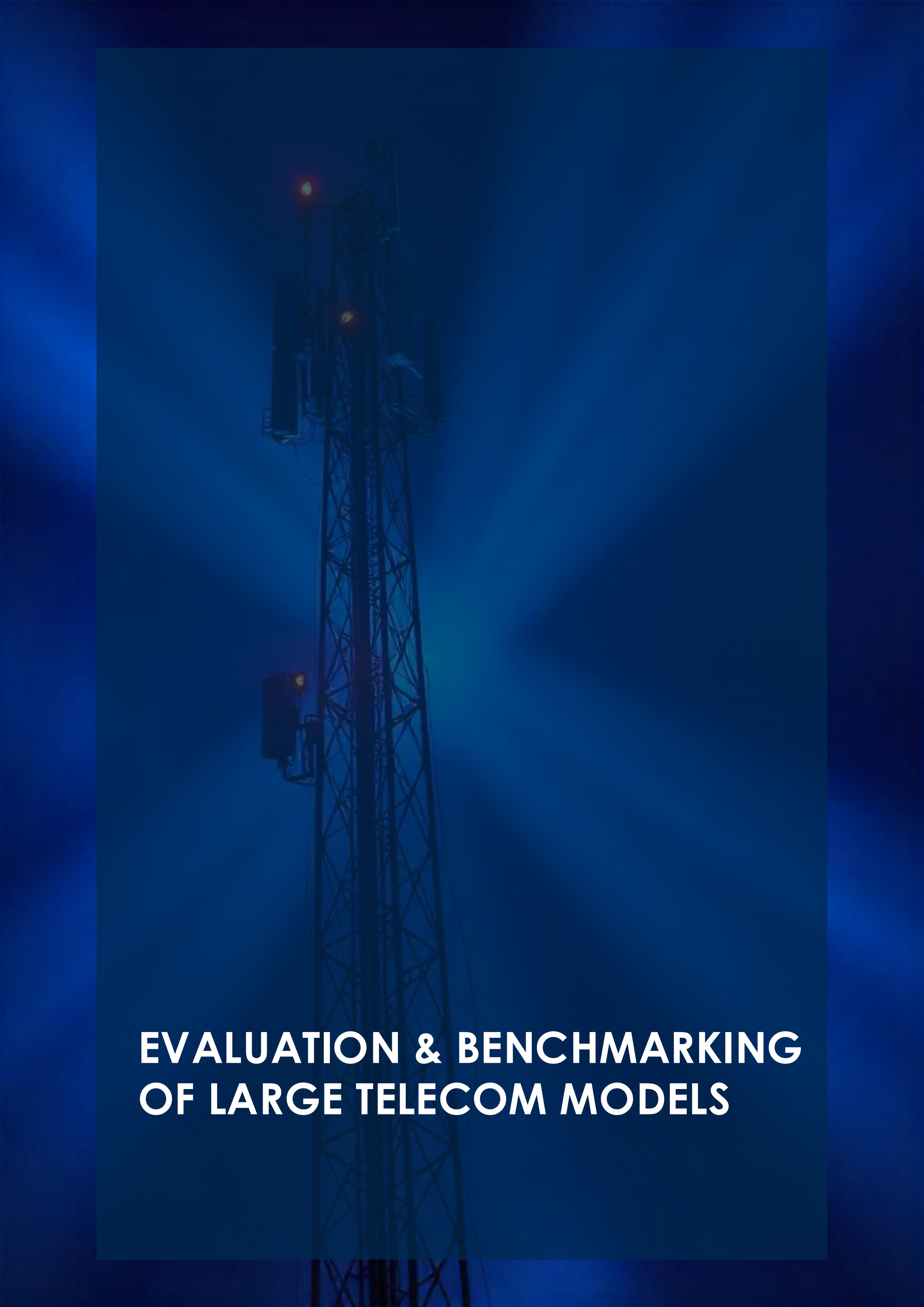

# EVALUATION & BENCHMARKING OF LARGE TELECOM MODELS

# 6 Evaluation & Benchmarking of Large Telecom Models

To harness the full potential of LTMs, it is imperative to establish robust guidelines and frameworks for their evaluation and benchmarking. This chapter addresses the need for systematic assessment methods tailored specifically for the telco domain. As we look toward the advent of 6G, which is expected to be AI-native, we will increasingly rely on AI. Consequently, we anticipate the deployment of a diverse array of models, each tailored to distinct aspects of the industry.

This is particularly the case for emerging Generative AI models, especially Large Language Models. Despite their potential, current state-of-the-art models like GPT-4 encounter significant challenges in the telecom domain. For instance, GPT-4 fails almost half of the specification-related problems in Telecom Question Answering (TeleQnA). This shortfall delays the potential deployment of such models in operational networks, highlighting the necessity for enhancing the telecom knowledge embedded within LLMs.

As we are already witnessing significant commitment and R&D efforts in developing telco-specific GenAI models, it is essential to develop models that can be broadly categorized into several types, each serving unique purposes and addressing specific challenges within the telecom domain. Such domain-specific LTMs are designed to handle specialized tasks within the telecom sector, leveraging domain-specific knowledge to enhance performance and accuracy.

This chapter begins with an overview of the key benchmarking metrics that are pertinent to LTMs in Sec. 6.1. Sec. 6.2 describes well-known evaluation frameworks developed by the AI community to test LLM capabilities. Sec. 6.3 present LLM requirements and capabilities to realize the most promising use cases for GenAI in telecom networks. Sec. 6.4 explores the role of digital twins in the evaluation of LTMs. Sec. 6.5 and Sec. 6.6 discuss the performance of LLMs on telecom knowledge and telecom math modelling, respectively. Sec. 6.7 present the evaluation results of a prompting solutions for LLM-based automation of generating commit message in 5G networks. Finally, Sec. 6.8 describes the evaluation methodology built to assess two LLMs developed by MNOs and the associated results.

## 6.1 Overview of benchmarking metrics

Today there are three main ways to assess a GenAI model: human evaluation, using a second model as judge, or running a benchmark test using well established metrics [170]. In human evaluation, users interact with the LTM across various domains and real-world telecom scenarios. The LTM's output is then manually scored based on predefined evaluation rules. These scores are weighted according to the importance of each evaluation criterion, and a final composite score is generated. Using human labelers to judge the GenAI outputs is very time-consuming and costly. Also, this approach lacks flexibility as when the model or its task are updated a new evaluation process is required. Replacing the human judge with a GenAI model is promising as it reduces the cost and time constraints of human evaluation [170]. By using automated tools to call the LTM's interface, results across various domains and evaluation tasks can be obtained. A third-party model or tool, acting as the GenAI judge, can then compare the LTM's outputs with reference datasets to compute objective metrics by evaluating the differences between the LTM's predictions and the ground truth. However, the GenAI judge may not surpass human evaluation in accuracy and quality. A possible approach to test the GenAI judge is to create a small human evaluation dataset, which can test the accuracy of the GenAI judge. Table 1 shows standard benchmark tests for LLMs, i.e., GLUE (General Language Understanding Evaluation), SuperGLUE, HellaSwag, TruthfulQA, and MMLU (Massive Multitask Language Understanding). These standard benchmarks typically use well-known evaluation metrics based on the evaluation task, e.g., accuracy [171] or F1-score [172] for classification tasks or question-answering tasks, and Bilingual Evaluation Understudy (BLUE) [173] or Recall-Oriented Understudy for Gisting

Table 11: Standard benchmark test for LLMs.

| Benchmark | Explanation | Metrics | Reference | URL |
|---|---|---|---|---|
| General Language Understanding Evaluation (GLUE) | Standardized set of diverse Natural Language Processing tasks | Correlation coefficients, accuracy, and F1 score | [175] | [176] |
| SuperGLUE | More difficult language understanding tasks with respect to GLUE | Accuracy, exact match, and F1 score | [177] | [178] |
| HellaSwag | Benchmark for commonsense natural language inference | Accuracy | [179] | [180] |
| TruthfulQA | Benchmark of questions designed to cause imitative falsehoods | Human evaluation | [181] | [182] |
| Measuring Massive Multitask Language Understanding (MMLU) | Multiple-choice questions related to 57 tasks including mathematics, history, computer science, and law | Accuracy | [183] | [184] |

Evaluation (ROUGE) [174] for evaluating text similarity.

As LLMs continue to play a vital role in both research and industrial use cases, their evaluation becomes increasingly critical, specially in specific domain such as Telecom. There are many choices LLMs both from commercial and open-source models and it is important to understand what models to choose to serve an objective for a given use case. IT needs to be evaluated systematically. It is important practice to measure LLMs performance on benchmark datasets with metrics that are aligned with a given objective. Benchmarking LLM performance requires creation of evaluation datasets based on the specific domain and use-cases. The initial point can be using open source datasets that is widely , However, many of use cases within telecommunication domain require domain adaptation and as result to evaluate those use cases creating a domain and task specific datasets are must. The important factors to consider on creating domain/task related datasets are their coverage in terms of data sources, their modalities and variant tasks. Some of the challenges involved in creating such a favorable datasets is creating them manually with human in the loop, which make time consuming and biased. In case of evaluating a task in hand, we need to pay special attention to the metrics considered. There are many suggested metrics ranging from classical statistical metrics and model based metrics. In order to have a comprehensive evaluation for each use case, couple of metrics may be required and it is important to measure. Some of this metrics needs to be domain adapted as well. Example of this is a metric called BERTScore which measures cosine similarity of embeddings of candidate and reference answers. In case of telecom texts it is important to use embeddings to compute similarity metrics. Similarly if LLM used as a judge it is vital to use telecom adapted LLM aware as judge. The most common way to evaluate the LLM is the to measures if the output is accurate, clear, and informative by checking if the output from the language model includes any fabricated or incorrect information (i.e., hallucination). We should also consider aspects of the model output such as determining if the language model's output doesn't contain potentially harmful or offensive content. Another critical metric is functionality, which measures the number of operational scenarios the LTM can support. These metrics assess the model's ability to handle various oper-ational contexts in different network domains. The more business scenarios the LTM can support, the stronger its knowledge base and generalization capabilities. A scoring mechanism can be applied based on the percentage of test cases the model successfully handles within these scenarios to evaluate its functionality. Additionally, we can apply classic statistical metrics for evaluating machine learning models such as accuracy, precision, recall and F1 score. Other Natural Language programming (NLP) metrics could be applied such as BLEU and ROUGE score. Another way to evaluate the output of the LLM model is using stable and performant Language Models to compare with the output of our model. We cite the Bert-score and GPT-Score as shown in Figure 33.

| Classic Statistical Metrics | Classic NLP Metrics | Model Based Metrics | Frameworks |
|---|---|---|---|
| Accuracy | BLEU | Bert score | HELM |
| Precision | ROUGE | G-Eval | DeepEval |
| Recall | Mean Reciprocal Rank (MRR) | GPT-score | PromptBench |
| F1 score... | Edit Distance... | NLI-Score | LM Evaluation Harness |
| | | BLEURT... | LMSYS Chatbot Arena... |

Figure 30: Metrics and frameworks for LTM evaluation.

In addition to the metrics presented in Figure 33, we can apply the Metric for Evaluation of Translation with Explicit ORdering (METEOR). METEOR is a flexible metric originally created to evaluate machine translation quality, but it can also be effectively utilized for assessing LLMs [185]. It emphasizes the alignment between reference and generated words, providing flexibility in matching synonyms and grammatical variations. The scoring system ranges from 0 to 1, where higher scores reflect better alignment. In the context of LLM evaluation, METEOR can measure the coherence of generated responses and enable comparisons across different models. Although it provides a detailed evaluation, its complexity and reliance on the reference corpus can pose challenges. Overall, METEOR is a valuable tool for conducting in-depth assessments of language models.

In the process of deploying LTM models within network operations, it is essential to establish a robust evaluation framework to comprehensively assess model performance. Without an effective evaluation mechanism, the model may encounter issues during real-world applications, increasing operational risk. The evaluation of LTM can be tiered across different stages of deployment, with varying requirements for performance metrics at each stage: pilot application stage, expanded application stage, and Full application stage, the details are as follows:

(1) Pilot application stage: In the early stages of LTM deployment, the model is typically tested in pilot operations or for performing auxiliary tasks such as fault diagnosis and performance optimization. A comprehensive evaluation plan is necessary to assess the model's accuracy, computational efficiency, and security. This ensures that the LTM can operate within a physical network and supports its transition to the next phase. Given the the relatively low risk and limited scope of application, the accuracy requirements for model predictions are relatively lower, focusing more on aiding network tasks and learning from data patterns. Computational efficiency may not be critical at this stage, but as data volumes increase, this becomes a more significant factor. Importantly, deploying the LTM in a system introduces risks related to security, stability, and data privacy. These must be carefully evaluated to ensure a safe and reliable deployment.

(2) Expanded application stage: Following a successful pilot, the LTM's scope of use expands to higher-risk operations, such as network security. In this stage, performance expectations are elevated, and the model faces increasingly complex scenarios. The accuracy of its predictions becomes more crucial as the model supports decision-making and optimization efforts. To meet operational needs, real-time processing and response capabilities are required, and model explainability becomes a key factor for gaining trust from network operators. As the application expands, the evaluation becomes more rigorous to ensure the model's predictions and operations align with the growing complexity of tasks.

(3) Full application stage: Once the LTM has demonstrated stable performance at larger scales, it can be deployed across the entire network lifecycle, from planning and design to deployment, operations, and maintenance. At this stage, the risks associated with the LTM are highest, necessitating the strictest evaluation mechanisms. The primary evaluation criteria include the model's ability to make highly accurate inferences, strong generalization across diverse, evolving scenarios, and rapid updates to meet new requirements. As the model handles increasing amounts of data and computation, its performance must remain optimal despite resource constraints. Degradation in performance could lead to reduced processing speeds or excessive memory consumption. Additionally, the model must be resilient to security threats and system disruptions, as any failure could severely impact the network. Furthermore, explainability becomes a critical requirement, ensuring the LTM provides clear, consistent, and interpretable decisions to network operators.

## 6.2 Evaluation Frameworks

This section presents a summary of popular evaluation frameworks developed by the AI community to test LLM capabilities.

### 6.2.1 HELM

The HELM also known as the Holistic Evaluation of Language Models framework is designed to provide a comprehensive assessment of LLMs by evaluating multiple dimensions of their performance, including accuracy, fairness, robustness, and efficiency [186]. HELM was developped by the collaboration of Center for Research on Foundation Models (CRFM), Institute for Human-Centered Artificial Intelligence (HAI) and Stanford University. The idea behind HELM is to include multiple metrics and not only centered in the accuracy of the model. It incorporates a diverse set of metrics to capture various aspects of model behavior, such as linguistic quality, factual accuracy, and ethical considerations. By offering standardized benchmarks, HELM facilitates comparisons between different models across various tasks and datasets. Additionally, it emphasizes user-centric evaluation, taking into account user experience and satisfaction. The framework encourages iterative improvement, allowing developers to refine models based on evaluation results.

Standardization benchmarks in the context of the HELM framework for evaluating language models typically include a set of established datasets and tasks that allow for consistent comparison across different models. Common types of benchmarks include:

- **Natural Language Understanding (NLU)**: GLUE and SuperGLUE.
- **Natural Language Generation (NLG)**: already addressed in the previous Section, it includes the BLEU and ROUGE metrics.
- **Factual Consistency**: datasets that assess the factual accuracy of generated text, such as **FEQA** (Factual Evaluation for Question Answering).
- **Ethical and Fairness Metrics**: benchmarks that evaluate bias and fairness, such as those assessing gender or racial bias in generated outputs.
- **Robustness**: tests that evaluate how well models perform under adversarial conditions or with noisy inputs.
- **User-Centric Metrics**: surveys or user studies that gauge user satisfaction and experience with model outputs.

### 6.2.2 PromptBench

PromptBench was developed by researchers in academia and Microsoft Research [187] in 2024. It is a framework designed to evaluate the performance of LLMs based on their responses to various prompts. It includes a diverse array of prompt types that test different capabilities of LLMs, such as reasoning, creativity, and factual accuracy. By providing a standardized methodology for assessing model performance across different tasks and datasets, PromptBench ensures consistency in evaluation. The framework emphasizes user-centric evaluation, incorporating prompts that reflect real-world applications and user needs. This focus allows for comparative analysis, enabling the assessment of multiple models based on their responses to the same set of prompts. Additionally, PromptBench encourages iterative improvement, promoting continuous refinement of models based on evaluation results and user feedback.

### 6.2.3 LLM Evaluation Harness

The LLM Evaluation Harness is a framework designed to facilitate the evaluation of LLMs across various tasks and benchmarks [188]. was developed by researchers at Hugging Face and was introduced in 2022. Its modular design allows for easy integration of different evaluation metrics and datasets, making it adaptable to a wide range of research needs. This flexibility ensures that researchers can tailor the evaluation process to suit specific objectives and contexts.

One of the key features of the LLM Evaluation Harness is its inclusion of standardized benchmarks, which ensure consistent evaluation across different models. By supporting a variety of tasks, such as text generation, question answering, and summarization, the harness enables comprehensive assessments of model performance. It also provides various metrics for evaluating aspects like accuracy, fluency, and relevance, offering a well-rounded view of each model's capabilities.

Additionally, the LLM Evaluation Harness is designed with a user-friendly interface, making it accessible for both researchers and developers. This simplicity simplifies the evaluation process, allowing users to focus on analyzing results rather than navigating complex tools. Overall, the LLM Evaluation Harness serves as a valuable resource, offering detailed insights into model capabilities and limitations while promoting best practices in model evaluation and development.

### 6.2.4 LMSYS Chatbot arena

LMSYS Chatbot Arena is a platform where you can engage with and evaluate different large language models. You can pose questions or provide prompts, and then compare the responses from various models side-by-side [189]. This allows you to get a sense of their strengths, weaknesses, and overall capabilities. Note that LMSYS chatbot is used to test Chatbots based on LLMs with prompting technique.

### 6.2.5 deepval

DeepVal is a tool developed by Microsoft that focuses on evaluating and analyzing the safety and reliability of LLMs [190]. It uses a technique called *meta-evaluation* which leverages LLMs themselves to assess the quality and potential harms of other LLMs' outputs.

## 6.3 Capabilities and Requirements of Large Language Models applied to the Telecom domain

Telecom use cases have diverse and complex needs spanning from energy consumption to knowledge of standard documents, which makes challenging for identifying the right model to be used for each specific telecom use case.

In this section, the requirements and capabilities to realize the most promising use cases for GenAI in telecom networks are introduced.

### 6.3.1 Capabilities of Large Language Models applied to the Telecom domain

The following capabilities allow to realize the most promising use cases for LLMs in telecom networks.

**6.3.1.1 Software development** LLMs should write, optimize and maintain software code, based on natural language descriptions, helping developers rapidly prototype or implement solutions. LLMs should also automate tasks such as debugging, refactoring, and providing suggestions for performance improvements. LLMs should be also capable of generating regular expressions (regex), helping users create complex pattern-matching rules from natural language descriptions.

**6.3.1.2 Compliance with output format** LLMs should generate content, whether code, text, or structured data, that adheres to specific formatting standards, conventions, or templates indicated by the user.

**6.3.1.3 Supports structured data as input** LLMs should be able to process, interpret, and transform structured formats like JSON, Extensible Markup Language (XML), Comma-Separated Values (CSV), and database tables. By understanding the relationships within structured data, LLMs can perform tasks such as data extraction, transformation, validation, and even synthesis of new structured data.

**6.3.1.4 Mathematical and logical reasoning** LLMs should be able to realize mathematical and logical reasoning, allowing it to perform problem-solving, calculations, and algorithmic thinking.

**6.3.1.5 Tool calling** LLMs should have tool invocation capabilities, enabling it to autonomously interact with various software tools, Application Programming Interfaces (APIs), and libraries based on natural language instructions. By recognizing user intent, LLM can trigger specific actions—like fetching data, performing calculations, or automating tasks—through external tools without requiring direct manual input.

### 6.3.2 Requirements of Large Language Models applied to the Telecom domain

Telecom operators operate in a consumer-driven, energy-intensive environment. For LLMs to be widely adopted, the telecom industry must consider their cost, energy consumption, and safety impacts alongside technical evaluations discussed in the previous section.

**6.3.2.1 Energy Consumption** LLMs demand substantial energy resources, especially during inference, which occurs continuously in telecom applications like customer service. Research indicates that inference consumes significantly more energy than training, as it runs in real-time and scales with user demand [191]. For telecom operators aiming to optimize energy use, efficient benchmarking should consider the model's architecture, deployment configuration, and optimizations such as batching and parallel processing [192]. Recent studies reveal that hybrid GPU-CPU configurations can reduce energy use by up to 7.5%, underscoring the importance of adaptive infrastructure design [193]. Although energy consumption is often secondary to performance in evaluations, there is a growing shift toward prioritizing energy efficiency to align with industry sustainability goals and reduce operational costs [194]. By implementing energy-aware strategies and infrastructure adaptations, telecom companies can balance real-time performance with energy demands, ultimately supporting a more sustainable AI deployment strategy.

**6.3.2.2 Safety First** With telecoms managing vast amounts of sensitive data, robust safety benchmarking is essential to maintain compliance and uphold customer trust. LLMs must be rigorously tested for their handling of personally identifiable information (PII), financial data, and potential safety risks. Frameworks like SafetyBench and TRUSTLLM provide comprehensive benchmarks to evaluate privacy, bias, and response appropriateness, ensuring that LLMs align with telecom's high standards for data security and ethical AI usage [195, 196]. Meanwhile, tools like WALLEDEVAL apply mutation-based testing to assess model consistency across varied customer interactions, simulating diverse scenarios to uncover response deviations and potential vulnerabilities [197].

Additionally, telecom-specific red-teaming exercises are instrumental in identifying industry-specific risks. For example, The Global Telco alliance's TelcoLLM (a joint venture of five telecom's operators, SK Telecom, Deutsche Telekom, e& Group, Singtel and Softbank) incorporates real-time detection filters, prompt engineering, and iterative red-teaming practices to enhance robustness, effectively safeguarding LLMs against adversarial inputs in high-stakes customer support environments [198]. These targeted measures help ensure that LLMs can responsibly manage sensitive telecom data and provide accurate, consistent support, reinforcing both operational safety and customer confidence.

## 6.4 Digital Twins for evaluation of LTMs

Digital twins act as sandboxes that accurately capture the features and properties of cellular networks [199]. They can reproduce a variety of channel conditions in real time, scale to a large number of nodes and run the cellular stack on real hardware. Such platforms include the Keysight Propsim channel emulator [200] and Colosseum [201, 202, 203], the world's largest wireless network emulator with hardware in the loop.

In the vision of 6G and beyond, LTMs have been identified as suitable tools to convert high-level intents from operators into control policies and reconfiguration actions that update the network configuration to match intents. While this will undeniably streamline network provisioning, monitoring and control, the concern is that policies generated by LTMs might be inaccurate or inefficient. This becomes especially problematic in the case where the control policies generated by the LTMs (e.g., starting from high-level intents expressed via natural language or high-level configuration files) are not optimal and/or could cause improper configurations that might result in performance degradation and even outages.

In this context, it is important to provide LTMs developers with a development and testing platform that can accurately replicate RF and network conditions, while offering a sandbox environment that prevents wrong control policies from affecting user performance.

Digital twins a versatile and robust framework for evaluating and benchmarking LTMs especially in the aforementioned cases. Indeed, the deployment of digital twins in this context enables a comprehensive, risk-free environment for testing and optimizing these LTMs in a repeatable way that not only allows to identify issues, but also to benchmark different LTMs against the same network conditions and configurations. This approach of offline training and testing of AI/ML models currently is the consensus the O-RAN alliance has agreed upon as the best practice [204]. Hence, digital twins for LTM benchmarkings are a key enabler for O-RAN compliant LTMs.

In the following, we identify three use cases that would not only benefit from digital twins, but also require their use before deploying the LTM on production networks.

**Policy Deployment and Effectiveness** One of the primary applications of LTMs in telecommunications is generating policies and network configurations. Digital twins allow us to create a virtual replica of one or more cellular network deployments where these policies can be implemented and evaluated against a set of target KPMs. By simulating various network conditions, topologies, traffic profiles and user profiles, we can assess how well the LTM-generated policies perform. This includes evaluating their impact on network efficiency, reliability, and overall performance. The digital twin enables continuous monitoring and provides detailed feedback, allowing us to refine the LTMs' policy generation algorithms to ensure they deliver optimal performance in real-world scenarios. Some works have already been done for extensively evaluating policy optimization algorithms on Colosseum, for problems such as slicing [205] and spectrum efficiency [206]. We argue the architecture demonstrated in those works could pave the way to evaluating LTM policies on such problems.

| | Mistral 7b | Mixtral (MoE) | GPT-3.5 | GPT-4 | Humans |
|---|---|---|---|---|---|
| Lexicon (500) | 56.80 | 83.80 | 82.20 | 86.80 | 80.33 |
| Research overview (2000) | 51.60 | 70.70 | 68.50 | 76.25 | 63.66 |
| Research publications (4500) | 49.27 | 70.20 | 70.42 | 77.62 | 68.33 |
| Standards overview (1000) | 46.20 | 66.73 | 64.00 | 74.40 | 61.66 |
| Standards specifications (2000) | 35.60 | 55.85 | 56.97 | 64.78 | 56.33 |
| Overall accuracy (10000) | 47.07 | 67.74 | 67.29 | 74.91 | 64.86 |

Table 12: Illustration of Mistral 7b, Mixtral (MoE), GPT-3.5, GPT-4, and active professionals accuracy (%) across the various TeleQnA categories.

**Misconfiguration Detection and Correction** LTMs can also be used for identifying misconfigurations and issues in telecom networks thanks to their ability to process large log files and generate summaries that highlight errors and warnings that could point at identifying misconfigurations. By leveraging digital twins, we can create controlled environments where we introduce misconfigurations intentionally to generate testing scenarios as well as labeled training datasets. Then, we can use the LTM's ability to detect these issues to process log files and look for errors and misconfigurations. Once identified, the LTM can propose corrective actions, which can be implemented and tested in the digital twin. This process ensures that the recommended fixes are effective and do not inadvertently introduce new problems. In this way, we can generate a closed-loop testing process where the digital twin facilitates the training and testing of LTMs capable of automatically identifying and correcting any improper network configuration.

**Performance Testing of Intent-based Network Configurations** Another critical application of LTMs is reconfiguring networks based on specific intents, such as optimizing for higher throughput or lower latency. Digital twins provide an ideal sandbox for testing the ability of the LTM in generating network configurations that meet intents but are general and vary depending on the specific network deployment, channel conditions, traffic load and user behavior. By simulating different network topologies, user densities, and traffic patterns, we can evaluate the LTMs' reconfiguration strategies for a variety of intents, and make sure the LTM is able to generalize and produce policies that are effective independently of the deployment scenario. The digital twin's feedback loop allows us to measure the performance improvements and potential trade-offs of these reconfigurations, and can help in fine-tuning generated policies to better match intents and adapt to changing operational conditions.

## 6.5 Assess Large Language Models Telecommunications Knowledge

We evaluate the performance of GPT-3.5 and GPT-4 across the various categories of the TeleQnA dataset. For comparison, we also evaluate the performance of two open source models, a small model Mistral 7b, and a medium model, Mixtral 8x7B mixture of experts (MoE), comparable to GPT 3.5. The results are reported in Table 12. As anticipated, GPT-4 consistently outperforms GPT-3.5, demonstrating around 7% improvement across all categories. We remark that the open-source alternative Mixtral MoE performs similarly to GPT-3.5, while Mistral 7b significantly underperforms, which is expected given its small size. Notably, LLMs exhibit exceptional performance in the lexicon category, which encompasses general telecom knowledge and terminology, achieving approximately 87% accuracy for GPT-4. Conversely, these models face challenges when confronted with more intricate questions related to standards, with the highest performing model, GPT-4, achieving a modest 64% accuracy in this domain. In summary, GPT-3.5 averaged an accuracy of 67%, while GPT-4 achieved an accuracy of 74%. These results demonstrate that these models possess a solid foundation in general telecom expertise. However, to attain higher accuracy in responding to complex inquiries, further adaptations to the telecom domain are necessary.

Figure 31: Accuracy (%) comparison among GPT-3.5, GPT-4, and GPT-3.5 with context in the standards specifications category.

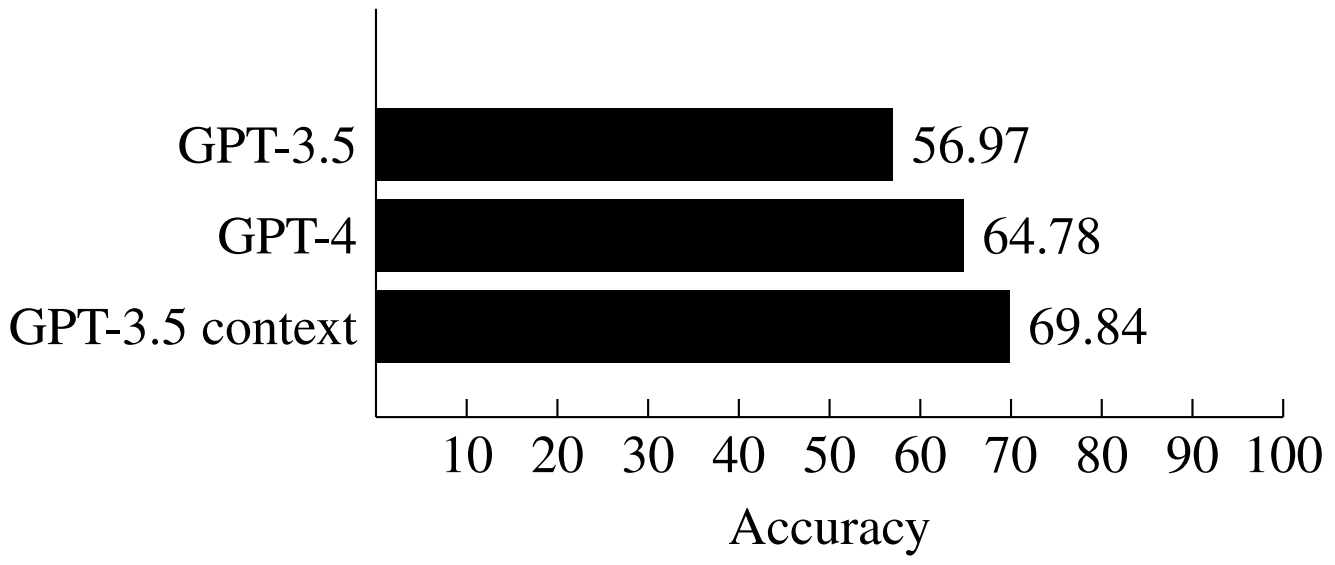

In our final step, we conducted a performance benchmark comparing active professionals to LLMs. The results reveal that LLMs and active professionals exhibit comparable performance in general telecom knowledge. However, in the case of intricate questions related to research and standards, LLMs demonstrate the capability to rival these professionals. This is attributed to LLMs' ability to digest and memorize complex and intricate documents. Furthermore, it is crucial to recognize the challenge faced by professionals when responding to these questions, as they encompass a broad range of telecom subdomains that these individuals may not be necessarily actively engaged with in their work. Considering all factors, our results underscore the significant promise that LLM hold within this domain, as demonstrated by their competitiveness within this extensive and comprehensive dataset.

### 6.5.1 Influence of Context

Until this stage of our benchmarking process, we have been querying the LLMs without accompanying contextual information for the questions. Nevertheless, in this subsection, our goal is to investigate how supplementing questions with additional context affects the accuracy of these models.

To accomplish this, we have focused on the standards specifications sources, encompassing thousands of technical standards pages. Our selection of this category is driven by the fact that this is where LLMs have exhibited the lowest performance. With this in mind, we segmented these pages into approximately 500-word segments before generating embeddings for each segment using OpenAI's Ada v2 text embeddings. Moreover, we employed the same OpenAI model to create embeddings for the questions and corresponding options belonging to this category. Following that, we constructed a distance matrix between the embeddings of each question-options pair and those of each segment. The next step involved querying the LLMs by supplying batches of five questions-options pairs, and additionally, as context, the top-3 closest segments to the five question-option pairs based on the distance matrix. The outcomes of these experiments are illustrated in Fig. 31.

The LLMs supplemented with contextual information reached an accuracy level of 69.84%, which translates into a relative accuracy enhancement of 22.5% compared to the scenario lacking context. This gain highlights the significant enhancement in performance achieved by incorporating contextual information, demonstrating how even less advanced models like GPT-3.5 can match the performance of state-of-the-art GPT-4 model. This underscores the necessity for a specialized telecom language models, fine-tuned or trained specifically on telecom-related data.

Developing such a foundation model has the potential to push the boundaries of LLMs performance in the telecom domain, paving the way for a wide range of use cases that demand telecom knowledge and expertise.

## 6.6 Evaluation of Telecom Math Modeling

Math modeling is a critical application in telecom domain. This includes accurately model the telecom environment or problem using telecom terminology and formulate the task as rigorously defined problems expressed in mathematical equations. LLM pre-trained or fine-tuned on math corpus has shown remarkable performance in math reasoning. such as DeepSeek Math, AlphaProof, and AlphaGeometry 2. A number of mathematical benchmarks have been used to evaluate the math capability of LLMs, from GSM8K, MATH, to IMO, AIME. However, using LLM to perform telecom domain system modeling and problem formulation requires in-depth telecom knowledge associated with factual reasoning capability. Therefore, a benchmark capable of evaluating the math modeling capabilities of LLMs in telecom domain is critical for using LLMs on telecom math problems.

In order to have a simple but effective benchmark for telecom math modeling, we developed a masked equations in-filling task. Specifically, we first extract the system modeling and problem formulation paragraphs from telecom domain research papers or technical documents. Then, we mask some crucial equations representing the modeling process, such as system model, channel model, modulation and coding. The masked equations will be replaced by a special placeholder. Meanwhile, we try to avoid masking equations related to definitions or referring to other documents, since they are unique to the paper rather than general telecom knowledge. Finally, we ask the LLM to recover these masked equations, which will be compared with the ground-truth to obtain a quality score. The LLM will predict merely one equation at a time in order to ensure the efficiency of running the benchmark. To further simplify the task, we include the ground-truth equations in the paragraph before the masked equations.

We utilize MathBERT, a variant of BERT fine-tuned on a large dataset of math equations, to evaluate the cosine similarity between the embedding of the predicted equations and the ground-truth. The cosine similarity of MathBERT is adapted to the operation trees of the equations and thus can be used as a semantic similarity in the sense of math structure. For a given equation represented by $y$ and a predicted equation $\hat{y}$ given by an LLM, the MathBERT score is defined as:

$$\text{score}(y, \hat{y}) = \max\left\{\frac{\cos(e(y), e(\hat{y})) - \cos(e(y), e(\text{Ø}))}{1 - \cos(e(y), e(\text{Ø}))}, 0\right\} \times 100\%$$

where $e(\cdot)$ is the embedding output of the MathBERT; cos denotes the cosine similarity between two embedding vectors; Ø represents an empty equation where LLM returns nothing. Furthermore, we normalize the cosine similarity to the range of $[0, 100]$ with regards to the difference between the ground-truth and the empty answer. Since the raw cosine similarity between them is usually greater than 0.7, the normalized score better aligns with the human common sense.

Example results of this benchmark on typical SOTA LLM is shown in Table 13.

## 6.7 Generating Commit Messages for Configuration Files in 5G Network Deployment Using LLMs: Evaluation

This use case was described in more details in Section 8.7. In this Section, the aim is to present the evaluation of the proposed prompting solutions for the automation of generating commit messages. The primary objective of the study in paper [207] is to examine the ability of LLMs to automatically generate meaningful commit messages that are not only technically accurate but also contextually appropriate for network automation. As presented in Section 8, we applied five types of prompting:

| Model | Average Score | $\geq$ 90% | $\geq$ 50% |
|---|---|---|---|
| GPT-4 | 49.38 | 3.77 | 50.35 |
| GPT-3.5 | 43.53 | 1.81 | 40.44 |
| Llama3-8B-Instruct | 40.78 | 2.51 | 34.45 |
| Mistral-7B-Instruct | 35.54 | 1.53 | 29.43 |
| Mistral-8x7B | 43.62 | 2.51 | 41.98 |

Table 13: Performance of Telecom Math Modeling Benchmark of SOTA LLM.

- **Prompt 1**: Basic Prompt with a background field to explain the role we want the model to consider when answering, the task, the input, and finally the output that contains the desired output format of the commit message as showcased in Figure 49.

- **Prompt 2**: Prompt with Negative Instruction

  This prompt adds a negative instruction to Prompt 1. (*Your output must be strictly in one line and in the format '<type>[optional scope]: <description>' without any extra text like 'this is the commit message:' etc., neither before, nor after*). The purpose of the negative instruction is to guide the model to avoid generating commit messages that do not meet the task requirements, such as avoiding unclear or overly simplistic statements.

- **Prompt 3**: Prompt without Repository Tree

  Prompt 3 removes information related to the code repository tree from Prompt 2. This change aims to test the model's performance without specific code organizational structure information, thereby assessing the model's sensitivity to environmental dependencies.

- **Prompt 4**: Prompt with one-shot

  Prompt 4 adds a specific commit message example (one-shot) to Prompt 2. This method helps the model learn how to construct commit messages through a concrete example, potentially improving the accuracy and relevance of the generated information.

- **Prompt 5**: Prompt with RAG

  Prompt 5 integrates RAG into Prompt 2. through RAG technology, the model queries related documents or existing data before generating commit messages, enhancing the accuracy and richness of the generated content.

We also considered three models for prompting: Mistral, Llama3 and GPT4. Two methods of evaluation was applied: automated evaluation with metrics and human evaluation. in the automated evaluation three metrics were considered: METEOR, BLEU and ROUGE.

This manual evaluation involved four human evaluators, all of whom are project staff members. They reviewed a random sample of 10% of the total commits, amounting to 50 distinct commits. Each commit was assessed 11 times, leading to a total of 550 individual evaluations. Among these, 30 commits were authored by humans, while 20 were generated by machines. Since the large-scale GPT series developed by OpenAI is not open-source, we opted to manually utilize the GPT-4 model for our studies to maintain information security. Ultimately, we only conducted the experiment for prompt 1. Due to the limitations of the chat platform, we were unable to automate the retrieval of commit messages; instead, we had to input them manually and generate multiple messages simultaneously. Consequently, our experimental approach differed from that of other models and was quite time-consuming.

The evaluators rated the commit messages generated under various prompts using a scoring scale from 1 to 3, defined as follows:

- **1 Point**: Does not meet standards.
- **2 Points**: Meets basic standards.
- **3 Points**: Fully meets standards.

Ratings were based on five criteria:

1. **Accuracy**: Determines if the generated commit message accurately reflects the code changes.
2. **Integrity-What**: Assesses whether the commit message fully describes the changes made.
3. **Integrity-Why**: Evaluates if the commit message explains the reasons behind the changes.
4. **Readability**: Checks if the language of the commit message is clear and correctly formatted.
5. **Applicability**: Evaluates whether the commit message is relevant in a real software development context.

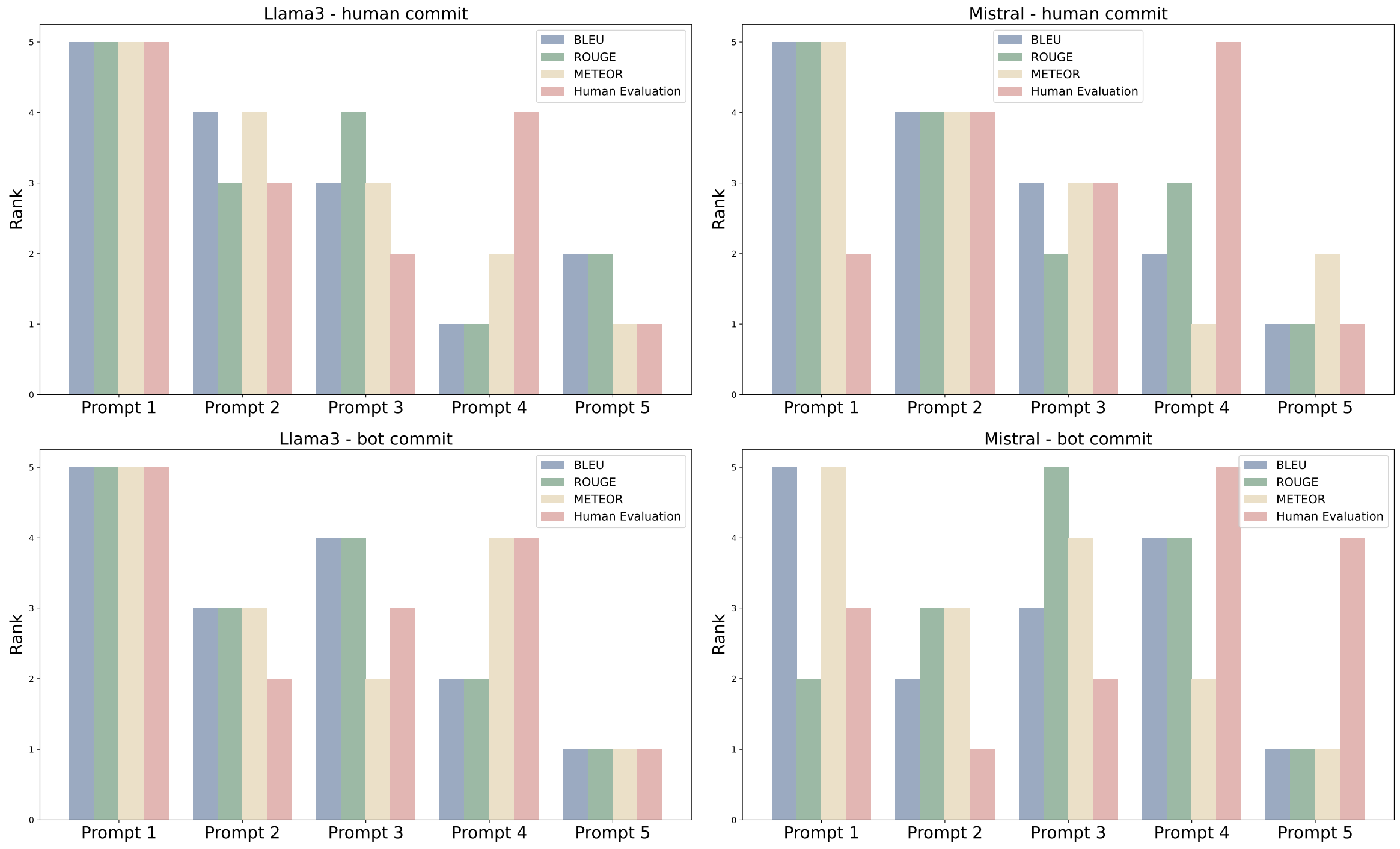

Figure 32: Comparing human evaluation and objective metrics

An interesting result to mention about the evaluation is the comparison between the human and metrics evaluation as illustrated in Figure 32. In Fig. 32, we compare human evaluation results with objective metrics through ranking. The Llama3 model exhibits relatively stable scores in both human evaluations and objective metrics, suggesting a degree of consistency between human preferences and automated scoring. For human commits, prompt5 or occasionally prompt4 yield the best results, while for bot commits, prompt5 consistently outperforms the oth-ers. This indicates that prompt5, and sometimes prompt4, is the top performer across all metrics. Mistral reveals some inconsistencies between human evaluations and objective metrics, especially in bot commits. Although objective metrics show very high scores for prompt5, human evaluations, while still high, are not as extreme. This implies that Mistral may excel in certain automated metrics, but human evaluators perceive its outputs differently, possibly due to subtleties that metrics like BLEU, ROUGE, and METEOR may not fully capture. The *Integrity why* score is consistently around 2, indicating that the model struggles to fully understand the context of changes. Additionally, the *Applicability* score hovers around 2, except for bot commits with Llama3, suggesting that the models are not yet fully ready for use without human oversight. In this use case, we demonstrated the importance to use both evaluation metrics and human evaluation to benchmark models. However, the human evaluation is laborious and takes time. One way to resolve this, is to consider the definition of use case specific metrics for evaluation.

## 6.8 Large Model Evaluation System from Telecom Operators

With the rapid development of big model technology, telecom operators begin to rely more and more on big models to improve the intelligence of their network operation and maintenance, customer service and business processes. In order to ensure the efficient application of big model in telecom business, operator big model evaluation system emerges as The Times require. This system not only focuses on the intelligence level and task execution ability of the model, but also systematically investigates its performance optimization, security assurance and application scenario adaptability.

The design of the operator big model evaluation system follows the four core dimensions of "ability, task, performance, security", and each dimension revolves around the actual needs of the telecom operation. Firstly, the ability evaluation mainly investigates the comprehensive performance of the model in terms of knowledge coverage, cognitive reasoning, interaction generation, and agent scheduling. Telecom operators need large models that can quickly respond to complex tasks such as network status query, fault diagnosis, and resource scheduling. Therefore, knowledge and reasoning ability are crucial. In addition, large models need to have good interaction capabilities to ensure that they can generate accurate responses or suggestions when interacting with users, operations personnel, or other systems. The ability of Agent scheduling is the key for large models to perform tasks automatically in a multi-task environment, especially in the multi-dimensional optimization and resource scheduling scenarios involved in telecom networks.

Performance evaluation plays an important role in operator large-scale model evaluation, mainly investigating the inference speed, resource consumption, throughput and concurrent processing ability of the model. Telecom operators have high network scale and real-time requirements, so large models need to have efficient inference speed and reasonable utilization of computing resources. The evaluation of throughput and concurrency is aimed at the processing ability of the model in the face of large-scale user requests or network operation and maintenance tasks, to ensure that it can work stably and efficiently in practical applications. The performance evaluation also covers the adaptability of the model in different hardware environments, such as whether the model can maintain excellent performance under resource constraints with the support of high-performance processors (e.g., NVIDIA, Huawei, etc.).

Security evaluation is the basis to ensure the safe operation of large models in the telecom field. The application of big models in the telecom industry involves the processing of a large amount of sensitive information, so data privacy protection and content compliance have become important evaluation contents. Through the security test of the model, the evaluation system ensures that the content generated does not involve the disclosure of sensitive information or violate industry standards. In addition, the adversarial attack defense ability is also included in the evaluation scope, which mainly investigates the robustness of the model in the face of malicious inputs or attacks.

The security evaluation system of the operator's large model also involves the detection and early warning of potential security vulnerabilities to ensure that the model will not cause network security incidents when dealing with complex tasks.

### 6.8.1 TianGong Large Model Evaluation System

China Telecom TianGong Large Model Evaluation System provides a comprehensive framework for evaluating large models from the dimensions of capability, tasks, performance, and security. Its multi-layered, modular design ensures that the model performs well in all key areas, offering robust support for large model applications in the telecom industry. Through the TianGong system, the telecom industry can better assess and optimize the application of large models, promoting the intelligent upgrade of network management, customer service, and business operations.

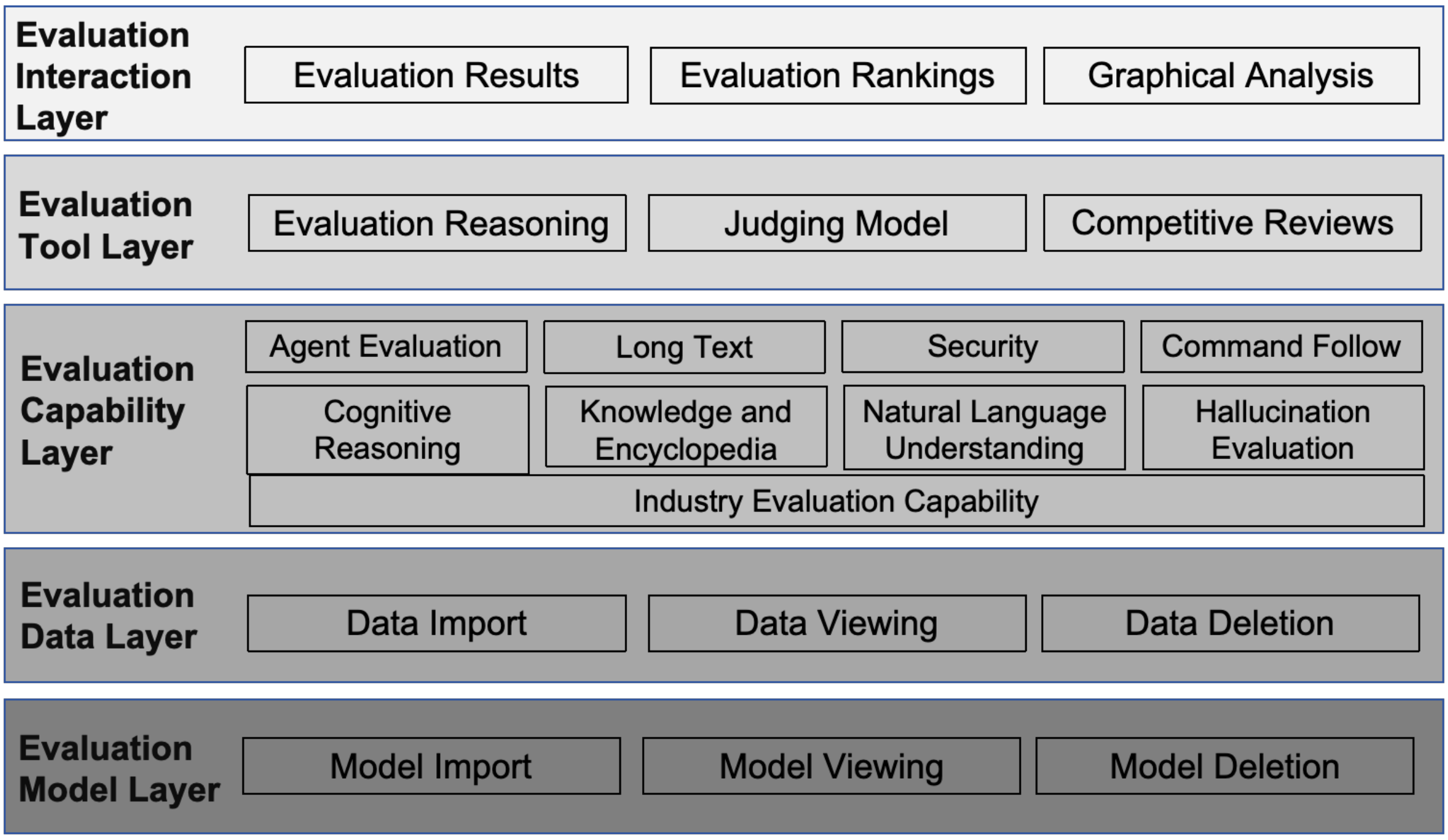

Figure 33: TianGong Large Model Evaluation System.

The system is divided into five layers: **evaluation interaction layer**, **evaluation tool layer**, **evaluation capability layer**, **evaluation data layer**, and **evaluation model layer**. These layers work together to complete a comprehensive assessment of the model. The evaluation interaction layer is the core module for presenting evaluation results, generating rankings, and producing graphical analysis. By providing visualization and result display, the evaluation system not only offers comprehensive data presentation but also makes the complex evaluation results easier for users to understand, facilitating the optimization and improvement of the model's application.

The evaluation tool layer provides automated reasoning functions and scoring mechanisms to standardize and automate the evaluation process. Through the integration of automated tools, large-scale evaluations can be conducted rapidly, reducing human intervention and ensuring fairness in the assessment. The competitive review mechanism enables different models to be fairly compared under the same tasks, evaluating their performance in a parallel context. This mechanism is particularly useful in the telecom industry, where large-scale, multi-scenario business tasks need to be processed simultaneously.

The evaluation capability layer is the core of the TianGong system and performs detailed grading and assessment of various model capabilities. The capability evaluation includes modules for agent evaluation, long-text process-ing, security evaluation, command-following capabilities, and reasoning abilities. Notably, agent evaluation is critical because it assesses the model's performance in automated task execution, especially in telecom network management, where agent efficiency directly impacts network optimization and resource scheduling. Given the telecom industry frequent need for handling extensive textual data, such as generating reports or providing network configuration recommendations, the TianGong system includes a long-text processing evaluation module to ensure that the model can maintain logical consistency and accuracy when generating long-form content. Additionally, the security evaluation module is a key feature of the TianGong system, testing the model's ability to handle sensitive information and protect against malicious attacks, ensuring that the model effectively safeguards network security and data privacy during actual use.

It is also worth noting that cognitive reasoning evaluation is highly emphasized in the TianGong system. This capability serves as the foundation for the model's ability to perform complex reasoning and decision-making, particularly in tasks such as telecom network fault diagnosis and traffic prediction. The strength of the model's reasoning ability directly impacts its performance in these complex tasks, so the evaluation system conducts in-depth assessments to ensure that the model is capable of handling high-demand reasoning tasks in real-world scenarios.

Unlike general telecom operator large model evaluation systems, the TianGong system specifically includes an industry task evaluation module. This module focuses on optimizing the model for key tasks in the telecom industry, ensuring that it excels in industry-specific tasks. These evaluation tasks include network traffic prediction, news content generation, and customer service interaction. For example, in the network prediction task, the model must predict future network traffic and resource demand based on historical data, helping telecom operators preemptively allocate resources and optimize network performance. In customer service interaction tasks, the model is tested on its ability to interact with customers, assess response speed, understand customer intentions, and generate accurate responses. Through these task evaluations, the system ensures that the large model performs optimally in the core business of the telecom industry.

In terms of security, the TianGong Large Model Evaluation System has established detailed security evaluation standards to ensure that the model does not pose security risks when processing sensitive information and generating content. Through content compliance testing, the system ensures that the model can filter and identify illegal information and sensitive content, preventing the generation of policy-violating content. Especially within the legal framework of the telecom industry, the model must meet strict compliance requirements to ensure that its output does not violate regulations. The evaluation system also fully tests the model's defense capabilities, ensuring that it can effectively defend against malicious input, network attacks, and other threats, thereby safeguarding the telecom network's security.

### 6.8.2 TelcoLM Large Model Evaluation System

TelcoLM, developed by Orange, is a language model adapted to the telecommunications domain. The adaptation process started from the Llama-2-7B base model and followed two main steps: (1) a domain adaptation phase using a corpus of 800M tokens collected from telecommunications technical documents (including 3GPP, ITU, and ETSI standards, research papers from arXiv's Networking and Internet Architecture category, and filtered web content from telco-related domains), and (2) an instruction-tuning phase using 80,000 domain-specific instructions. These instructions were generated through a combination of automatic transformation of technical documents and human-curated question-answer pairs from telecommunications forums. All fine-tunings are full fine-tunings (as opposed using LoRA).

The evaluation of TelcoLM was conducted using multiple test sets, divided into domain-specific and general-purpose evaluations. This section describes the evaluation protocol and presents the results obtained by Orange.

**Telecom-Specific Tests:** Telecom-specific test set have been designed to evaluate the telecom knowledge integrated in TelcoLM.

- TeleQnA Test Set: A collection of 900 multiple-choice questions developed by Huawei to evaluate telecommunications knowledge. Questions cover various aspects of telecommunications technology, with 4-5 options per question and the possibility of multiple correct answers.

- Nokia Certification Exams: A set of 632 multiple-choice questions extracted from Nokia's official practice exams. These questions specifically come from the NRS I, NRS II, and SRA certification programs, covering topics in signal processing and network engineering.

- Domain-Generated MCQs: Questions automatically generated from technical specifications using GPT-4. The source materials included 3GPP, ATIS, and ETSI standards, focusing on formal technical knowledge assessment.

**General-Purpose Tests:** The goal of the general-purpose tests is to make sure that the adapted LM has not significantly forgotten general knowledge.

- OpenBookQA: 500 questions from the test set of the official OpenBookQA dataset, focusing on elementary-level science.

- TruthfulQA: The multiple-choice subset of the TruthfulQA dataset, comprising 817 questions on general knowledge.

- Big-bench: 600 questions from the "abstract narrative understanding" subset of the validation set.

The evaluation employed different metrics depending on the task type:

- For multiple-choice questions: accuracy (percentage of correct answers)

- For question-answering tasks: Meteor scores and LLM-based scoring using GPT-4 as a judge (from 1, bad, to 5, perfect).

**Results:** Tables 14 and 15 present the results obtained for different versions of TelcoLM using continued pre-training and instruction-tuning, or instruction-tuning only. The baseline model is Llama-2-7b. TelcoLM versions are compared to this baseline as well as the chat version of Llama-2-7b, GPT3.5, and GPT4. On telecom-specific tasks, TelcoLM achieves scores closer to GPT-3.5, better than the base Llama2-7B model or its chat version. On general-purpose tasks, TelcoLM maintained performance. The question-answering evaluation leads to similar conclusions. The LLM-based scoring shows that telecom-specifc questions without questions are difficult for general LMs, even GPT4.

Table 14: Performance of TelcoLM on MCQA tasks: telecom domain (top); general domain (bottom).

| Model | Cont. pre-tr. | Inst. tuning | Telco domain | | | | | |
|---|---|---|---|---|---|---|---|---|
| | | | ATIS | 3GPP | ETSI | TeleQnA | Nokia | Avg. |
| Llama-2-7b | | | 0.62 | 0.46 | 0.41 | 0.56 | 0.34 | 0.48 |
| Llama-2-7b | | yes | 0.69 | 0.50 | 0.52 | 0.61 | 0.32 | 0.53 |
| Llama-2-7b | yes | yes | 0.72 | 0.52 | 0.48 | 0.60 | 0.35 | 0.53 |
| Llama-2-7b-chat | | | 0.65 | 0.45 | 0.47 | 0.45 | 0.35 | 0.47 |
| GPT3.5 | | | 0.71 | 0.57 | 0.61 | 0.61 | 0.45 | 0.59 |
| GPT4 | | | 0.85 | 0.65 | 0.64 | 0.72 | 0.63 | 0.70 |

| Model | Cont. pre-tr. | Inst. tuning | General purpose | | | |
|---|---|---|---|---|---|---|
| | | | OpenBookQA | TruthfulQA | BigBenchNarr.QA | Avg. |
| Llama-2-7b | | | 0.38 | 0.30 | 0.27 | 0.32 |
| Llama-2-7b | | yes | 0.43 | 0.33 | 0.28 | 0.35 |
| Llama-2-7b | yes | yes | 0.48 | 0.21 | 0.31 | 0.33 |
| Llama-2-7b-chat | | | 0.50 | 0.30 | 0.22 | 0.34 |
| GPT3.5 | | | 0.79 | 0.64 | 0.34 | 0.59 |
| GPT4 | | | 0.83 | 0.78 | 0.59 | 0.73 |

Table 15: Performance of TelcoLM of QA without context converted from a subsample TeleQnA.

| Model | Cont. pre-tr. | Inst. tuning | METEOR | MOS |
|---|---|---|---|---|
| Llama-2-7b | | | 0.11 | 1.41 |
| Llama-2-7b | | yes | 0.20 | 2.18 |
| Llama-2-7b | yes | yes | 0.20 | 2.28 |
| Llama-2-7b-chat | | | 0.18 | 2.10 |
| GPT3.5 | | | 0.21 | 2.50 |
| GPT4 | | | 0.24 | 2.64 |

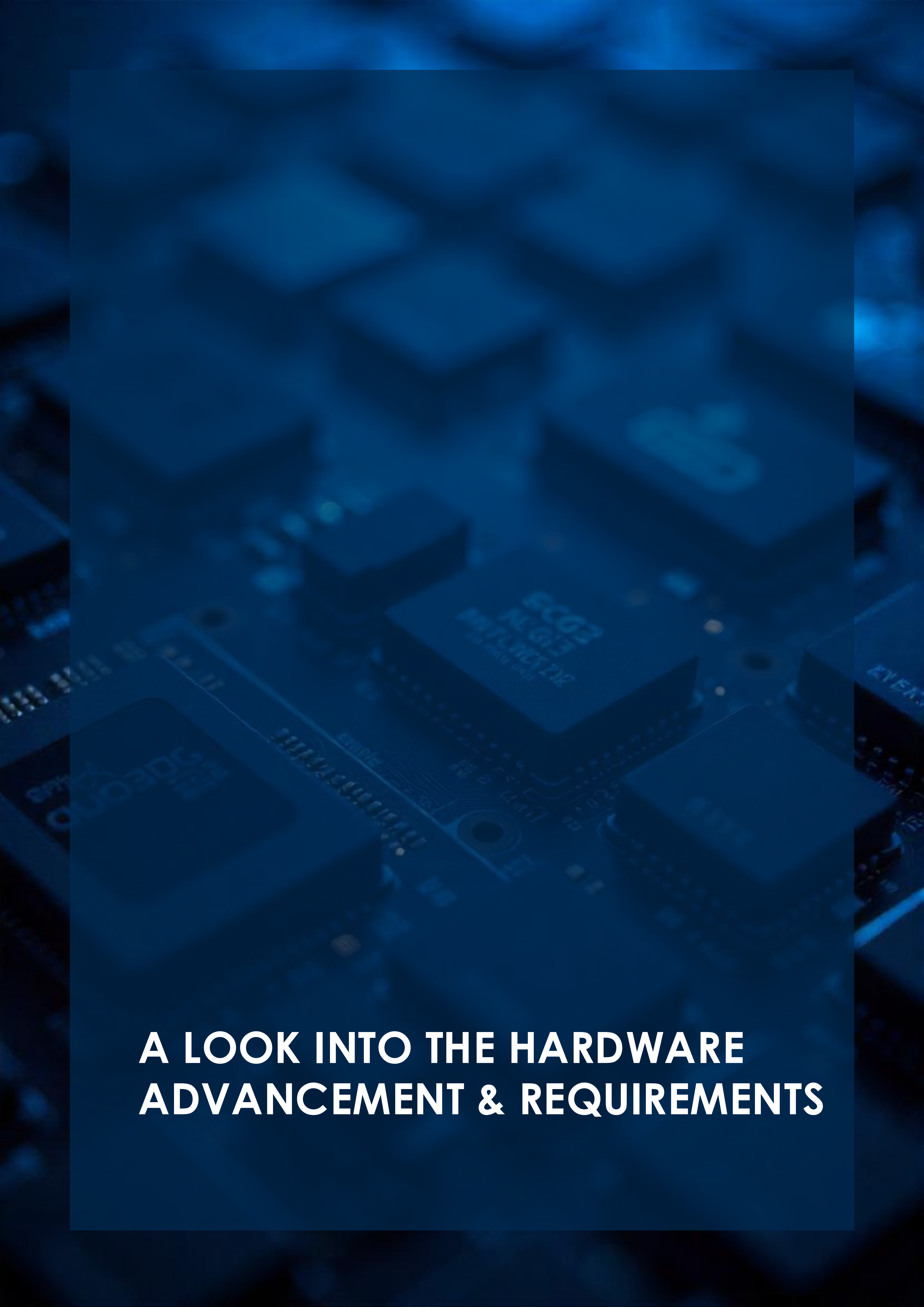

# A LOOK INTO THE HARDWARE ADVANCEMENT & REQUIREMENTS

# 7 A Look into the Hardware Advancement & Requirements

## 7.1 Background

The rapid evolution of AI, particularly in the domain of generative AI (GenAI), has created great demands on computational infrastructure. This shift is particularly evident in the telecommunications industry, which finds itself at the intersection of data management, network operations, and AI deployment.

The telecom industry generates and processes an enormous volume of data, far surpassing many other sectors. Annual mobile data traffic has been on the rise historically [208], with data coming from various sources, including call detail records, network performance logs, customer interactions, and Internet of Things (IoT) devices.

This exponential growth in data volume is accompanied by rising customer expectations, resulting in increased pressure on customer service agents to provide faster, more accurate, and personalized support. Concurrently, novel use cases are emerging that leverage AI to enhance network operations, such as "network agents" that autonomously manage and optimize network performance, alongside predicting and preventing outages [209]. The ongoing research on Telecom Foundational Models also present an opportunity to simplify network operations, enhance development productivity, aid trustworthy networks, and boost business profitability [210]. These AI-driven solutions are becoming crucial for telecom companies to maintain a competitive edge and meet the growing demands of 5G and beyond.

Convergence of massive data scales, heightened customer expectations, and emerging AI applications present both challenges and opportunities for the telecom industry, necessitating rapid advancements in hardware infrastructure to support these data-intensive and computationally demanding workloads.

## 7.2 Telecom Companies as AI Factories

Telecommunication companies, having an extensive array of centralized datacenters and distributed sites close to and at the edge of their networks, are at a pivotal moment when they can transform from being merely data conduits and cloud computing consumers to being AI-driven service providers, and even compute providers in their own right. As trusted national technology providers, telecoms in many countries are in an ideal position to provide sovereign AI for regional governments, enterprises, and startups, enabling them to build, customize, and deploy GenAI applications [211].

By transforming into an "AI Factory," telecoms can leverage accelerated computing infrastructure, software, and services in their existing datacenter footprint to deliver AI intelligence at a national scale. In addition to running their own AI workloads for internal operations and customer services, the AI Factory approach enables telecoms to evolve from being merely an IT consumer to becoming a provider of infrastructure and services themselves. This change in roles creates a new business opportunity where customers run their own AI workloads on telecom infrastructure on a subscription basis. This approach gives telecom companies a competitive edge, as it moves them from providing only commodity-based connectivity, to being a modern AI content, application, and services provider.

## 7.3 The Rise of HPC-AI

The introduction of the transformer neural network architecture by a team at Google [48] enabled the rapid growth and adoption of LLMs. Correspondingly, the demands for the growth in scale, capability, and speed of computational resources have skyrocketed.

Relatively flat growth in CPU computational power compared to the growth in LLM size has created a need for hardware accelerators that can keep pace with the demand. It was realized in the 2000s that the graphics processing unit (GPU), a processor originally designed to accelerate real-time computer graphics, could also be used to parallelize the matrix multiplications used in neural networks [212]. GPUs found a role as a dedicated "vector" processor for other computing workloads and gradually became an increasingly used accelerator in high-performance computing (HPC) sites throughout the world. With the advent of LLMs, GPUs and other matrix multiplication accelerators evolved to become indispensable tools for AI.

### 7.3.1 LLM accelerator chip technology

The burgeoning demand for LLM training and inference has fueled rapid advancement of GPU technology in addition to a host of competing accelerators to challenge the dominance of the GPU as the leading LLM training and inference engines. Originally developed for rendering 3D graphics in gaming, GPUs have evolved into versatile processors widely used in artificial intelligence (AI), machine learning (ML), data science, video editing, and scientific research. For example, NVIDIA has developed several lines of datacenter GPUs with varying capabilities, size, and energy consumption that are designed to accelerate intensive computing tasks that can benefit from parallel processing. At the time of this writing, the current NVIDIA Blackwell GPU architecture introduces specific Transformer Engines [213] to perform LLM processing using 8-bit floating point (FP8) precision [214], with the ability to quantize LLMs down to the FP4 level. In addition to NVIDIA, AMD, Intel, and Google also produce GPU accelerators for AI. IBM has developed accelerators for their own Z Series mainframe platforms. In addition to GPU-based accelerators, many alternative AI accelerator architectures are also under development [215].

For software developers to leverage the AI acceleration afforded by GPUs, or any other accelerator hardware, requires a translation layer that makes it easy to add directives to traditional CPU-based code that can offload parallelizable work to the GPUs in a transparent fashion. Because of this requirement, NVIDIA has developed an extensive set of drivers, libraries and AI frameworks that can be used in a stand-alone solution or to accelerate other popular AI frameworks. At a lower level, CUDA (Compute Unified Device Architecture) allows developers to harness GPU power for parallel computing. The NVIDIA Collective Communication Library (NCCL) [216] enables high-performance inference and training over multiple GPUs across networks, which is essential for training the largest LLMs in use today in a feasible length of time. The NVIDIA AI software stack provides a hierarchy of abstraction layers that make GPU acceleration readily consumable and can be used to create a complete GenAI ecosystem [217]. Correspondingly, any AI accelerator entering the market needs a mature software framework to facilitate its adoption by telecoms.

### 7.3.2 The Role of High-Performance Networking in LLM Processing

The demand for high-performance computing and accelerators for LLM workloads is met by an equal demand for high-performance networking. The need for many highly coordinated and low-latency computing resources for LLM training is driving the evolution of cloud-based resources, and a movement from general commodity public-cloud consumption to the use of dedicated "private clouds" that adopt the architecture, scalability, and performance of HPC systems typically found at academic and government-sponsored supercomputing centers. Training LLMs also requires using multiple accelerators in parallel to reduce training times. This is achieved at the server level, with multiple AI accelerators communicating over the PCIe bus or using dedicated high-bandwidth connections such as NVIDIA® NVLink® [218].

NVLink Switch technology extends NVLink to provide a topology connecting multiple GPUs. Starting with NVIDIA Grace Blackwell GB200-based systems, up to 72 GPUs residing in multiple trays of a compute rack may be connected [219]. As the scale of LLM models grows and high-bandwidth, low-latency cross-server data movement becomes increasingly critical to AI training performance. Fabrics based on the InfiniBand standard [220] have long led performance in latency and effective bandwidth, providing lossless data transfer via RDMA (direct memory access bypassing CPU buffers), advanced congestion control, and adaptive routing between fabric endpoints [221, 222]. The InfiniBand Trade Association (IBTA) also promotes RDMA over converged Ethernet (RoCE) as a means of using RDMA over Ethernet fabrics [223].

Due to the prevalence of Ethernet and the level of familiarity consumers have with it, there has been a large incentive for vendors to develop Ethernet with true HPC-level performance approaching that of InfiniBand. NVIDIA, the main InfiniBand developer, has developed NVIDIA Spectrum™-X, an Ethernet solution for high-performance AI training, that delivers RDMA over converged Ethernet (RoCE), adaptive routing, and the same telemetry-based congestion control using endpoint telemetry found in InfiniBand [224, 225]. Spectrum-X requires a specific combination of hardware on both the switch side (select models with the Spectrum-4 ASIC and later) and on the SuperNIC side (specific NVIDIA® BlueField®-3 and ConnectX®-8 network interface card (NIC) models validated and enabled for Spectrum-X) [226, 227]. In addition, the Ultra Ethernet Consortium, with over 55 members, is currently developing a specification to ”deliver an Ethernet based open, interoperable, high performance, full-communications stack architecture to meet the growing network demands of AI & HPC at scale” [228].

Over the years, demand has increased for many organizations to move toward running workloads in public clouds, even computation-intensive scientific and AI workloads that have traditionally been hosted in shared or dedicated supercomputing centers. One prominent example is NOAA’s Cloud-based Warn on Forecast system (Cb-WoFS) [229]. The telecom industry has also made inroads into public clouds. Joint research by SoftBank and the University of Tokyo has created a “stateless” 5G core network on Amazon Web Services (AWS) [230]. Deutsche Telekom and T-Mobile have collaborated with cloud providers to provide cloud-based 5G and AI services with 5G Advanced Solutions [231] and Magenta Cloud [232, 233, 234]. This performance demand has driven the large cloud service providers (CSPs) to provide co-located compute instances that run on entire, dedicated bare-metal servers with a high-performance interconnect for scalability. An additional imperative for CSPs is to avoid network fabric congestion from multiple tenants. Microsoft Azure instances [235] use InfiniBand with partition keys [236] and advanced congestion control to enable multiple tenant workloads to run in parallel without interfering with each other. Amazon AWS uses its own high-performance Elastic Fabric Adapter (EFA) for high-performance workloads [237].

## 7.4 Building Blocks for AI Infrastructure Deployment

As telecoms are relatively new to the concept of building HPC systems, hardware vendors have started to provide roadmaps. They work with ecosystem partners to provide certified reference architectures (RAs) for building optimized and scalable AI infrastructure and cloud services for training of and inference on GenAI models.

The evolution of HPC in the era of GenAI is that of increasing scale and complexity of infrastructure building blocks. For architectures to effectively address the scale of LLMs, the data center has become the new unit of compute rather than individual servers. Interconnected GPUs, CPUs, memory, storage, and other resources across multiple compute nodes orchestrate large-scale AI workflows. This infrastructure requires not only high-performance compute and networking previously discussed, but carefully designed storage, cooling technologies, and power delivery to sustain optimal performance and efficiency for each data center environment. For companies that do not have expertise in designing, procuring, and running HPC systems, moving into the realm of large-scale AI training and inference can be a daunting, high-risk endeavor.

To lower the barrier to entry, several vendors provide foundational building blocks to build, customize, and deploy rapidly evolving generative Al and large language models (LLMs), both on-prem (Dell, HPE, NVIDIA, Supermi-cro, Deloitte [238] and others) and as hosted private clouds (e.g. CoreWeave [239]). Turnkey data center solutions accelerate time-to-delivery and eliminate the complexity of building a large compute cluster. Such compute infrastructure was previously only achievable through intensive design, tuning, and time-consuming optimization of supercomputing resources.

An example of this building-block approach are Supermicro SuperCluster system design [240]. In these systems, the core compute component for AI training is a server with 8-way NVIDIA HGX H100, H200, or Blackwell GPUs. Each GPU is in a dedicated PCIe 5.0 slot and paired with NVIDIA Quantum-2 400Gb/s InfiniBand networking. NVIDIA ConnectX-7 network interface cards (NICs) enable NVIDIA GPUDirect RDMA and storage for direct data flows to GPU memory to minimize latency. The systems are available in both air-cooled and liquid-cooled variants. NVIDIA® NVLink® interconnects the GPUs for high GPU memory bandwidth and enables capacity to run LLMs cost-effectively. These solutions are designed in a building-block approach, starting from a given "pod" or scalable unit. For scalability, the largest training jobs leverage InfiniBand or Spectrum-X interconnects for high-speed data passing between these pods. For example, the leaf-spine network topology in a SuperCluster allows it to scale from 32 8-way GPU nodes to thousands of nodes. An added advantage of this building-block approach is that the vendor has already validated performance at scale and has made component and design selections that ensure scalability of performance.

Along with compute and networking, high-performance storage is also an integral part of an AI computing environment. Storage is required for storing model weights, inference query history, checkpoints, and for caching data. The storage needed for AI factories differs from typical enterprise storage. Fortunately, several vendors have developed scalable solutions that are designed for AI and certified with AI infrastructure providers, such as DDN [241], VAST Data [242], WEKA [243], Dell [244], Pure Storage [245], and IBM [246]. These vendors offer ongoing support and performance tuning of those solutions once deployed. Such storage solutions mesh ideally into the building-block approach for AI factories.

## 7.5 Convergence of AI and RAN

The advent and adoption of 5G technology has introduced a new level of speed and capability in networks, enabling them to deliver not only voice, but also Internet-based services, data, and streaming. This generalization of telecom networks into a data service network allows the introduction of LLM-driven services. A further benefit comes from the flexibility built into 5G network architecture, allowing different components to be strategically distributed in different locations, from the radio unit (RU) site at the very edge, to DU sites (such as MSOs), to the CU which has more space and power available. This flexibility also means that components can be decoupled, such that following Open RAN (O-RAN) specifications enables a merger of the best solutions for both RAN and AI processing while meeting latency demands [247]. This flexibility allows smaller footprints for AI inference to be located closer to the network edge, where response time is at a premium, while larger-scale computation and training can be done in centralized datacenters.

The merger of AI with 5G/6G RAN technology, referred to as "AI-RAN," opens a broad array of possibilities for telecoms, with key avenues being defined by the AI-RAN Alliance [248]. The AI-RAN Alliance outlines three key synergies between AI and RAN:

1. **AI-for-RAN**, which focuses on embedding AI into radio signal processing to improve RAN performance metrics such as spectral efficiency;

2. **AI-and-RAN**, which enables a common accelerated infrastructure to host 5G RAN software and AI workloads concurrently (as first demonstrated by a SoftBank and NVIDIA collaboration [249]), maximizing platform utilization and creating new AI monetization opportunities; and

3. **AI-on-RAN**, which considers the requirements to run AI applications over 5G and 6G infrastructure.

A recent example of this AI-RAN synergy is demonstrated by NVIDIA AI Aerial™ platform, [250] which includes hardware and software components to enable AI-RAN development and deployment. NVIDIA's Aerial RAN Computer-1 [251] provides a common GPU-accelerated infrastructure to host 5G radio processing and AI workloads at the edge, while NVIDIA Aerial CUDA-accelerated RAN provides the software libraries to build a full stack 5G virtual RAN. This capability enables LLM inference to be served closer to the edge and provides the ability to dynamically orchestrate RAN and AI workloads as demand changes over the course of a day, maximizing infrastructure utilization [252].

This convergence of AI and RAN also means that both the RAN and AI infrastructures can be designed as one platform that works together seamlessly, not two separate platforms that need to be co-housed and made to interoperate [253]. SoftBank has already completed the first live field trial of AI-RAN [254, 255] showcasing its superior performance and economics and is also driving the commercialization of AI-RAN with the launch of their own product – AITRAS [256, 257]. T-Mobile and NVIDIA have recently announced a joint AI-RAN innovation lab with Ericsson and Nokia [258, 259] to define a scalable commercial AI-RAN solution built on accelerated infrastructure.

## 7.6 Conclusion

LLM-driven services show much promise in the telecom industry in delivering content, providing AI-driven services, facilitating the internal operations, and optimizing network performance and maintenance. While telecoms have traditionally focused on delivering connectivity and content over wired and wireless networking, LLM-driven applications and advanced computing technologies create a new opportunity for telecoms to also play a role in the AI value chain [260].

LLMs have been enabled by key advancements in GPUs and other computing accelerators, as well as advances in networking that enable the performant resource scaling and parallel processing required to make LLM training feasible. Along with high-performance storage, LLM training and large-scale inference systems are very similar to a typical HPC supercomputing environment. Such complex environments take special design and operational expertise not typically extant in enterprise computing datacenters. Fortunately, key hardware advancements in compute, networking, and storage that drive LLM training and inference today are no longer exclusively the purview of large-scale HPC facilities but are being offered as certified reference architectures that can be delivered in a building-block approach, providing telecoms an onramp to LLM-driven transformation. Telecom operators now have the opportunity to become national AI infrastructure and service providers by building an AI grid spanning across both centralized AI factories and distributed AI-RAN datacenters providing LLM inference at and near the network edge.

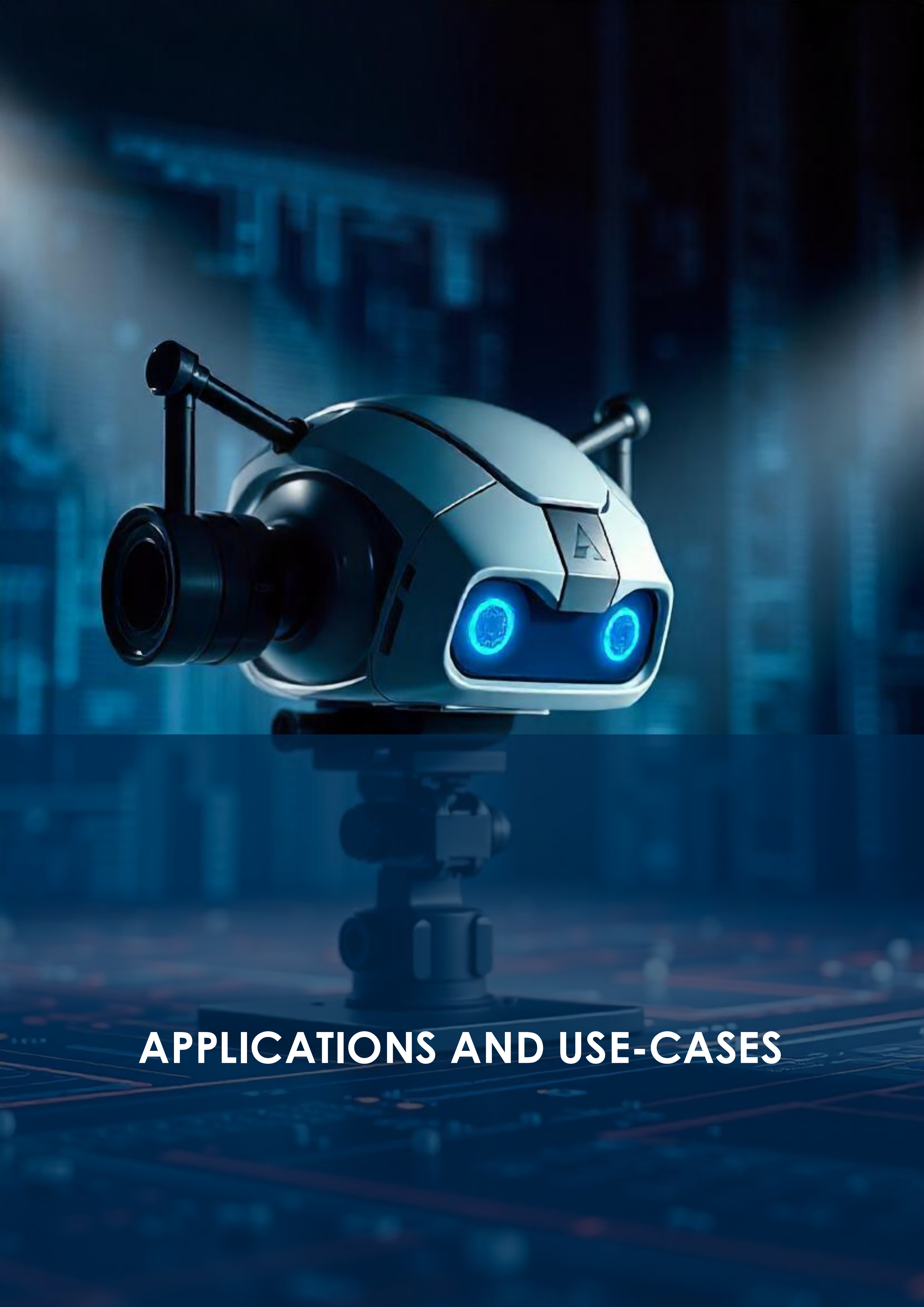

# APPLICATIONS AND USE-CASES

# 8 Applications and Use-Cases

## 8.1 LTMs at the Edge

### 8.1.1 The Synergy of LTMs and Edge Computing

The integration of large telecommunication models and edge computing represents a paradigm shift in computational and networking technologies. Traditional cloud-centric frameworks are increasingly supplemented by edge computing, which processes data closer to the source. This evolution is particularly vital for modern applications like autonomous systems, industrial IoT, and smart cities, where real-time processing, low latency, and privacy are crucial. The fusion of these two domains utilizes the immense power of large AI models while addressing the resource constraints of edge environments. By blending computational advances with distributed networks, this synergy promises to transform industries while overcoming significant technical challenges [261, 262].

Large language models are foundational components of modern AI, with architectures like GPT-4 and LLaMA employing billions of parameters. These models excel in tasks requiring contextual understanding, such as natural language processing, multimodal reasoning, and downstream task generalization. Their power derives from the transformer architecture, which uses self-attention mechanisms to capture long-term dependencies in data [261, 262].

Despite their capabilities, deploying LLMs poses significant challenges. Their enormous size and resource-intensive nature often exceed the computational, memory, and energy capacities of edge devices. Techniques like model pruning, quantization, and parameter-efficient fine-tuning have been developed to address these constraints. These innovations enable LLMs to be tailored for specific applications, maintaining performance while minimizing resource usage [263, 264].

- **The Role of Edge Computing**
  Edge computing decentralizes computational power, placing processing closer to data sources. Unlike cloud-based systems, edge computing reduces latency by processing data locally. For applications like autonomous driving, where split-second decisions are critical, this capability is indispensable. Additionally, edge computing minimizes bandwidth usage, as only essential data is transmitted to centralized systems, and enhances privacy by keeping sensitive information localized [262, 265].

  A growing area within edge computing is collaborative edge systems, where multiple edge devices work together to share computational loads. This approach transforms previously idle resources into active participants in data processing, significantly improving efficiency [263, 265].

- **Integration of LLMs with Edge Computing**
  The fusion of LLMs and edge computing addresses the trade-off between the high performance of large models and the constrained resources of edge environments. Innovations in model optimization have made it feasible to deploy LLMs at the edge. For instance, model quantization reduces precision requirements, while pruning removes redundant parameters. Techniques like EdgeShard and collaborative edge frameworks allow LLMs to be distributed across multiple devices, balancing workloads and ensuring faster processing [263, 264].

  Moreover, a hybrid cloud-edge synergy enables the most resource-intensive computations to be offloaded to the cloud while retaining real-time, latency-sensitive tasks at the edge. This division of labor ensures optimal performance without overburdening edge devices [266, 265]. For example, applications such as video analytics for public safety or predictive maintenance in industrial settings rely on cloud-edge collaboration to analyze massive datasets in real-time.

- **Applications and Use Cases**
  The integration of LLMs and edge computing opens doors to numerous applications. In autonomous systems like robotics or smart factories, edge computing facilitates rapid decision-making, ensuring that tasks such as object recognition or movement coordination occur without delays. For instance, edge-deployed LLMs allow real-time adjustments in robotic systems, enhancing both precision and reliability [265].

  In healthcare, edge computing provides real-time analytics for diagnostics while preserving patient privacy. Hospitals and clinics can deploy edge-based LLMs for tasks such as analyzing medical images, processing patient histories, or delivering tailored treatment plans, reducing the need to transmit sensitive data to external servers [264, 265].

  Smartphone-based virtual assistants illustrate the consumer-facing potential of this integration. Lightweight LLMs deployed at the edge enable context-aware services, including personalized recommendations, natural conversation capabilities, and localized processing of user data. These features are vital for ensuring a seamless user experience while adhering to strict privacy regulations [262, 263].

  In industrial IoT, predictive maintenance powered by edge computing and LLMs optimizes equipment performance. By processing sensor data locally, companies can identify anomalies, forecast failures, and initiate preventive measures in real time, minimizing downtime and operational costs [263, 265].

- **Challenges and Future Directions**
  Despite its promise, the integration of LLMs with edge computing faces several challenges. Resource allocation is a primary concern, as edge devices must balance their limited processing power with the high demands of LLMs. Efficient algorithms for task scheduling, dynamic resource management, and real-time offloading are critical for overcoming these limitations [266, 265].

  Energy consumption poses another significant challenge. Training and deploying LLMs at the edge require innovative solutions to manage power demands sustainably. Research into green AI practices, such as energy-efficient hardware designs and adaptive inference techniques, is essential for scaling these technologies [265].

  Security and privacy are equally important. Protecting sensitive data during edge-cloud interactions and safeguarding models against adversarial attacks are critical for maintaining trust in these systems. Advanced encryption methods, secure communication protocols, and robust model defense mechanisms are areas of active research [264, 266].

  Future directions include the development of adaptive LLMs capable of real-time learning in dynamic environments. These models could adjust to changing conditions, such as variations in network traffic or user behavior, without requiring complete retraining. Additionally, integrating 6G networks with edge computing could further enhance the bandwidth and connectivity available for deploying large telecommunication models [262, 264].

### 8.1.2 Edge-Enhanced TinyML for Beyond 5G Networks

The growing interest in the TinyML Foundation, recently renamed the EdgeAI Foundation, and the rapid progress of 3GPP toward AI-enabled beyond 5G networks highlight a major shift in communication systems. The parallel efforts of the TinyML Alliance and 3GPP standardization in both RAN (Radio Access Network) and SA (System Architecture) illustrate how edge AI is transforming the design of devices, networks, and their operation.

This transformation is evident in recent advancements such as the emergence of reduced-capability (RedCap) de-vices, progress in integrating sensing capabilities into wireless devices, and the democratization of neural network-based algorithms for processing not only sensor data but also communication signals. These advancements necessitate the adoption of native edge-enabled schemes.

In this framework, edge LLM (Large Language Models) and edge LTM (Large Transformer Models) play a crucial role. Models embedded on devices must be trained and updated using techniques like knowledge distillation, over the air model transfer [267, 268], enabling devices to seamlessly access highly capable computing frameworks to support their native operations. These mechanisms ensure efficient adaptation to changing environments and enhance the performance of resource-constrained devices.

### 8.1.3 Fine-tuning LLMs using Federated Multi-task Learning

The rapid advancement of Large Language Models (LLMs) has revolutionized natural language processing, enabling applications ranging from conversational agents to sophisticated content generation thanks to the ability of LLMs to be multi-task models [269]. However, customizing these powerful models to specific domains or tasks often necessitates extensive fine-tuning, which can be computationally intensive and raise significant privacy concerns. Traditional fine-tuning approaches typically require centralized access to vast amounts of labeled data, which may be impractical due to data ownership restrictions, privacy regulations, or the sheer volume of data involved.

Federated Learning (FL) [270] offers a paradigm shift by enabling multiple clients to collaboratively train a global model without sharing their local data. This approach inherently preserves data privacy and reduces the risks associated with centralized data aggregation. However, traditional FL algorithms are designed to optimize a single global model, which may not perform optimally across diverse tasks or heterogeneous data distributions inherent in different clients. This limitation becomes particularly pronounced when fine-tuning LLMs, which are often deployed across varied domains with distinct tasks and requirements.

Federated Multi-Task Learning (FMTL) [271] extends the FL framework by allowing each client to train a personalized model tailored to its specific task while still benefiting from the collective knowledge of the entire network. By modeling the interactions among clients as a graph where nodes represent clients and edges quantify task similarities, FMTL facilitates the learning of individualized models that are both specialized and informed by related tasks. This approach not only enhances model performance on individual tasks but also ensures scalability and adaptability in heterogeneous environments.

One of the key challenges in fine-tuning LLMs is the computational and communication overhead associated with these models, which can have millions or even billions of parameters. Existing federated fine-tuning of LLMs leverages parameter-efficient fine-tuning (PEFT) methods to enhance communication efficiency and reduce trainable parameters, as seen in frameworks that utilize prompt tuning [272], zeroth-order optimization [273], and Low-Rank Adaptation (LoRA) [274]. However, these approaches typically learn a single global model through averaging, which may be suboptimal in heterogeneous environments. To implement FMTL for fine-tuning LLMs, we introduce MIRA [275], a parameter-efficient algorithm that integrates LoRA with an FMTL paradigm to enable efficient and effective fine-tuning of LLMs in a federated manner. LoRA decomposes the weight matrices of LLMs into lower-dimensional subspaces, significantly reducing the number of trainable parameters and mitigating the computational and communication overhead typically associated with LLM training. In MIRA, each client maintains its own set of low-rank matrices, which are iteratively updated based on local data and regularized to align with similar tasks through a centralized parameter server. The MIRA algorithm operates through iterative communication rounds. In each round, a subset of clients is selected to perform local updates on their respective low-rank matrices using their private data. After completing the local training steps, clients transmit their updates to the server, which then applies a regularization step to harmonize the models based on the similarity graph. This process ensures that clients with related tasks influence each other's model parameters, promoting collaboration and knowledge sharing among clients while preserving the task-specific adaptations of each client.

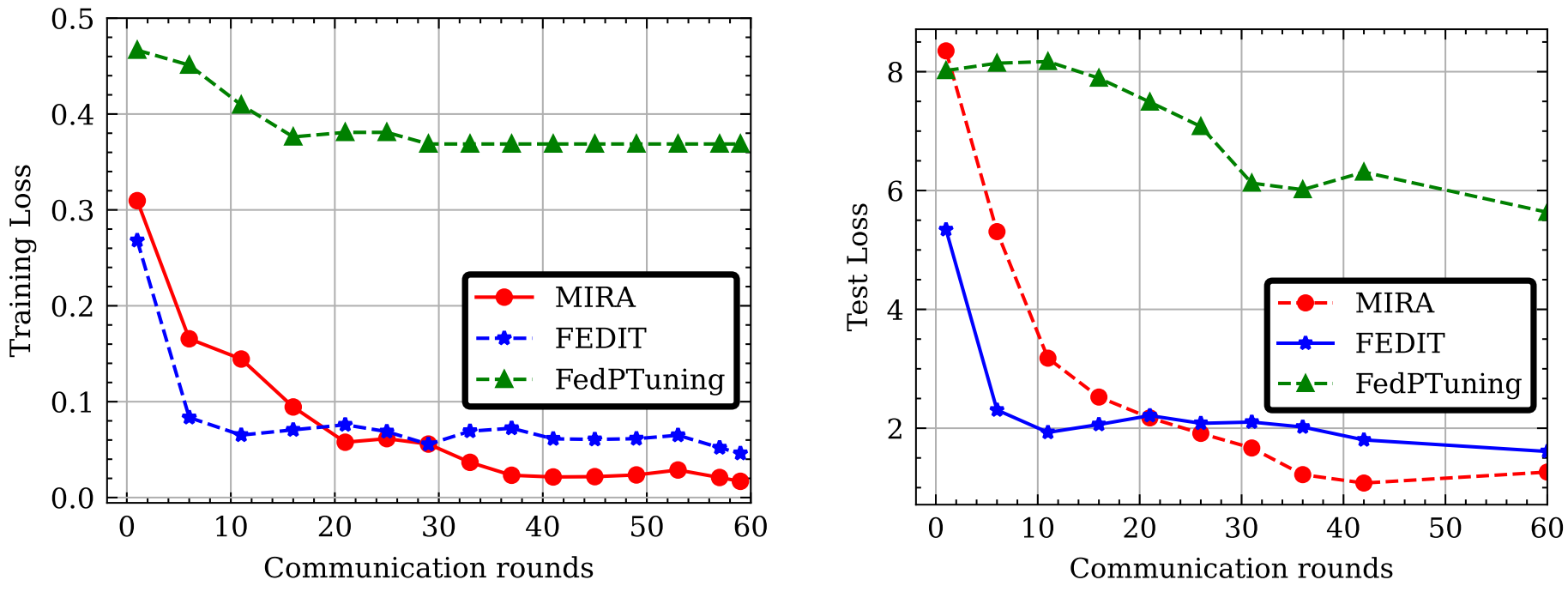

(a) (b)

Figure 34: Performance comparison of the proposed method and the baselines on Data-Juicer using the Natural Instruction dataset.

Next, we detail our setup and assess the performance of our proposed method, MIRA, against existing federated fine-tuning approaches for large language models (LLMs), specifically comparing it to FedIT [276] and FedP-Tuning [277]. We employed Data-Juicer, a Llama-based LLM model with 1.3 billion parameters on the Natural Instruction [278] dataset. Following [273], we pre-processed the data by sampling around 20% of the training set and 2% of the test set. The federated setup included 80 clients, each with a unique local task, with 10% of the clients randomly selected to participate in each communication round for 60 communication rounds. We measured performance using average training and testing losses and the Rouge-L score, which evaluates text generation quality by identifying the longest common subsequence between the model's output and reference texts. All experiments were conducted on an NVIDIA A100 GPU with 40 GB of VRAM, ensuring that the best hyper-parameters were selected for all methods to maintain a fair comparison. In Figure 34, we demonstrate how our proposed method, MIRA, compares to the baselines. Specifically, when fine-tuning the Data-Juicer model using the Natural Instruction dataset, MIRA shows better performance compared to both baselines. In fact, Figure 34(b) shows that MIRA surpasses FedIT, the closest baseline, after approximately 20 communication rounds. In Table 16, we evaluate the impact of FMTL on the performance of individual tasks to assess how effectively MIRA adapts to each task compared to baseline methods. We randomly selected four clients/tasks and recorded the average test loss throughout the training period. We can see that MIRA achieves a lower average loss for three out of the four clients/tasks, highlighting the effectiveness of FMTL and its ability to accommodate task-specific requirements, in contrast to the model averaging approach of the other baselines.

## 8.2 The interplay between LLM/Foundation models and federated learning

The integration of Foundation Models (FMs) and Federated Learning (FL) [270] presents a transformative approach to addressing challenges in training large-scale models while maintaining data privacy. The constrained communication resources and inherent data/system heterogeneity in real-world networks can impede performance and scalability when deploying FL. By incorporating FMs as a vital element of intelligent network infrastructure, we can utilize FMs to improve FL training efficiency and enable new application scenarios beyond the capabilities of conventional AI models. Hence, here is no "one-size-fits-all" scheme. The integration of FMs into an FL system should be tailored to align with the system's characteristics, wherein FMs conceptually function as tailored service providers [279]. In other words, we can think of the usage of FMs into federated edge learning systems in terms of "*Foundation Model as a Service*" (FMaaS) principle. In the following, we provide a comprehensive discussion on the potential and limitations of this interplay, illustrated through typical use cases.

Table 16: Local test loss per client for some selected subset of clients on Data-Juicer using the Natural Instruction dataset.

| Client Task | Algorithm | Test Loss |
|---|---|---|
| Question Answering | MIRA | 2.7 |
| | FedIT | **1.8** |
| | FedPTuning | 9.55 |
| Program Execution | MIRA | **0.23** |
| | FedIT | 0.37 |
| | FedPTuning | 0.87 |
| Speaker Identification | MIRA | **1.87** |
| | FedIT | 7.01 |
| | FedPTuning | 11.98 |
| Explanation | MIRA | **1.99** |
| | FedIT | 2.55 |
| | FedPTuning | 3.08 |

#### 8.2.1 Mutual Promotion between FL and FMs

As FM grow in scale, the next step after utilizing all publicly available data is to leverage personal data, which is inherently distributed across wireless networks. However, domain-specific FMs in fields such as law, healthcare, and finance face significant challenges in accessing the proprietary data due to stringent privacy regulations. FL provides a strategic solution by enabling FMs, including LLMs, to be trained directly on decentralized edge devices, without the need to centralize sensitive data. This collaboration between FMs and FL harnesses the strengths of both approaches to overcome their respective limitations, fully unlocking the computing potential of edge networks and establishing a synergistic relationship [280, 281].

**FL expands data availability for FMs.** By allowing data to remain on edge devices, FL facilitates training on sensitive data while ensuring privacy compliance. This privacy-preserving approach empowers FMs to enhance their training using a wide variety of real-world data that would otherwise be inaccessible. This allows FMs to maintain adaptability and robustness across different tasks and user-specific applications, such as healthcare diagnostics or personalized financial recommendations [282, 283]. Additionally, Real-world data, such as from edge devices and IoT sensors, continuously expands, posing challenges for updating FMs. FL enables seamless fine-tuning of FMs with new data, integrating it incrementally into existing models, ensuring the model stays up-to-date and adapts effectively to changing environments.

**FMs boost FL with advanced features and few-shot learning.** Pre-trained FMs provide FL with advanced feature representations and few-shot learning capabilities. By leveraging the foundational knowledge embedded in FMs, FL can accelerate its learning process, allowing models to adapt quickly and efficiently to specific downstream tasks with minimal training. Moreover, FMs' generative capabilities can assist in addressing data heterogeneity challenges in FL, such as by synthesizing additional data to improve model convergence and overall performance.

### 8.2.2 Empowering FedFMs Deployment: Challenges and Solutions

Despite the promising advantages of integrating FMs in federated settings, several significant challenges remain. These include managing the substantial computational requirements and huge communication overhead of training and sharing large models, ensuring efficient communication between edge devices and the central server, and addressing the heterogeneity of devices, data, and models across the network. Overcoming these obstacles is essential for realizing the full potential of FL with FMs.

**High demand for training and sharing FMs.** The significant resource demands of FM training, contrasted with the limited and heterogeneous resources in FL systems, such as communication bandwidth, computational power, and memory[284], pose major challenges to the efficiency of Federated Edge Fine-tuning. The frequent exchange of training data over limited bandwidth channels [285] creates a communication bottleneck, and the large number of FM parameters further compounds this issue, hindering the overall training process. To mitigate these challenges, parameter-efficient training methods have been developed to adapt FMs to specific domains or tasks. These methods typically freeze the majority of FM parameters and fine-tune only a small adapter, thus reducing both the computational load and communication overhead. Techniques such as BitFit [286], adapter tuning [128], prompt tuning [287], and Low-Rank Adaptation (LoRA) [130] are examples of these parameter-efficient approaches that facilitate efficient fine-tuning in resource-constrained environments. In addition to these, model compression techniques like model pruning [288], sparsification [289], and quantization [290] are also employed to further enhance resource efficiency. These methods reduce the size and complexity of FMs, enabling more efficient storage and transmission across FL clients with limited computational and memory capabilities. Together, these strategies make it feasible to deploy and fine-tune FMs in federated learning settings while addressing the constraints imposed by device resources and communication bandwidth.

**Adaptability** Adaptability challenges stem from the need to adjust a FM to specific downstream tasks in FL settings, particularly due to heterogeneity in models, data sources, and system resources across clients. Different devices and environments in FL systems often possess varying computational capacities, storage, and data characteristics, which makes it difficult to uniformly adapt and fine-tune large-scale FMs across diverse nodes. To address these challenges, techniques such as Knowledge Distillation and Mutual Learning are employed. Knowledge distillation [288] helps by transferring knowledge from a larger, more complex FM to a smaller, more resource-efficient model, which can be better suited for deployment on resource-constrained devices. Mutual learning [291] allows for bidirectional knowledge exchange between models across clients, improving both model adaptation and performance consistency in heterogeneous settings. In response to the challenges posed by resource heterogeneity, Split Learning has also been proposed. This technique, such as in FedBERT [292], allows the model to be partitioned between clients and servers, where each client processes only part of the model locally, and the rest of the model computation is offloaded to the server. This not only reduces the computational burden on individual clients but also minimizes the communication overhead by reducing the amount of model information that needs to be exchanged during the training process. These methods together improve the adaptability of FMs in FL environments, ensuring that models can effectively handle the varying constraints of different clients while maintaining robust performance on specific tasks.

**Other potential issues.** In addition to the aforementioned cost-intensive properties of FMs, the hallucination of FMs serves as another non-negligible concerns. Hallucination presents a significant challenge when using FMs in FL systems. In this context, hallucination refers to the generation of inaccurate or nonfactual information by the FM, which can be particularly problematic in critical automated decision-making scenarios, such as specific tasks in autonomous driving. Such inaccuracies could result in severe consequences, highlighting the need for robust mechanisms to detect and mitigate hallucinations in FMs.

### 8.2.3 The role of FMs in FL-enabled networks

Under the principle of FMaaS, FMs can be utilized in existing FL-enabled networks to deliver various services at different stages, such as data pre-processing, training, and calibration of the collaboratively trained model, as outlined below.

**FM in pre-processing stage.** The imbalanced data commonly present in real-world wireless networks and the resulting data heterogeneity across clients pose significant challenges that limit the training performance of FL systems. Leveraging one of the well-known capabilities of certain FMs, "their ability to generate data", can be adopted to enhance model training. Possible integrated systems include "Data Augmentation at the Edge" and "Synthetic Data at the Server." In such scenarios, FMs' role as data augmenters by incorporating public or locally generated synthetic data into global distillation or local training processes [279]. This approach allows the trained model to learn more balanced representation knowledge from combined datasets rather than relying solely on local private data. As a result, it significantly enhances privacy protection and improves robustness against adversarial attacks, such as gradient inversion.

**FM in the training.** In conventional deep learning, a well-trained model can serve as a teacher model to transfer knowledge for training smaller models (i.e., student models). Given that pre-trained foundation models (FMs) have acquired extensive knowledge from massive training datasets, it is logical to design an integrated system that retrieves and transfers this knowledge from FMs to enhance small model training within FL-enabled networks.

**FM for model evaluation** FMs at the edge server could be equipped with enhanced functionality by accessing the updated local models or aggregated model, rather than merely participating in the training process. Given that pre-trained FMs demonstrate exceptional performance across various downstream tasks, their performance can be used as a benchmark for evaluating smaller models. Currently, performance evaluation and validation rely on limited validation and test datasets. While a trained model may show strong generalization performance on these test sets, the risk of overfitting still exists. Therefore, the output of an FM could serve as a performance evaluation criterion for smaller models by comparing their outputs against the FM's corresponding outputs. Additionally, in scenarios involving malicious clients, FMs could be utilized to identify such clients through output verification, distinguishing between malicious and benign clients. With pre-trained FMs integrated into FL-enabled networks, the edge server could leverage these capabilities for complementary model evaluation services.

### 8.2.4 FedFM-Empowered User Cases

**Speech.** With the rapid advancements in AI, there has been substantial progress in the development of speech-related FMs, such as wav2vec 2.0 [59] and Whisper [293]. These models are increasingly paired with Federated Learning (FL) to manage privacy-sensitive audio data. FL's decentralized nature makes it particularly suited for speech applications where on-device processing and user-specific customization are critical. Audio data is continuously generated by end-user devices like smartphones, making it essential to keep this data local for privacy reasons rather than transmitting it to external servers. Additionally, while FL enables collaborative model training across users to improve general accuracy, a universal model may not always meet the unique requirements of individual users [294], highlighting the need for personalization. Key applications of FL in this domain include Automatic Speech Recognition (ASR) [295] and Speech-to-Text (S2T) [296], where FL ensures privacy while enhancing user-specific model performance.

**Recommendation.** Federated Recommendation (FR) aims to provide personalized content to users while ensuring data privacy by leveraging decentralized learning [297]. Recently, the use of LLMs in recommendation systems has gained significant traction [298] due to their ability to understand complex language inputs and generalize across domains. One common approach to adapting FMs for FR involves fine-tuning them with historical user-item interaction data. To optimize resource usage in such settings, techniques like FedPEFT, including adapter tuning [299] and split learning [300], are often employed.

Beyond parameter fine-tuning, LLMs can also be used in a zero-shot manner through prompt engineering to assist in recommendation tasks [301]. For instance, GPT-FedRec [302], a two-stage FR framework, utilizes ChatGPT's zero-shot generalization capabilities. In the first stage, it collaborates to train both ID-based and text-based retrievers, which are then used to generate prompts for GPT. In the second stage, these prompts are processed by GPT to re-rank the retrieved results. Additionally, [303] explored using pre-trained BERT models to create item description vectors, which are then incorporated into a recommender system as enriched input data for improved recommendations.

**Healthcare.** FMs, LLMs, have demonstrated remarkable success in healthcare, excelling in various tasks such as mental health assessments [304], disease diagnosis [305], and drug discovery [306]. However, privacy concerns arise when sensitive patient data is uploaded to centralized commercial servers hosting these models. Federated Learning (FL) addresses this challenge by enabling decentralized model training without exposing private data, offering a secure way to harness FMs' potential in healthcare.

One such example is FedTherapist [307], a mobile mental health monitoring system that fine-tunes FMs using FL based on user-generated speech and keyboard inputs. This system has demonstrated high accuracy in predicting mental health conditions, including depression, stress, and mood fluctuations. Another study [308] explored the application of FL in improving MRI reconstruction. The researchers fine-tuned an FM, pre-trained on public datasets, using visual prompts from decentralized clinical data. This personalized FL approach reduced communication overhead while achieving competitive performance with limited local data, highlighting the viability of the FM-FL combination in medical imaging and other healthcare applications.

## 8.3 Reinforcement Learning with LTMs Interaction (9.4.1)

### 8.3.1 RL-Empowered Communication

Reinforcement Learning (RL) has achieved significant applications in communications, including network access and rate control, caching and offloading, network security, and connectivity preservation.

- **Network Access and Rate Control**
  In network access and rate control, RL is utilized for dynamic spectrum access, where users, such as sensors in the Internet of Things (IoT), select channels based on their states to maximize throughput. It also addresses joint user association and spectrum access, optimizing data rates and service quality by determining the best base station and channel for users in heterogeneous networks[309]. Additionally, RL aids in adaptive rate control, such as in Dynamic Adaptive Streaming over HTTP (DASH) systems, where clients select video segment bitrates to enhance the Quality of Experience (QoE) by balancing average bitrate and minimizing buffering.

- **Caching and Offloading**
  In the realm of caching and offloading, RL is employed for wireless proactive caching, where base stations pre-cache popular content to reduce transmission duplication, access delays, energy consumption, and overall traffic[310]. This involves making decisions on what content to cache and when to replace it based on user requests and content popularity. RL also facilitates data and computation offloading, enabling IoT devices to offload computational tasks to nearby Mobile Edge Computing (MEC) servers, thereby reducing processing delays, saving battery energy, and enhancing security. Mobile users, for instance, decide whether to offload data to cellular networks or WLAN and select the appropriate MEC server based on network conditions and their specific needs.

- **Network Security and Connectivity Preservation**
  Network security and connectivity preservation are other critical areas where RL is applied[311]. In network security, RL helps combat jamming attacks in cognitive radio networks by enabling users to learn optimal strategies for channel selection, transmit power adjustment, or using UAVs as relays to avoid interference. It also aids in responding to cyber-physical attacks in autonomous systems, such as self-driving cars, where vehicles learn to adjust their speed based on the attacker's activities to maintain safe spacing. For connectivity preservation, RL is used in multi-robot systems, like multi-UAV networks, where each robot dynamically adjusts its speed and direction to maintain communication range and connection stability.

- **Other Applications**
  RL is also instrumental in traffic engineering and routing, optimizing data traffic paths to maximize network utility and minimize delays[312]. It enhances resource sharing and scheduling in multi-user massive MIMO systems and cloud radio access networks, ensuring efficient resource allocation and service quality. Furthermore, RL supports power control and data collection in non-cooperative cognitive radio networks, massive MIMO networks, and wireless sensor networks, optimizing power allocation, sensing, and control to improve network energy efficiency and user fairness. Additionally, RL techniques are used for direction of arrival (DoA) estimation, signal detection, user association, load balancing, user localization, and access device detection, leveraging deep learning to improve accuracy and performance in various communication scenarios.

### 8.3.2 LLM-Empowered 6G Communication

- **Optimization and Resource Management**
  - **Resource Allocation**
    LLMs have significant potential applications in maximizing spectrum efficiency and energy efficiency in wireless communication resource management. By analyzing vast amounts of communication data, LLMs can optimize spectrum allocation, ensuring that frequencies are used more effectively and reducing interference. This leads to improved spectrum efficiency, allowing networks to handle more data and users simultaneously. Additionally, LLMs can predict traffic patterns and user demand, enabling dynamic resource allocation that adapts in real-time to changing conditions, thus maximizing energy efficiency. By automating network configuration and fault detection, LLMs can also reduce the energy consumption associated with manual network management tasks. Furthermore, in the context of intelligent edge computing, LLMs can support real-time decision-making and resource optimization, further enhancing both spectrum and energy efficiency in wireless networks[313].
  - **Protocol Understanding**
    In the field of wireless communication, LLMs are proving to be invaluable for protocol understanding and implementation. These models can analyze and interpret complex wireless communication protocols, such as 5G, LTE, and Wi-Fi standards, facilitating the development and optimization of communication systems. By leveraging their advanced natural language processing capabilities, LLMs can assist engineers in decoding protocol specifications, identifying potential issues, and ensuring compliance with industry standards. Additionally, LLMs can help in automating the generation of protocol documentation, streamlining the design and testing processes, and enhancing overall system performance. Their ability to understand and process vast amounts of technical information makes LLMs a powerful tool in advancing wireless communication technologies[314].

- **Generative and Design Solutions**

  - **Automated Code Design**
    LLMs have demonstrated significant potential, particularly in generating hardware description language (HDL) code. Research indicates that LLMs can not only produce HDL code for simple computational tasks but also play a crucial role in developing more complex wireless networking prototypes and products. Case studies have shown that LLMs can substantially enhance productivity in code refactoring, reuse, and validation for researchers and developers. Furthermore, LLMs have proven effective in generating HDL code for advanced wireless signal processing algorithms, such as successfully creating a 64-point Verilog Fast Fourier Transform (FFT) module. These achievements highlight the broad applicability of LLMs in automated code design, showcasing their ability to handle complex task decomposition and multi-step reasoning, thereby driving innovation in wireless networking systems development[315].

  - **Generate Network Configurations**
    LLMs are increasingly being utilized to automate the generation of network configurations, offering significant benefits to telecom network operators. By translating natural language requirements into formal specifications, LLMs can create high-level and low-level device configurations without the need for extensive manual input[316]. This capability allows for automatic network provisioning, optimization, and performance tuning, as well as security and compliance configuration[317]. Additionally, LLMs can assist in fault diagnosis and troubleshooting, and support network virtualization efforts. By reducing manual effort and improving the efficiency, reliability, and security of network management, LLMs are transforming how network configurations are generated and maintained in complex telecom environments.

  - **Recommend Troubleshooting solutions**
    LLMs have shown significant potential in recommending troubleshooting solutions for complex telecom networks, which require resolving both software and hardware faults known as trouble reports. Research has demonstrated the effectiveness of LLMs in this domain, with models being used to generate and rank multiple possible solutions for system faults [318]. By incorporating transfer learning and non-task-specific telecom data, these models have enhanced their ability to handle unseen trouble reports [319]. The LLM-enabled method uses trouble report observations, headings, and fault areas as input to generate the top-K possible solutions, significantly improving efficiency and enabling faster response and repair times in telecom networks [320].

- **Predictive Analytics**

  - **CSI Prediction**
    Channel State Information (CSI) is vital for optimizing wireless communication systems as it provides detailed insights into channel quality and characteristics. LLMs can predict CSI by analyzing vast amounts of historical communication data to identify complex patterns and trends that affect channel conditions. By incorporating environmental factors, user behavior, and device characteristics, LLMs can generate more accurate CSI predictions. This predictive capability not only enhances network performance but also optimizes resource allocation and reduces channel estimation errors, thereby improving overall communication efficiency. Additionally, the natural language processing capabilities of LLMs can help engineers better understand and interpret prediction results, facilitating more effective decision-making[321].

  - **Prediction-based beamforming**

Beamforming is a wireless communication technique that enhances signal quality and reduces interference by directing transmissions towards specific users. LLMs can significantly improve prediction-based beamforming by analyzing user mobility patterns, environmental changes, and historical signal data to forecast optimal beam directions. This allows networks to dynamically adjust beamforming strategies to meet changing user demands and environmental conditions. By leveraging multimodal data, LLMs can capture dynamic network features and predict the best beam for current and future conditions[322]. This not only improves connection stability and signal quality but also reduces energy consumption and interference. Additionally, LLMs can process large datasets in real-time, making the prediction and adjustment process more efficient and precise. This flexibility enhances high-mobility and latency-sensitive applications, paving the way for more efficient and intelligent wireless communication systems.

– **Traffic load Prediction**
LLMs have the potential to revolutionize Traffic Load Prediction by leveraging their advanced natural language processing and reasoning capabilities. In the context of traffic management, LLMs can interpret and analyze vast amounts of traffic data, including historical traffic patterns, real-time flow information, and various environmental factors. By doing so, they can identify underlying trends and patterns that may not be easily detectable through traditional statistical methods. The integration of LLMs into traffic prediction systems allows for more accurate and nuanced forecasts, which can inform urban planning, traffic signal control, and driver navigation systems. Moreover, LLMs can adapt to changing traffic conditions and learn from new data, continuously improving their predictive accuracy over time. This human-mimetic approach to Traffic Load Prediction not only enhances the efficiency of traffic management but also contributes to the development of smarter and more responsive urban transportation ecosystems[323].

#### 8.3.3 LLM-Enhanced RL

- **Information Processor**
LLMs can serve as information processors in the field of reinforcement learning. They can extract feature representations or handle natural language-based information to accelerate RL learning[324]. For example, pre-trained LLM models can be used as feature representation extractors, either by directly using the frozen pre-trained model or by fine-tuning it with contrastive learning to increase sample efficiency and generalization. LLM can also act as a language translator, converting diverse and informal natural language information into formal task-specific information to assist the learning process of the RL agent.

- **Reward Designer**
LLMs have the potential to act as reward designers in RL. They can leverage pre-trained common-sense knowledge, code generation, and in-context learning ability to design or shape reward functions. There are two ways for LLMs to serve as reward models: one is to be an implicit reward model that directly provides auxiliary or overall reward value based on the understanding of task objectives and observations, either by direct prompting with language descriptions or by scoring the alignment between the feature representation of the visual observations and language-based instructions[325]; the other is to be an explicit reward model that generates executable codes of reward functions to transparently specify the logical calculation process of reward scalars[326].

- **World Model Simulator**
LLMs can be applied as World Model Simulators in the context of RL. They can be trained to 1) act as a trajectory rolloutor, auto-regressively generating accurate trajectories for the agent to learn and plan[327], or 2) function as a dynamics representation learner, predicting the latent representation of the world using representation learning[328]. For example, pre-trained large-scale models have been used to synthesize trajectories in games, and those based on action-free video representations have shown improved performance in visual RL tasks.

- **Decision Maker**
  LLMs can play the role of a decision-maker in RL, either as a direct decision-maker or an indirect decision-maker. In direct decision-making, LLM enhances the Decision Transformer-based methods with a more powerful pre-trained model and prior world knowledge to solve sparse-reward and long-horizon tasks more efficiently[329]. For example, by framing the learning process of offline RL into a supervised learning problem, LLM can predict future actions based on sequence modeling, improving the sample efficiency and generalization of the model. In indirect decision-making, LLM instructs the action selection by either generating a set of action candidates or providing a reference policy. This helps address the challenges posed by large action spaces and natural language in applications such as instruction-following and text-based games[330]. For instance, when generating action candidates, RL agents can further re-rank them based on the value function to maximize cumulative rewards.

## 8.4 Distributed LTMs

### 8.4.1 Integrating Big Data Analytics and LLMs

Distributed large-scale telecommunication systems are essential for handling the massive amount of data generated by modern communication networks. These systems may utilize distributed architectures, big data analytics, and artificial intelligence to efficiently manage, optimize, and secure telecommunications infrastructures. In this section, we aim to highlight key concepts, challenges, and applications based on recent research, particularly in big data analytics, large-scale wireless networks, and the application of LLMs in telecommunications.

- **Big Data Analytics in Telecommunications**
  The integration of big data into telecommunications has revolutionized the way data is managed and processed. Telecommunication networks generate an enormous volume of data from various sources, such as sensor nodes, mobile networks, and customer interactions. Efficiently analyzing this data is crucial for optimizing network performance, reducing latency, and improving user experience.

  A significant aspect of big data analytics in telecommunications involves the so-called *Lambda architecture* [331], which is designed to handle massive data streams in real-time and batch processing modes. This architecture enables the simultaneous analysis of historical data for long-term insights and real-time data for instant decision-making. The need for such architecture arises from the ever-increasing complexity of telecommunications networks, where real-time decisions are critical for functions like network load balancing and congestion control [332].

  Moreover, the ability to predict user behavior and network traffic patterns is one of the key applications of big data analytics. This allows telecommunications operators to anticipate network loads, allocate resources efficiently, and avoid service disruptions. The challenge, however, lies in processing the large volume of structured and unstructured data generated by these networks in real time, especially with the growing prevalence of IoT and 5G technologies [332, 333].

- **Large-Scale Wireless Networks: Challenges and Solutions**
  Wireless networks, especially large-scale ones, present unique challenges due to the constantly changing environment in which they operate. Factors such as user mobility, interference, and network topology shifts make it difficult to maintain stable, high-performance communications. For instance, the signal quality in a wireless network can be affected by physical barriers, such as buildings or even people, and by environmental factors like weather or radio interference [334].

One key approach to managing these networks is *real-time monitoring and topology discovery*. Wireless access points (APs) must be continuously monitored to detect fluctuations in the network, such as changes in user density or interference levels. Collecting beacon data (information transmitted by beacon devices in wireless communication that enables devices to discover, identify, and interact with the network or service) from APs allows network administrators to map the network topology and identify areas where the signal may be weak or congested [335]. This data can also be used to optimize power configurations and minimize interference between neighboring access points [334].

Furthermore, large-scale wireless networks face the challenge of efficiently distributing users across access points. Traditional methods rely on signal strength for user association with APs, which often leads to inefficient distribution and network overloading. More advanced techniques, such as those incorporating *Software-Defined Networking (SDN)* [336], can dynamically adjust user associations based on multiple factors like channel occupancy and network load, improving network efficiency and enhancing user experience [334].

- **Large Language Models (LLMs) in Telecommunications**
  A more recent advancement in the field of telecommunications is the application LLMs. While LLMs were initially developed for natural language processing (NLP), their potential in telecommunications networks has become increasingly evident. LLMs can be applied to a wide range of telecommunications tasks, such as network optimization, traffic prediction, and troubleshooting.

  One of the major uses of LLMs is in *network traffic prediction*. Predicting traffic patterns in telecommunications networks is a critical task, as it allows operators to optimize network resources, avoid congestion, and provide better service to users. LLMs can learn from large volumes of historical network data to predict future traffic loads with high accuracy. This predictive capability is especially valuable in the era of 5G and IoT, where the volume and variety of data streams have increased dramatically [333].

  LLMs can also assist in *troubleshooting network issues*. By analyzing network logs and customer service interactions, LLMs can identify common network problems and suggest solutions. This can significantly reduce downtime and improve the efficiency of network operations. Furthermore, LLMs can aid in automating routine tasks, such as network configuration and load balancing, freeing up human operators to focus on more complex issues [333].

  However, deploying LLMs in telecommunications networks presents several challenges. One of the main concerns is the computational cost of training and running these models. Telecommunications networks often operate at the network edge, where resources like storage and processing power are limited. Techniques such as *parameter-efficient fine-tuning* and *split edge learning* have been proposed to mitigate these challenges by reducing the computational load required for training LLMs at the edge of the network [333].

- **Future Directions and Challenges**
  As telecommunication networks continue to evolve, the role of distributed architectures, big data analytics, and AI models like LLMs will only grow. However, there are several challenges that need to be addressed to fully realize the potential of these technologies.

  One major challenge is the *scalability of distributed architectures*. Telecommunications networks are con-stantly expanding in size and complexity, with the introduction of new technologies like 5G and edge computing. Managing these networks requires scalable solutions that can handle vast amounts of data and support real-time decision-making across multiple nodes [332].

Another challenge lies in the *privacy and security* of telecommunications data. With the increasing use of AI models and distributed architectures, ensuring the privacy and security of user data has become more difficult. LLMs, in particular, are susceptible to issues such as data leakage and model inversion attacks, where malicious actors can reverse-engineer sensitive information from the model's output. Addressing these security concerns is critical for the widespread adoption of AI in telecommunications [333].

Additionally, the integration of *multi-modal data sources* is an important area for future research. Telecommunications networks rely on a wide variety of data types, including text, image, and sensor data. LLMs and other AI models need to be capable of processing and integrating this multi-modal data to provide more accurate predictions [333].

### 8.4.2 Communication-efficient Fine-tuning in Decentralized Settings

While the use of Large Language Models (LLMs) covers a wide range of applications across various fields, scaling these models and adapting them to specialized tasks and domains often require fine-tuning of pre-trained models. Motivated by the distributed nature of the data and the efficiency of training models using distributed computing over many devices, distributed or semi-distributed fine-tuning is an appealing approach in decentralized settings. While standard fine-tuning approaches rely on first-order (FO) optimization methods, such as Stochastic Gradient Descent (SGD) and Adam, these methods suffer from limitations that hinder their implementation in this context. In fact, FO optimization methods rely heavily on backpropagation, which significantly increases memory overhead. Furthermore, the devices are supposed to exchange high dimensional vectors of gradients, which presents a critical challenge, particularly in resource-constrained environments like edge devices.

Zeroth-order (ZO) optimization represents a potential approach to overcome these challenges. ZO optimization belongs to the wider field of gradient-free optimization. It is based on estimating gradients using finite difference approximations. These techniques rely solely on function evaluations - ZO information - rather than explicit gradient information, yet their algorithmic frameworks closely resemble those of FO gradient-based methods.

While several strategies exist for calculating ZO gradients, the most prominent one is based on randomized directions, which estimates gradients based on finite differences in function values evaluated along random directional vectors. Given a scalar-valued loss function $L(x)$, where $x$ resides in a $d$-dimensional space, the ZO gradient estimate is computed using the central difference formula:

$$\hat{\nabla} L(x) = \frac{1}{q} \sum_{i=1}^{q} d \frac{L(x + \delta u_i) - L(x - \delta u_i)}{2\delta} u_i,$$

where $u_i$ represents a random direction vector usually sampled from a normal distribution $\mathcal{N}(0, I)$, $2q$ is the number of function queries which generally improves the estimation, and $\delta > 0$ is a step size parameter (often referred to as a smoothing parameter). The idea of ZO estimates derives from directional derivatives, where, as $\delta \to 0$, the finite difference of function values in the $u_i$ direction, denoted as $L'(x, u)$, approximates the directional derivative $\nabla L(x)^\top u$.

ZO optimization offers several key advantages, making it a valuable tool in various fields. Here are some of its primary benefits:

- ZO methods do not require explicit gradient information, relying solely on function value evaluations. This makes them suitable for problems where gradients are difficult, expensive, or impossible to compute, such as black-box optimization scenarios where only input-output relationships are accessible.This is crucial for applications involving proprietary systems or experimental setups where internal computations are not visible [337, 338].
- Improved Computational, Communication, and Energy Efficiency: Computing ZO gradients requires only querying the objective function and applying finite difference calculations, which significantly reduces computational overhead compared to methods that rely on explicit gradient computation. Additionally, ZO optimization does not require storing large amounts of intermediate data, making it memory-efficient [339]. Moreover, innovative algorithms can be designed to exchange only scalar values in distributed settings, making use of the structure of ZO gradient estimates [340, 341, 342], which allows for high communication efficiency. This simplicity in computation and communication not only speeds up the optimization process -despite the slower convergence rates of ZO vs FO- but also reduces energy consumption [342], which is particularly advantageous in resource-constrained settings, such as embedded systems or on-device machine learning.
- Reduced Dependency on Model Structure: Unlike gradient-based methods, ZO optimization does not rely on the internal structure or specific details of the model. This model-agnostic property is particularly useful in machine learning tasks like adversarial testing [343, 344] and interpretability [345].

Due to these advantages, ZO can be seen as an appealing approach for fine-tuning in decentralized settings. In [339], a memory-efficient zeroth-order optimizer fine-tunes LLMs using only forward passes. The algorithm uses two forward passes to calculate the loss functions with the two perturbations, which are then used to estimate the gradient. To ensure consistency in perturbations when estimating each gradient, the method employs random seed resetting. This approach eliminates the need to store random perturbation vectors, thus avoiding additional memory overhead. This method demonstrates comparable performance to fine-tuning with backpropagation across multiple tasks, including classification and text generation, with memory usage reduction to the level of inference. While naturally, ZO methods need more iterations to converge than their FO counterparts, the authors show that the per-step speedup in their ZO algorithm can often make fine-tuning run faster than a standard implementation of fine-tuning with backpropagation. In addition, its potential extends to effectively optimizing non-differentiable objectives while also maintaining compatibility with both full-parameter and parameter-efficient tuning techniques such as LoRA and prefix tuning.

[340] extends the work in [339] to decentralized settings and proposes a communication-efficient decentralized fine-tuning framework that leverages shared randomness to minimize the bandwidth requirements for distributed model training. In this approach, multiple edge devices collaborate by first synchronizing their initial model states and agreeing on a shared mechanism to generate random perturbations. Each device independently estimates gradients using the method mentioned previously and shares only scalar gradient projections, minimizing thus the communication overhead. Devices also share the corresponding seed values, allowing others to reconstruct the perturbations locally. This ensures consistent gradient aggregation and synchronized model updates, all while minimizing communication overhead by only transmitting scalars. For training the 6.7 billion parameter OPT model, a naive baseline of sharing FO gradients between 4 machines would require 100s of terabytes, communicating the gradients of LoRA would cost 100s of gigabytes, whereas the scalars of the ZO gradients only require communicating a few 10s of kilobytes. This constitutes a huge saving in communication resources.

In parallel, ZO methods have also been explored in [341, 342] to train and fine-tune models over wireless systems. In [341], a novel two-step ZO-based federated learning (FL) has been proposed. This method has the advantage of including the wireless channel in the learning itself, avoiding thus decoding and equalization at the receiver, and allowing a high number of devices to participate simultaneously in the training. The use of ZO significantly reduces the communication overhead to two scalars per device, improving the communication efficiency of the system substantially. In [342], the impact of quantization and wireless errors on ZO-based FL has been studied. It has also been shown that ZO-based FL allows for the achievement of high energy and communication energy savings, making it a promising approach for edge devices. Even for the slower convergence rate of ZO methods, they show that there is an important saving in convergence time needed to compute and transmit the ZO gradients for the whole method's iterations compared to the standard FL method. Similarly, for energy consumption, ZO requires a negligible amount of computation/transmission energy compared to its FO counterpart.

## 8.5 LTMs in Network Optimization

### 8.5.1 Reinforcement Learning with LTMs Interaction

In modern telecommunications networks, optimizing user QoE is significant. Traditional methods that rely mainly on objective metrics, such as latency, data rate, and packet loss, often fall short of capturing the detailed, subjective experiences of users. This gap motivates us to design a more comprehensive approach that integrates subjective QoE assessments to truly enhance user satisfaction. LTMs have emerged as powerful tools for this purpose. These models can act as intelligent agents for users, simulating their interactions within network management scenarios and bridging the gap between objective performance metrics and subjective user experiences.

In this case study, we consider a network-aided image generation service where different users may have vastly different perceptions of the same image based on their personal preferences and experiences. For example, even when generating images of *"a dog on a table"*, the resulting images can vary significantly in style. Different users may have distinct preferences for these styles. Capturing and responding to these user preferences is vital for optimizing service delivery in next-generation network services. However, collecting and analyzing extensive user behavioral data to understand these preferences raises significant privacy concerns.

To address these challenges, Reinforcement Learning with LLMs Interaction (RLLI) is a novel approach to optimize QoE in AIGC services [346]. RLLI leverages LLM-empowered generative agents deployed on edge devices, such as smartphones, to simulate diverse user preferences without compromising privacy. These agents are initialized with prompts based on the Big Five personality traits, allowing them to mimic a wide range of user preferences and behaviors. The RLLI system operates by presenting generated images to the LLM agent, which evaluates them based on the simulated user's personality. These agents provide subjective QoE scores that serve as rewards for DRL algorithms, e.g., Proximal Policy Optimization (PPO). The DRL algorithm then learns an optimal policy for selecting AIGC service providers (ASPs) to maximize the overall QoE across all users. This innovative approach enables personalized service delivery without the need to transmit sensitive user data to central servers, effectively addressing both QoE optimization and privacy concerns. By integrating subjective QoE assessments through LLM-powered agents, RLLI bridges the gap between objective network performance metrics and users' subjective experiences. Experimental results demonstrate RLLI's superiority over benchmark methods such as random selection and DQN-based algorithms in maximizing user QoE, showcasing its potential to enhance service delivery in next-generation network services.

### 8.5.2 Traffic prediction for network optimization

The optimization of mobile networks is a process that aims at determining a network configuration (parameters, resource allocation, etc.) that induces the minimal cost, under QoS and energy performance constraints. The performance being highly dependent on the traffic, a precise knowledge of traffic is essential for an effective optimization process.

A plethora of optimization schemes have been proposed in the literature, most of them considering a perfect knowledge of the traffic and an instantaneous adaptation of the network to it. However, some optimization scenarios involve reconfiguration of network elements, such as numerology modification for network slice launch/adaptation or the wake-up of resources in deep sleep mode. Accurate traffic prediction is then needed for anticipating optimization actions on different time scales, integrating the spatial dimension.

LLMs have recently been extended to time series prediction, with promising results [347]. The authors of [348] claim that current LLMs have the potential to revolutionize time series analysis, facilitating more efficient decision-making and evolving towards a more universal form of time series analytical intelligence. Furthermore, authors of [349] showed that models pre-trained on natural language processing or image analysis tasks can achieve comparable or state-of-the-art performance in various time series analysis tasks, including prediction. This result indicates that LLMs stand out for their ability to transfer knowledge from one domain to another.

The main challenge in developing pre-trained models for time series analysis is the lack of large amounts of training data. Yet, a team at Amazon pre-trained a family of models for time series forecasting called Chronos [350] built on a T5 architecture [351] by gathering a large database and performing data augmentation. These models outperform traditional and deep learning techniques on observed datasets. The idea behind Chronos is that a language model that predicts the next token should not be so different from a model that seeks to predict the next value of a time series. It simply requires moving from an infinite continuous domain, that of the numerical values of time series, to a finite domain — more precisely, a dictionary of tokens used by LLMs. In other words, Chronos tokenizes time series into discrete bins through a simple scaling and quantization of the real values, thus creating a "time series language".

We propose to evaluate the predictive performance of Chronos on a dataset that was not part of the training data. This is called "zero-shot" prediction. More precisely, we have the traffic in Mbit/s aggregated per week from about twenty base stations spread over La Rochelle (in France), from which we seek to predict the future values or each site.

Before looking at the simulation results, a common practice is to define a baseline model, which helps establish a minimum level of performance that more sophisticated models will encounter. A widely used statistical technique for predicting network traffic is the integrated moving average autoregressive model (ARIMA) [352], and a variant model that captures seasonality: SARIMA[353]. The ARIMA model makes predictions by considering lagged values of a time series while taking into account the non-stationarity of the data. The simulation results are given Table 17. One can notice that the number of parameters does not always seem to lead to better performance. Indeed, the Base model (200M) does better than the Large model (700M). This is probably due to the fact that a larger model with more parameters may have a greater ability to learn from the training data, including noise or small fluctuations. This leads to overfitting, where the model performs well on the training data, but less well on data it has never seen.

A major drawback of this solution is that it only models the temporal dimension of the data and it has been shown by the authors of [354] that removing the LLM component or replacing it with a basic attention layer does not negatively affect the prediction results in the case of an univariate time series. For the moment, this study does not address the case of multivariate time series. In our case, this corresponds "intuitively" to the fact that if a base station sees its traffic increase or decrease sharply, it is very likely that neighboring stations will experience the same situation.

Thus, we propose to focus on a Spatio-Temporal LLM (ST-LLM), presented first by the authors of [355] to anticipate the flow of bicycles and taxis in a city. Although we are rather looking to predict data traffic, it is interesting to evaluate the performance of such a model on our time series. We have represented (Figure 35 and Figure 36) the superposition of the traffic observed and predicted by ST-LLM for sites 1 and 20; the choice is completely arbitrary.

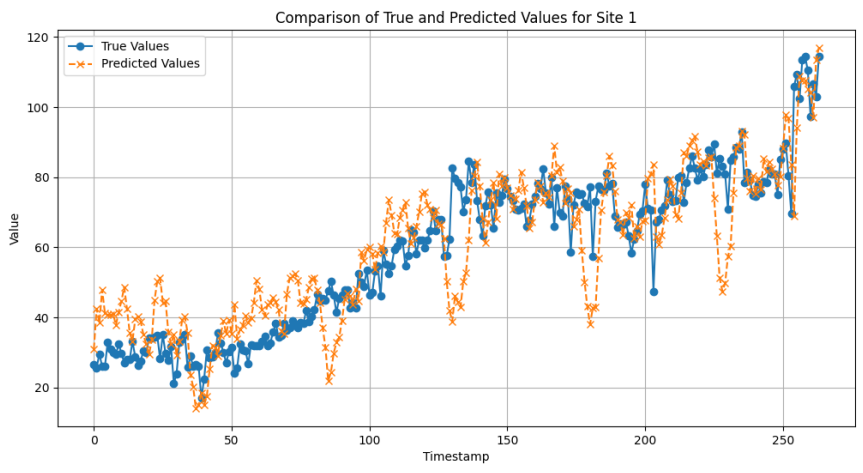

Figure 35: Traffic prediction of site 1.

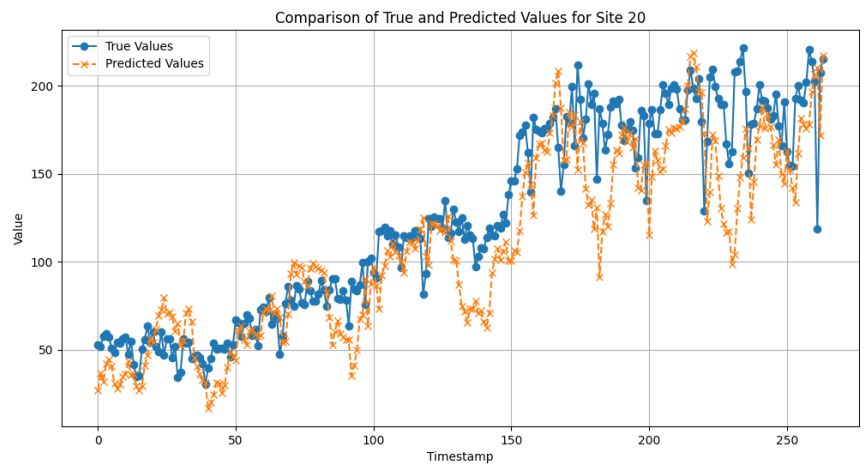

Figure 36: Traffic prediction of site 20.

By browsing the prediction curves, we noticed that the prediction was bad for some sites. By looking at the geographical position of this site, we see that it is "isolated" (Figure 37), in the sense that it has few neighbors, and that the distance separating it from its neighbors is large. One hypothesis is that the model, built to capture spatial dependencies, adapts poorly when there is precisely not much to capture.

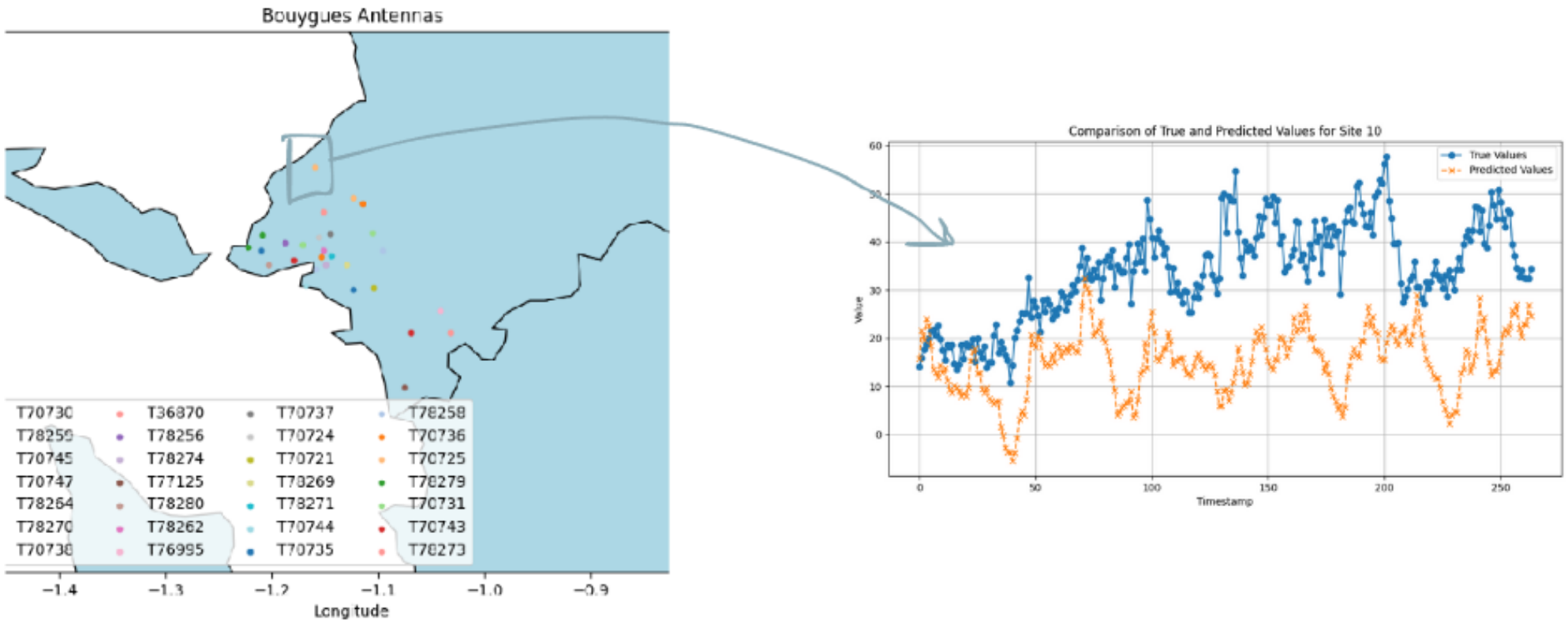

Figure 37: Prediction for an isolated site.

Even though it appears that Chronos performs better than ST-LLM, it should be kept in mind that the model was pre-trained on more than 84B observations, compared to barely 1M for ST-LLM.

Table 17: Comparison in terms of RMSE of different models

| Model | RMSE |
|---|---|
| SARIMA | 29.2108 |
| Base | 25.500391 |
| Large | 26.406957 |
| ST-LLM | 27.4306 |

We now move to the ability of LLM to propose accurate predictions far in the future. Looking at the performance of ST-LLM for different time horizons (Table 18), we notice that it performs worse than SARIMA as soon as the horizon is set to 2. The decrease in performance is expected, but it should be kept in mind that a horizon set to 2 is equivalent to a prediction for the traffic in 2 weeks, where seasonality plays an important role. Thus, it can be expected that the RMSE would decrease less quickly for a small time horizon, e.g. on the second scale. Retraining ST-LLM on network traffic data, instead of the current zero-shot approach is also a promising way for increasing accuracy.

Table 18: Comparison of ST-LLM predictions over different horizons.

| Horizon | RMSE |
|---|---|
| 1 | 27.4306 |
| 2 | 35.2581 |
| 3 | 44.8849 |
| 4 | 53.9882 |

### 8.5.3 Complex Optimization Problem Solving with Generation AI

The impressive multi-modal data generation capabilities demonstrated by generative models in the AI community have enabled their application to tasks such as wireless channel coding [356] and data feature extraction [357] in the telecom community. While these models have achieved remarkable success, they often overlook the key distinction between generative and discriminative models. Generative models learn the joint probability distribution of data, $P(x,y)$, whereas discriminative models focus on learning the conditional probability distribution, $P(y|x)$. This difference makes generative models particularly adept at addressing the multi-modal nature of high-quality solution spaces in network optimization problems, allowing them to generate better predictions than discriminative models.

Network optimization is a frequent challenge in telecom, particularly in scenarios that involve joint sensing, communication, computing, and control [358]. These tasks often require developing optimal resource allocation strategies to either maximize or minimize an objective function within specific network parameters and constraints. Due to the complexity of such problems, the high-quality solutions form a probabilistic multi-modal distribution within the solution space, for which the probability density function is typically unknown. Discriminative models, due to their deterministic nature, are limited in their ability to learn such complex distributions.

On the other hand, learning to describe the solution space based on input data offers a broader global perspective, leading to improved solutions. Notably, the distribution of high-quality solutions in the solution space is structured so that the difference in performance between these solutions and the optimal solution is minimal. In this distribution, the optimal solution holds the highest probability, followed by other high-quality solutions, while the probability of non-high-quality solutions approaches zero. The concept of transforming a single output solution into a distribution of high-quality solutions has been successfully demonstrated in recent research [359, 360] on two classic combinatorial optimization problems: the Traveling Salesman Problem (TSP) and the Maximum Independent Set (MIS) problem. Significantly, [361] was the first to define the parametrization of the solution space distribution, providing a continuously differentiable output target for neural network learning.

There have been few works that directly employ generative models as optimization solvers. For example, large language models (LLMs) have been explored to tackle differentiable simple constrained optimization and linear optimization problems [362], where feasible solutions are generated iteratively, and the optimal solution is approximated based on human feedback regarding the quality of each generated solution. However, LLMs currently face challenges in handling high-dimensional problems, and their performance often fails to justify the practical costs of training and inference. Meanwhile, diffusion generative models have also been studied as optimization solvers. For instance, [359] and [360] apply graph diffusion generative models to address TSP and MIS problems. However, these problems feature relatively simple objective functions and constraints, leaving the application of such models to more complex network optimization problems yet to be fully explored. Additionally, other researchers [357] have used diffusion models to generate solutions for purely convex optimization problems. Most of these works do not approach optimization from the perspective of learning high-quality solution distributions, and none fully realize the potential of diffusion generative models as standalone solvers for network optimization problems.

Diffusion models are a type of generative model that gradually adds noise to real data and learns to denoise it at each step [363]. The result of the noising process is to convert the data into completely noisy data, such as by continuously adding standard Gaussian noise until the data becomes a pure Gaussian distribution. The model learns to denoise data at various noise levels, continuously refining the corrupted data until clean data is obtained. In the context of our network optimization problem, the data being processed in the noising and denoising stages of the diffusion model represent the solution.

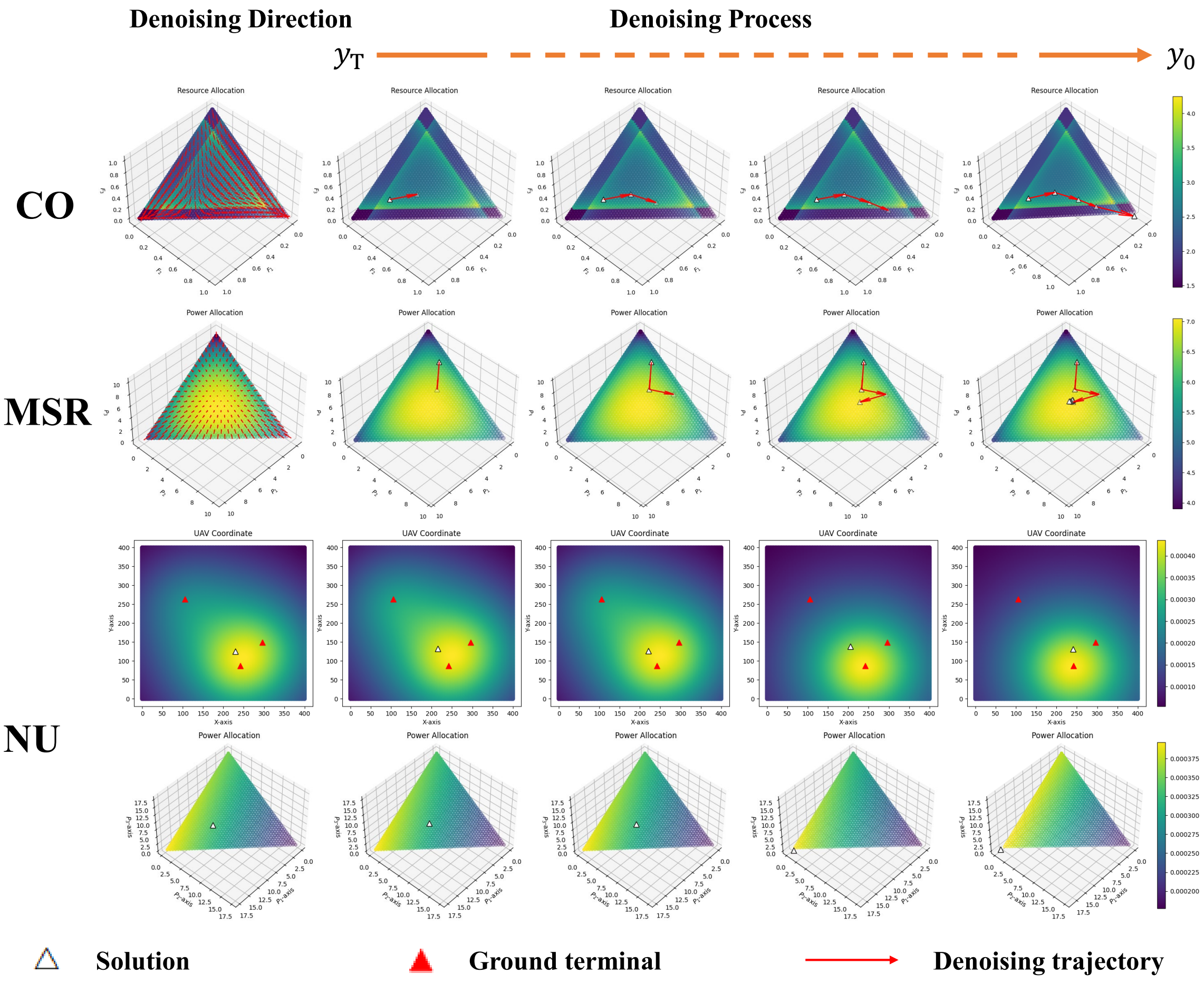

Figure 38: The denoising process for the CO, MSR, and NU problems, where each axis of the sub-figure represents a one-dimensional optimization variable, and the color of each solution point corresponds to its objective function value. In the first column, CO and MSR show the initial denoising direction, while NU does not, due to its hierarchical non-convex nature.

A multi-modal distribution has a probability density function with multiple distinct peaks (e.g., a mixture of Gaussian and Laplace distributions) and its exact probability density function is often impractical to derive. Existing research indicates that in both theoretical evaluation and engineering verification, the generalization error of diffusion models is polynomially small rather than exponentially large with respect to the number of training samples and model capacity [360]. This avoids the curse of dimensionality and supports the application of diffusion models to a wider range of problems.

To explore and verify the solution generation capability of diffusion models for network optimization problems with complex objectives and constraints, we designed a new framework, Diffusion Model-based Solution Generation (DiffSG) [364], and conducted exploratory experiments on several typical network optimization problems. Specifically, we considered three cases: computation offloading (CO) [159] to minimize the total weighted cost of task latency and power consumption, maximizing the sum rate of multiple channels (MSR) [357], and maximizing the sum rate in a NOMA-UAV system (NU) [365]. We followed the original models for these three network optimization problems.

We use the classic Denoising Diffusion Probabilistic Model (DDPM) [363] for model implementation. From Fig. 38 we have demonstrated that the proposed DiffSG effectively converges across various optimization problems, transforming the goal of directly inferring the optimal solution into fitting a high-quality solution distribution. Also, the optimization performance exceeds the original works [159, 357, 365], see [364] for details.

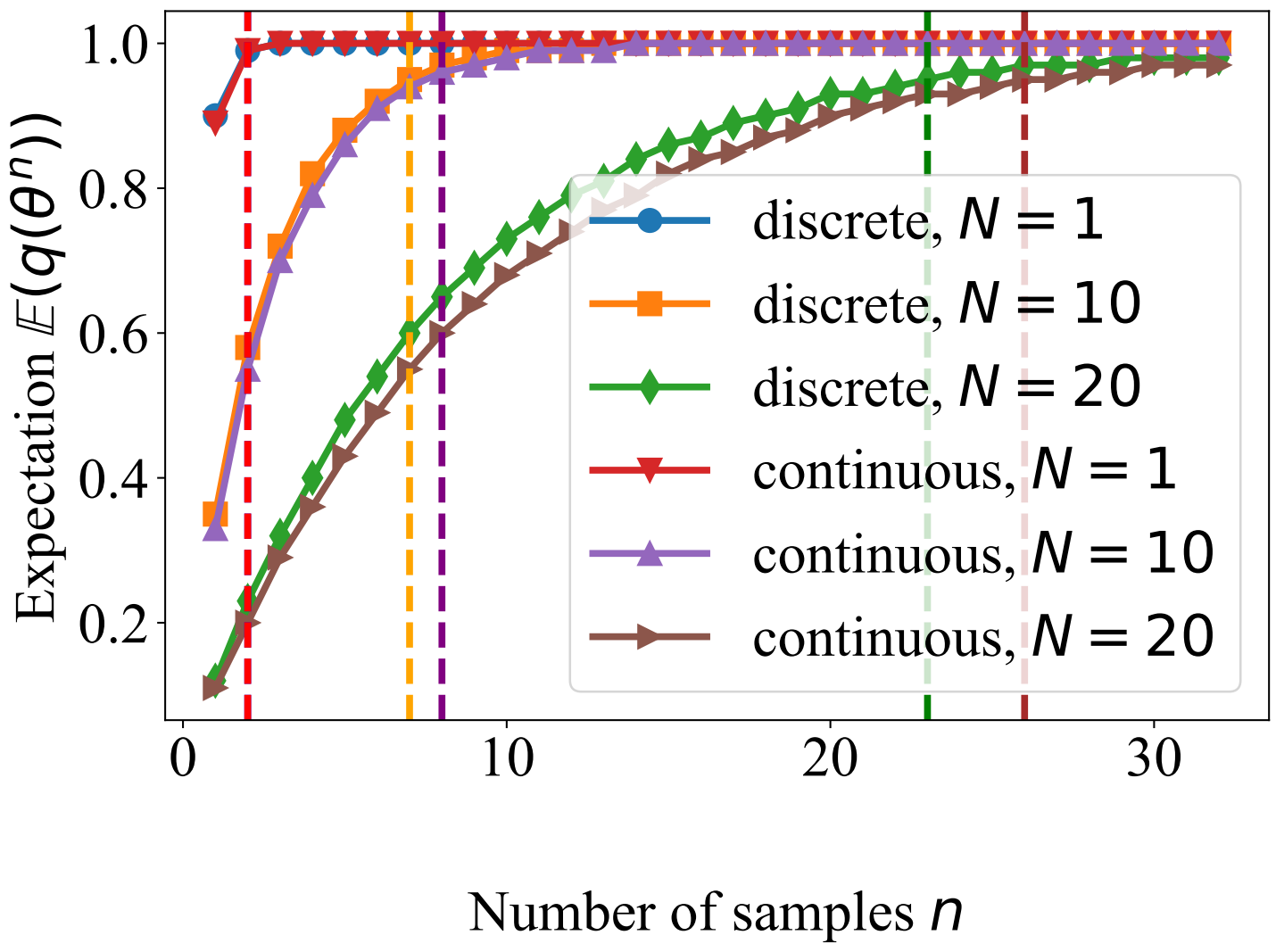

Figure 39: The expectation of hitting $y^*$ for different number of samples $n$ and variable dimension $N$. The dotted lines represent the lower bounds on $n$, where the red and blue dotted lines coincide.

After demonstrating the effectiveness of the diffusion model in generating solutions to complex network optimization problems, we further explored the underlying theory and design of the model. We have addressed the key questions of how and why to learn a high-quality solution distribution and provided a theoretical lower bound on the number of times this distribution should be sampled to reach the optimal solution. This theoretical foundation enables the generative model to learn sub-optimal solutions and sample toward the optimal solution. Building on this, we proposed a problem modeling method that reformulates a broad class of network optimization problems as graph optimization tasks, and designed a Graph Diffusion-based Solution Generation (GDSG) model. For experimental evaluation, we applied it to the multi-server multi-user (MSMU) computation offloading problem, which is a multi-task NP-hard problem involving both classification and regression.

As shown in Fig. 39, we have demonstrated that increasing the number of sampling iterations, $n$, brings the expected probability of hitting the optimal solution, $E(q(\theta^n))$, closer to 1, ensuring convergence. Parallel sampling significantly reduces the cost compared to serial sampling by accelerating neural network matrix operations, mak-ing it practical for real-world applications. Additionally, our implemented GDSG model exhibits nearly 100% inter-task orthogonality between edge classification and edge weight regression, meaning the loss gradients of both tasks on the same parameter are orthogonal during training, benefiting from the diffusion model's loss function setting. In terms of optimization performance, GDSG approximates the optimal solution with less than 10% error in fewer than 20 samplings (see [366] for details).

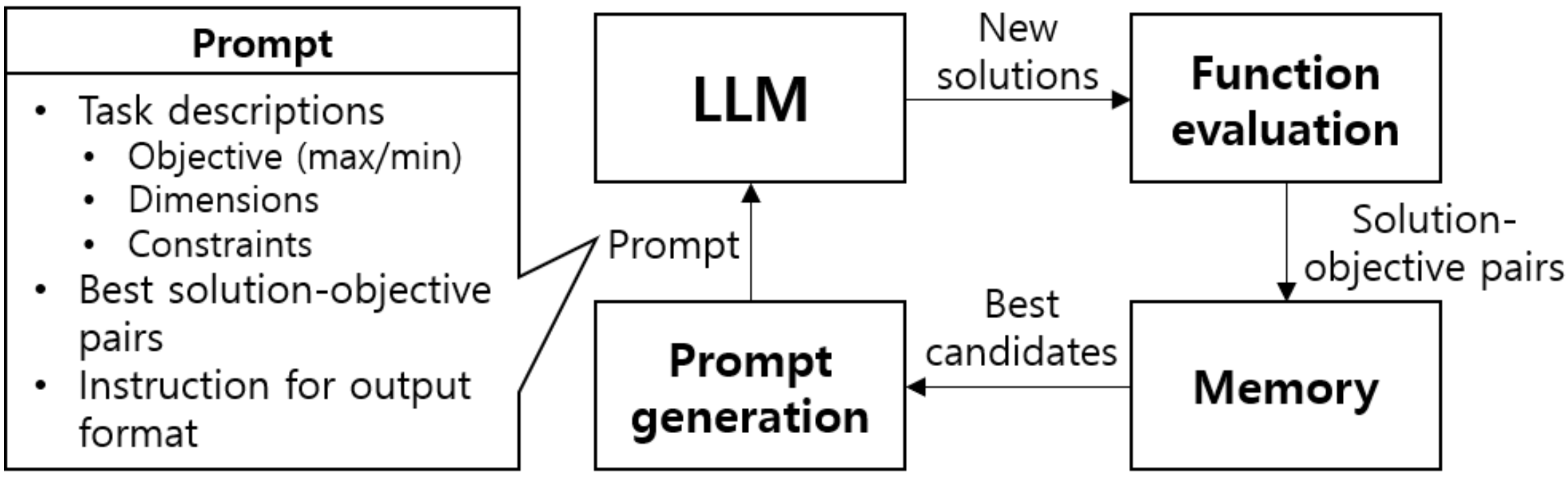

Figure 40: Knowledge-free LLM optimizer

Therefore, the diffusion generation model is applied to solution generation for complex network optimization problems. We have provided not only theoretical convergence guarantees but also empirical validation through various practical implementations, demonstrating the significant potential of this approach.

### 8.5.4 Knowledge-free LTM for network optimization

Conventional network management algorithms heavily depend on prior knowledge of system models and specific networking scenarios. However, there is a growing need for a universal optimization framework where a single optimization module can be seamlessly adopted to diverse network management tasks without relying on system-specific information. This requires knowledge-free optimization techniques that operate independently of scenario-specific details such as objective functions, system parameters, and network configurations. The core challenge of this approach lies in developing a hyper-intelligent black-box optimizer capable of formulating efficient decision-making policies through its own reasoning mechanisms.

Potential candidates for the knowledge-free optimization method include genetic algorithms (GA) [367] and reinforcement learning (RL) approaches. Rather than relying on mathematical models, these methods utilize optimization agents based on stochastic search mechanisms or neural networks (NNs). However, developing efficient agents in these approaches requires significant human involvement, including hyperparameter tuning and the design of training procedures, which must be meticulously customized for each unique network configuration and performance indicator. As a result, traditional methods struggle to generalize effectively, limiting their ability to accommodate the diverse and dynamic characteristics of future wireless networks.

Such challenges can be addressed through recent large language model (LLM) optimizer techniques [368] illustrated in Fig. 40. A pretrained LLM is utilized as an optimization module that generates new solutions based on its decision history. For knowledge-free operations, an input prompt is designed to exclude any system-specific knowledge such as mathematical models, channel state information, and application scenarios. Instead, it involves simple task descriptions, e.g., objective (maximization or minimization), solution dimension, and constraints, past decisions and their objective values, and desired output formats in natural language. Outputs of the LLM are assessed via a function evaluator, and the resulting solution-objective pairs are stored in a memory unit. The best candidates are then sampled from the memory and are utilized for generating the input prompt. Such a procedure

```
Now you will help me maximize a function with input
variables x1 and x2 which are between 0 and 1. I have
some [x1, x2] and the function values at those points.
The pairs are arranged in ascending order based on their
function values, where higher values are better.

Input :  [0.204 0.878]
Output :  1.512
Input :  [0.419 0.685]
Output :  1.452
....

Give me at most 8 new pair [x1, x2] that is different from
all pairs above, and has a function value higher than
any of the above. Do not write code. The output must
end with a pair [x1, x2], where x1, x2 are numerical values.
```

Figure 41: An example input prompt.

is repeated until the convergence.

The few-shot learning ability, which aligns pretrained LLMs to unseen tasks by prompting several examples [369], is a crucial component of the knowledge-free LLM optimizer framework. In the input prompt, the best solution-objective pairs act as few-shot examples for solving target network management problems. This helps the LLM to grasp the underlying characteristics of the optimization problem without invoking any prior information. Through this few-shot learning process, the LLM gradually refines its decisions, generating increasingly effective solutions that lead to improved objective values.

The most effective actions are retrieved from the memory and reintroduced into the input prompt in subsequent iterations. This self-feedback loop allows the LLM to iteratively enhance its decisions by learning from historical objective values [370]. The multi-step reasoning mechanism implemented here embodies the chain-of-thought process [371], where the LLM is guided through a sequence of intermediate reasoning steps. This approach improves the decision-making capabilities of the LLM optimizer without requiring any problem-specific information.

The viability of the LLM optimizer techniques has been examined in addressing traveling salesman problem (TSP) [368], multi-objective optimization [372, 373, 374], and wireless resource management tasks [375, 376]. In particular, the knowledge-free optimization ability of the LLM optimizer has been investigated in [376]. Without mathematical models of objective functions and channel state information, wireless resource allocation solutions obtained by using the LLM optimizer exhibit identical performance to model-based optimization algorithms.

The performance of the GPT-enabled LLM optimizer is assessed for solving transmit power control tasks in a two-user interference channel to maximize the minimum rate. The corresponding max-min rate problem is formulated as

$$\text{maximize} \min \left\{ \log\left(1+\frac{h_{11}x_1}{1+h_{21}x_2}\right), \log\left(1+\frac{h_{22}x_2}{1+h_{12}x_2}\right) \right\} \tag{37}$$

$$\text{subject to } 0 \le x_1 \le 1 \text{ and } 0 \le x_2 \le 1 \tag{38}$$

where $h_{ij}$ stands for the channel gain from transmitter $i$ $(i = 1,2)$ to receiver $j$ $(j = 1,2)$ and $x_i$ indicates the transmit power of transmitter $i$. The input prompt is designed as in Fig. 41. GPT-4 is tasked to generate 8 new solution vectors based on few-shot examples of past solution-objective pairs sorted in the descending order of objective values. No explanations about network configurations, channel gains, and closed-form objective functions are included in the input prompt.

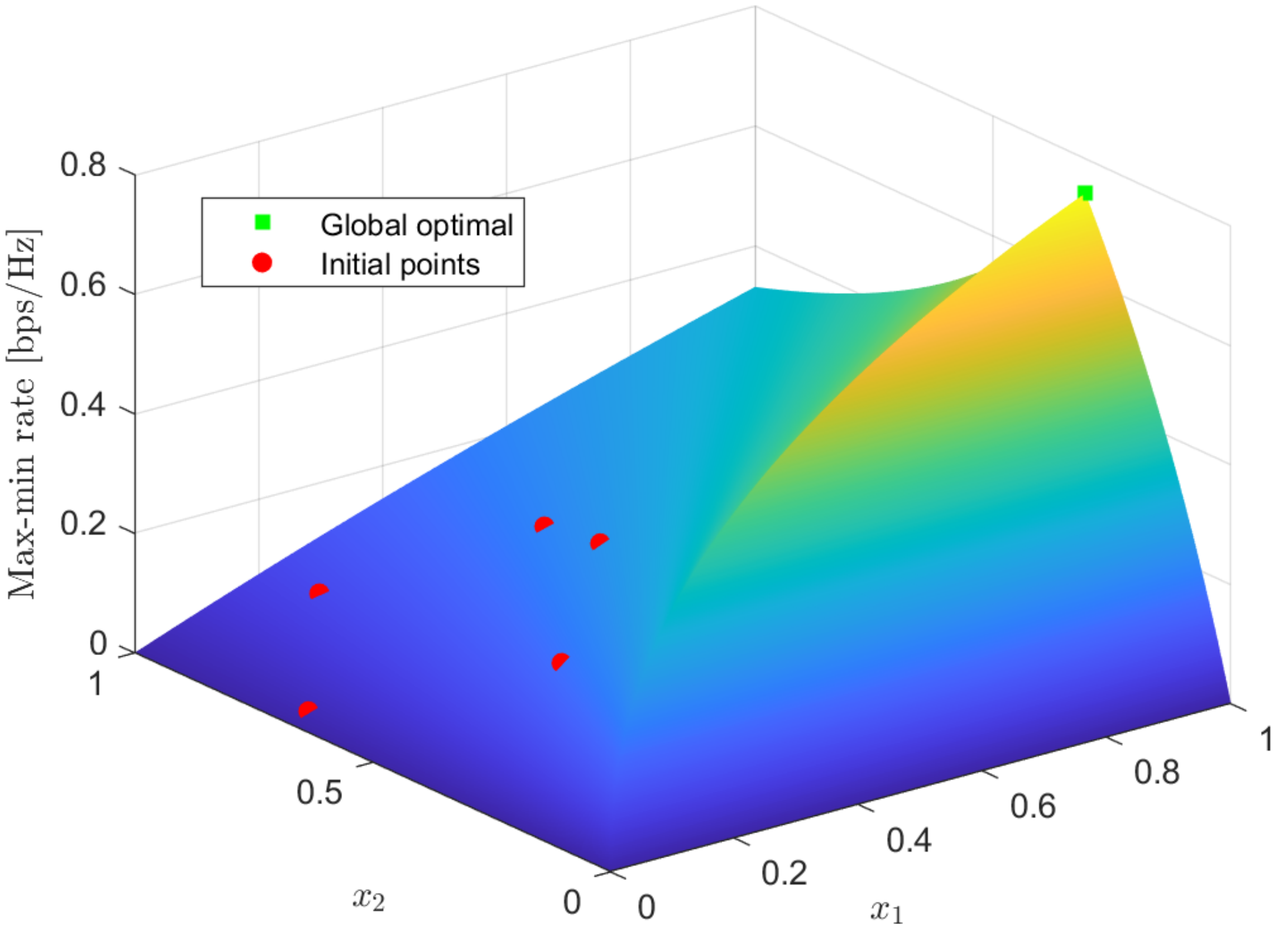

Figure 42: Max-min rate objective function.

Fig. 42 shows the max-min rate objective function for certain random channel gains. The global optimal solution for this task is obtained at $[x_1^\star, x_2^\star] = [1, 0.306]$, and the corresponding optimal value is given as 0.742 bps/Hz. For the initialization, 4 candidate solutions that are distant from the global optimal solution are prompted to GPT.

Fig. 43 plots transmit power control solutions generated by the LLM optimizer approach at different iterations. At the first iteration, the LLM optimizer deploys new solution vectors in a trapezoid shape near the initial points. By doing so, it can infer the ascending direction of the objective function. In the subsequent iteration, new solutions are located within a line corresponding to the gradient ascending direction. The LLM optimizer further exploits this search direction to identify the global optimal solution. Solutions obtained at the 10th iteration are closely located to the global optimal.

Fig. 44 depicts the max-min rate performance with respect to the iterations. For the LLM optimizer, the best objective value over 5 independent runs is plotted. It is observed that the max-min performance of the LLM optimizer gradually increases with the iterations. In particular, the performance is quickly improved at early iterations. However, after the 6th iteration, it gets stuck to a certain point and the performance is no longer enhanced. Such premature behavior is a primary challenge of the LLM optimizer technique, which can be tackled by employing multiple LLMs simultaneously [373, 376].

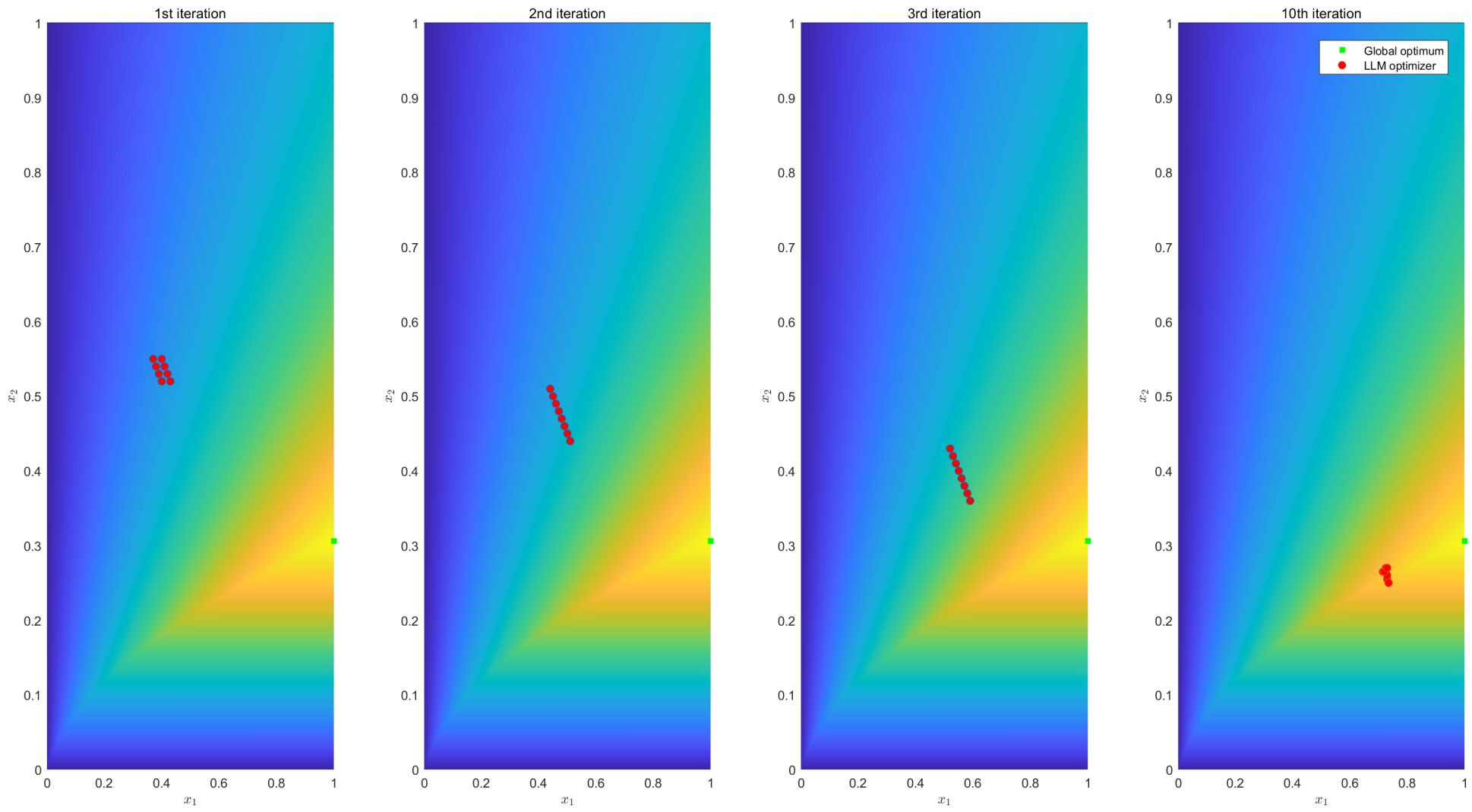

Figure 43: Solutions generated by LLM optimizer.

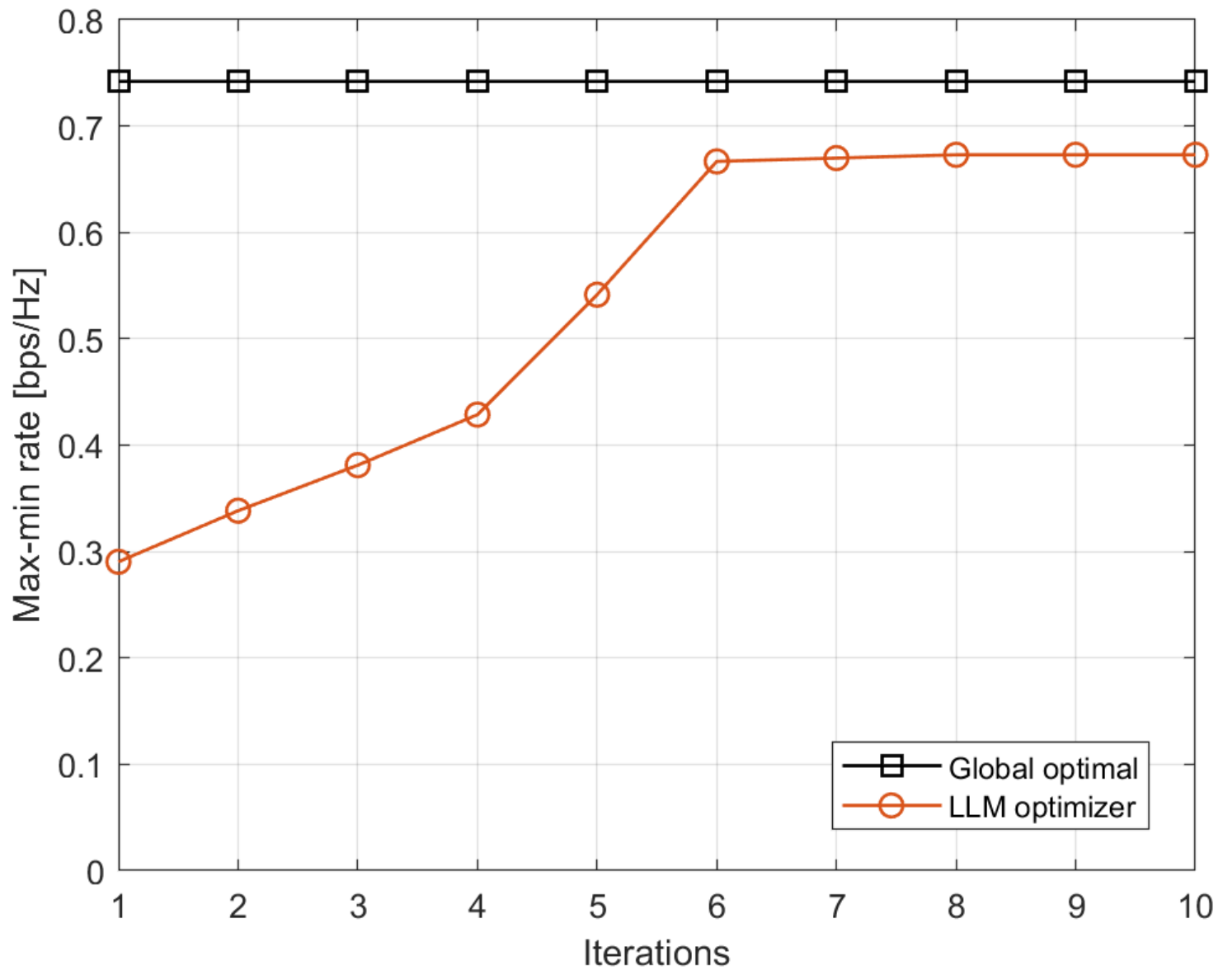

Figure 44: Max-min rate performance with respect to iterations.

The LLM optimizer remains in its early research stage, requiring further investigation to fully understand its optimality and advantages across diverse network scenarios and complex optimization problems. In particular, addressing limitations such as premature convergence and systematically validating the effectiveness of multi-LLM approaches are critical next steps. These efforts will pave the way for the LLM optimizer to emerge as a practical and scalable alternative or complement to traditional knowledge-based optimization methods.

## 8.6 Network automation and intent-based management with LTMs

Recently, Mobile Network Operators (MNOs) are considering to fully automate their networks to reduce management costs and deployment and maintenance time. Figure 62, showcases the level of network automation as defined by TMforum. MNOs aims to reach the level 5 for full autonomous networks. In order, to reach this autonomy, the concept of “intent” was introduced in standardization bodies.

| Level Definition | L0: Manual Operation & Maintenance | L1: Assisted Operation & Maintenance | L2: Partial Autonomous Network | L3: Conditional Autonomous Network | L4: High Autonomous Network | L5: Full Autonomous Network |
|---|---|---|---|---|---|---|
| Execution | P | P/S | S | S | S | S |
| Awareness | P | P | P/S | S | S | S |
| Analysis | P | P | P | P/S | S | S |
| Decision | P | P | P | P/S | S | S |
| Intent/Experience | P | P | P | P | P/S | S |
| Applicability | N/A | Select scenarios | | | | All scenarios |

**P:** Personnel, **S:** Systems

Figure 45: TMForum automation levels [377].

In the context of the standard organization of telecommunication and technologies as 3GPP, GSMA, TMForum and ETSI. The “*intent*” and “*intent based*” concepts often refer to a method of defining and configuring communications system based on their intended purpose or outcome. The concept of intent and intent-based system is an important part of work of those organizations to create more efficient, flexible, and responsive telecommunications systems and services that meet the needs of users and businesses. For instance, TMForum defines intents as the formal specification of all expectations including requirements, goals, and constraints given to a technical system. Intent is therefore purely an expression of what needs to be achieved rather than indicating how this can be done [378]. Intent based networking (IBN) was mostly applied in Software Defined Networks (SDN) [379]. SDN is a concept that involves separating the control and data plane of a network, allowing for quick and automated network reprogramming. 5G allows for the introduction of various new services and enables different industries to deploy their own customized networks using network slices. As a result, GSMA has proposed a template-based approach. In this approach, IBN users can personalize their intent for creating a network slice by specifying a set of attributes through a Generic Network Slice Template (GST). After the introduction of intents in the context of SDN controllers. Standardisation bodies provided multiple definitions of what is an ”*intent*”. For instance, the most recent definition of intent by IETF: “… a set of operational goals that a network should meet and outcomes that a network is supposed to deliver, defined in a declarative manner without specifying how to achieve or implement them” [380]. The purpose of intent is to define and communicate knowledge about expectations to a system in a way that allows automated processes to reason about it and derive suitable decisions and actions. Intent for network automation are managed by intent management functions. Intent serves the purpose of defining and conveying information about expectations to a system. This enables automated processes to analyze and make appropriate decisions and take actions accordingly. In the context of network automation, intent is managed by intent management functions or closed loops as chown in Figure 46. However, in order to achieve the level automation defined in Figure 62, LLMs are used as a key component in enabling autonomous networks. LLMs are responsible for learning from data, making decisions, and executing actions in an autonomous manner. They play a crucial role in enhancing the capabilities of autonomous networks and enabling them to adapt and evolve based on changing conditions. Intents interpretation is crucial to achieve fully autonomous networks. The intents are introduced via multiple channels as shown in figure 46: customer portal, business portal, order management and operations portal. For instance, a customer portal such as a chatbot enables the client to order a private 5G slice for its specific needs (for example to offer 5G connectivity for a stadium event). The clients intents are expressed with simple Natural language and should be translated into a description of a 5G service that answers the client's requirements. LLMs such as ChatGPT are very efficient with text analysis, summarising, answering questions, providing explanations, or engaging in interactive conversations.Therefore, LLM started to be applied for intent interpretation and management in networks. We can cite multiple tasks where LLMs could be applied for intent management:

- Business and service Intent resolution: A Business Intent Resolver transforms a business request from simple service questions into a Product (or a package of Products). The service resolver acts as the Communication Service Management Function (CSMF) among the slicing management entities specified by 3GPP [381].
- Intent conflict management: first step of intent management consists of gathering intents from various source (i.e., portals). Next step is to aggregate the similar intents to simplify management. However, before the implementation step, we need to apply solutions to detect potential conflicts between intents. For example, an intent to maximize bandwidth might conflict with an intent to minimize costs.

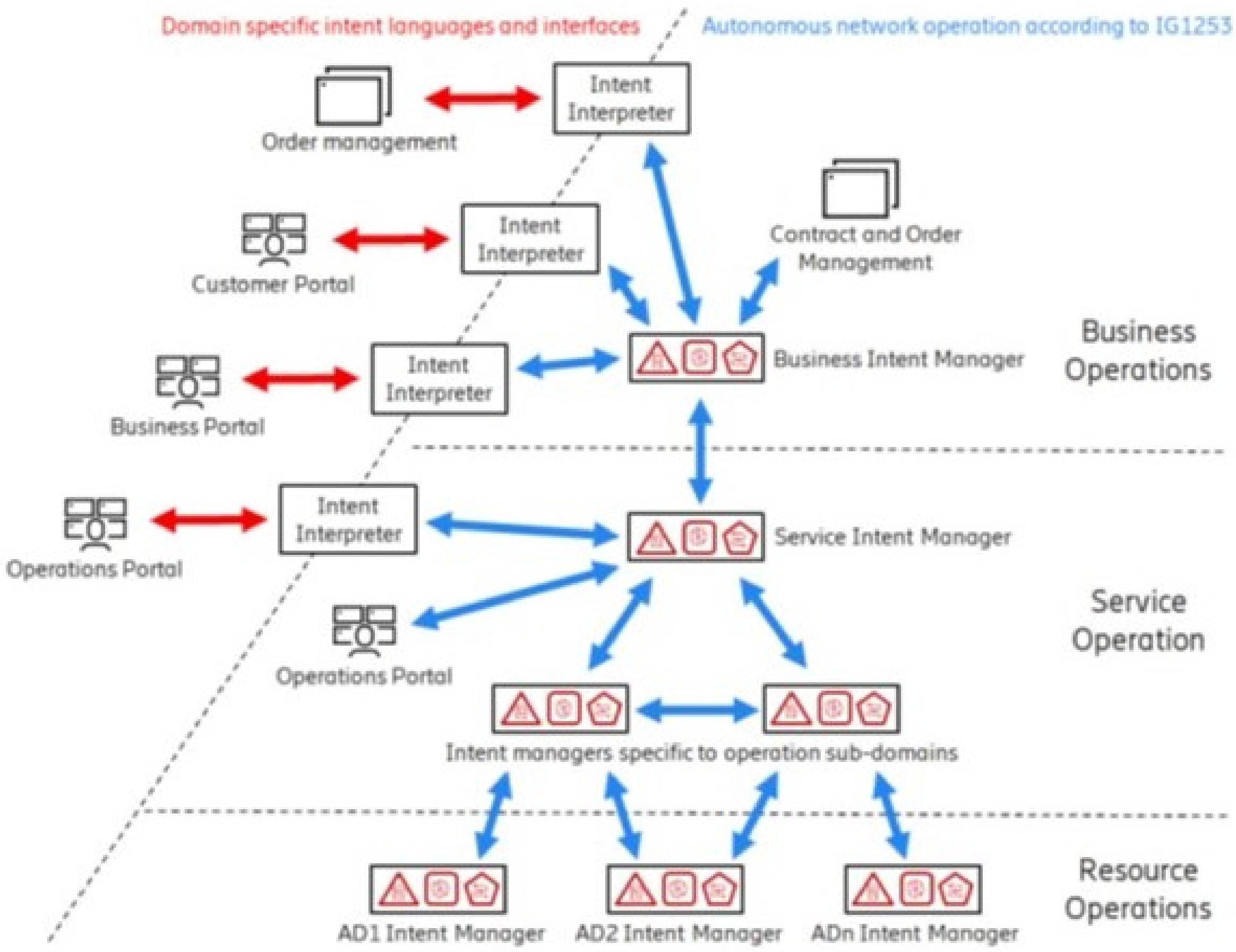

Figure 46: Intents interpretations in autonomous networks [378].

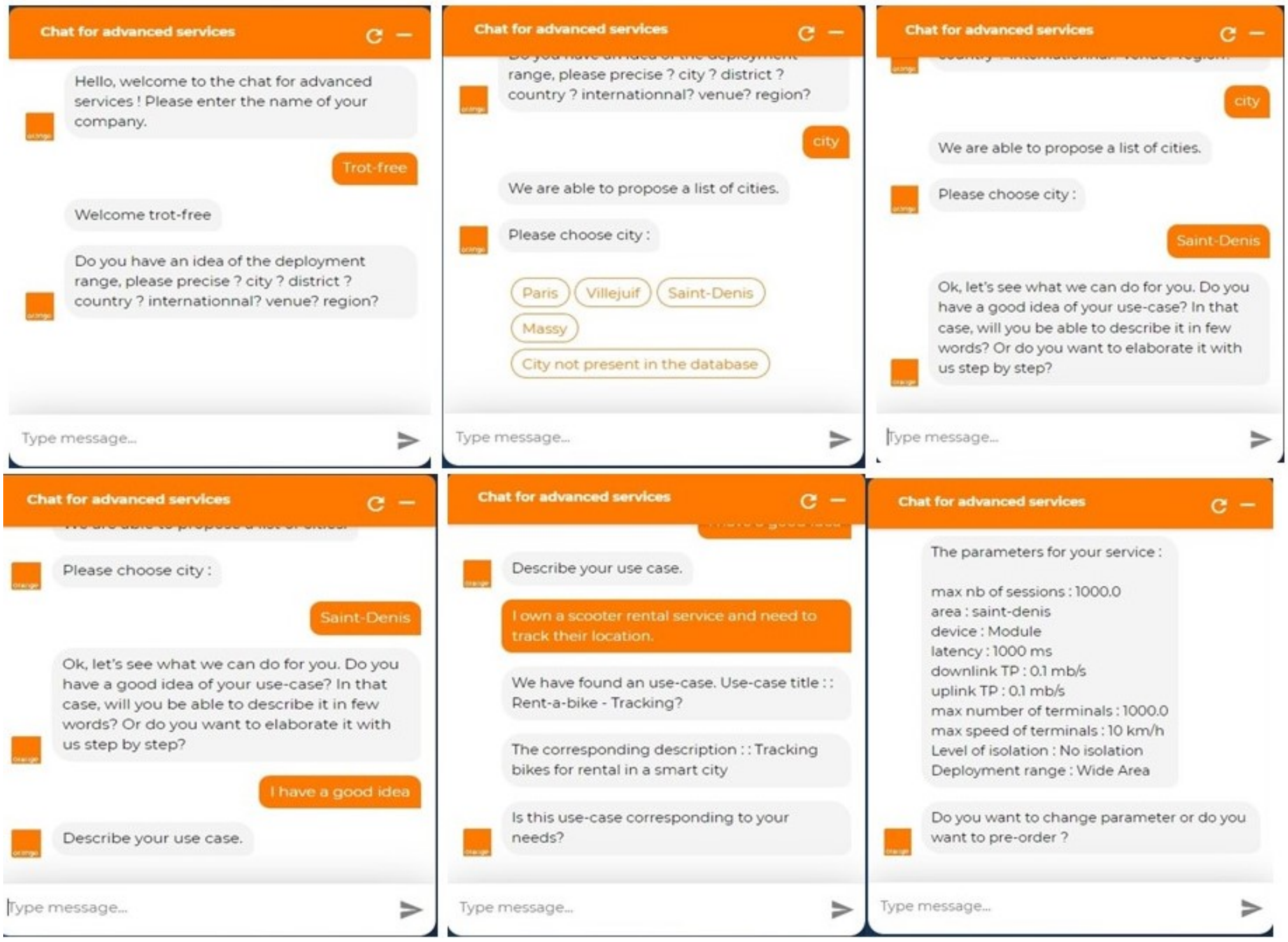

Figure 47: An example of scooters renting company.

- Intent assurance: its a number of procedures to ensures that the system behaves according to the specified intents (i.e., SLA or clients requirements). For assurance of intents, we should detect the root cause of violation and propose reconfiguration actions to come back to the desired state that respects the clients intents.

Authors of [382] developed a user friendly channels (i.e., Chatbot and web UI) for clients to automate 5G service ordering using LLM. Th authors used bert, a small masked language model developed by Google. The model was finetuned using the network entity recognition NLP task and the product catalog data in order to find a match between a client intents described in natural language and a product in the catalog. Figure 47, showcases an example of company that wants a 5g connectivity service to track its scooters. The company describes in simple natural language the need *I own a scooter rental service and need to track their location* that is matched behind using the LLM API to a nearest product in the catalog. The clients enters also other intents such as the supported latency and the maximum number of terminals.

## 8.7 Generating Commit Messages for Configuration Files in 5G Network Deployment Using LLMs

Network automation plays a vital role in enhancing network performance. Commit messages detail the various actions involved in modifying network configuration files and deployments. This study presents experiments and research on the automated generation of commit messages in the context of 5G network deployment.

In contemporary network management, the concept of Network as Code (NAC) is gaining traction. NAC fundamentally applies software development methodologies to manage and configure network devices and services. It enables the management of network configurations through version control systems and the implementation of automated tools and processes, similar to software development operations. This methodology not only boosts operational efficiency and minimizes human errors but also speeds up the deployment of network configurations. The adoption of NAC greatly enhances the consistency, traceability, and replicability of network configurations, providing exceptional flexibility and control in network operations.

Our research on NAC focuses on the leftmost section of Figure 48, which integrates vendor intentions and operations into a network configuration system that incorporates the Open Container Initiative (OCI) registry and Git for source control. This segment manages "Vendor Intents," which are specific configuration needs from external suppliers, and "Operator Intents," which represent internal configurations established by the network management team. We develop and manage these intents through code, aiming for systematic, traceable, and efficient deployment and modification of network configurations, thereby improving efficiency, consistency, and flexibility in network operations.

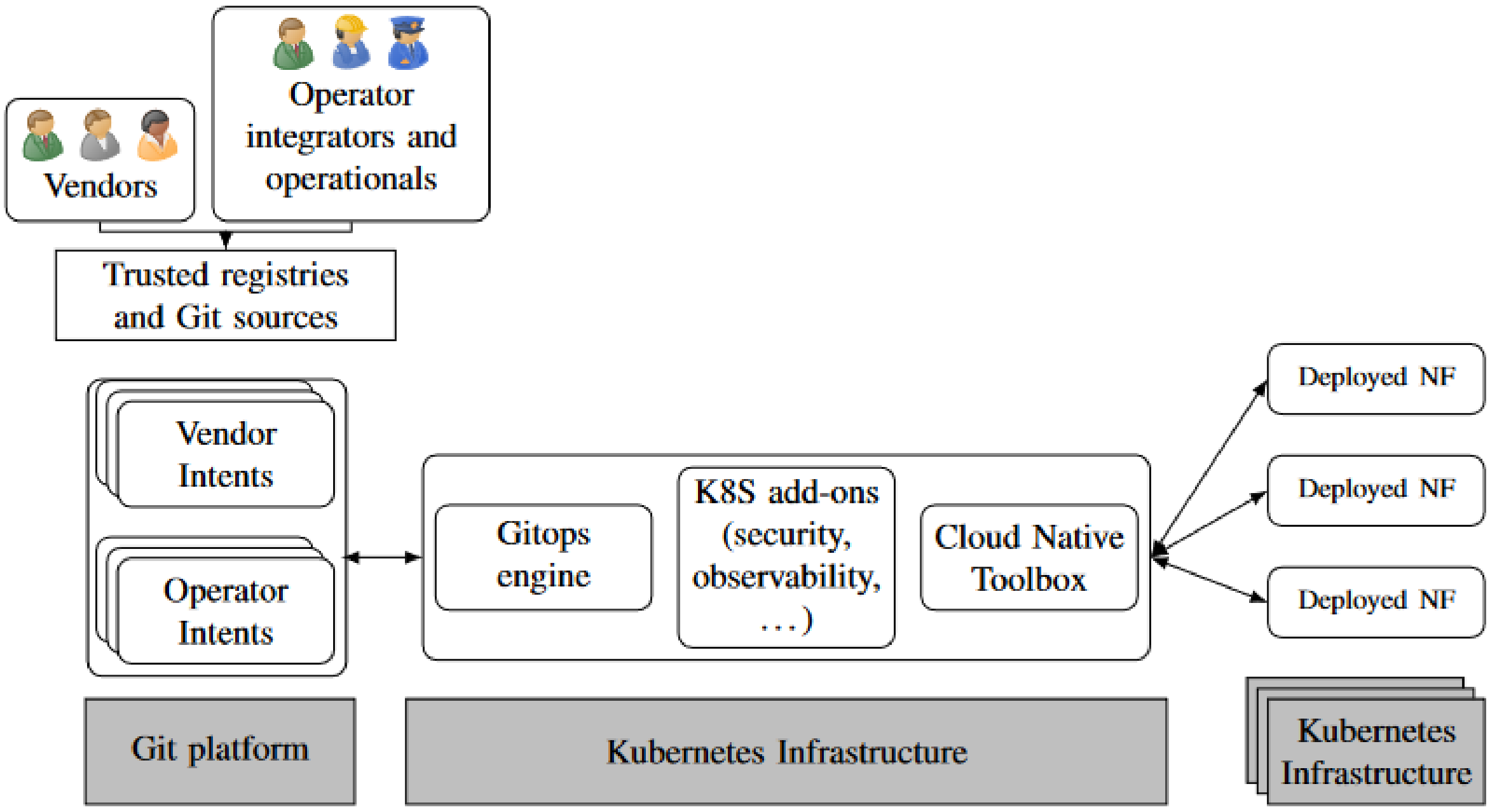

Figure 48: Network as a code. [207]

Large Language Models (LLMs) utilize deep learning techniques to automate and enhance various computational processes, including network management. These models are built on transformer architecture, which employs an attention mechanism to comprehend contextual relationships in text, essential for interpreting complex network configurations and assisting in coding tasks for NAC practices.

In network configuration management, precise commit messages are essential for version control and audit trails. However, creating these messages manually can be laborious and prone to errors. LLMs can automatically generate descriptive and accurate commit messages based on changes in configuration files, thereby enhancing documentation quality and operational transparency.

In our study [207], we investigated the use of prompt engineering for the automatic generation of code commit messages by designing and implementing five distinct prompts. Each prompt included unique elements and instructions to assess their impact on the quality of the generated results. In our use case, we considered the

following prompts:

```
Prompt1:  background + instruction + input data + output indicator
└─Prompt2:  Prompt1 + negative prompting
   ├─Prompt3:  Prompt2 - Repository tree
   ├─Prompt4:  Prompt2 + one-shot
   └─Prompt5:  Prompt2 + RAG
```

> Background: You are a bot producing commit messages.
>
> Task: Write a commit message that follows the Conventional Commits format ('<type>[optional scope]: <description>') based on the provided changes description as Input.
>
> Input:
>
> Change details:
> [Repository Tree] + [DIFF]
>
> Output:
>
> Using the Conventional Commits as a format('<type>[optional scope]: <description>')

Figure 49: Prompt1 structure [207].

In the following, we describe each Prompt in more details:

- **Prompt 1**: Basic Prompt with a background field to explain the role we want the model to consider when answering, the task, the input, and finally the output that contains the desired output format of the commit message as showcased in Figure 49.

- **Prompt 2**: Prompt with Negative Instruction

  This prompt adds a negative instruction to Prompt 1. (*Your output must be strictly in one line and in the format '<type>[optional scope]: <description>' without any extra text like 'this is the commit message:' etc., neither before, nor after*). The purpose of the negative instruction is to guide the model to avoid generating commit messages that do not meet the task requirements, such as avoiding unclear or overly simplistic statements.

- **Prompt 3**: Prompt without Repository Tree

  Prompt 3 removes information related to the code repository tree from Prompt 2. This change aims to test the model's performance without specific code organizational structure information, thereby assessing the model's sensitivity to environmental dependencies.

- **Prompt 4**: Prompt with one-shot

  Prompt 4 adds a specific commit message example (one-shot) to Prompt 2. This method helps the model learn how to construct commit messages through a concrete example, potentially improving the accuracy and relevance of the generated information.

- **Prompt 5**: Prompt with RAG

  Prompt 5 integrates RAG into Prompt 2. through RAG technology, the model queries related documents or existing data before generating commit messages, enhancing the accuracy and richness of the generated content.

To test the performance of the different AI models GPT-4, Llama3 and Mistral, we meticulously extracted data from a total of 581 commits, out of which 168 were authored by human developers, and 413 were generated by automated systems (i.e., bots). This dataset encompassed a variety of information, including the differences between commits (i.e., diffs), commit messages, authorship details, and unique SHA identifiers for each commit. Such comprehensive data capture facilitates an in-depth analysis of the project's evolutionary development and the incremental modifications made over time. For the evaluation of the different prompt using the different models, we used both automated and human evaluation methods, selecting BLEU, ROUGE, and METEOR as our metrics for automated assessment [173, 174, 383].

We present multiple evaluation results in [207]. as results we noticed that overall, LLMs generally excel in tasks involving the automatic generation of commit messages, especially with Llama 3 after applying RAG techniques. The bot-generated commits from Llama 3, enhanced by RAG, demonstrate strong performance in both automated and human evaluations. This success is likely attributed to Llama 3's improved ability to produce well-structured and formatted content. In contrast, the evaluation scores for human-generated commits are more balanced and consistent across various models and prompting methods.

## 8.8 Use cases in optical networks

Generative AI, and LLMs in particular, have garnered tremendous attention since 2023 in the optical network community. As an example, a complete workshop titled "How Can Generative AI be Used for Network Operations?" was organized at the March 2024 edition of OFC, the leading conference in the field.

As optical networks still rely on manual operation, large language models are seen as an opportunity to ease network operation, by acting as the human interface to the NMS (network management system), and abstracting complex concepts and tasks so that less skilled labor can operate a network, or operations can be shortened and/or automated. When combined with a digital twin, the output (i.e., suggestions of desired actions) of an LLM agent can first be tested within the twin, which acts as a sandbox, then be pushed to the field network. This combination AI agent (based on LLM)/digital twin has become popular in the research community, with several lab experimental demonstrations [384] and even field trials [385, 386].

As such, a large model can be used as a copilot for optical network management. As generalized LLMs have little optical communication/networking background, they need to be given additional information on one or more of the following: physics of optical networks; optical network management rules; and product information. This can be done through providing context/prompt engineering, RAG, and/or fine-tuning, as in [384, 385, 386, 387]. This allows an LLM-based management system to advise on certain problems faced during operation, and in certain cases, to interact directly with the equipment itself, when the network state and product information are given through one the techniques mentioned above.

Typical tasks envisioned to be delegated to LLMs mainly pertain to network management, including but not restricted to, network design (selection of equipment matching operator and physical constraints), resource allocation (e.g., routing and spectrum allocation), physical layer optimization (e.g., channel power equalization, setting of various equipment such as the optical amplifiers) and fault management (root cause analysis, suggestion for remediation) [388].

In most cases, the LLM is simply an intermediate layer between the human operator and the network management system, and its role is to interpret a human request and call the right network function, which implements some optimization algorithm. Correct interpretation of the human request, a language-oriented capability, and calling the right function, can be quantified in terms of API calling accuracy as in [389, 384, 390].

In some cases, the cognitive ability of LLMs is also leveraged to perform more advanced tasks. For instance, an LLM was able to output an algorithm and the associated code for a resource allocation problem (routing and spectrum allocation) in [391]. Reasoning ability is used in [392, 384] to analyze network logs, identify the root cause of a failure, and propose a solution.

Looking forward, possible and desired applications for large models include:

- Pro-active network operation, whereby a large model predicts issues in the network (typically, those are relatively slow degradations such as an aging board; the most common failures, fiber cuts, are impossible to predict until a few seconds before their occurrence.) This can be done through telemetry or log analysis. Pro-active maintenance then includes turning off and replacing the soon-to-be faulty board, reroute the traffic before the failure, etc.
- Security enhancement: detect network vulnerabilities through literature and code analysis.
- Various operation optimization, such as adjusting the operation mode of the equipment to tune its energy consumption to the (current or foreseen) demand.
- Finally, as large models progress, they could complement or even replace some of the physics that drive the optical physical layer equipment. The behavior of most equipment is well-modeled with physics, but some models can be improved as the main underlying assumptions are limited, and others rely on heavy computations such as nonlinear differential equations solving. It is possible that large models eventually find better approximations, or even new (more accurate or faster) models.

# 9 LTMs for Network Planning

Part of the following has been reported in [393].

Optimization tasks are critical in identifying the most effective solutions within a complex decision space, and they have become increasingly important in the evolving landscape of wireless communication. Moving from second-generation networks, supporting voice calls and text messages, to fifth-generation and beyond (B5G) networks, there has been a giant leap in capacity and capability. These networks are expected to provide an umbrella to the Internet of Things, machine-to-machine communications, virtual reality, and other emerging applications [394]. Commensurate progress has been observed in network planning techniques that have evolved to address increasing network complexity and the diverse needs of our ever-increasing digitalized society [395].

In wireless network planning, the traditional approach has been heavily dependent on the experience of network engineers, especially in the crucial task of selecting positions for the installation of Access Points (APs). In more recent times, their strategic decisions have been supported by radio propagation models, for example, the empirical Okumura-Hata model [396] or deterministic ray-tracing algorithms [397] to predict the received signal strength and so confirm the suitability of chosen sites.

The advent of advanced computational capabilities has shifted network planning towards an algorithmic-based approach, often supplementing or replacing human expertise with optimization algorithms. These methods, particularly metaheuristic algorithms such as evolutionary strategies [398, 399], optimize AP placement and network coverage using detailed radio propagation models. Successful network planning relies on the optimization algorithm's ability to identify optimal deployment configurations and the optimization performance is confirmed through propagation models. Recent innovations, such as optimization algorithms integrating Large Language Models (LLMs) [362, 400, 401], show promising results in efficiently addressing optimization issues describable using natural language. However, little research studies their applications in complex scenarios such as wireless network optimization, where integration with expert models is required.

Against this above background, we aim to address the challenge of network deployment within the wireless communication sector by seamlessly incorporating an LLM-based framework with sophisticated propagation models. We introduce an LLM-based optimization framework, termed Large Language Model-based combinatorial optimization (LMCO), which to our knowledge is a unique implementation in wireless communications. This innovative framework demonstrates notable advantages over conventional optimization techniques. Specifically, our experiments suggest that LMCO not only surpasses traditional solutions in terms of performance but also reveals its adaptability to address an extensive range of analogous optimization challenges.

## 9.1 LLM as optimizer in network planning

It is plausible that LLMs inherently embed human-like experiences of continuous learning through reasoning and decision-making, offering significant potential as optimizers across various domains.

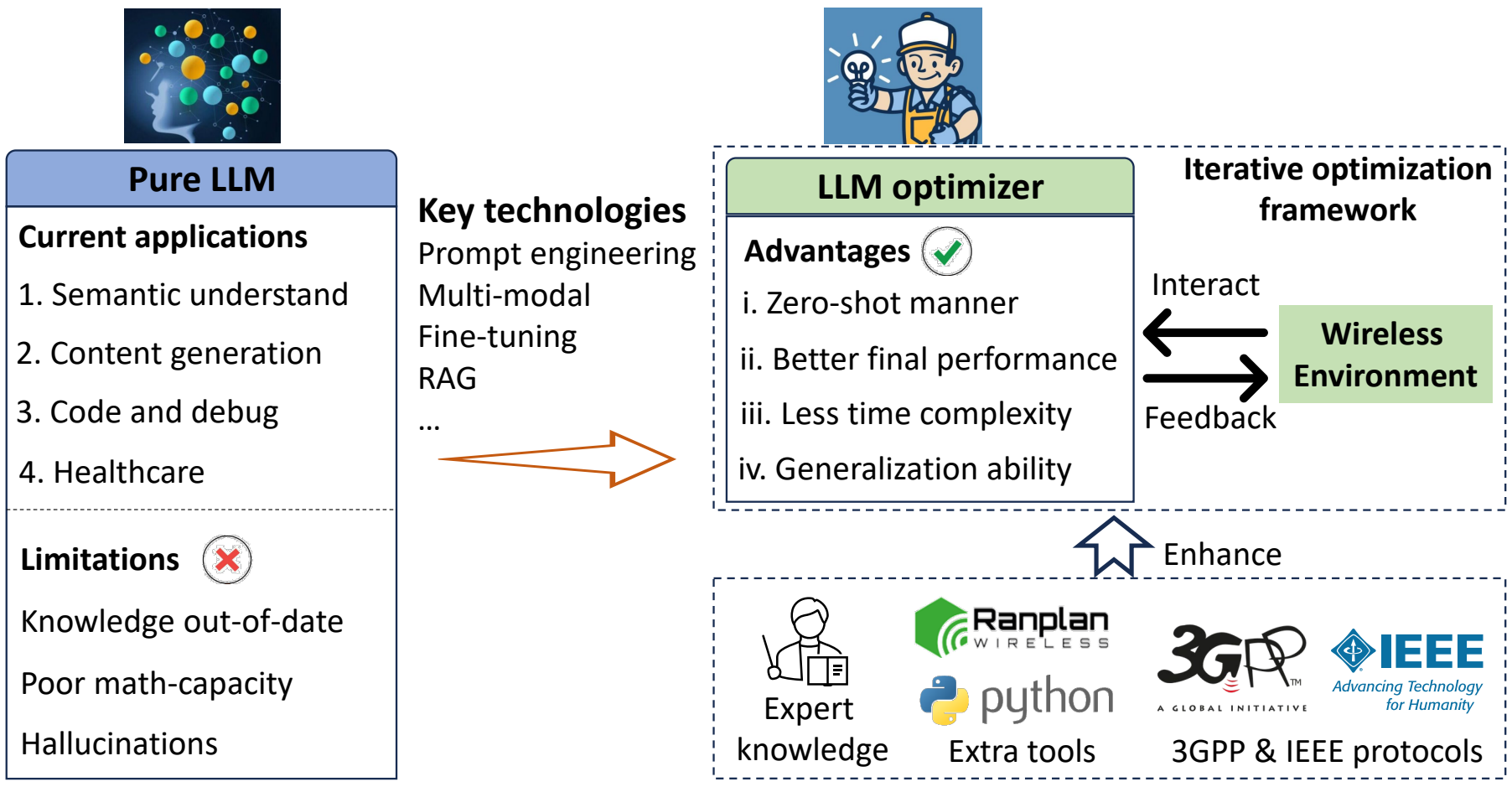

Figure 50: Overview of LLM-based optimizer.

While LLMs exhibit strong optimization capabilities, their limitations—such as hallucination, unstable knowledge, and challenges with complex mathematical problems necessitate careful prompt design [402]. Techniques like retrieval-augmented generation (RAG), and transfer learning can significantly enhance their performance and accuracy. By structuring prompts to include system parameters, problem descriptions, and expert knowledge, LLMs can dynamically interact with users to form robust optimization chains. LLMs have demonstrated impressive optimization abilities in diverse fields, including education, healthcare, and customer service [403, 404, 405, 406]. In wireless communication, frameworks such as WirelessLLM [314] and multi-agent LLMs [407] have been developed to address network challenges.

Unlike traditional optimization methods, LLMs leverage pre-trained knowledge and continuous learning through prompts, enabling efficient problem-solving without detailed step-by-step programming.

### 9.1.1 Preliminary

A key element of the LLM-based framework is the "prompt", which dynamically incorporates: (*i*) the user's query, (*ii*) brief examples to refine the model's response and (*iii*) instructions for processing input. LangChain [408] provides tools for constructing prompts using specialized templates that generate reproducible text strings, accepting user-defined parameters. Its modular design enables developers to build applications with multiple prompts, supporting complex and adaptive interactions with LLMs, simplifying development, and enhancing user experience.

The proposed LMCO framework [393], built on LangChain, aims to replace traditional single-target optimizers like evolutionary algorithms and ant colony optimization [409, 410] in wireless network planning. Unlike conventional methods requiring detailed step-by-step programming, LMCO leverages minimal domain expertise and relies on domain-specific information provided through prompts. Prompts also specify strict output formats to aid interpretation. Subsequent sections detail the AP placement optimization problem, LMCO's structure, inputs, outputs, and component implementation.

### 9.1.2 AP Placement Task for Network Optimization

The primary objective of network planning optimization frameworks is to determine the network topology, i.e., the number of APs and their locations, that optimize and meet some target key performance indicators (KPIs), e.g., coverage, delay, power supply, or installation and administration costs [411]. In this work, we consider the optimization task of meeting a target coverage level, $\phi$, while minimizing the installation costs, i.e., the number of APs, $N$. The corresponding optimization problem $T$ can be formulated as follows:

$$\begin{aligned} & \min_{\{x_n,y_n\}} N \\ \text{s.t.} \quad & \sum_{k=1}^{N} R_k \geq \phi, \\ & x_{\min} \leq x_n \leq x_{\max}, \forall n \in \mathcal{N}, \\ & y_{\min} \leq y_n \leq y_{\max}, \forall n \in \mathcal{N}, \\ & \min\left\{ \left|x_i - x_j\right|, \left|y_i - y_j\right| \right\} \geq L, \forall n,l \in \mathcal{N} \end{aligned} \tag{39}$$

where $(x_n, y_n)$ indicates the location of the $n$-th AP, $R_k$ is a function that evaluates the coverage for each AP, and we use 3D-ray-tracing software from Ranplan [412] for the evaluation of coverage in this work. Specifically, $x_{\min}$, $x_{\max}$, $y_{\min}$, $y_{\max}$ refer to the boundary constraints of AP locations in a square scenario, $L$ is the minimum distance between APs. It is set to 1 meter in the following experiments, ensuring that APs do not overlap, and the constraint $\sqrt{(x_i - x_j)^2 + (y_i - y_j)^2} \geq L, \quad \forall i \neq j$ guarantees that the distance between any two APs is greater than or equal to $L$. Next, we will show how we can use the proposed LLM-based framework, i.e., LMCO to address the above problem.

### 9.1.3 LMCO

A block diagram depicting the LMCO framework [393] is presented in Fig. 51, showing the workflow and the prototype prompts crafted for the utilization of an LLM in the context of network planning. The framework entails two modules, leveraging the in-context learning potential of the LLM, orchestrated by strategically formulated prompts: (*i*) an initialization and (*ii*) an LLM-driven deployment optimization module, whose details are presented in Algorithms 1 and 2, respectively. The initialization module is used to automatically determine the maximum number, $N_{max}$, of APs to be deployed, i.e., its output is $N_{max}$. Then, given this upper bound, the LLM-driven optimization module outputs the required number of APs, $N$ ($N \leq N_{max}$), and their respective locations, $S_{locations}$, resulting in a network deployment that meets the target coverage levels.

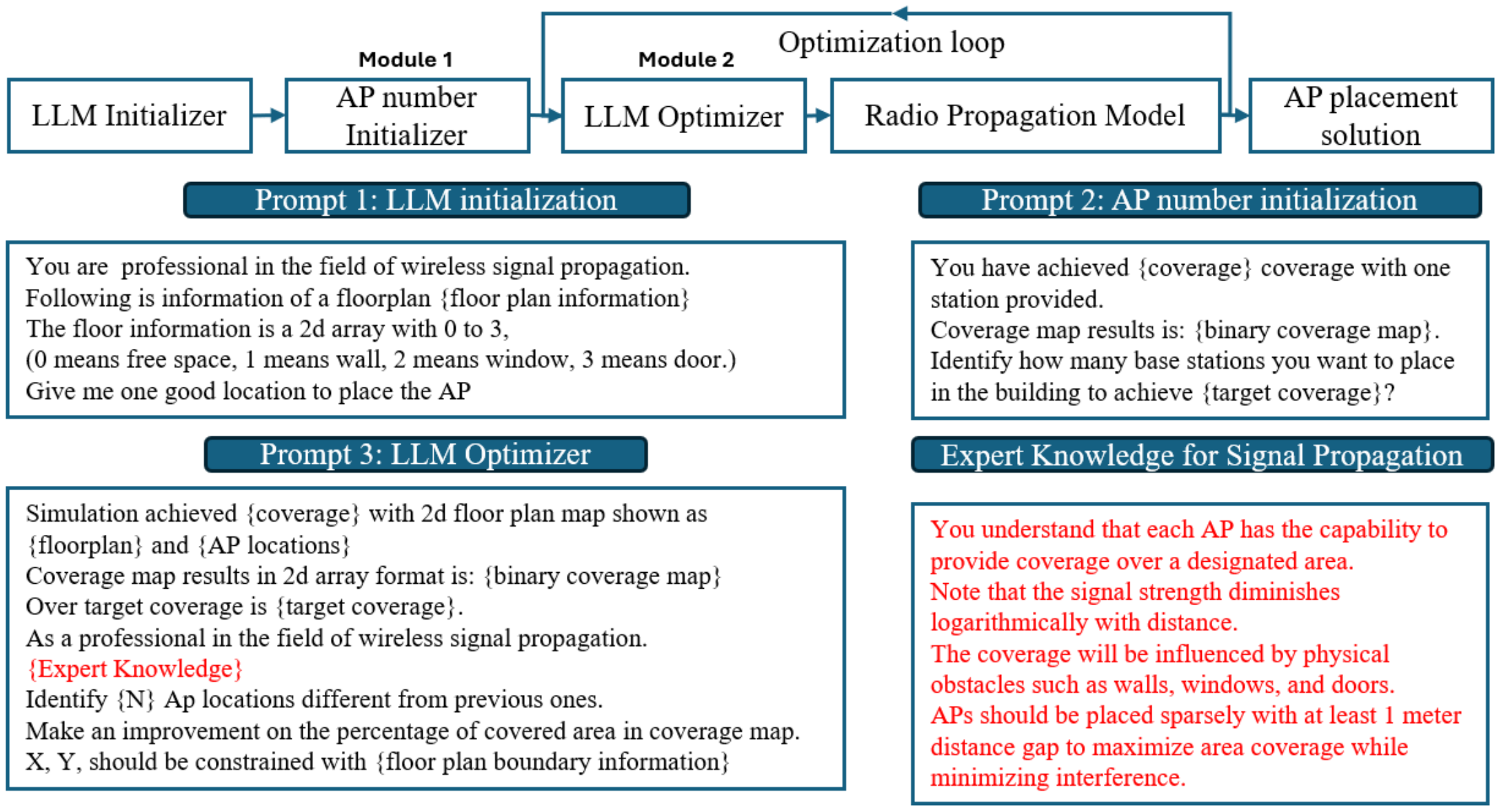

Figure 51: The LMCO framework has 3 groups of crucial prompts that are employed to guide and instruct LMCO in resolving wireless optimization issues. The segments enclosed in "{}" within the prompt are placeholders that will be substituted with the relevant content when communicating with the LLM, it also has a block providing expert knowledge to the LLM [393].

The floor plan comprises a grid of points depicting the wall layout, and the construction materials used, and it is represented as a two-dimensional (2*D*) array with different numbers indicating the use of a different material. Finally, the target coverage refers to the percentage of grid points at which the received signal strength (RSS) is larger than a threshold.

The functionality of the LLM for each module is configured through appropriately designed prompts. The suitably selected prompts for the other two modules, along with their implementation details are discussed in the following subsections.

**Module 1: AP number Initializer**

As mentioned previously, the goal of the AP number initializer is to determine the minimum number of APs to deploy without human intervention. Consequently, in the first instance, a designed prompt for the initialization module is provided to the LLM. Based on this prompt, the LLM will generate a location $s_loc$, where the AP will be deployed. Then a ray tracing simulator is employed to simulate the RSS distribution in the indoor environment and calculate the coverage. This output, in turn, is utilized by LMCO to ascertain a preliminary AP count, setting the stage for subsequent optimization processes.

**Module 2: LLM Optimizer**

The second module integrates an LLM as a combinatorial optimizer, operating in a zero-shot fashion. Again, a designed prompt informs the LLM about (*i*) the geometry layout, (*ii*) the initial number of APs indicated by Module 1, (*iii*) the network topology, (*iv*) the attained coverage, through a binary coverage heatmap representation (with 0 and 1 indicating above and below the coverage threshold, respectively) and an aggregated coverage percentage. Note that the latter constitutes the objective function of (39) that the LMCO aspires to maximize the coverage. In addition, an *expert knowledge prompt* is embedded in the LLM optimizer prompt, to induce common network engineering knowledge to the LLM and orchestrate its actions. Given these pieces of information, the LLM is asked to provide the number of APs, $N$, and their 2D locations (different from the previous ones) such that the coverage is improved. As shown in Algorithm 2, this process is repeated iteratively until the desired coverage level is reached, and at each iteration, the RSS and the coverage are evaluated with a ray-tracer for the network deployment indicated by the LLM. During this iterative process, the initial number of available APs is increased by one, if the attained coverage is not improved for the 6 consecutive iterations.

### 9.1.4 Coverage Evaluation via Ray Tracing

Our target is to identify an optimized configuration of AP locations such that when they are deployed in our target environment and the path loss is calculated using the propagation model, the resulting coverage fulfills our specified criteria. Figure 52 showcases a scenario with a floor plan and includes a path loss map derived from the strategic positioning of APs. We begin by setting a path loss threshold—any location with path loss surpassing this threshold is classified as an uncovered zone. We then measure coverage as the percentage of the area that achieves satisfactory signal strength within the established path loss threshold.

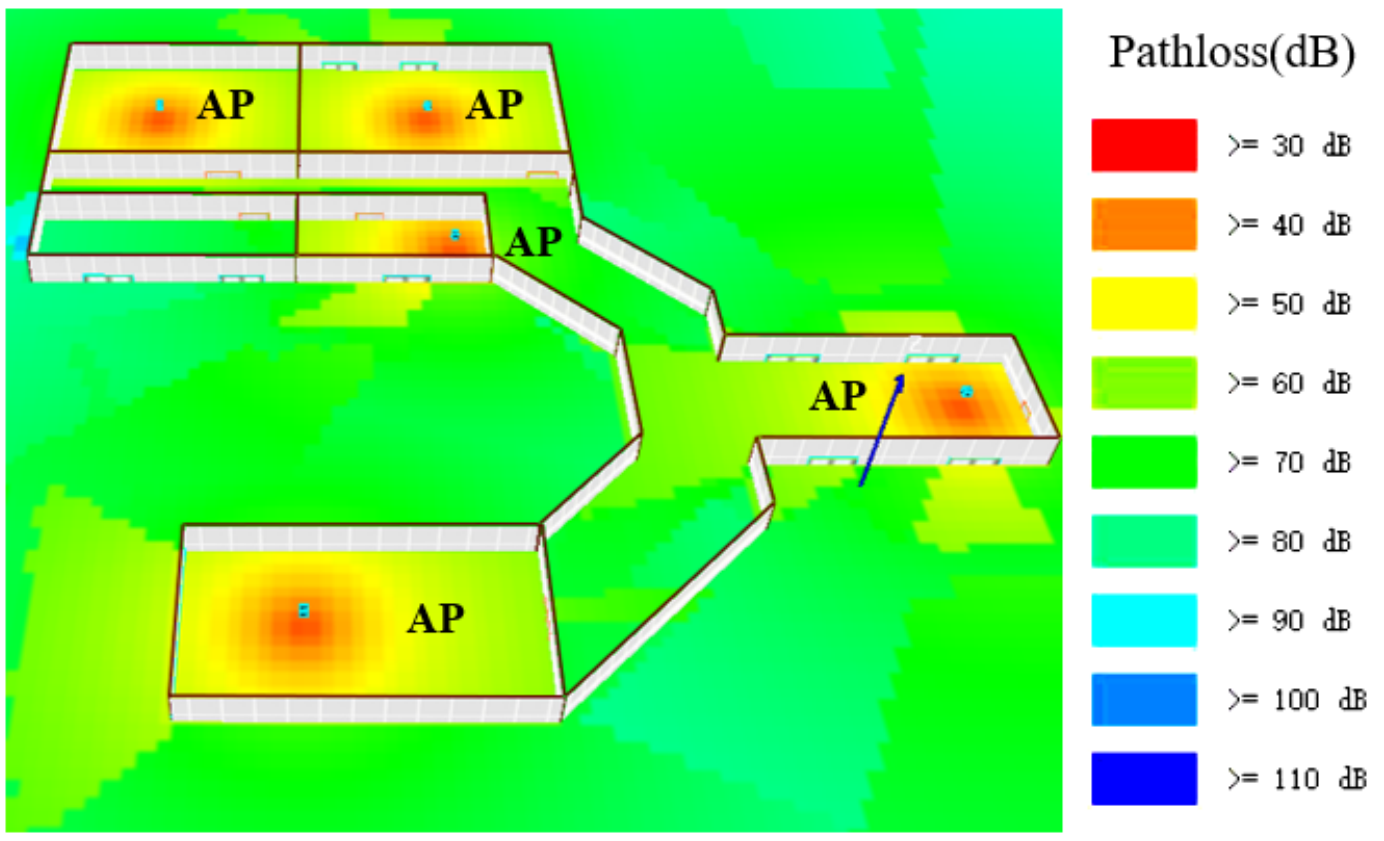

Figure 52: A sample scenario encompasses details of the floorplan and the path loss map computed with ray-tracer

### 9.1.5 Baseline Metaheuristic Optimization Algorithm

The performance of LMCO will be compared to Ant Colony Optimization (ACO), a widely used metaheuristic inspired by the foraging behavior of ants, known for its effectiveness in combinatorial optimization problems. ACO works by simulating "artificial ants" that explore the optimization space to identify near-optimal solutions. Starting with random initial solutions, each ant evaluates its results to determine a weight, analogous to pheromone levels used by real ants to mark efficient paths. Subsequent solutions are chosen probabilistically based on these weights, and the process repeats until the objective function converges. The weight calculation and path selection probabilities depend on the specific ACO implementation; this paper adopts a greedy ACO approach similar to [413].

### 9.1.6 Evaluation in practical use cases

To showcase the potential of LMCO to solve combinatorial problems and assist network planning we consider two use cases. The first experiment is conducted in a controlled environment, where a fixed number of APs, $N$, is predetermined to achieve an optimal coverage metric. In the second experiment, we remove the constraint on the number of APs and seek solutions that would meet our predefined coverage requirements, regardless of the number of APs deployed. This approach allows us to evaluate the models' adaptability and efficiency in achieving the target coverage levels, which can entail a varying number of APs. In both experimental setups, our goal is to ensure that at least in 90% of the building area, i.e., $P = 0.9$, the RSS is above -90 dBm. Each instance in which the optimizer calculated coverage metrics using the proposed AP configurations was considered a single iteration.

To ensure a fair evaluation, we design two experimental protocols, each representing a common scenario in wireless network design. These scenarios are selected to assess different aspects of performance. For each experimental setup, we conducted 20 tests using OpenAI's gpt-4-turbo-preview iteration as the LLM within the LMCO strategy. We compare the effectiveness of the solutions on the basis of the number of iterations needed to satisfy the established coverage criterion.

**Experiment to find a solution in simple indoor scenarios**

In this experiment, we first skip the initialization phase in the LMCO algorithm due to the use of a predetermined number of APs and we conduct tests in two indoor environments. The first one assumes a simple geometric space measuring 23.8m by 20.2m and a more complex configuration with dimensions of 58.8m by 63m. In both cases, the number of APs used in each scenario was determined by prior experimentation, which established the number of APs required to meet the coverage criteria.

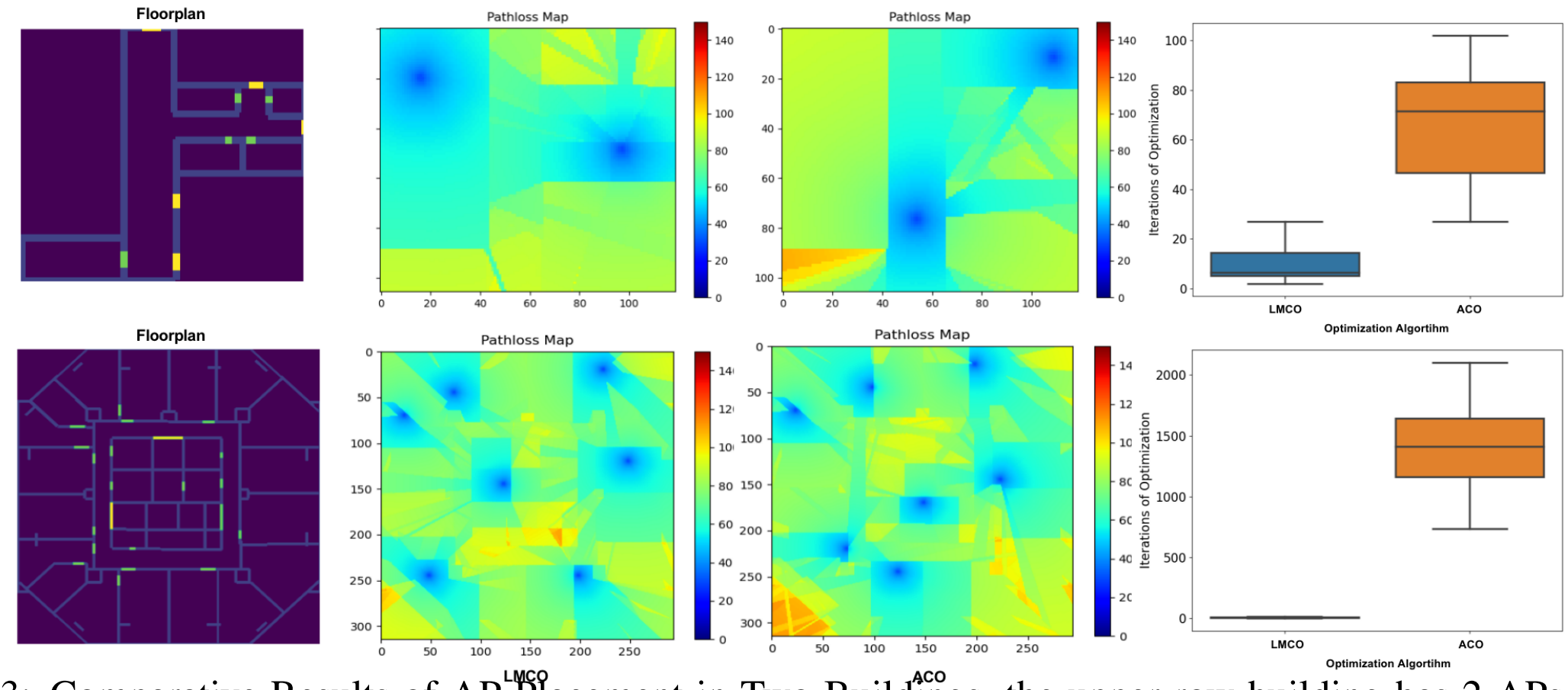

Figure 53: Comparative Results of AP Placement in Two Buildings, the upper row building has 2 APs and the lower row building has 7 APs: Iteration Efficiency of LMCO versus ACO

Figure 53 illustrates the results of this experiment. In both scenarios, the LMCO strategy significantly outperforms the ACO method. On average, LMCO requires only 9.2 iterations to meet the coverage criteria in the simple scenario with 2 APs, compared to the 63.4 iterations needed by ACO.

Similarly, in the more complex scenario with 7 APs, LMCO averages 10.9 iterations, while ACO requires a substantial 1394 iterations to achieve comparable coverage levels. Due to the inherent stochastic nature of both algorithms, the results are presented as a bar plot that reflects the distribution of outcomes across 20 independent tests for both scenarios.

In an advanced experiment, we utilized the initialization phase and did not provide either algorithm with information on the number of APs needed to meet coverage requirements. LMCO's average iteration count of 9 versus ACO's 63 in the simpler geometric space underscores its superior convergence. The difference becomes more pronounced in the complex environment, where LMCO's iteration count was two orders of magnitude lower than

ACO's (11 compared to 1394).

**Experiment on finding a solution in real-world scenario**

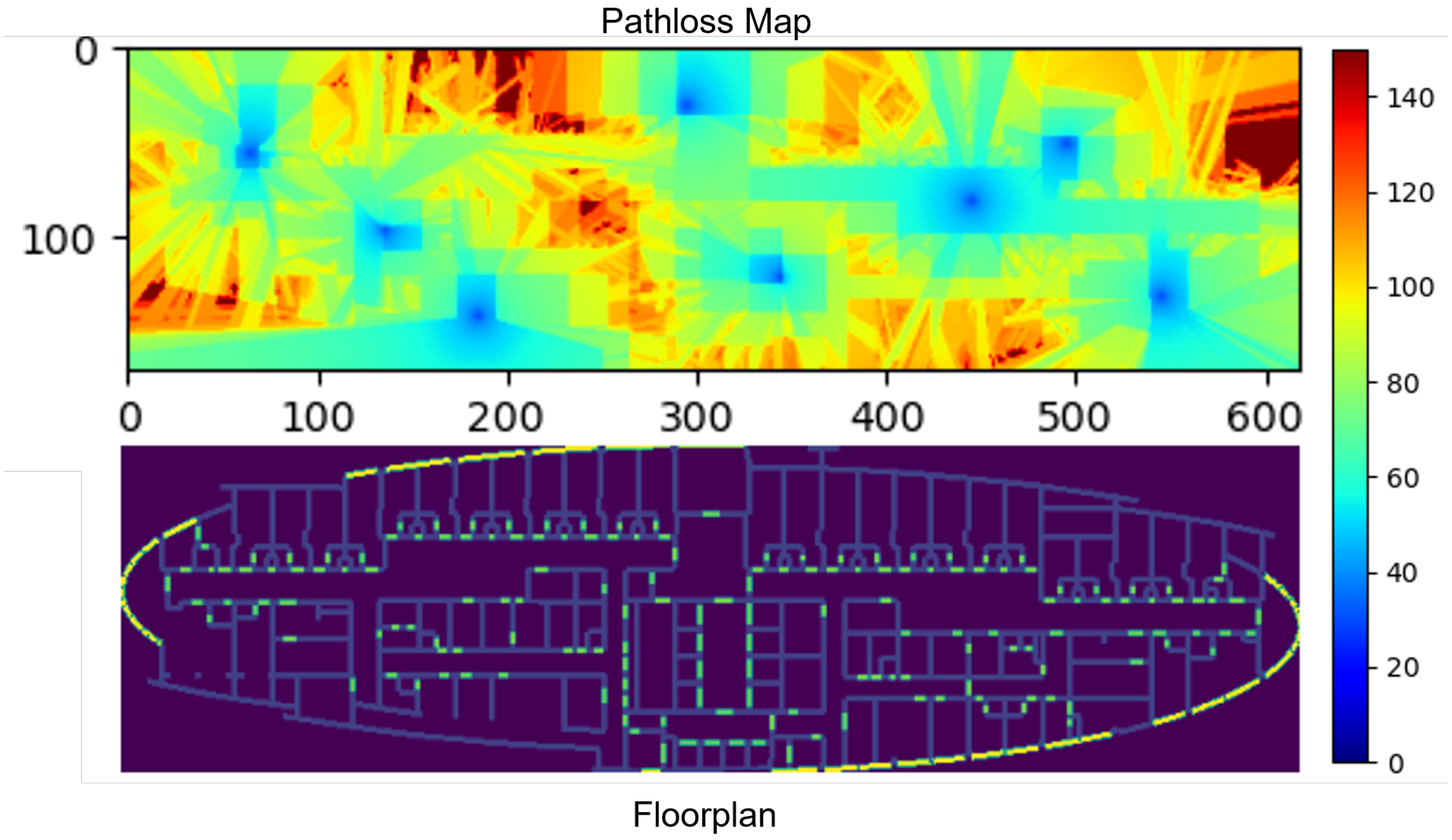

Figure 54: Result of Real-world Scenario Optimization for AP Placement by LMCO without Prior Knowledge of the Number of APs

Table 19: Comparison of Average Iterations Required for LMCO and ACO in a real-world complex scenario.

| Algorithms | **LMCO** | **ACO** |
|---|---|---|
| Iterations | 16 | 2275 |
| Time used(s) | 197 | 11802 |

An experiment in a real-world scenario, as depicted in Figure 54, provides evidence of LMCO's adaptability and further confirms its robustness. Table 19 presents the average number of iterations required by both optimization algorithms in this scenario, and the results indicate a clear difference in performance between the LMCO and ACO algorithms for AP placement.

The result shows that the LMCO algorithm consistently required fewer iterations and time to achieve the desired coverage criterion in both simple and complex environments.

### 9.1.7 Conclusion

In conclusion, LMCO introduces a novel framework for Large Language Model (LLM) based optimization in wireless communications. It presents a flexible and generalized LLM-based optimizer that incorporates expert knowledge, enabling its application across various domains to address increasingly complex problems. The LMCO algorithm demonstrates significant improvements in iteration efficiency and robustness, which are crucial for the requirements of large-scale wireless network deployments and real-time applications. The comparison highlights LMCO's superiority in handling complex and dynamic network configurations, substantially reducing time and computational overhead. This establishes LMCO as the preferred choice for optimizing wireless communication systems, with potential benefits extending to other optimization scenarios.

Continuation of this work will not only strengthen the existing capabilities of LMCO within the field of wireless network optimization but also explore its scalability and adaptability to challenges across diverse domains.

## 9.2 LTMs for Immersive communication

With the advancement of XR headsets and omnidirectional locomotion platforms enabling 6 Degrees of Freedom (DoF) [414], the concept of the Internet of Senses (IoS), which offers immersive multisensory experiences, is becoming increasingly achievable.

Additionally, advancements in cloud continuum architectures have made ultra-low latency communication possible, allowing telepresence systems to operate seamlessly. These technologies enable users to interact with teleoperated environments in ways that simulate the experience of physical presence, providing a heightened sense of immersion in remote locations. Moreover, the rapid progress in robotics and UAVs has opened up a wide range of innovative applications. UAVs, in particular, are being employed to improve immersion in VR through beyond visual line of sight (BVLOS) control [415] [416]. This creates a sense of teleportation, enabling virtual tourism to distant or inaccessible locations, and facilitating safe exploration of hazardous environments. However, achieving sub-millisecond delays with multisensory media is still a challenge. This is due to various limitations related to bandwidth constraints, network delays, and the synchronization of multiple media streams beyond audio and visual media. Additionally, the accurate representation of multiple sensory modalities presents significant difficulty due to the current limitations in sensor technology specifically for scent and taste as well as data processing capabilities. Furthermore, UAVs face additional challenges when operating at higher altitudes. The beams from base stations (BS), typically designed for ground-level communication with downtilted antennas, result in suboptimal coverage and increased interfering noise beams for airborne vehicles. This scenario leads to frequent handovers and connectivity issues for flying vehicles, thereby exacerbating the difficulty in maintaining seamless communication and low latency required for high-fidelity.

Therefore, integrating LTMs with finely tuned LLMs is expected to become one of the foundational paradigms for the IoS. This combination is set to pave the way for enabling semantic communication and synchronizing multiple sensory modalities [417]. By leveraging the capabilities of LLMs to understand and generate nuanced context, alongside LTM's proficiency in minimizing latency, this approach promises to significantly enhance the fidelity and immersion of multisensory experiences.

The architecture depicted in Figure 55 represents an optimal approach to deliver multisensory experiences and create digital twins (DTs) with low latency, while also saving bandwidth. The system will include 4 main components that are detailed below :

**Teleoperated Vehicle:** The teleoperated Vehicle in this usecase a UAV is equipped with a 360° camera, an onboard compute and a 5G modem. The UAV streams its position, and IMSI (International Mobile Subscriber Identity) to the cloud server. Other sensorial data such as Video frames, Object annotations from frames, vibration and position are sent to the Edge server.

**Cloud Server:** The Cloud server hosts a LTM fine tuned on 3GPP data. Once the Cloud server receives IMSI from the UAV, Network metrics and position, the LTM can act on the QoS by selecting the opimal profile from the Core network leveraging Network APIs such as the one developed by GSMA in the project CAMARA [418]. Furthermore, through the Network API, the position of the attached BS can be retrieved, thus the LTM can decide on routing the UAV to the nearest edge cloud server.

**Edge cloud servers:** The selected edge server hosts a fine-tuned LLM that specializes in WebXR code frameworks, specifically React360 and A-Frame. The LLM generates code for 3D representations of annotated objects received from the UAV, which are updated in real-time for the end user. The generated code is also saved in a database, which can serve as a 3D map representing a DT of the environment [419].

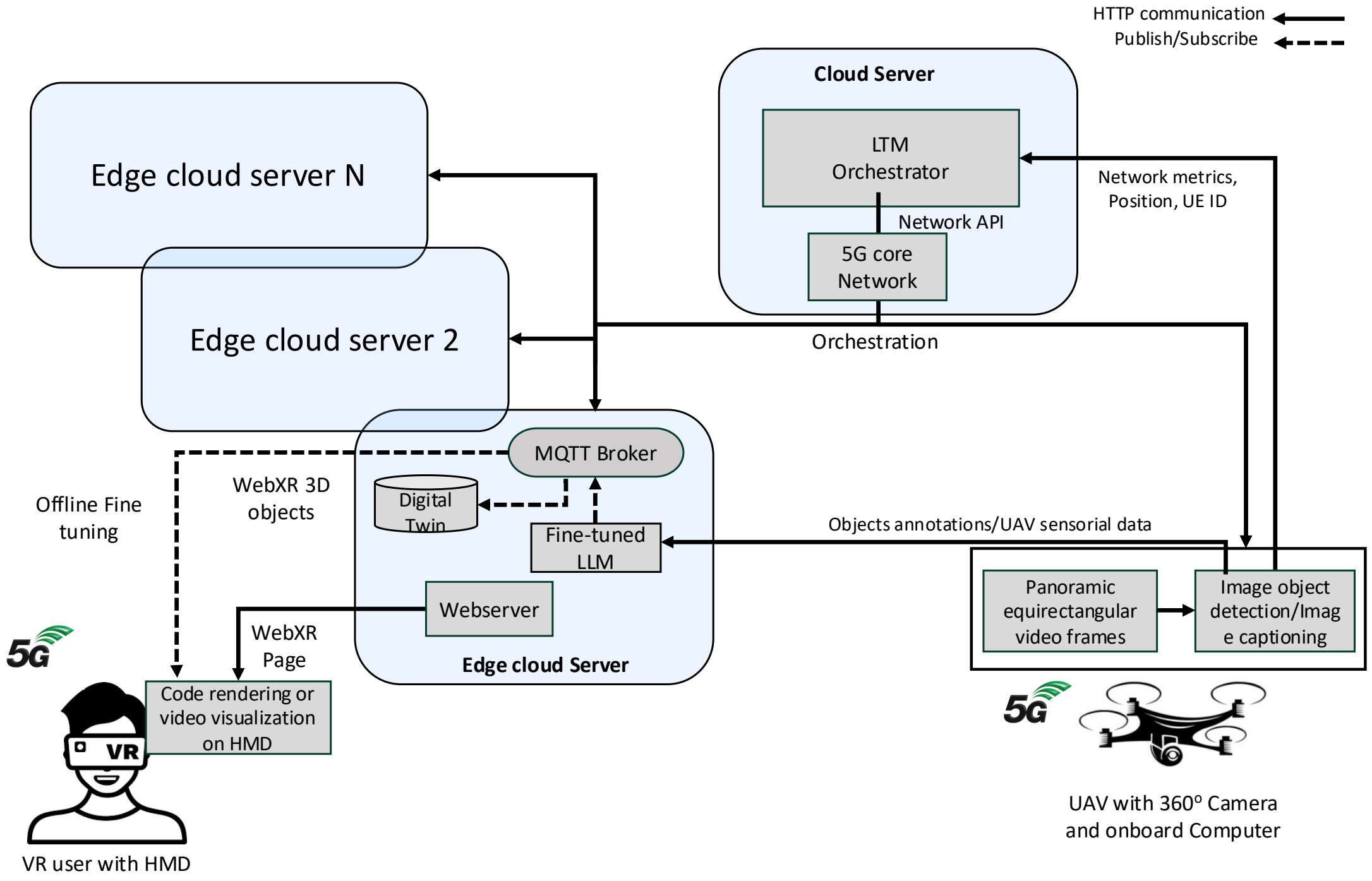

Figure 55: Proposed Architecture for providing real-time multisensory experience through LLMs and LTMs

**End User** The end user wears a HMD and can view the surroundings in 6DoF using a locomotion platform. By teleoperating the remote UAV with their movements, they experience a sensation of flying.

In conclusion, by leveraging LTMs and network APIs, we can enable intelligent orchestration that significantly reduces network delays and enhances communication efficiency. LLMs will play a crucial role in the creation of DTs and support multisensory streaming by generating additional sensory modalities. For instance, LLMs can estimate parameters like wind speed and temperature from images, while also optimizing bandwidth consumption by converting visual and sensor data into code. Furthermore, by integrating edge cloud servers and teleoperated UAVs, we can provide immersive, low-latency experiences, ensuring real-time updates and interactions in remote environments. This approach not only enhances the fidelity and immersion of the IoS but also offers scalable solutions for creating rich digital representations of physical spaces.

## 9.3 Towards Sustainable, Intelligent, and Autonomous Data Centers Enabled by Large Models

Data centers, as critical infrastructure for telecommunications, play a vital role in interconnection, storage, and computing, forming the backbone of modern telecommunications. As large-scale and complex systems, they face significant challenges, including high energy consumption, reliance on manual operations, and limited intelligence. Despite considerable research efforts to improve data center operations, challenges such as data scarcity, generalization, and adaptability remain. The rise of large models presents both new challenges and opportunities for optimizing data center operations. In this section, we provide an overview of the background, limitations, and challenges in existing research on data centers, and propose future directions for large model empowered data centers. Specifically, we discuss time series forecasting, PUE optimization, the development of intelligent and autonomous data centers, and present a case study. Through this work, we aim to advance the development of more sustainable, autonomous, and intelligent data centers.

### 9.3.1 Introduction

Data centers are significant contributors to energy consumption and carbon emissions. The emergence of large models such as ChatGPT and Llama has led to a dramatic surge in demand for computational power within these centers. As a result, energy consumption in data centers is expected to increase substantially in the coming years, making it a key focus for both industry and government efforts aimed at enhancing energy efficiency and reducing carbon emissions. Consequently, research into data center operations has attracted considerable attention, and governments, along with data center operators, are actively implementing policies and initiatives to curb energy consumption.

A review of the development history and future directions of data center operations reveals three distinct generations, based on technological advancements and the degree of reliance on manual operations: **Theoretical and Experience-Based Models Generation:** Currently, most data centers depend on expert-driven manual control for operations. However, as systems and equipment become increasingly complex, traditional approaches—relying on theoretical models and experience—fail to fully realize the energy-saving potential of data centers and are also prone to risks due to their heavy dependence on human expertise and judgment [420]. **Data-Driven Models Generation:** AI-based approaches for optimizing Power Usage Effectiveness (PUE) have shown considerable potential. Companies such as Baidu, Google, and Facebook have successfully implemented AI technologies to reduce PUE, and a variety of AI-driven techniques have been adopted in data centers to improve energy efficiency [421, 422, 423]. Despite their promise, current AI methods often depend on large-scale data collection and training. However, in real-world production environments, obtaining comprehensive datasets requires ongoing data accumulation, facilitated by the widespread deployment of sensors and other hardware, which can make data inherently scarce. This limitation slows the deployment and practical application of traditional AI approaches. As a result, the operations and management of data centers continue to rely heavily on human expertise for fault diagnosis, monitoring, management, and analysis, lacking reliable, automated, and intelligent AI-based solutions. **General Intelligent Large Models Generation:** The success of large models such as ChatGPT in various domains is expected to bring new opportunities for the development of data centers, ushering them into a new phase.

Although large models have demonstrated remarkable success across various tasks, their application in specialized industries, such as the data center domain, still faces significant challenges. Balancing generalization, specialization, and cost-effectiveness in large model deployment has become a critical obstacle to their industrial application. Moreover, research into data centers remains largely unexplored in this context. Therefore, this section aims to provide a roadmap for large model powered data centers, covering areas such as time series forecasting, PUE optimization, and autonomous data centers. It highlights state-of-the-art research, identifies the challenges associated with large models in these areas, and seeks to inspire further discussions and research toward achieving a sustainable, autonomous, and generally intelligent data center.

### 9.3.2 Large Models for Time Series Forecasting in Data Centers

As the demands for future intelligent data centers, time series analysis plays an important role. Empowered by time series analysis, accurately forecasting for data center opeartion status can be achieved, which help to monitoring and PUE optimization; anomaly detection can help operators timely discover anomaly status in data center that can greatly helpful to data center safety; predictive maintenance would help operators find possible fault in advance, preventing security incidents. Besides, there are many other crucial time series tasks in data centers, including generation, imputation and denoising, event detection, trend extraction, etc., which will greatly enhance data center level of intelligence.

As intelligent data centers evolve, time series analysis plays a crucial role. Empowered by time series techniques, accurate forecasting of data center operational status can be achieved, aiding in monitoring and optimizing PUE. Anomaly detection allows operators to quickly identify irregularities in the data center, significantly enhancing safety. Predictive maintenance helps operators anticipate potential faults, preventing security incidents. Furthermore, other important time series tasks in data centers, such as generation, imputation, denoising, event detection, and trend extraction, can substantially enhance the overall intelligence of data center operations.

Recently, there has been growing interest in developing large time series models, with the aim of making significant advancements in universal models for time series forecasting. Similar to the development of large language models (LLMs), foundation models for time series forecasting have emerged. The first foundation model, TimeGPT [424], based on transformer architecture and trained on 100 billion data points, demonstrates high accuracy and strong zero-shot capabilities. In a similar vein, Liu et al. [425] developed a GPT-style architecture trained on 1 billion time points, showing promising performance across diverse time series applications. Lag-Llama [426], designed for probabilistic univariate time series forecasting, has demonstrated superior performance across various time series datasets.

Studies have highlighted the remarkable capabilities of LLMs, prompting exploration into their application in time series forecasting. Consequently, this has led to more direct approaches, specifically the transfer of LLM capabilities into the time series domain for forecasting. Xue et al. [427] directly employed LLMs for forecasting by transforming numerical inputs and outputs into language prompts, demonstrating superior generalization compared to traditional numerical forecasting methods. To apply LLMs to time series, Jin et al. [428] aligned time series and language modalities by training text prototypes with frozen LLMs, leveraging the internal capabilities of LLMs. Chang et al. [428] pursued a similar approach but proposed a deeper and more flexible fine-tuning method to transfer LLM abilities from natural language to time series. This approach used a two-stage fine-tuning process: supervised fine-tuning for LLMs followed by task-specific downstream fine-tuning.

**Remarks:** Both time series foundation models and large models transferred from other domains represent promising approaches for constructing large models for time series forecasting in data centers. For time series foundation models [427, 428], despite task differences between current foundation models and data center applications, both fall within the domain of time series. This inherent alignment gives these models a distinct advantage in transferring to, or even being directly applied in, data center domains. In contrast, large models such as LLMs or other domain-specific models [424] operate on different modalities compared to time series data, posing challenges in modality alignment and effective transfer of domain-specific capabilities to time series tasks. To address these challenges, techniques such as parameter-efficient fine-tuning may be required to adapt these models with minimal or no additional training. Nevertheless, knowledge from other domains can provide valuable assistance in building more interpretable models, integrating external knowledge relevant to data centers [429]. Furthermore, LLMs have significant potential to function as agents, enabling broader, more advanced roles. This versatility may pave the way for achieving general intelligence in time series applications.

**Challenges and Future Directions:** Despite recent advances in large models for time series forecasting, significant challenges and opportunities remain in their application to data centers. The limitations of current research and potential future directions can be summarized as follows:

- **Develop Tailored Foundation Models for Data Centers:** Construct tailored foundation models specifically for data centers, enabling emergent capabilities such as multi-task adaptability, generalization across different scenarios, and exceptional few-shot and zero-shot performance. However, these efforts face significant challenges, including the collection and processing of massive datasets, ensuring data privacy and secure sharing, and addressing the high costs and computational demands. These issues must be carefully addressed to realize the full potential of foundation models in this domain.

- **Efficient and Fast Application to Data Centers:** Investigating efficient methods for transferring the capabilities of existing large time series models or large visual models to data centers is highly valuable. This approach leverages the powerful inherent abilities of these models, enabling faster deployment compared to training models from scratch. Additionally, it facilitates more diverse functionalities, such as integrating text-based auxiliary information. Therefore, designing efficient and lightweight transfer learning or fine-tuning strategies for these large models is critical to their successful application in data centers.

- **Unified Time Series Large Models Framework for Data Centers:** Time series tasks in data centers extend beyond forecasting to include anomaly detection and other applications. The ultimate goal is to develop a unified large model framework capable of addressing all time series tasks comprehensively. One promising approach is to establish an agent-based framework where diverse large models act as specialized agents, each responsible for a specific task. These agents collaborate through well-organized workflows to handle the entire range of time series tasks effectively.

### 9.3.3 Large Models for PUE Optimization in Data Centers

Data centers play a pivotal role in global energy consumption and carbon emissions, with their impact projected to grow significantly, particularly with the increasing deployment of large-scale models such as ChatGPT and Llama [430, 431]. To address these concerns, governments and data center operators have introduced various policies and initiatives aimed at reducing energy consumption [432]. PUE is a widely recognized metric for assessing the energy efficiency of data centers. Optimizing PUE has become a primary objective for data center operators, focusing on minimizing energy consumption and carbon emissions [422]. Recent advances have highlighted data-driven approaches as effective alternatives to traditional methods based on expert knowledge and thermodynamic principles. Among these, deep reinforcement learning (DRL)-based solutions, such as the DQN-based chiller energy optimization [420], event-driven deep reinforcement learning [433], and branching double-dueling deep Q-network [422], have demonstrated significant potential.

Large models, with their extensive knowledge, exceptional generalization capabilities, and superior reasoning abilities, show great potential for advancing optimization. Current research on this topic can be divided into two main approaches: using LLMs as optimizers and optimizing through LLMs. The approach of using LLMs as optimizers involves leveraging LLMs to analyze, reason, and optimize problems step by step. This method allows LLMs to handle optimization tasks in an intelligent, human-like manner. Yang et al. [434] explore the use of LLMs as optimizers by prompting them to address scenarios involving the absence of gradients, where optimization problems are presented in natural language. They demonstrate the effectiveness of the OPRO framework on tasks such as linear regression and the traveling salesman problem, and show that LLM-based solutions outperform human-designed ones. A similar study in [435] focuses exclusively on prompt-based optimization problems.

Recent works also explore the approach of optimizing through LLMs, where LLMs are integrated with evolutionary algorithms to enhance traditional optimization methods. Lehman et al. [436] find that LLMs trained to generate code can significantly enhance the effectiveness of mutation operators in genetic programming. LLMs can generate hundreds of novel examples not seen during pre-training, demonstrating their considerable potential for optimization. Meyerson et al. [437] exploit the natural in-context learning abilities of LLMs to create variation operators in evolutionary algorithms. Their experiments, which include sentences, equations, and code, lead them to conclude that LLM-based crossover is a flexible and effective method for optimization. While neural architecture search through prompt-tuning may be challenging for LLMs [438], Chen et al. [439] demonstrate that LLMs, when used as general adaptive mutation and crossover operators, can consistently identify diverse and high-performing models by directly generating code. Additionally, Nasir et al. [440] combine LLMs with quality-diversity algorithms to generate code for neural network search, incorporating an evolutionary process. Their results demonstrate the competitive performance of LLMs, even without prior domain-specific knowledge. Google DeepMind [441] has developed an evolutionary procedure for searching programs by pairing a pre-trained LLM with a systematic evaluator. This approach has led to genuinely novel discoveries—solutions that were not present in the training data—surpassing existing methods in the Cap Set and Online Bin Packing problems. Their work highlights the exceptional potential of LLMs to explore uncharted areas, showcasing their creativity.

**Remarks:** Large models possess superior capabilities in knowledge, reasoning, and generalization, positioning them as promising candidates for either acting as optimizers or collaborating with traditional optimization methods. This evolution in optimization represents a new generation of intelligent solutions. Both approaches offer viable paths toward leveraging large models for general optimization tasks. When large models are used as optimizers, they can fully exploit the powerful abilities of LLMs, enabling them to address optimization problems in a more intelligent and interpretable manner, akin to human experts. This approach is the most direct way to create general intelligent optimizers. On the other hand, when large models collaborate with traditional optimization methods, the combination leverages the strengths of both. Traditional optimization techniques are particularly advantageous for tasks requiring specially designed strategies, such as fast convergence, learnable policies, and tailored solutions for specific classes of problems. Meanwhile, large models contribute their vast knowledge and expertise, leading to more effective and intelligent solutions.

**Challenges and Future Directions:** Despite the success of applying large models to optimization tasks, several challenges remain in realizing a truly general intelligent optimizer enabled by large models. These challenges and potential future research directions can be summarized as follows:

- **Enhancing the Mathematical Abilities of Large Models:** Existing studies highlight some limitations of LLMs in optimization, such as difficulties in tackling large-scale problems [434] and inconsistent performance for the same problem when presented with different prompts [442]. Consequently, further investigation is needed to strengthen the mathematical reasoning capabilities of large models. Chain-of-Thoughts [443] is a promising approach that has shown significant improvements in arithmetic and symbolic reasoning tasks.

- **Synergy between Large Models and Traditional Optimization Methods:** Although large models alone are capable of making a significant impact, the advantages of traditional optimization methods should not be overlooked. Collaboration between large models and these methods can yield superior performance. Previous research has explored the integration of large models with evolutionary algorithms. Furthermore, LLMs can be combined with other classic optimization techniques, such as Bayesian optimization [444] and DRL [436], unlocking additional potential based on the specific characteristics of the optimization algorithms.

- **Integrating Different Modalities with Large Models:** Current research, whether using LLMs as optimizers or combining them with other algorithms, primarily relies on natural language inputs. This reliance limits the ability of large models to address a wider array of optimization problems. For instance, LLMs struggle with tasks involving large datasets or graph-structured information. These input limitations can significantly hinder the optimization process. Therefore, more efforts are needed to integrate various modalities within large models, enhancing their capabilities and enabling them to handle a broader range of optimization tasks more effectively.

- **Incorporating Data Center Knowledge into Large Model-Enabled Optimizers:** One of the core ideas and key strengths of utilizing LLMs in optimization is their ability to leverage powerful internal knowledge. However, the integration of domain-specific knowledge, such as that required for data center optimization, presents significant challenges. Training large models for data center optimization from scratch, fine-tuning existing models, or enhancing them with retrieval-augmented generation are promising directions for incorporating this expertise into optimization tasks.

- **General Optimizer for Data Centers Driven by Large Models:** Data centers require optimization across various facets, including PUE, carbon emissions, water utilization, operational efficiency, and fault rates. Optimizing these aspects necessitates integrating expert knowledge, diverse document sources, and specialized tools into large models. The challenge lies in effectively combining these large models with diverse inputs and tools to create a general and intelligent optimizer. This area remains largely unexplored and requires significant further investigation.

### 9.3.4 Large Models for Intelligent and Autonomous Data Centers

AI Operations for data centers is a broad concept that refers to utilizing AI technologies to automate various aspects of data center operations. This includes not only time series-related operations and optimization but also data analysis, root-cause analysis, daily monitoring, and maintenance. Despite the development of advanced technologies for data center operations, these operations still heavily rely on manual processes and expert knowledge, which are time-consuming, labor-intensive, and prone to errors. The advent of large models, which aim to achieve artificial general intelligence, brings new opportunities for developing generally and intelligently autonomous data center operations. These models are expected to enable self-optimization, self-healing, and self-management, leading to more efficient, reliable, and intelligent data center operations.

Several studies have explored the application of large models in AI operations [445, 446, 447]. Log analysis is a crucial task in AI operations. Considering the challenges posed by semi-structured log data with limited grammatical structure, Gupta et al. [445] propose BERTOps, the first LLM that can effectively generalize to multiple downstream tasks of log analysis. They build on BERT-BASE, pre-train on large-scale public and proprietary log data (41.6 million public data and 1.9 million proprietary data), and fine-tune on specific log tasks. The proposed BERTOps achieves superior performance compared to existing methods in log format detection, golden signal classification, and fault category prediction. Qingzhi operation large model [446] is proposed to realize more intelligent AI operations. It utilizes several effective technologies targeting various scenarios, including log and alert analysis, automatic fault report generation, alert understanding, and detection of logs, alerts, and events. Specifically, they incorporate fine-tuned LLMs for specific AI operation tasks, construct an operations knowledge database, and leverage Retrieval Augmented Generation (RAG) to enhance the capabilities of LLMs. They also improve the reasoning ability of LLMs through prompt engineering and Chain of Thought (CoT) techniques to better analyze alerts and logs, and utilize various tools such as databases, algorithms, and code to enhance general abilities. To accurately and fairly evaluate the performance of large models on AI operation tasks, comprehensive benchmarks are needed. Liu et al. [447] propose OpsEval, which evaluates LLMs' performance across three pivotal scenarios: wired network operation, 5g communication operation, and database operation. It considers varying ability levels encompassing knowledge recall, analytical thinking, and practical application.

**Remarks:** Large models have significant potential for AI autonomous operations in data centers. AI operations are closely related to natural language processing, a domain where large models have achieved remarkable success. This success can drive advancements in AI across various domains. In data centers, logs and alerts are often in natural language, making large models well-suited for AI operations. Additionally, LLMs can receive, process, and generate natural language content, facilitating more user-friendly operations and lowering the expertise threshold required for data center management. Beyond natural language processing, large models, including LLMs, have demonstrated exceptional performance in various areas that can support AI operations. For instance, their code generation capabilities can be crucial tools in developing autonomous systems. Their strong reasoning abilities help identify root causes, understand and analyze operational status, and manage complex tasks. Furthermore, by leveraging advanced tools, we can enhance the capabilities of large models in AI operations. Through effective prompt engineering, we can improve large models for data center operations without additional training. Techniques such as RAG and the use of external resources like knowledge bases, knowledge graphs, and documentation can incorporate specialized data center knowledge, transforming large models into proficient data center operators. Additionally, efficient fine-tuning methods can adapt large models to the specific needs of data center operations at a low cost.

**Challenges and Future Directions:** Although there are significant potentials and advantages in applying large models for autonomous operations, existing works still face several deficiencies. More efforts are needed to address these challenges and make steady progress toward intelligent autonomous data center operations.

- **Develop Tailored Large Models for Autonomous Data Centers:** Large models are promising solutions for AI operations; however, there is no research conducted on AI operations for data centers. Due to the specialized nature of data centers, previous studies cannot be directly applied. Therefore, developing well-tailored large models for autonomous data centers remains largely unexplored. Techniques such as prompt engineering, RAG [448], training from scratch, and fine-tuning pre-trained models each have their own strengths and weaknesses, and the choice should be based on specific circumstances.

- **Comprehensive Abilities of Large Models are Indispensable:** Autonomous data centers require no manual interventions and should be capable of handling all aspects of data center operations. This includes not only prediction capabilities but also optimization, analysis, reasoning, and data management. Large models must process diverse input modalities, including text, logs, and data. These diverse abilities are indispensable for large models in achieving autonomous data centers. However, developing such comprehensive abilities in large models remains a significant challenge.

- **Operation Reliability of Large Models:** Data centers store and process large amounts of data and support numerous critical services, including communication and computing. Therefore, the reliability of data center operations is crucial. Current investigations show that larger models may become less reliable [449]. Large models may generate content that deviates from facts, known as factual hallucination. More efforts and investigations are needed to mitigate such unreliability; otherwise, these large models cannot be applied in real production environments.

- **Autonomous Data Center Driven by Large Model-based Agents:** While large models are incredibly powerful, a pure large model may struggle to realize autonomous data centers. *The fundamental distinction between humans and other animals lies in the ability to create and use tools.* This description, though not entirely accurate, highlights the importance of tools for human progress. Each industrial revolution has been driven by new tools and technologies. Inspired by this truth, autonomous systems should utilize diverse external tools [450] to overcome limitations in pre-training data and address specific downstream tasks. Enabling agents with large models will greatly enhance their potential and capabilities, representing a promising avenue toward achieving Artificial General Intelligence (AGI) for autonomous data centers.

### 9.3.5 Case Study

We have done initial works towards intelligent and autonomous data centers empowered by large models, we introduce them as a case studies in this section.

Considering that conventional small models for predicting data center status often face data scarcity issues in practical deployment. While large models show promise in addressing this challenge, they encounter obstacles such as multivariate tasks, computational intensity, and ineffective information capture. Moreover, their applications in data centers remain largely unexplored. In this case studies, we investigate local semantic capture empowered large model for multivariate time series forecasting in data centers. We first introduce time series tasks within data centers and propose the Point Lag (Plag)-Llama framework with the Lag-Llama backbone to support zero-shot forecasting and fine-tuning for multivariate point time series forecasting. To address computational intensity and enhance the capabilities of multivariate forecasting, we propose the **Lo**cal **Se**mantic **C**apture (LOSEC) for adapter fine-tuning, which captures local semantic information across time and channel dimensions alternately with low-complexity. Specifically, time series are segmented into tokens, and channels are clustered together, forming local semantic information that can be captured more effectively. Extensive experiments demonstrate that Plag-Llama exhibits superior zero-shot capability and that the LOSEC empowered adapter fine-tuning achieves state-of-the-art performance on real-world datasets collected from data centers, with ablation studies further validating the effectiveness of each module within the proposed models.

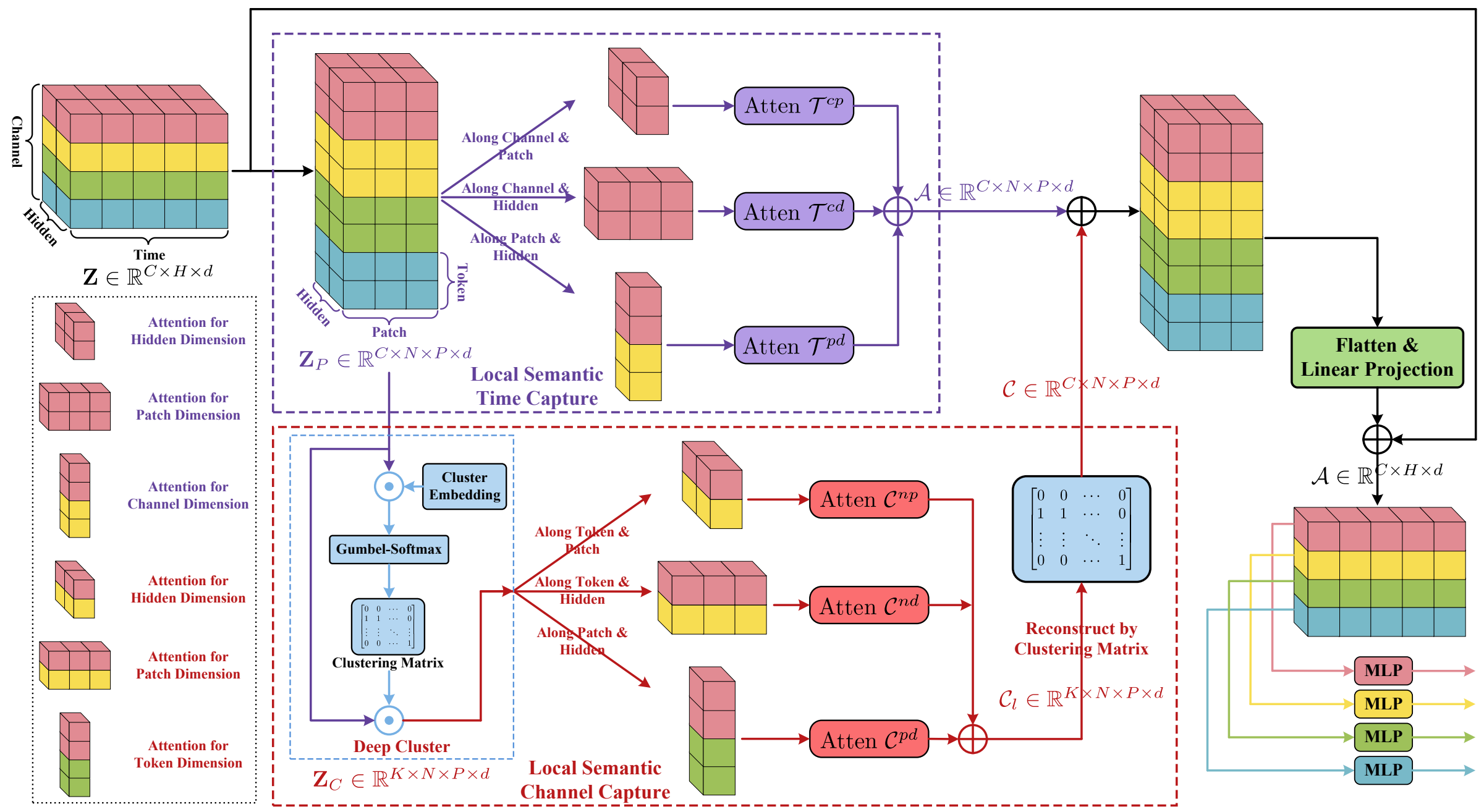

Figure 56: Architecture of LOSEC: Capturing local semantic information alternately across time and channel dimensions.

**Plag-Llama, A Multivariate Time Series Point Forecasting Framework:** We introduce Plag-Llama, a framework crafted for multivariate time series point forecasting, which extends the capabilities of the pre-trained Lag-Llama. The overall architecture is depicted in Fig. 57. In Lag-Llama, historical inputs $\mathbf{X}_{t-L+1:t} \in R^{C\times L}$ undergo sequential processing involving a projection layer, $N$ masked transformer decoder layers, and a distribution head, yielding a probabilistic distribution for each time series. To adapt the probabilistic forecasting capability to point forecasting, the pre-trained Lag-Llama, excluding the distribution head, serves as the backbone of Plag-Llama. Furthermore, Plag-Llama integrates supplementary blocks, such as the transfer block and adapter block, incorporates a revised loss function, and employs fine-tuning techniques. The transfer block corresponds to the full fine-tuning method, while the LOSEC adapter block implements the adapter fine-tuning method. As shown in Fig. 57, both methods share a forward pass process, but differ in the backward pass. Specifically, full fine-tuning involves backpropagation through both the pre-trained Lag-Llama model and the transfer block, whereas LOSEC adapter fine-tuning involves backpropagation only through the adapter block and a backward pass through the pre-trained Lag-Llama. This distinction arises from their differing operational mechanisms. In this work, we propose an intuitive and zero-shot block that leverages the average operation across feature dimensions. The details of LOSEC adapter fine-tuning are introduced in the following sections.

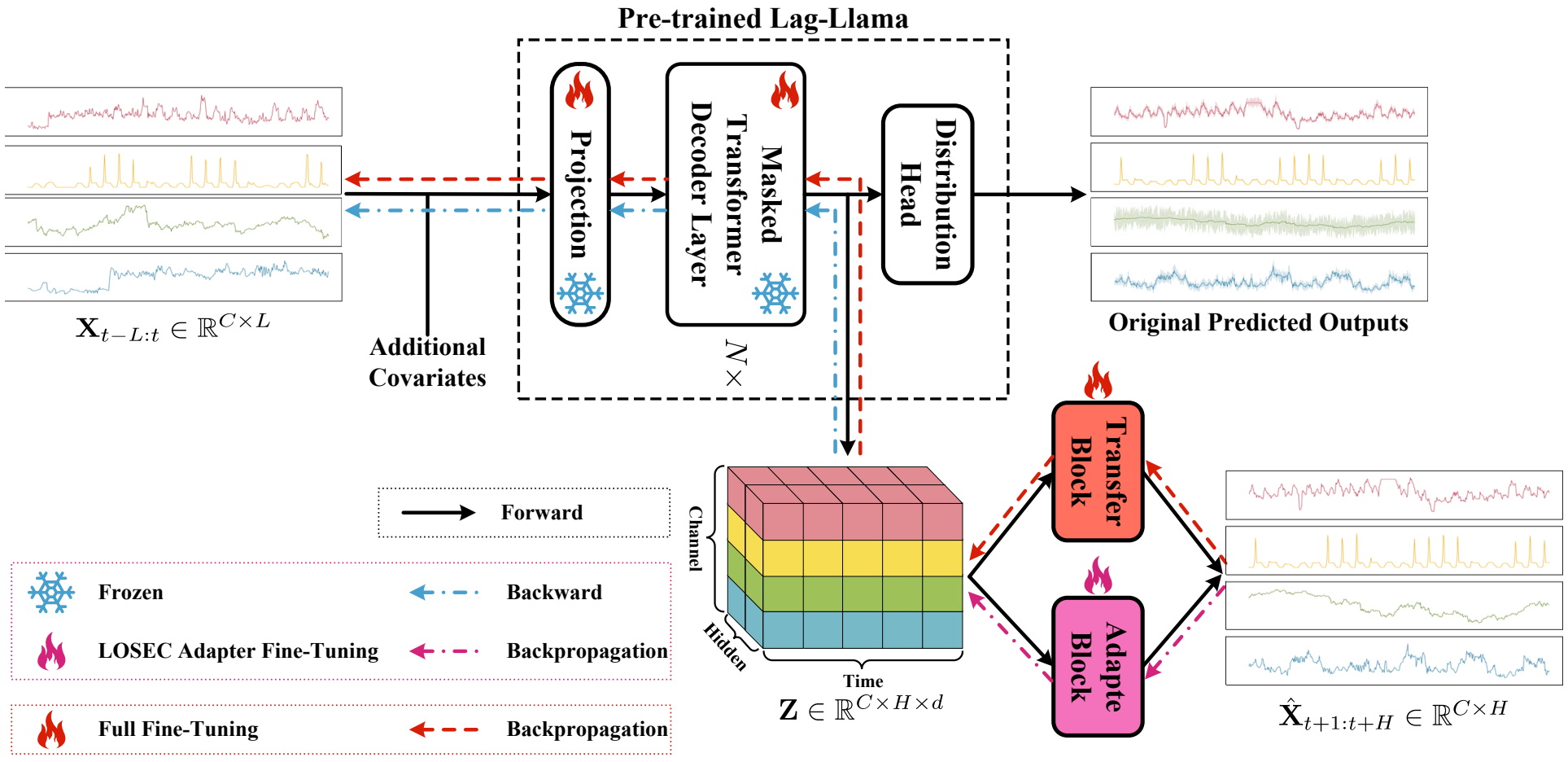

Figure 57: Plag-Llama: Overview of the architecture.

**Local Semantic Capture Empowered Adapter Fine-Tuning:** In the full fine-tuning approach, all parameters $\boldsymbol{\omega}$ and $\boldsymbol{\theta}$ must be adjusted for each task, leading to significant computational intensity and resource demands. Despite implementing early stopping and random sampling strategies, full fine-tuning remains prone to overfitting. Adapter fine-tuning presents a promising method for capturing new correlations beyond those established by original models, thereby enhancing the capabilities of Lag-Llama in multivariate time series forecasting. However, existing studies [451, 452, 453, 454] fail to adequately capture channel-dependent and other critical information, which would results in a significant degradation of performance. Therefore, we propose a novel model, LOSEC, aiming capturing local semantic information to compensate drawbacks of base models and achieve SOTA performance.

To empower Lag-Llama with the ability to perform multivariate forecasting, we seek to understand the correlations between different channels. However, simple attention-based methods struggle to capture this information due to irrelevant and even interfering data among channels [455]. Original attention mechanisms are designed to capture correlations between words in a sentence; however, individual channels in a multivariate time series are more analogous to letters in a sentence, and therefore do not inherently carry semantic meanings. Moreover, this issue extends beyond capturing correlations between channels, affecting the temporal relationships between time steps as well [456]. To address this, we propose the LOSEC model, which constructs and extracts semantic information for both the channel and time dimensions. In this context, certain individuals (channels or time steps) are grouped together to form meaningful semantic units, which we refer to as ”local semantics” due to their localized nature. To efficiently capture local semantic information while maintaining complexity, we first extract local semantics along the time dimension, followed by the channel dimension. This results in two distinct processes: local semantic time capture and local semantic channel capture. This ensures that the attention mechanism can effectively perform its role in capturing these relationships. These local semantic captures across both dimensions are central to the LOSEC model, with its overall architecture depicted in Fig. 56. The LOSEC model is designed to be low-complexity, featuring three key strategies to reduce computational load: 1) the formation of local semantic information across the time and channel dimensions, 2) alternating captures across the time and channel dimensions, and 3) dimension-specific attention mechanisms. Detailed discussions of the complexity reduction and design philosophy behind these modules are provided in [457].

**Experimental Results:** We compare our proposed methods with SOTA models in the data center multivariate time series forecasting task. The results for three categories: supervised, zero-shot, and fine-tuning models are summarized in Table 20. Leveraging the powerful forecasting capabilities of Lag-Llama and the proposed transfer block, Plag-Llama demonstrates strong zero-shot performance, achieving the second-best MAE and third-best SMAPE, along with an average rank of 5.75. Although it falls short of the best performance, it still ranks third among the supervised SOTA models. This superior zero-shot performance indicates that the Plag-Llama framework is a promising solution to address data scarcity challenges and expedite deployment within data centers. It is important to note that the original Lag-Llama aims to align the overall distribution with the ground truth; however, the simple average layer effectively transfers these probabilistic capabilities to point forecasting. This is attributed to the symmetric Student's t-distribution adopted in Lag-Llama, where learning optimal mean values is beneficial for probabilistic forecasting. However, full fine-tuning significantly undermines the performance of Plag-Llama due to overfitting, resulting in an average rank of 9.25, which is even worse than its zero-shot performance. Empowered by the proposed local semantic information capture framework, the LOSEC adapter consistently demonstrates superior performance compared to other SOTA models, exhibiting an average improvement of 3.34% over the second-best model.

Table 20: Multivariate time series forecasting performance: The context length is set to 8, and the prediction length is set to 4 (best results in **bold**, second best in underlined, and third best in dotted underline).

| MODEL | MAE | MSE | SMAPE | MASE | ARk |
|---|---|---|---|---|---|
| **SUPERVISED** | | | | | |
| Autoformer | 0.139 | 0.180 | 26.244 | 0.184 | 8.75 |
| Crossformer | 0.137 | 0.190 | 23.863 | 0.181 | 8.75 |
| DLinear | 0.135 | 0.179 | 24.786 | 0.178 | 6.50 |
| FEDformer | 0.141 | 0.182 | 26.858 | 0.186 | 9.75 |
| Informer | 0.128 | 0.179 | 22.939 | 0.169 | 3.50 |
| iTransformer | 0.135 | 0.245 | 19.451 | 0.178 | 6.25 |
| LightTS | 0.168 | 0.204 | 34.194 | 0.222 | 12.50 |
| NS-Transformer | 0.130 | 0.182 | 22.971 | 0.171 | 5.75 |
| PatchTST | 0.136 | 0.249 | **19.424** | 0.179 | 7.25 |
| TimesNet | 0.130 | 0.187 | 22.580 | 0.171 | 5.00 |
| Transformer | 0.353 | 0.397 | 67.239 | 0.466 | 14.00 |
| **ZERO-SHOT** | | | | | |
| Plag-Llama | 0.127 | 0.194 | 20.579 | 0.180 | 5.75 |
| **FINE-TUNING** | | | | | |
| Plag-Llama (Full) | 0.153 | **0.176** | 28.643 | 0.201 | 9.25 |
| Plag-Llama (LOSEC Adapter) | **0.123** | 0.177 | 21.980 | **0.162** | **2.00** |

More experimental results, detailed analyses on few-shot learning, visualization of local semantic information capture, ablation studies, and other aspects, along with comprehensive technical details of this case study, can be found in our work [457].

## 9.4 LLM-enabled Semantic Communication

Underwater communication plays a vital role in environmental monitoring, marine biology research, and underwater exploration [458]. Efficient and reliable transmission of multimodal data, including images and sensory information, is essential for tracking marine ecosystems, studying marine life, and ensuring the success of exploration missions [459]. Traditional methods like acoustic, optical, and RF communications dominate underwater communication [460]. Acoustic communication, the most practical for long distances, suffers from limited bandwidth (kbps-level), high latency due to the slow speed of sound, and vulnerability to noise, attenuation, and multipath effects [461, 462]. These limitations—low bandwidth, high latency, and poor robustness—pose significant challenges in complex underwater environments.

Semantic Communication (SC) based on artificial intelligence has been proposed to address challenges in low-bandwidth, high-attenuation scenarios [463, 464, 465]. SC introduces a semantic channel to extract, encode, and transmit semantic information, optimizing bandwidth usage and achieving efficient transmission [466]. Unlike traditional systems that focus solely on bitstream accuracy, SC emphasizes meaningful content transmission [467, 468]. However, SC faces key challenges in underwater environments. First, multipath propagation and Doppler shifts lead to semantic mismatches, distorting signals and disrupting coherence, which traditional SC struggles to handle effectively. Second, SC lacks flexibility in prioritizing critical information for diverse underwater tasks, such as distinguishing between images for environmental monitoring and marine species detection. This limitation hinders efficient communication, potentially delaying critical decision-making in underwater operations.

To tackle these challenges, we propose a novel SC framework using Large Language Models (LLMs). Unlike traditional models, LLMs excel at generalization and can prioritize and filter data based on task-specific needs. This allows for more efficient data management and transmission. Our framework uses visual LLMs to process image data, performing semantic compression and prioritization by encoding key elements and applying higher compression to less critical areas. On the receiver side, a text LLM recovery mechanism and two ControlNet networks help reconstruct the data and improve semantic coherence, mitigating information mismatches. This approach reduces data size to 0.8% of the original and enhances system resilience against noise and signal loss.

The framework, specifically designed for underwater image transmission, is proposed. It begins with a query sent by individuals above the water surface, which is transmitted to the underwater environment. Upon reception, the underwater transmitter uses a semantic encoder combined with an LLM-based prioritization mechanism. This mechanism identifies and ranks critical visual information based on the context of the query. The prioritized data is then compressed for transmission through the underwater communication channel. At the receiver's end, the information is decoded using a semantic decoder, supported by a diffusion model and LLM recovery mechanism. The decoded visual content is then reconstructed and delivered to the individuals above the water. This approach optimizes the transmission of visual data in underwater communication systems. The framework serves multiple purposes:

1. **LLM-based Semantic Compression and Prioritization Framework:** Introduced a novel semantic communication framework leveraging Large Language Models (LLMs) for underwater image transmission. By understanding user queries and identifying key semantic elements in images, the framework performs semantic compression and prioritization, significantly reducing the transmission of non-critical data and enhancing communication efficiency and adaptability.

2. **Integration of Newly Designed ControlNet Networks:** Developed and integrated two specialized ControlNet networks: Key Region ControlNet and Global Vision ControlNet. These networks, combined with a diffusion model, enhance semantic coherence and effectively address semantic mismatches in complex underwater environments.

3. **Achieved Significant Data Compression:** Achieved a remarkable data size reduction, compressing transmitted data to only 0.8% of the original size. This approach maintains high-quality semantic information reconstruction, even in high-noise conditions, while significantly reducing bandwidth requirements.

4. **Robust Multimodal Recovery Mechanism:** Incorporated an LLM-based textual recovery mechanism on the receiver side, combined with triple guidance signals in the diffusion model. This ensures high-precision and semantically consistent image reconstruction, leveraging joint optimization of text and visual information to improve recovery robustness under challenging communication conditions.

## 9.5 LTM-enhanced Data Augmentation for Spectrum Sensing in Cognitive Radio Networks

Spectrum sensing is a critical function in cognitive radio networks, enabling devices to detect unused frequency bands and avoid interference with Primary Users (PUs). One effective approach to spectrum sensing is *automatic modulation classification* (AMC), which is based on the idea that by detecting the modulation schemes used by PUs, spectrum occupancy can be accurately determined. Once the modulation schemes are identified, the corresponding frequency bands can be marked as unavailable for Secondary Users (SUs), ensuring efficient spectrum utilization and minimizing interference.

AMC can be performed using large-scale vision models that have demonstrated remarkable success in image recognition tasks. However, the accuracy of these models in the context of modulation classification heavily depends on large-scale labeled datasets. Acquiring such datasets is challenging and resource-intensive. Additionally, if a new modulation class is introduced, the existing model requires retraining to accommodate the new class, which is both time-consuming and computationally demanding.

To address these challenges, the proposed methodology involves transforming raw in-phase and quadrature (IQ) time-series data into image representations using techniques like Recurrence Plots (RP), Markov Transition Fields (MTF), and Gramian Angular Fields (GAF). These transformations serve multiple purposes:

1. **Recurrence Plots (RP):** Recurrence plots (RPs) are a powerful tool for analyzing non-linear time series data, allowing for exploring hidden patterns and structures not easily detected by conventional methods. RPs provide a way to observe a system's behavior over time and identify recurring states or cycles within the data. The construction of a recurrence plot is based on the concept of recurrence in phase space. Given the time series $\{x_i\}_{i=1}^{N}$, the first step is reconstructing the phase space using delay embedding, as stated by Taken's theorem [469]. This is done by constructing delay vector $X_i$ in an $m$-dimensional space:

$$X_i = (x_i, x_{i+\tau}, x_{i+2\tau}, \ldots, x_{i+(m-1)\tau}) \tag{40}$$

In Equation (1), $m$ is the embedding dimension and $\tau$ is a time delay. The choice of $m$ and $\tau$ can be optimized using methods like the False Nearest Neighbors (FNN) algorithm [470] for $m$ and the first minimum of the mutual information function for $\tau$. The recurrence matrix $R$ is defined to quantify the recurrences of states in phase space.

$$R_{i,j} = \Theta(\varepsilon - \|X_i - X_j\|) \tag{41}$$

In Equation (2), the recurrence matrix $R_{i,j}$ is defined using the Heaviside step function $\Theta$ [471] to quantify whether the distance $\|X_i - X_j\|$ between state vectors $X_i$ and $X_j$ is within a threshold $\varepsilon$.

$$\|X_i - X_j\| = \sqrt{\sum_{k=1}^{m} (x_{i+k\tau} - x_{j+k\tau})^2} \tag{42}$$

In Equation (3), the distance $\|X_i - X_j\|$ between state vectors $X_i$ and $X_j$ is the Euclidean distance in $m$-dimensional phase space, summing the squared differences of their time-delayed components.

$$\Theta(x) = \begin{cases} 1 & \text{if } x \geq 0 \\ 0 & \text{if } x < 0 \end{cases} \tag{43}$$

The Heaviside step function $\Theta(x)$ in Equation (4) is used to return 1 if $x \geq 0$ and 0 if $x < 0$, effectively determining if the distance between vectors is within the threshold $\varepsilon$.

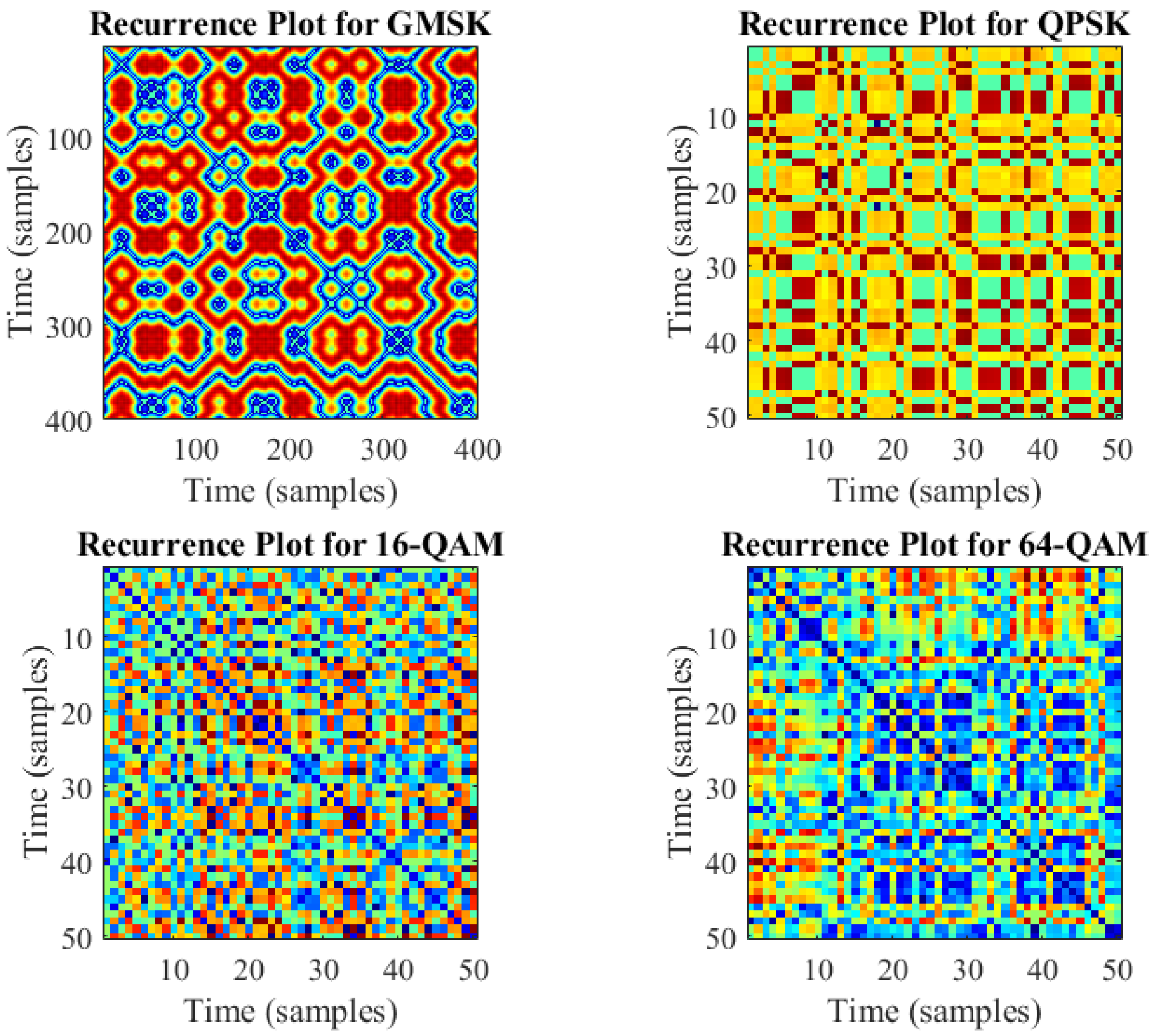

Figure 58: Recurrance Plots for various modulation schemes

Figure 58 displays RPs for various modulation types: GMSK, QPSK, 16-QAM, and 64-QAM. These plots reveal the periodicities and repetitive patterns within the signals, highlighting differences in signal dynamics across the modulation schemes.

2. **Markov Transition Fields (MTF):**

Markov Transition Field (MTF) is another technique that transforms time series data into images. This method was introduced by Wang and Oates [472]. The central idea is to evaluate how likely the time series will transition from one value to another within a given timeframe. The given time series is divided into a finite number of non-overlapping intervals acting as the states. For each pair of states $s_i$ and $s_j$, the transition probability of moving from $s_i$ to $s_j$ in one-time step is computed. Lastly, an MTF matrix is constructed where each element $M(i, j)$ corresponds to the transition probability from state $s_i$ to state $s_j$. This matrix represents the Markov Transition Field.

Once the MTF is constructed, it can be visualized as an image, where each pixel intensity is proportional to the transition probability between states. Figure 59 shows MTF plots for different modulation schemes: GMSK, QPSK, 16-QAM, and 64-QAM. Each plot visualizes the transition probabilities within a signal, highlighting how different modulation techniques affect signal characteristics.

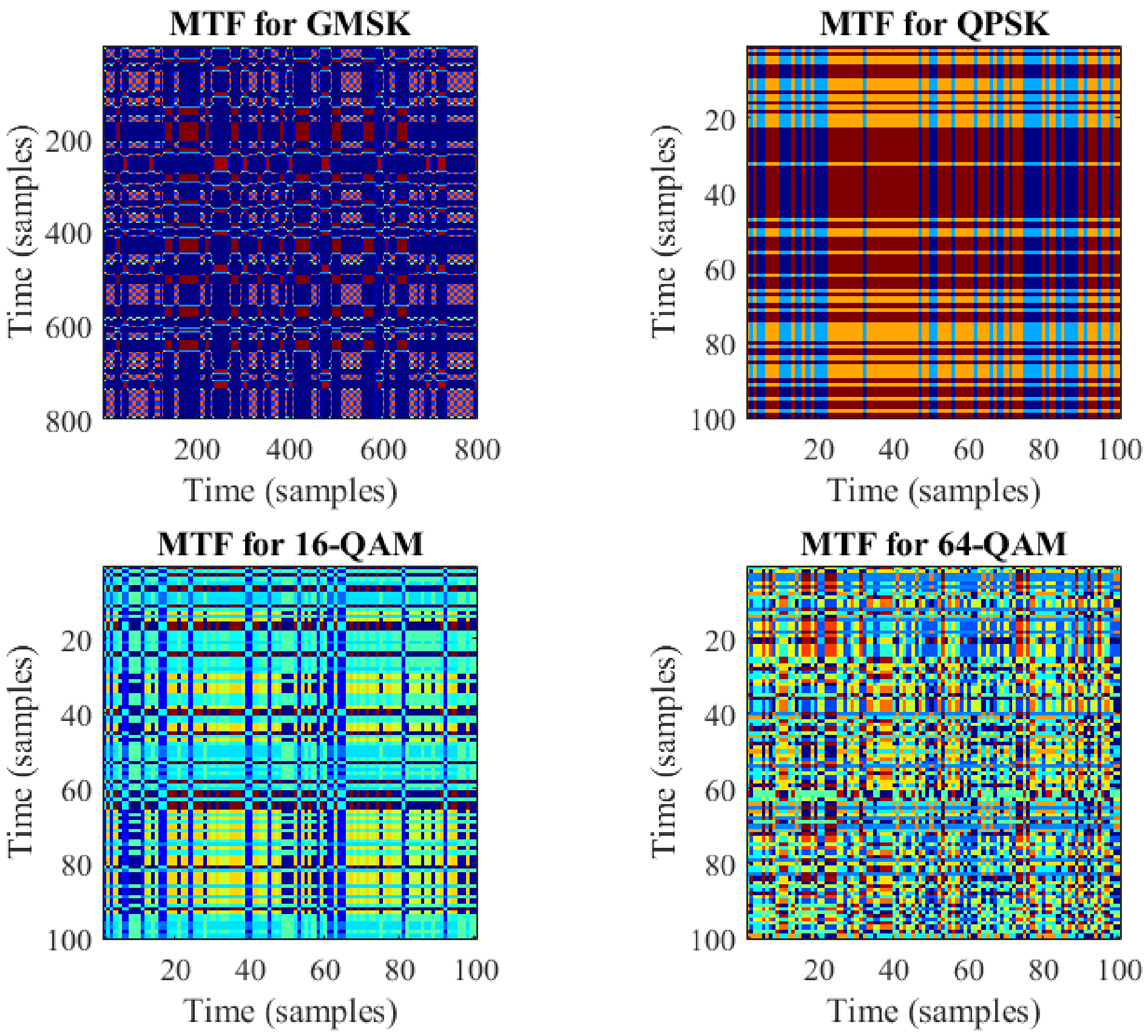

Figure 59: MTF Plots for various modulation schemes

3. **Gramian Angular Fields (GAF):** Gramian Angular Field (GAF) captures the relationships between each pair of points in a time series in a polar coordinate space. While MTF focuses on transition probabilities, GAF captures temporal correlations between values in a time series. The given time series is first normalized to have values in the range $[-1,1]$. The normalized time series is then transformed into polar coordinates. GAF values are calculated based on the type of GAF being used. There are two types of GAFs: *Summation* and *Difference*.

   - **Gramian Angular Summation Field (GASF):** This type captures the temporal correlations between values in the time series.

   - **Gramian Angular Difference Field (GADF):** This type captures the temporal anti-correlations between values in the time series.

   The GAF is defined as follows:

$$\text{GASF} = \cos(\theta_i + \theta_j) \tag{44}$$

$$\text{GADF} = \sin(\theta_i - \theta_j) \tag{45}$$

   where $\theta_i$ and $\theta_j$ are the angles corresponding to the $i$-th and $j$-th points in the normalized time series transformed into polar coordinates.

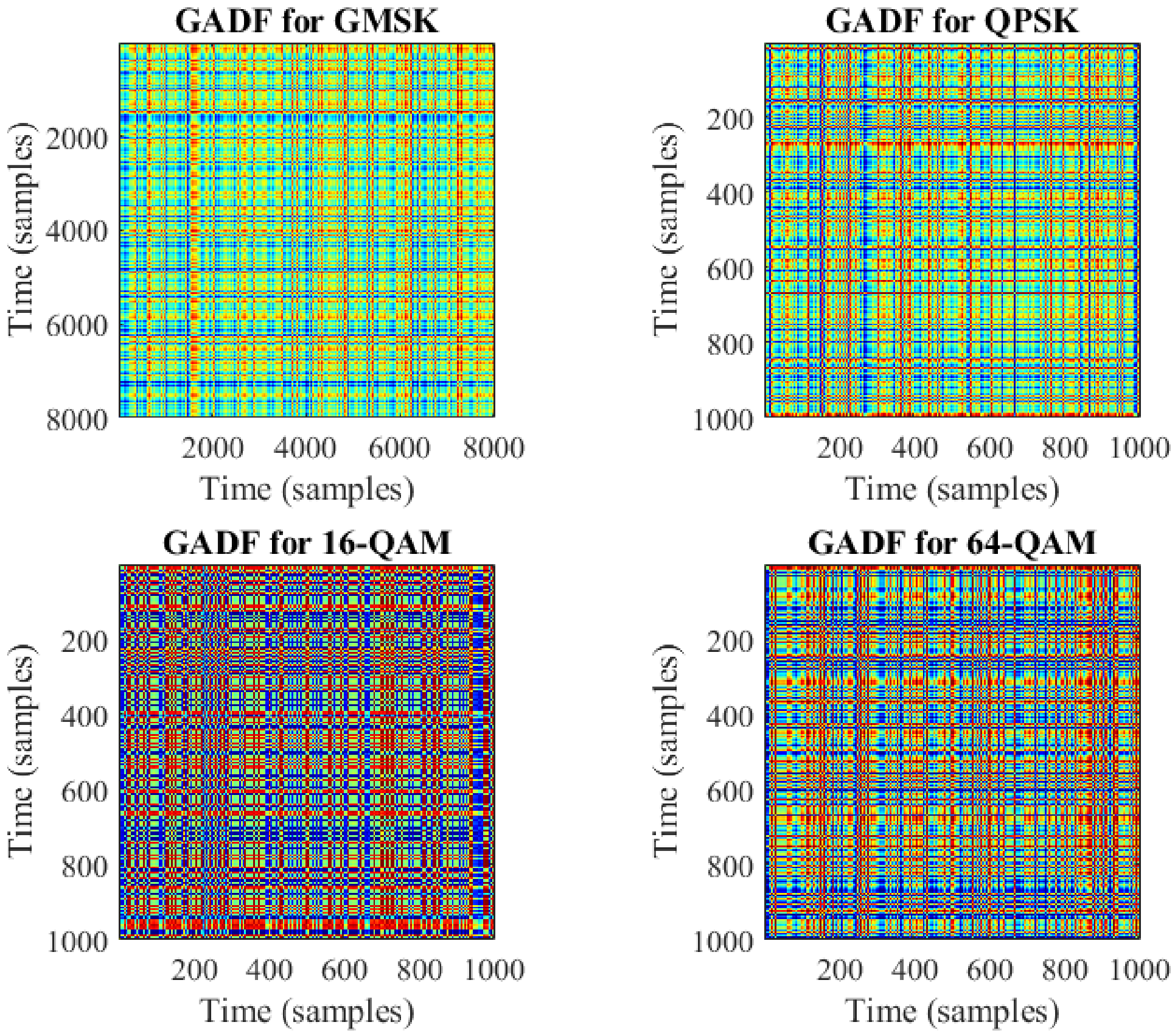

Figure 60: GADF Plots for various modulation schemes

Figure 60 shows GADF plots for four different modulation schemes: GMSK, QPSK, 16-QAM, and 64-QAM. GADF transforms visualize the relative angular movement between points in a time series, providing a detailed view of the temporal patterns and structural differences in signal behaviour across various modulation types.

Large-scale AI models can efficiently generate these augmented datasets. Figure 61 shows an example of this, where RP plots were generated using OpenAI's ChatGPT.

Once the augmented Image dataset is generated, the Few-Shot classification technique can be used for AMC. Few-shot classification involves using a (pretrained) model to classify images with only a limited number of examples per class. OpenAI CLIP (Contrastive Language-Image Pre-Training) model [473] offers a powerful approach to this problem by learning to associate text with images through embedding both in a shared space. A classifier can then be created by providing captions describing the possible labels for generated modulation images and selecting the caption with the highest similarity. CLIP can be fine-tuned in various ways, either by training the entire network end-to-end or focusing on the image encoder alone.

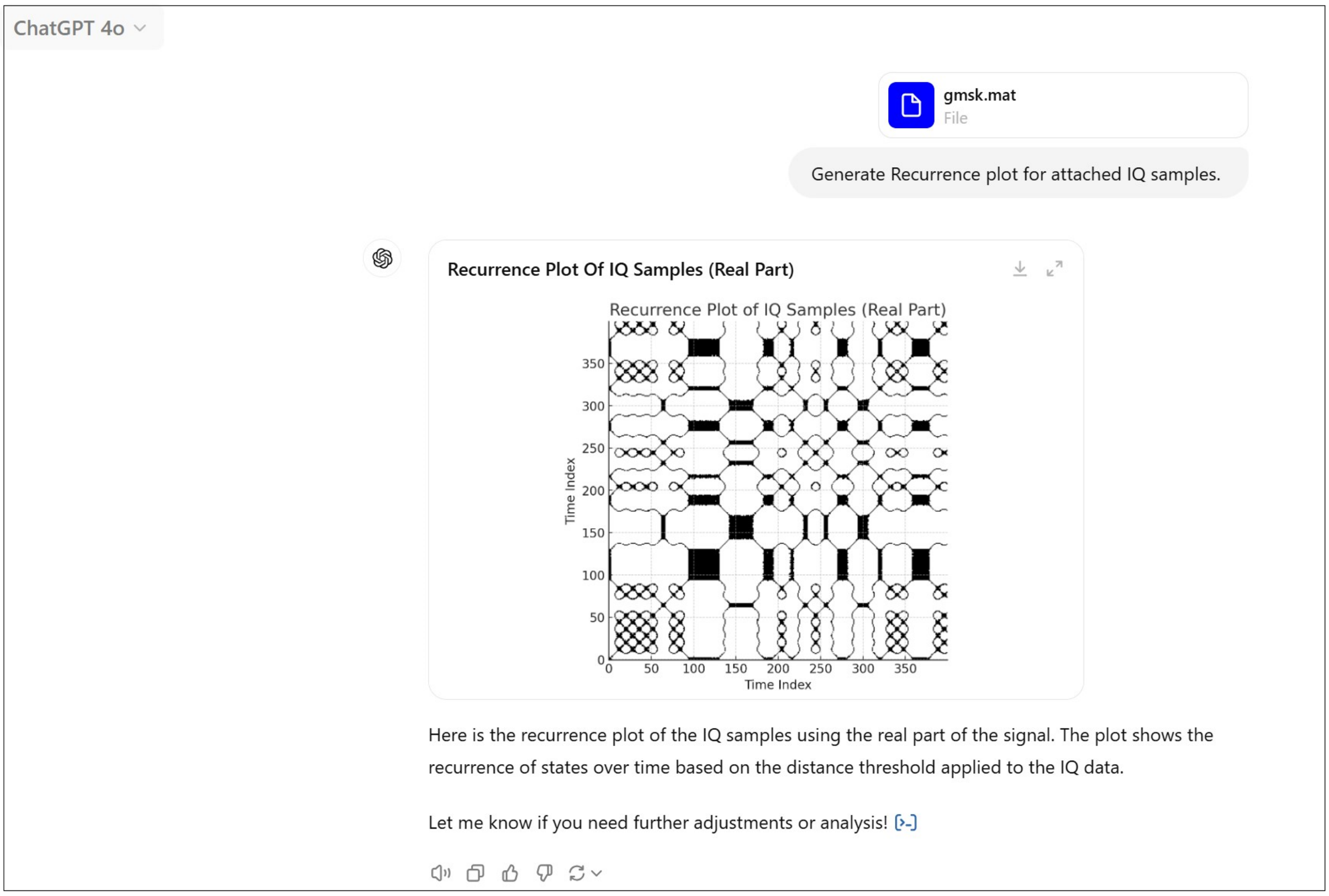

Figure 61: RP plot generate by OpenAI ChaGPT

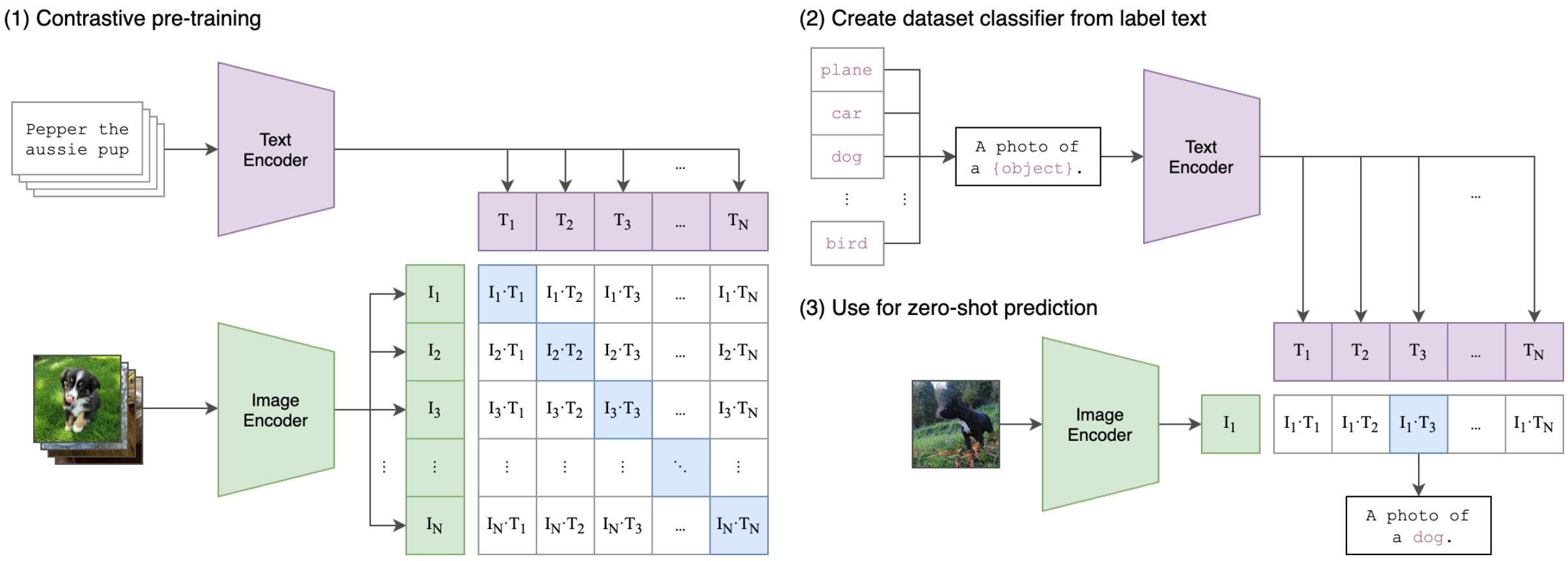

Figure 62: OpenAI CLIP model [473]

## 9.6 Agents for Telecommunications (Telecom-Copilots)

The process of understanding, developing, and researching modern wireless communication technologies is often lengthy and demanding. It requires systematically exploring through numerous web pages and technical specification documents to gather essential information and synthesize it into actionable insights. This section explores Telecom-Copilot, a Gen-AI tool designed to efficiently synthesize and process information from wireless communication specifications, simplifying complex data for streamlined analysis and understanding.

The complex nature of technical specifications poses a significant challenge for understanding and refining modern wireless communication technologies. Researchers, engineers, and students often struggle with overwhelming acronyms and complex terminology, compounded by information scattered across numerous documents. This shattering of content makes it a labor-intensive and time-consuming endeavor. The dawn of foundation models, such as ChatGPT, offers a promising solution to this challenge. These models represent a substantial leap forward in delivering synthesized, easily understandable responses to user queries about wireless communication specifications and technologies, streamlining information retrieval and comprehension. While state-of-the-art foundation LLMs are sometimes effective in addressing several queries related to modern wireless communication technologies, they often provide responses that are irrelevant or inaccurate, limiting their reliability in this domain [141]. Much like existing conversational copilots such as ChatGPT, the telecom industry stands to gain significantly from a specialized conversational AI tool designed for the synthesis of wireless communication specifications—a Telecom-Copilot. A Telecom-Copilot provides a question-and-answer interface specifically tailored to the telecom domain, offering enhanced capabilities to deliver more accurate and contextually relevant answers on topics related to the technical specifications of modern wireless communication technologies. As described in [141], a Telecom-Copilot can be built upon foundation LLMs and features three key additional components:

- Domain-Specific Database: Foundation models are typically trained on extensive web-based datasets. However, technical specifications and documents related to modern wireless communication technologies, while publicly available, are often not easily accessible due to their niche and specialized nature. This lack of representation leads to gaps in the ability of state-of-the-art foundation models to recognize and generate responses with patterns relevant to wireless communication systems. To overcome this limitation, the Telecom-Copilot system supplements foundation models by extracting and segmenting text from various technical specifications, including 3GPP releases, WiFi standards, and O-RAN documents, ensuring more accurate and context-aware outputs.

- Context Extractor: The Telecom-Copilot employs a context extractor (i.e., a RAG framework) to identify and retrieve the most relevant text samples from a domain-specific database. These extracted samples are provided as contextual input to either a base or fine-tuned foundation model. The framework combines this context with the user's query to generate precise and informed responses. To ensure traceability, the specification document identifier is stored for citation purposes.

- Feedback Mechanism: The copilot may incorporate a robust feedback feature, enabling users to interactively refine their experience. Users can like or dislike responses or seek clarification from an expert. When expert feedback is requested, the system automatically generates an issue in a designated repository. This issue includes the user query, the provided context, and the generated response. Experts can then address the issue by offering feedback or contributing additional data, enhancing the system's accuracy and reliability over time.

## 9.7 LTMs in Network Operations and Maintenance

With the continuous progress of mobile communication technology and the continuous growth of network demand, the network structure is becoming more and more complex. However, traditional network management can no longer meet the needs of future development. In the future, intelligent autonomous networks based on AI-driven automated analysis and multi-dimensional data perception will provide more flexible and efficient network strategies, but this will also require smarter network management methods. Large language models (LLMs), will play an important role in promoting the intelligent autonomy of communication networks. At the same time, they will also provide new application scenarios for the future development of LTMs in the telecommunications field.

Typical application scenarios of network intelligent operations and maintenance (O&M) include anomaly detection, fault diagnosis, event warning, and performance optimization. In traditional network O&M, O&M personnel need to obtain network state information through manual inspection and data analysis, which is inefficient. By introducing LLM technologies, network state information can be monitored in real-time and efficiently, and the network can be analyzed and processed through automated O&M, thereby effectively improving the stability and reliability of the network. The simplified architecture of network operation and maintenance assisted by LTMs is shown in Figure 63.

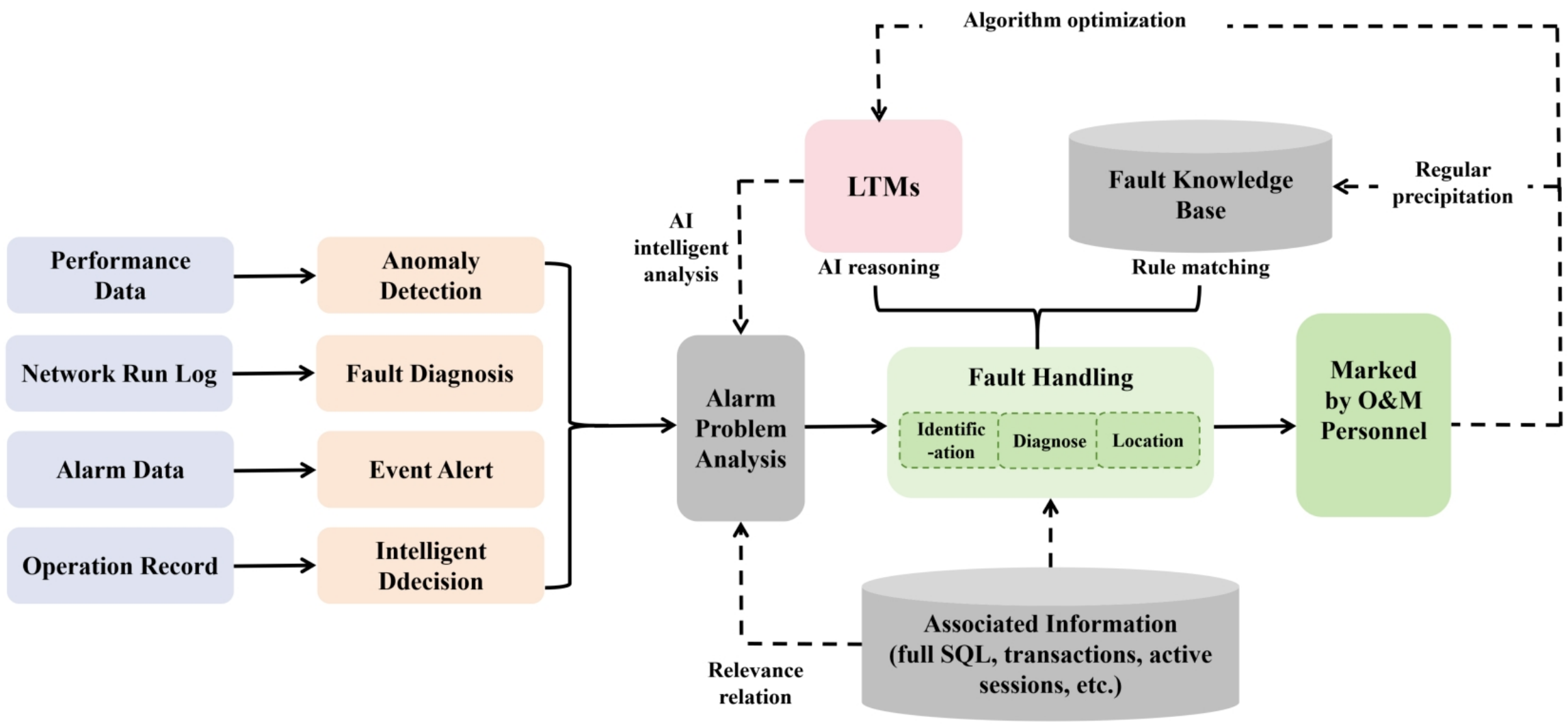

Figure 63: LTMs simplify network operations and maintenance.

By collecting various performance data, network congestion information, and network operation logs in the network, and sending real-time network information to LTMs, the complete process from monitoring to processing can be completed, enabling intelligent network operation and maintenance. Specifically, LTMs will quickly perform statistical analysis, then predict the network in combination with various network service scenarios, perform fault diagnosis, and provide corresponding operation and maintenance decisions, which will then be sent to the database for storage. In addition, LTMs can also visualize the current and predicted network conditions, better displaying network status and trends to operation and maintenance personnel, helping them to perform network operation and maintenance more efficiently and intelligently.

Furthermore, in combination with LLM technologies, the underlying components can be comprehensively analyzed using intent-driven, communication-network-specific ML models and advanced policies. This approach achieves an autonomous loop of “fault localization-policy generation-policy verification.” Unlike traditional intelligent O&M methods assisted by manual decision-making, this method uses spatiotemporal representation learn-ing for knowledge inference in network operation state detection, automatically generates and verifies fault recovery and multitask management policies, and ensures service bandwidth and network performance through techniques such as bypass routing and resource orchestration. Moreover, it can autonomously perform fault repairs based on learning results to support network management and state adjustments, thereby promoting autonomous network O&M.

## 9.8 Large telecom foundation model for the physical-layer

Since LLMs lack a profound understanding of physical-layer wireless signals, there is a need to develop foundation models specifically trained to handle any type of data that comprises such signals. Typically, wireless use cases provide this data in the form of (IQ-based) time series, sampled at varying lengths and sampling rates. However, such a model should generally also understand derived statistics such as channel impulse responses, chirps, power spectral densities, FFT statistics, etc. To obtain a semantic understanding of wireless signals a wireless foundation model can be trained through the following steps *next-sample prediction*, *masking* and *denoising* as highlighted in Figure 64. In contrast to current machine learning approaches, that focus on specific wireless use cases such as, e.g. sensing using UWB radar, spectrum management of overlapping technologies, health care sensor data, etc, the goal of such a foundation model is to be task agnostic and offer universal knowledge of the heterogeneous time series in these use cases. To do so, the model must transform the input data into a common embedding and / or tokenization space. Time series can be divided along the time dimension into patches of size $p$, preserving the two I and Q channels. Determining $p$ can be challenging when targeting multiple use cases [474]. For wireless use cases, the tokenization strategies investigated for natural language processing cannot be directly applied because the entropy of different wireless use cases will be different from text and even across multiple use cases. These patches can be embedded using a linear projection layer together with a positional embedding dimension. Next, these samples can be given e.g. to a transformer foundation model which needs to be pre-trained first on different pre-training tasks from various use cases. These tasks can include next-sample prediction, masking, denoising, etc., and serve for generalized feature learning in a self-supervised fashion. After pre-training, downstream tasks can be adapted using finetuning with supervised, few- or zero-shot learning or reinforcement learning. Concerning the use cases (sensing using UWB radar, spectrum management of overlapping technologies, and healthcare sensor data), the pre-trained foundation model can be fine-tuned to different tasks such as determining people walking, obstacle detection, interference detection, technology identification, and more. The output of these tasks can include classification, prediction or semantic context prediction. For example, in addition to spectrum classification, the system can predict environmental information such as room size, whether the setting is indoor or outdoor, and the presence of people. The integration of physical layer information from foundation models with LLMs enables network optimization and configurations. This integration is achieved by incorporating semantic context from the foundation model, as illustrated in Figure 64. Along with human prompting and wireless specification and standardization documents, LLMs can automatically adapt and optimize wireless networks in near real-time.

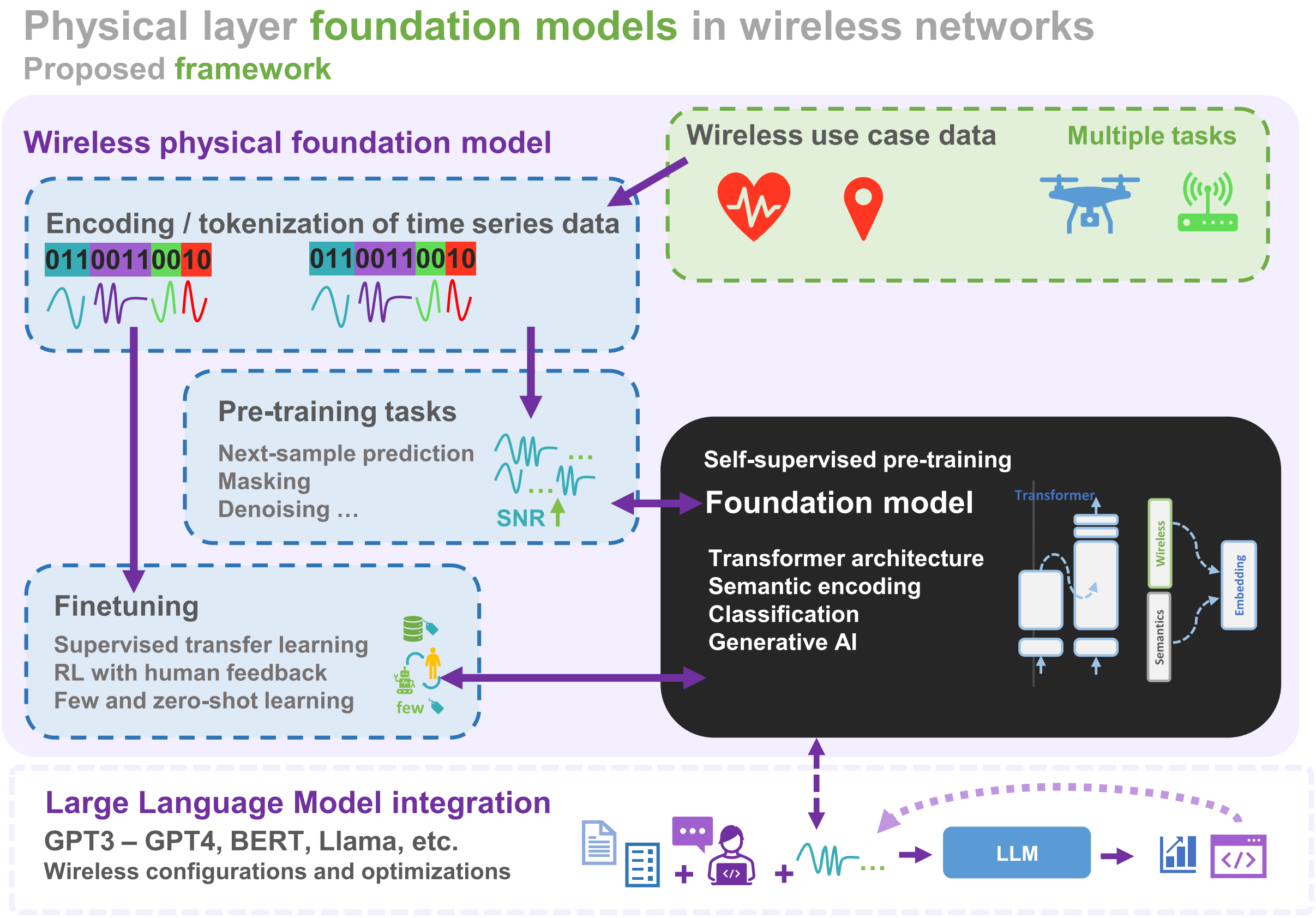

Figure 64: A wireless foundation model architecture for physical-layer wireless signals [475]

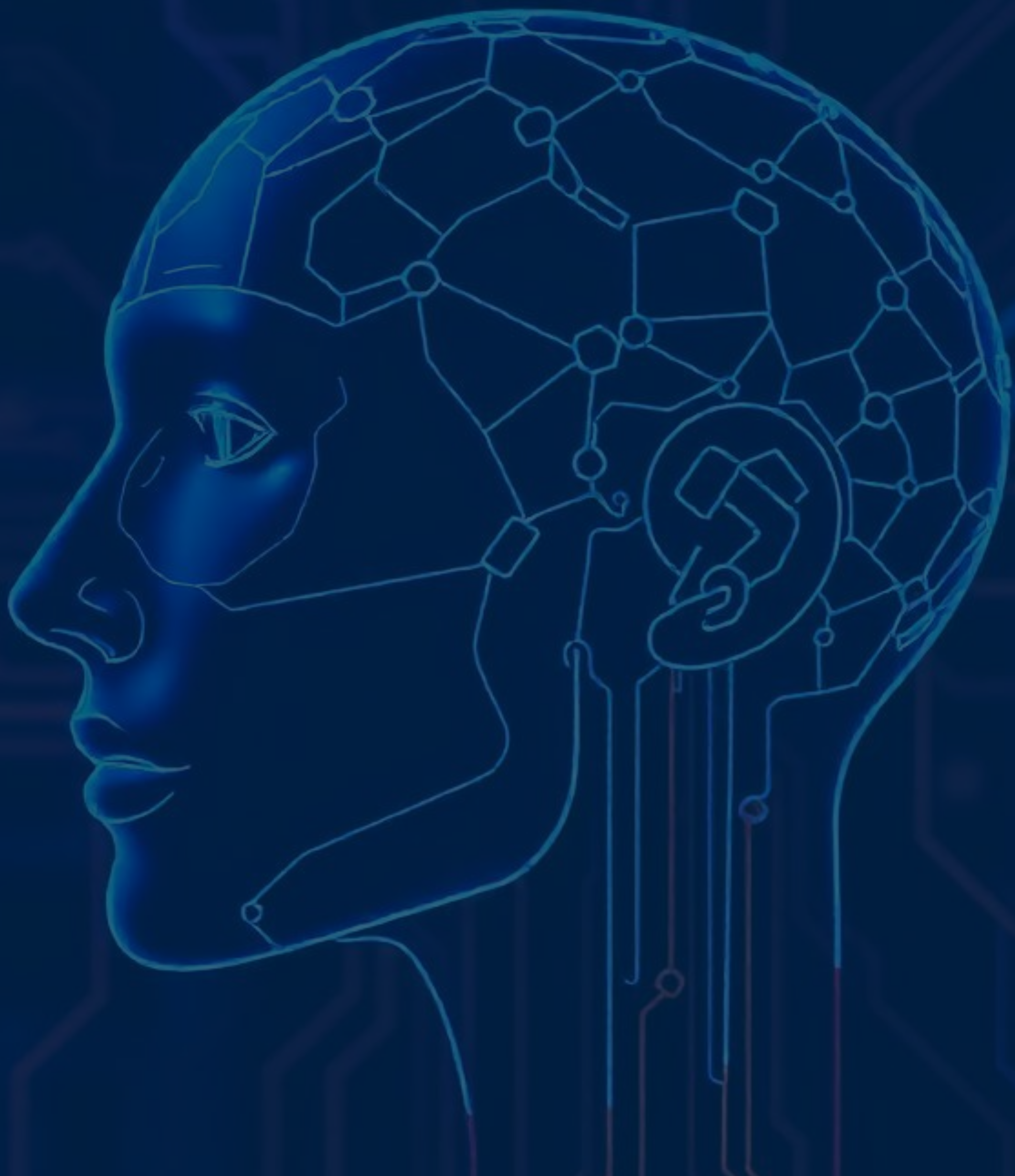

# LARGE TELCO MODELS: NAVIGATING REGULATORY AND ETHICAL COMPLEXITIES

## 10 Large Telco Models: Navigating Regulatory and Ethical Complexities

As large telco models become more embedded across telecom networks, they face significant regulatory and ethical scrutiny. These models, essential for powering advanced capabilities like predictive maintenance, personalized customer interactions, and network optimization, also bring challenges in data governance, accountability, and privacy. The regulatory landscape is complex, influenced by both regional laws and the broader global policy environment. With these models increasingly intersecting with sensitive personal data and critical infrastructure, establishing robust governance structures has become a necessity for telecom operators worldwide.

### 10.1 Data Governance and Accountability

Data governance in the telecom sector must be particularly rigorous due to the extensive volume and sensitivity of data that telecom networks manage. Establishing a robust governance framework involves implementing clear protocols for data collection, processing, and storage, especially for large AI models which often require vast amounts of varied data to function optimally. Telecom companies need to define roles, such as data stewards and compliance officers, who can oversee data quality and ensure that datasets are properly anonymized, retained, and utilized in a way that aligns with both operational needs and regulatory standards. Ensuring accountability across these models means building transparency into their processes, allowing regulators and internal stakeholders to understand decision-making, particularly in customer service automation, fraud detection, and network management.

In addition to data handling, telecom AI models must be developed with a commitment to transparency in outcomes. This transparency helps to build trust among stakeholders—be they customers, regulatory bodies, or internal users of the AI systems. Trust-building in AI decision-making is critical, especially when these systems interact directly with customers, such as in automated support interactions or in delivering targeted service recommendations.

### 10.2 The Regulatory Landscape: EU and U.S. Differences

The European Union, through its GDPR and the newly proposed AI Act, is taking a highly proactive regulatory stance that emphasizes user rights and stringent data protection requirements. GDPR's influence on telecom AI is significant, with its requirements for data minimization and explicit consent presenting unique challenges. These principles impact telco AI in ways that extend beyond the traditional data processing paradigms, given that telco models are often cross-border by nature and require continuous data to maintain and optimize network performance. Ensuring GDPR compliance means telecom operators need to implement complex processes for data anonymization and localization, potentially increasing operational costs and impacting the efficiency of large AI models.

The EU AI Act, if adopted, will further add to these challenges by classifying telecom applications under high-risk categories, mandating detailed audits, risk assessments, and transparency measures. For telcos, this will likely translate into extensive compliance obligations, particularly for customer-facing AI solutions or applications that have significant impacts on user privacy or service quality.

In contrast, the United States presents a more fragmented regulatory landscape, especially given the current political climate. At the federal level, initiatives like Executive Order 14110 set forth requirements for transparency, anti-discrimination, and explainability in AI systems. However, with the recent 2024 shift towards a Republican-led administration, there is a possibility that stringent federal oversight could be reduced or modified, particularly as it pertains to regulatory demands that may be seen as overly restrictive to innovation or economic growth. Federal AI regulation in the U.S. is also largely voluntary and lacks the uniform enforcement mechanisms found in the EU, placing more emphasis on industry self-regulation and state-level laws like the California Consumer Privacy Act (CCPA) and the California Privacy Rights Act (CPRA). Telecom companies operating in the U.S. face the unique challenge of balancing these diverse state requirements with their broader federal obligations.

Given this evolving landscape, telecom operators need to remain agile, prepared for potential shifts in federal regulations that could either relax or intensify compliance requirements. In an era where regulations may be overturned or significantly altered, especially as AI continues to advance and reshape industries, telecom operators will need to establish adaptable compliance structures that can swiftly respond to these political and regulatory changes.

## 10.3 Ethical and Operational Challenges in Telco AI

Beyond regulatory compliance, ethical considerations remain at the forefront of deploying large telco AI models. Given that these models directly affect consumer interactions and potentially influence customer trust, operators must address issues of bias, discrimination, and accountability. The scale of telecom operations adds a layer of complexity: while automated AI models can efficiently manage customer service and optimize network operations, they can also reinforce or exacerbate existing biases if not carefully designed and tested.

To mitigate these risks, telecoms need to embed ethical guidelines into the AI development lifecycle, ensuring that fairness, accountability, and transparency are prioritized from model conception through deployment. This approach is particularly essential in high-stakes applications such as emergency service routing, fraud detection, and automated dispute resolutions, where biased decision-making could have serious implications for user trust and regulatory standing.

## 10.4 Future Directions and Recommendations for Telco AI Governance

Looking ahead, telecom providers face both challenges and opportunities in aligning their AI models with evolving regulatory requirements while maintaining operational efficiency. Emerging trends in AI governance, including adaptive risk assessment, continuous auditing, and real-time monitoring, present avenues for telco providers to bolster their data governance practices in ways that proactively address both regulatory and ethical considerations. By embracing these practices, telecoms can not only improve their compliance posture but also establish a more resilient foundation for responsible AI deployment.

International cooperation will also play a critical role in harmonizing standards across borders, with telecom providers advocating for frameworks that account for the unique, data-intensive nature of their networks. Organizations such as the OECD and G7 are increasingly involved in fostering consensus on AI governance, offering telecoms the potential to contribute to globally recognized standards that bridge the divide between the EU's stringent requirements and the more flexible U.S. approach. For telecom operators, active participation in these international forums will be essential in shaping regulations that are both robust and feasible for their large-scale operations.

In summary, the deployment of large telco AI models brings a host of regulatory, ethical, and operational considerations that telecom providers must navigate thoughtfully. By establishing strong data governance frameworks, staying agile in response to shifting U.S. regulatory dynamics, and adhering to rigorous ethical standards, telecom operators can not only comply with current requirements but also set a foundation for sustained, responsible AI-driven growth.

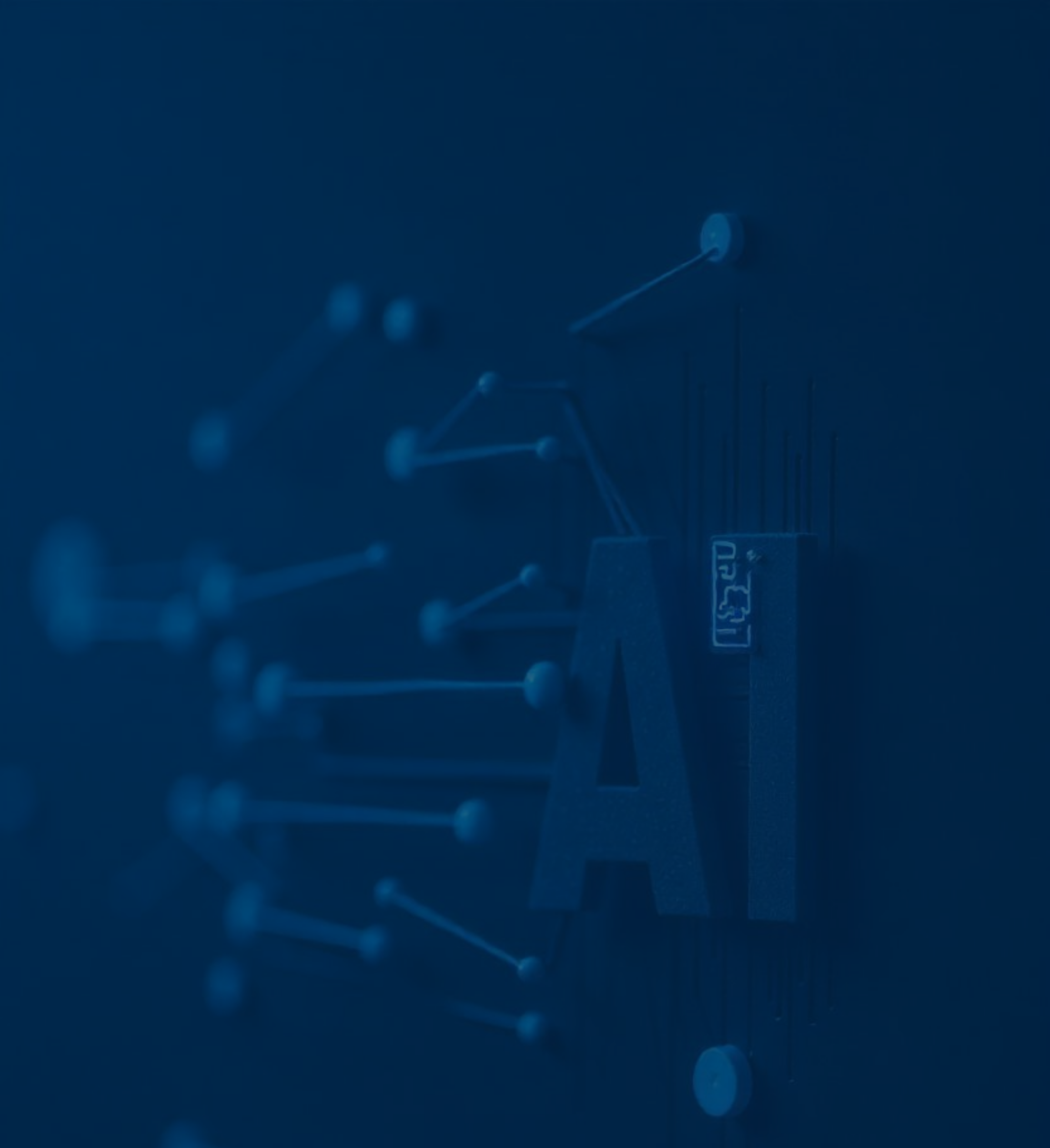

# STANDARDIZATION ACTIVITIES AND ROADMAP

# 11 Standardization Activities and Roadmap

The integration of Artificial Intelligence (AI) into the telecommunications industry is a rapidly evolving field, and standardization activities are crucial to ensure interoperability, security, and efficiency. Here are some key points about the standardization activities and roadmap for large-scale AI in telecom: Standardization of Large-Scale AI in Telecom

## 11.1 Key Standardization Activities

1. ITU-T Y.3000 Series:

    - Frameworks and Architectures: These provide the structure for how AI should be integrated into telecom networks. This includes defining the layers, components, and interactions necessary for AI to function within telecom systems.

    - Protocols: These are sets of rules that ensure AI applications can communicate effectively with each other and with existing telecom systems.

    - Use Cases: The ITU-T Y.3000 series includes detailed use cases that demonstrate practical applications of AI in telecom, such as network optimization, predictive maintenance, and customer service enhancements.

2. CEN-CENELEC Focus Group:

    - Accountability: Standards to ensure AI systems are transparent and that their decisions can be traced and explained. This is crucial for trust and compliance with regulations.

    - Quality: Ensuring the data used by AI systems is of high quality and appropriate for the tasks at hand. This includes data accuracy, completeness, and timeliness.

    - Security and Privacy: Developing standards to protect data used by AI systems from unauthorized access and breaches. This also includes ensuring that AI systems comply with privacy regulations.

    - Ethics and Safety: Guidelines to ensure that AI systems are used ethically and do not pose risks to users. This includes avoiding biases in AI decision-making and ensuring the safety of AI applications.

3. GSMA Responsible AI Maturity Roadmap:

    - Vision and Strategy: Helping organizations define their AI strategy and vision, ensuring alignment with ethical principles and business goals.

    - Operating Model: Developing an operating model that supports the deployment and management of AI systems. This includes roles, responsibilities, and processes.

    - Technical Controls: Implementing technical measures to ensure AI systems are secure, reliable, and perform as expected.

    - Third-Party Ecosystem: Managing relationships with third-party providers involved in AI development and deployment, ensuring they adhere to the same standards.

    - Change Management and Communications: Ensuring that changes brought about by AI adoption are managed effectively, with clear communication to all stakeholders.

## 11.2 Roadmap for Large-Scale AI in Telecom

1. Network Infrastructure

   - Performance Enhancement: Integrate AI algorithms to optimize network performance by dynamically managing resources, reducing latency, and improving data throughput.
   - Resource Management: Implement AI solutions to predict and manage network resources efficiently, preventing congestion and optimizing bandwidth usage.
   - Security Enhancements: Use AI-driven security systems to detect and respond to threats in real-time, enhancing overall network security.

2. Network Management

   - Operational Optimization: Deploy AI systems that monitor network operations continuously, identifying and resolving issues proactively to minimize downtime and maintenance costs.
   - Predictive Maintenance: Utilize AI to analyze historical data and predict when network components might fail, allowing for timely maintenance and reducing unexpected outages.

3. Business Operations

   - Customer Service Automation: Implement AI-powered chatbots and virtual assistants to handle customer queries efficiently, providing quick responses and reducing the workload on human agents.
   - Intelligent Applications: Enhance business support systems (BSS) and operation support systems (OSS) with AI to automate routine tasks, improve decision-making, and provide insights based on data analysis.

4. Vertical Applications

   - Healthcare: Apply AI for remote patient monitoring, predictive diagnostics, and personalized treatment plans, improving patient outcomes and reducing healthcare costs.
   - Automotive: Enable advanced driver-assistance systems (ADAS), predictive maintenance for vehicles, and smart traffic management systems with AI.
   - Smart Cities: Use AI to manage urban infrastructure efficiently, including traffic management, energy consumption, waste management, and public safety.

5. Future Networks

   - 5G and Beyond: Develop AI technologies to support the next generation of telecom networks, focusing on network slicing, edge computing, and massive IoT deployments.
   - Network Intelligence: Enhance network intelligence with AI to optimize performance, manage complex interactions between different network elements, and provide personalized services to users.
   - Intent-Driven Network Management: Utilize AI to understand and fulfill user intents, automating network configuration and management based on high-level user requirements.

## 11.3 Timeline and Milestones

- Short-term (1-2 years): Focus on integrating AI into existing network management systems and enhancing customer service operations.
- Medium-term (3-5 years): Expand AI applications to vertical industries and develop more advanced predictive maintenance and security solutions.
- Long-term (5+ years): Prepare for the full integration of AI in future networks, including 5G and beyond, with a focus on network intelligence and intent-driven management.

This roadmap ensures that AI technologies are implemented safely, ethically, and efficiently in the telecom industry. It outlines clear steps and milestones for integrating AI into various aspects of telecom operations, paving the way for a smarter, more efficient future.

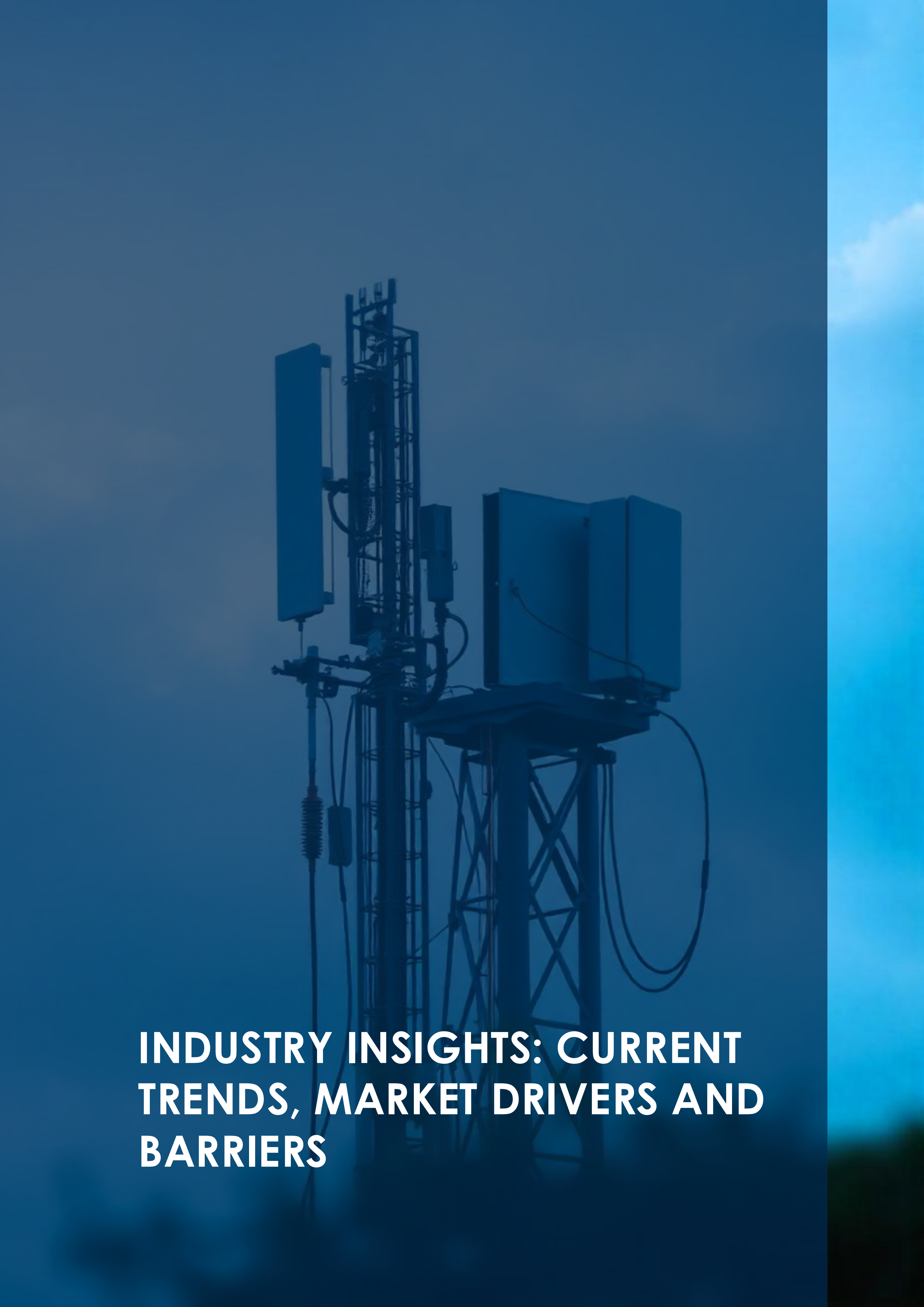

# INDUSTRY INSIGHTS: CURRENT TRENDS, MARKET DRIVERS AND BARRIERS

# 12 Industry Insights: Current Trends, Market Drivers and Barriers

In this chapter, we will introduce the initial applications of LLMs in the telecommunications industry. LLMs have found various use cases in fields such as medicine, education, and engineering, including tasks such as code generation [476]. In the telecommunication sector, LLMs have primarily been used for the generation of text and code. However, the telecommunication industry is now exploring the potential benefits of using LLMs in different use cases to optimize networks and improve performance. In addition, telecommunication companies are studying the potential revenue streams from LLMs, such as the application of chatbots that represent user-friendly interfaces to sell services. Customers are also using LLMs in their devices, which requires more data usage needs that should be considered in the design of the beyond 5G and 6G. Operators are actively addressing the barriers and drawbacks associated with implementing LLMs. In the following sections, we will dive into the current trends, market drivers, and barriers in more detail.

## 12.1 The application of LLMs in Telecommunication: Current Trends

Since the appearance of the first LLMs in the market such as GPT-1 by OpenAI and Bert by Google, multiple studies in the application Of LLM has emerged in the telecom sector. The first use cases considered the usage of LLM where the data is *text* type such as the development of Chatbots that understand clients intentions or requirements for telecommunication services ordering. In this case, LLMs were applied with fine-tuning technique to understand client's needs described in natural language [382, 477]. In the following, we will describe a number of trends and use cases that showcase the first application of LLM in the telecommunication sector.

### 12.1.1 Multimodal LLMs

Multimodal models, often referred to as MLLMs, XLMs and so on are models whose training has incorporated two or more domain-specific training datasets. resulting in a unified latent space combining representations from all modalities. It's important to distinguish between multimodal models typically employed for text-to-media generation tasks, and models typically integrating text and other modality domains, which are designed for *domain-domain(-domain...)* interaction tasks.

Underlying model architectures vary widely, dependent on task type (such as media generation, domain-domain interaction, or cross-modal reasoning). Stability AI's Stable Diffusion and Stable Audio families of models, as their names imply, are media (image and audio) generation models based on a diffusion architecture which for these tasks generate media by progressively refining noisy data into coherent image and audio content.

OpenAI's CLIP and Microsoft's LLaVA models, designed for image-text interaction, employ dual-encoder architectures. Separate encoders (i.e., based on transformers for text and Vision Transformers (ViTs) or CNNs for images) map each modality into a shared latent space without directly generating new content. They are oriented to cross-domain tasks such as media classification and cross-modal reasoning.

In terms of effects on cross-modal reasoning, model architectures can additionally be categorized by their use of early fusion or late fusion strategies for integrating modalities. Meta's Flamingo model, using early fusion, combines data from different modalities at the input stage, and the model processes them jointly throughout the model layers. CLIP employs the late fusion strategy and each modality is processed separately through distinct branches, for example using text-based encoders for language and image encoders for visual data, with the final combination of their latent representations occurring only after this isolated processing. Early fusion is more computationally expensive but offers more tightly integrated representations which could better encode more complex modal relationships. Late fusion is better suited to retrieval and mapping tasks where modalities can be processed independently.

The same architectures supporting models oriented to human-consumable media tasks are of course applicable to any modality. For example in the telecom space, a multimodal model oriented to baseband operation tasks could fuse signal characteristics, spectrum, and network configuration data in a shared latent space for subsequent prediction and dynamic configuration operations.

### 12.1.2 LLMs for network assurance

LLMs can play a crucial role in the evolution of network assurance, including anomaly detection, prediction and corrective and preventive measures and actions [478]. By leveraging their ability to process vast amounts of data and recognize patterns, LLMs can identify unusual behaviors in network traffic, predict potential failures, and suggest corrective measures in real-time. This integration not only improves network reliability but also optimizes resource allocation, ensuring a seamless user experience in the hyper-connected world of 6G.

Various works have shown the potential of LLMs at detecting multiple categories of anomalies:

- Logical anomalies: refer to errors in the logical flow or structure of data, where the sequence or relationships between data points do not follow expected rules [479]. For example, an LLM could identify logical anomalies in network logs where the sequence of events does not follow the expected logical order, such as a user logging out before logging in. These anomalies often indicate mis-configurations or potential security vulnerabilities that need to be addressed.

- Semantic Anomalies: involve inconsistencies in the meaning or context of data, where the content does not align with the expected semantics but where there isn't any problem with any individual component of a system [480]. Research have shown that LLM-based monitor can effectively identify semantic anomalies in a manner that shows agreement with human reasoning. In the context of network data, an LLM might detect a semantic anomaly in routing if a route appears valid syntactically but violates expected routing policies or patterns [481].

- Visual anomalies: are irregularities that can be identified through visual inspection of data representations. LLMs have demonstrated strong performances for detecting this kind of anomalies [482]. In network data, they are less common but can be relevant in the context of network monitoring dashboards and visualizations. For instance, an LLM can detect visual anomalies in a network traffic heatmap where certain regions show unexpected spikes or drops in activity [483]. Another example is identifying unusual patterns in network topology diagrams, such as a sudden change in the structure of connections that might indicate a network attack or failure. Visual anomalies help in quickly identifying and diagnosing issues that might not be immediately apparent through raw data analysis.

LLMs can offer significant advantages in analyzing network logs, system events, and contextual data. Their natural language processing capabilities allow for the identification of anomalies that traditional systems might overlook. By interpreting complex data patterns, LLMs can detect unusual behaviors indicative of potential intrusions or future anomalies and SLA violation, thereby improving the overall security and reliability posture of 6G networks.

LLMs can also provide valuable insights into the decision-making processes of AI models. They can present security alerts in an easily understandable format, enabling analysts to make informed decisions. This interpretability is crucial for identifying anomalies, and facilitate corrective procedures in automated systems.

LLMs have a wide range of applications in anomaly detection for networks. Below are some notable use cases [484]:

- Personalized Assistants: Intelligent assistants powered by LLMs can offer context-aware interactions, an-

ticipating user needs and responding proactively. This personalization enhances user experience while simultaneously monitoring for anomalies in user behavior.

- System Log Anomaly Detection: LLMs can analyze system logs to detect anomalies that may indicate security breaches or operational failures. Their ability to process large volumes of data quickly makes them ideal for real-time monitoring.

- Self-Healing Systems: Integrating generative AI, such as GPT-4, into self-healing systems allows for automated code generation and repairs. This reduces the need for human intervention, optimizing system functionality and efficiency.

- Event Sequence Prediction: LLMs excel in predicting event sequences, capturing complex behavioral patterns and dependencies. Their ability to understand contextual nuances enhances the accuracy of forecasts, making them invaluable in user behavior analysis and transaction sequences.

- Real-time Anomaly Detection: Edge-based LLMs facilitate predictive maintenance by identifying and mitigating anomalies in real-time. By making informed decisions, these models enhance the reliability and efficiency of industrial operations, ensuring seamless functionality in 6G environments.

While LLMs offer significant advantages in anomaly detection, several limitations and challenges must be addressed.

#### 12.1.3 Large Action Models (LAMs)

Large Actions Models (LAMs) represent a further evolution of model inference integration into code execution chains. In this context the continuum of AI models includes:

1. Purely text-generation models.

2. Function-calling models triggering predefined API calls.

3. Tool-calling models interacting more broadly with external systems and utilities.

4. LAMs orchestrating multi-step, goal-oriented workflows on multiple platforms. Not quite synonymous with agentic systems but with similar characteristics.

5. Direct binary output LAMs, generating executable in-memory program states for immediate execution. This technology represents a distant horizon.

Purely text generation LLMs and function/tool-calling models are both pre-trained using commonly available ”Big Text” corpora. Function and tool-calling models, in contrast to the instruction-following and chat fine-tuning of textgen LLMs, are fine-tuned using datasets which include API usage, code execution, and structured data (such as OpenAPI JSON API definitions) for external function-calling task capability. The model identifies specific function call features within natural language input and maps them to predefined, single-step API interactions.

Tool-calling models can be considered as extensions of function-calling models and are fine-tuned on more heterogeneous datasets covering a more extensive range of interactions, such as with application suites and multiple utilities (for example, the standard set of utilities available with a Ubuntu installation). The extended training process for tool-calling models often involves simulation environments involving complex tool use and error recovery scenarios.

LAMs extend the function/tool-calling model concept and support autonomous, multi-step process interactions integrating multiple systems and tools. Function/tool-calling models are capable of invoking a single function or tool in response to a user's request, while LAMs dynamically manage interaction patterns associated with complex workflows. Action models generate "command" sequences intended to trigger operations within other systems. For example, a LAM would determine actions necessary to satisfy a natural language request and then generate what are effectively API calls to carry out the previously determined actions. The calls are text-based (albeit in most cases comprising data structures appropriate to the target as opposed to free text) with direct binary output straight to an external app or even representing an app's in-memory state a likely capability.

LAMs are usually considered in the same breath as agentic systems, in that LAMs represent the actual task execution aspect of multi-task, goal-achieving agent reasoning solutions. SalesForce's xLAM model family serves as an early indicator of model/deployment signatures for agentic frameworks.

### 12.1.4 AI Native

The AI native concept can be defined as follows: "AI native is the concept of having intrinsic trustworthy AI capabilities, where AI is a natural part of the functionality, in terms of design, deployment, operation, and maintenance. An AI native implementation leverages a data-driven and knowledge-based ecosystem, where data/knowledge is consumed and produced to realize new AI-based functionality or augment and replace static, rule-based mechanisms with learning and adaptive AI when needed." [485] Defining AI native: A key enabler for advanced intelligent telecom networks[485]

AI Native presupposes the operation of an underlying model or system for the delivery of the exposed features, for example, VAE models for anomaly detection and auto-regressive models (with diverse latent spaces) for generative and control operations.

AI Native is typically considered in two contexts: architecture and Deployment Modes. AI Native Architecture intends the deployment of AI functionality in a flexible manner via widely distributed AI modules delivering self-contained capabilities, APIs exposed via AI backends realized in a range of functional and/or spatial centralization, in various arrangements suited to a particular stack configuration realizing some feature.

Distributed LLMOps is a key aspect of AI Native. The ubiquitous deployment of AI functionality must be matched by the ability to deliver associated training and lifecycle management operations regardless of an AI subsystem's functional and operational location. This carries an implication for the underlying data infrastructure in terms of both data and hardware availability. Ideally a physical AI Native deployment supports optimized training via duplicated or shared data distribution combined with optimal GPU farm location(s) providing a Pareto-optimal cost front (although this will certainly change over time as model efficiency/size and GPU technology maintain their improvement trends). The CAPEX and OPEX impacts stemming from this result in a somewhat narrow TCO forecast window (depending on a CSP's infra update strategy).

Deployment modes consist of legacy "hard-coded" component replacement and/or augmentation, introduction of completely new AI-based components and introduction of "shim" AI-based control components acting as an interface point for one or more legacy components. The latter approach represents a rationalized introduction of AI functionality, providing automation and function enhancements to an already-developed feature base. This brings with it LLMOPs considerations as described above but with lower CAPEX/OPEX impacts due to the comparatively limited scope of AI use.

### 12.1.5 LLMs for network automation with intents

The ambition of network automation in TMForum is to enhance the efficiency of service providers by reducing manual intervention and operational costs. It aims to improve agility, allowing for rapid deployment of services and quicker responses to market changes. Additionally, network automation seeks to increase reliability by minimizing human errors and enhancing service quality through automated processes. This initiative also supports innovation by facilitating the introduction of new services and technologies, such as IoT. Ultimately, the goal is to enable self-management of networks, allowing them to self-configure, self-heal, and self-optimize, thereby creating a more flexible, responsive, and efficient telecommunications ecosystem.

Moreover, a manual process of ordering, deployment and adaptation may not be sufficient to meet business demands. In particular, turnaround and assurance times may require further automation for self-adaptation. The implementation of machine learning techniques should help automate how the system will be able to adapt in order to guarantee the services of offers to the verticals. Nevertheless, the 5G/6G network management system can only adapt and keep up with the needs of the company if it knows them. This includes knowledge of expectations, including strict requirements, but also preferences and priorities. This can evolve dynamically as customer needs change. It is therefore the goal of an intention-based approach to define and communicate knowledge of expectations to a system in a way that allows automated processes to reason about them and derive appropriate decisions and actions. Initially, Intent-Based Networking (IBN) was essentially a commercial option for ergonomics and time saving in network management solutions. Early implementations focused on intentions to automate connectivity control via Software Defined networks (SDN) controllers and infrastructure configuration intentions such as Open Stack's open-source Group-Based Policy (GBP) solution. In standards bodies such as the Internet Engineering Task Force (IETF), IBN was equated with configuration management automation, as in the ANIMA group [486]. Early industry solutions include Cisco Digital Network Architecture, Nokia Altiplano, Apstra Operating System, Huawei Network Model, which focus on connectivity configuration and tracking intentions.

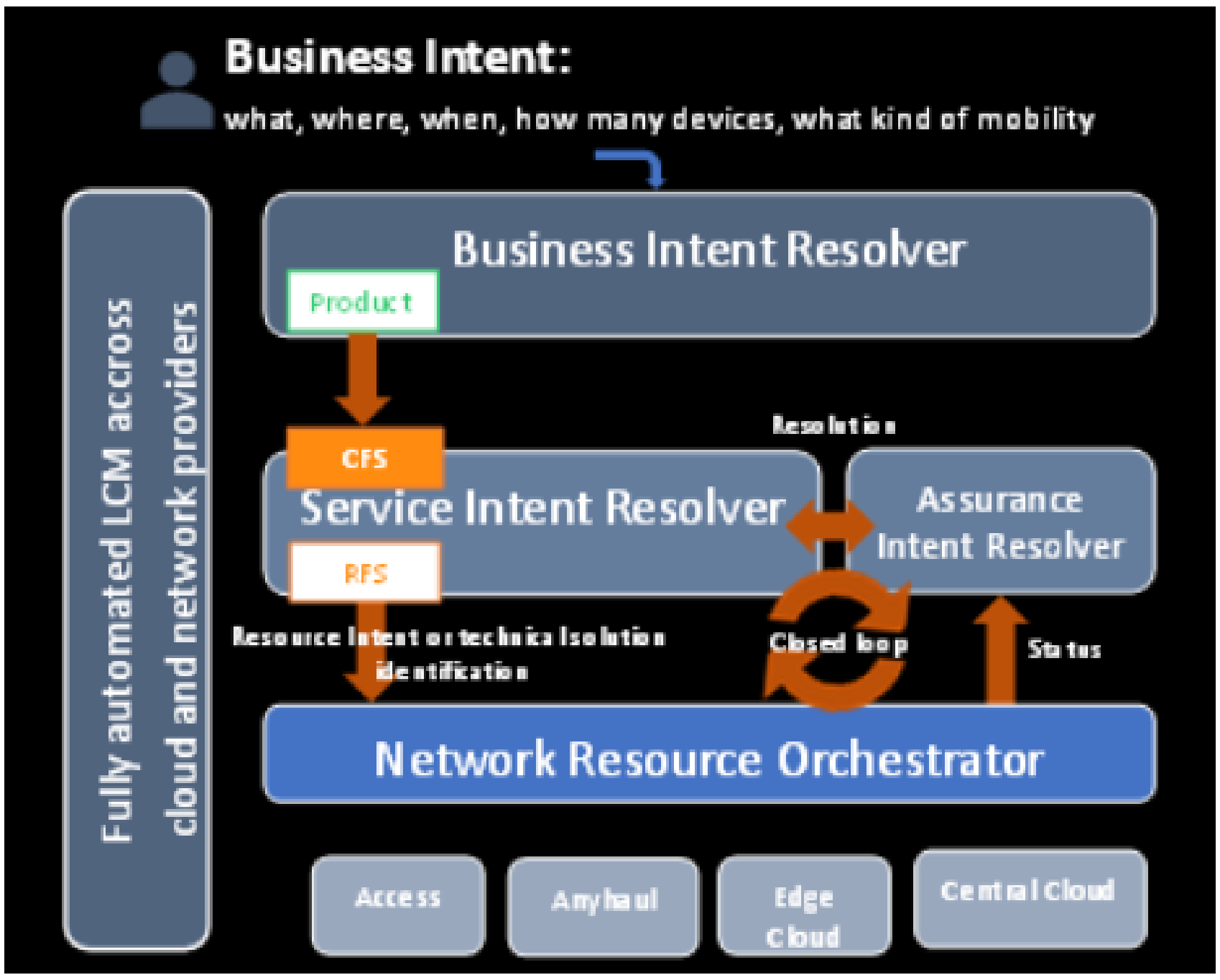

Figure 65: Intent based management Framework TMForum. [487]

Intent-based management is a key component of network automation. It focuses on defining desired outcomes or intents rather than specifying how to achieve them. This approach allows networks to automatically configure, manage, and optimize themselves based on high-level business objectives. By translating business intents into network policies, intent-based management enhances agility, reduces complexity, and improves overall network performance, making it an integral part of modern network automation strategies. From a very high-level point-of-view, the choice has been made to implement business intent handling through a hierarchy of decision engines from order intent to lower intent levels and service/slice deployment action (fulfillment) as an architecture in three layers illustrated in Figure 65. Two intent resolvers are defined for intent fulfillment as it will be further described below:

- Business intent resolver : from client requirements to products: A Business Intent Resolver transforms a business request from simple service questions into a Product (or a package of Products).
- Service intent resolver from Customer Facing Service (CFS) to Resource Facing Service (RFS): The Product can be defined in the catalog and configured to adress customer's needs. When the customer validates his order, the system identifies the CFS at the origin of the product (the product being a restriction of a CFS). A Service Intent Resolver (for simplicity frequently named "Service Resolver" in the remainder of the project's documentation) identifies the technical solution relevant for the context . This technical Solution (Resource Facing Service RFS in TMF) is the way the Know How (CFS) will be delivered.

The application of LLMs for intent translation involves leveraging their natural language processing capabilities to interpret and convert high-level business intents (e.g., "I want 5G connectivity for the 360 video service" or "I need a 5G coverage to track my company bikes in Paris") into actionable network configurations. The application of LLMs for intent translation plays a crucial role in modern network management by enabling the conversion of high-level business intents into configurations and actions. LLMs allow clients to express their requirements in simple human natural language, which can then be parsed and interpreted accurately. Once the intent is understood, LLMs can generate the necessary configuration policies or scripts. Additionally, these models can consider real time network data to ensure that the generated configurations align with current network conditions. They also enhance reliability by detecting potential issues or conflicts in the configurations. Furthermore, LLMs can continuously learn from user feedback and adjustments, improving their accuracy over time. This integration of LLMs into intent translation significantly enhances the agility and efficiency of network automation operations.

#### 12.1.6 LLMs for user friendly interfaces

User-friendly interfaces are crucial for enhancing user experience and engagement. They simplify interactions, making it easier for users to navigate and accomplish their tasks. Mobile Network Operators do not neglect the usage of user-friendly interfaces such as chatbots in order to reduce the waiting time for clients and to automate their service ordering process. The application of chatbots in telecommunication sector enable for operators to enhance customer experience, reduce support costs, collect users feedback's as insights for future improvements and services.

Chatbots are increasingly important for clients as they enhance customer service by providing instant, 24/7 support, which improves user experience and satisfaction. They can handle a wide range of inquiries, from answering frequently asked questions to assisting with transactions, thereby reducing wait times and freeing human agents to focus on more complex issues. Additionally, chatbots can gather valuable data on customer preferences and behaviors, enabling businesses to tailor their services and marketing strategies effectively. This not only boosts operational efficiency but also fosters stronger customer relationships, ultimately driving loyalty and revenue growth. Collecting the customers' expectations is the first step to end-to-end vertical network automation. Having an accurate representation of these expectations with a certain level of abstraction to hide the network complexity as well as a user-friendly interface to communicate them to the management system are the main challenges to face in order to make 5G/6G an automatable and affordable technology.

Before the emergence of LLMs, chatbots primarily relied on rule-based systems and simpler machine learning techniques. For instance, the first chatbot was developed in the mid-1960s by Joseph Weizenbaum at MIT called ELIZA [488]. ELIZA simulated conversation by using pattern matching and substitution techniques to mimic a psychotherapist's responses. It could engage users in dialogue by recognizing keywords and phrases, creating the illusion of understanding. For these first chatbots, Rule-based systems operated on predefined scripts and rules, where specific keywords or phrases triggered predetermined responses. While this approach was effective for straightforward queries, it often struggled with complex or ambiguous language, leading to limited user engagement and satisfaction.

Some chatbots incorporated basic machine learning techniques, such as decision trees or simple classifiers, to enhance response accuracy. However, these models lacked the depth and contextual understanding that LLMs provide. Overall, the earlier methods were less capable of handling the complexity and variability of human language, making them less effective compared to the sophisticated capabilities of modern LLMs. Nowadays, thanks to LLMs, a new modern era emerged for chatbots and operators are considering again the usage of chatbots to increase clients' satisfaction and reduce support cost. LLMs have significantly transformed chatbots by enhancing their ability to understand and generate human-like responses. With improved comprehension of context and nuances in language, LLMs enable chatbots to engage in more natural and coherent conversations. They draw from vast datasets, allowing them to provide relevant information across a wide range of topics.

Additionally, LLMs can be fine-tuned for specific industries, making them more effective in specialized applications. Their capability to maintain context over multiple exchanges facilitates complex, multi-turn dialogues, resulting in a more engaging user experience. Overall, LLMs have elevated chatbots from basic query-response systems to sophisticated conversational agents, greatly improving their utility and effectiveness. Nowadays, we have multiple famous and efficient example of chatbots such as ChatGPT from OpenAI and Gemini from Google. Moreover, for the telecommunication sector a number of chatbots was proposed such as TelecomGPT and standardGPT that we will explain in the next Section 12.1.7. In addition, authors of [382], for example propose the usage of a chatbot with LLMs to automate the clients' 5G services ordering. This use case was described in more details in Section 8.

### 12.1.7 StandardGPT and TelecomGPT

Most of LLMs was trained on diverse and extensive datasets that include books, articles, websites, and other text sources, allowing them to learn a wide range of topics. In addition, they utilize the transformers architecture, which enable them to capture complex patterns and relationships in text. This is thanks to the attention mechanism that allows the model to weigh the importance of different words in a sentence relative to each other, enabling it to understand context and relationships between words regardless of their position [3]. Several versions of GPT have been proposed or adapted for specific use cases, enhancing their effectiveness in targeted applications. One notable example is *Codex*, a variant of GPT-3 fine-tuned for programming tasks, which can understand and generate code in multiple programming languages, making it particularly useful for software development and coding assistance. In this Section, we discuss about two examples of finetuning ChatGPT: StandardGPT and TelecomGPT.

One of the first use cases raised by operators is to customize LLM models such as ChatGPT to include telecommunication knowledge and standards specifications. The authors of [489] proposed to fine-tune Llama2-7B, Mistral-7B, and Llama3-8B in order to include telecommunication knowledge, but also math calculation capabilities in the telecommunication sector. One example of the prompt presented in the paper is to ask the model to write a Python function to convert an IPv6 address from string format to an integer, or to develop a C function that updates the decrypt status flag based on the decryption result for a received 802.11 frame.

OpenAI also proposed a customized ChatGPT called StandardGPT accessible at this link: [490]. StandardGPT,

like other customized versions of ChatGPT, was developed by OpenAI, the organization behind the original ChatGPT models. OpenAI creates general-purpose AI models and offers tools for users or organizations to customize and optimize them for specific use cases. In this instance, the GPT was tailored with custom instructions to suit tasks related to standards, compliance, and technical queries. StandardGPT is designed to address a wide range of standards in various industries and sectors. These include international standards such as ISO and IEC, which cover areas such as quality management (ISO 9001), environmental management (ISO 14001), and functional safety (IEC 61508). It also addresses safety and regulatory standards, such as product safety regulations (IEC 60335, ISO 12100) and occupational health and safety standards (ISO 45001). Industry-specific standards are another focus, with examples like IATF 16949 for automotive quality, AS9100 for aerospace, and ISO 22000 for food safety. Standards for environmental and sustainability management, such as ISO 14001 and LEED certification, are also included. In addition, it addresses telecommunication standards that cover a wide range of areas, including network technologies, wireless communications, and telecommunication infrastructure. It includes ITU standards from the International Telecommunication Union, which set global guidelines for broadband, radio communications, and Internet protocols, as well as IEEE standards like the IEEE 802 series for Ethernet and Wi-Fi. Additionally, it covers 3GPP standards for mobile networks such as 4G LTE and 5G, which are critical for modern mobile communication. ETSI standards from the European Telecommunications Standards Institute for mobile and broadband networks, and IETF standards for internet protocols like TCP/IP, are also within its scope. Furthermore, StandardGPT can address 5G and next-generation network standards, along with network security standards (e.g., ISO/IEC 27033) and compliance with national regulatory frameworks like those from the FCC in the U.S. This broad coverage makes StandardGPT a useful tool for navigating both technical and regulatory aspects of the telecommunications industry.

OpenAI also proposed a more specific model for telecommunication: TelecomGPT [491]. TelecomGPT inludes telecommunication knowledge but it is also capable of doing tasks such as VoIP and analyzing SIP logs, help troubleshootubg voice-over-IP (VoIP) services, analyzing SMPP logs, troubleshooting SMS delivery, and understanding Short Message Peer-to-Peer (SMPP) protocol-related issues. TelecomGPt has also the capacity to fetch voice and SMS wholesale pricing via the TelecomsXChange Market. It could also perform Home Location Register (HLR) lookups to provide real-time information on phone numbers such as status, roaming, and porting data.

## 12.2 GenAI On-Device

Large scale generative AI models are rapidly evolving, driven by advancements in both cloud and on-device processing capabilities. The integration of hybrid AI architectures, which combine cloud and edge computing, is becoming increasingly prevalent. This approach enhances performance, reduces latency, and improves data privacy. Generative AI applications are expanding across various domains, including content creation, digital assistants, and autonomous systems. The trend towards more personalized and context-aware AI solutions is also gaining momentum, as these models become more sophisticated and capable of understanding and generating human-like text, images, and other forms of media.

With an installed base of billions of AI-capable phones, PCs, and other devices in users' hands today, the potential to leverage on-device AI processing for generative AI is already significant and poised to grow steadily in the coming years. A key question is which generative AI models can run on devices with appropriate performance and accuracy. The great news is that highly capable generative AI models are becoming smaller while on-device processing capabilities continue to improve. Fig. 66 illustrates a broad range of generative AI capabilities that can run on devices using models that range from 1 to 10 billion parameters. Models like Stable Diffusion, with over 1 billion parameters, are already running on phones with performance and accuracy levels comparable to their cloud equivalents. Additionally, many other generative AI models with 10 billion parameters or more are expected to run on devices in the near future.

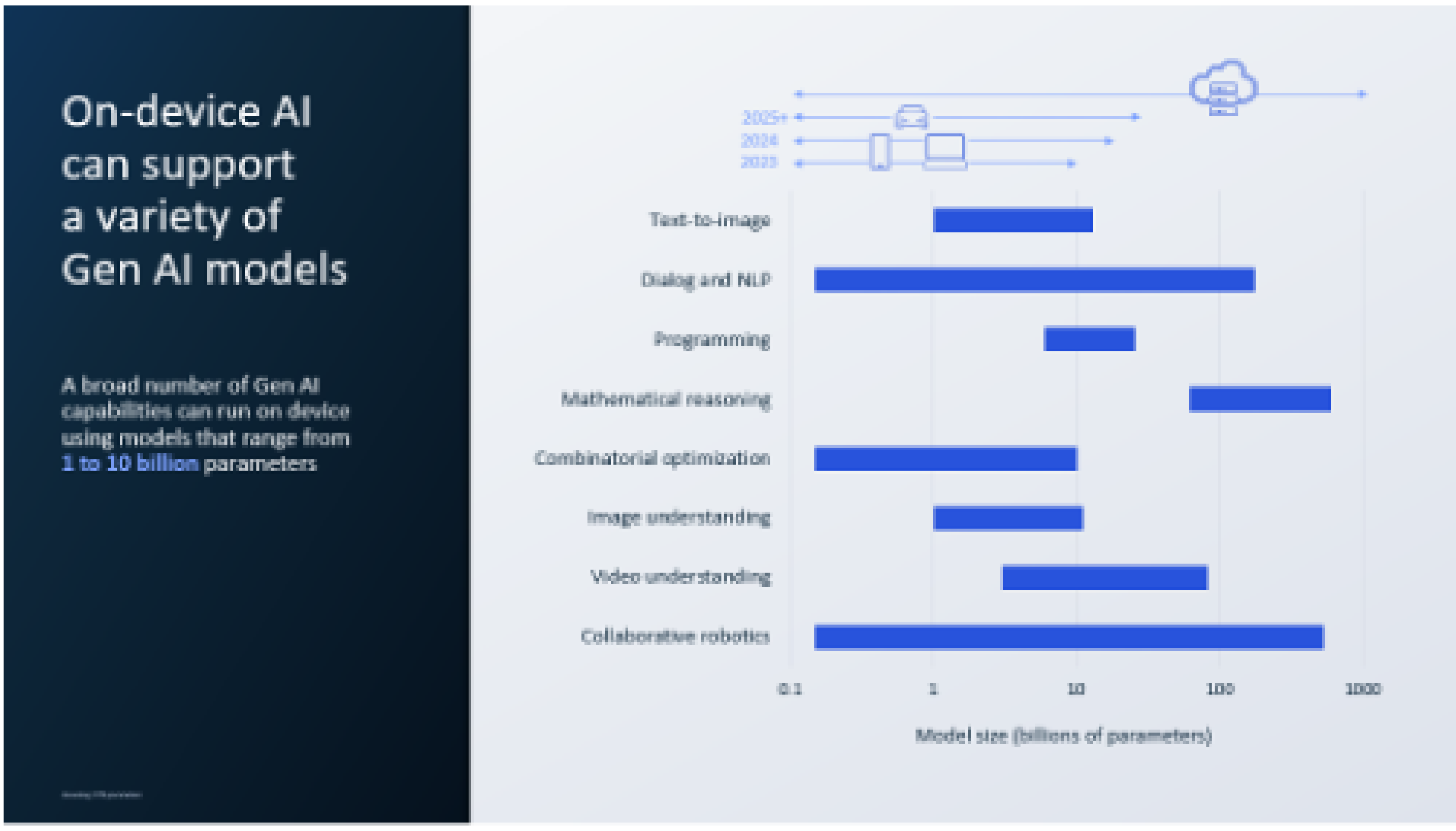

Figure 66: (Source: Qualcomm)

### 12.2.1 Generative AI use cases across device categories

The rise of generative AI with foundation models is driving a new wave of use cases around content generation, search, and productivity across device categories, including smartphone, laptop and PC, automotive, XR, and IoT. The hybrid AI architecture will enable generative AI to provide new and enhanced user experiences across these segments.

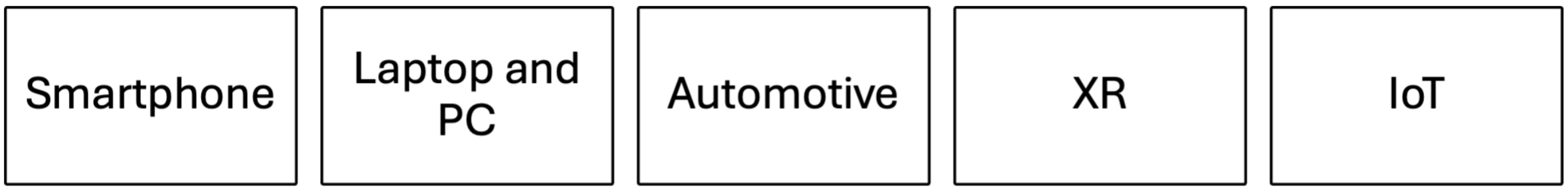

Figure 67: Impactful GenAI use cases across device categories

- In smartphones, generative AI enhances search capabilities and digital assistants, making interactions more intuitive and efficient. With over 10 billion searches conducted daily and mobile devices accounting for more than 60% of these searches, the adoption of generative AI is set to significantly increase the computing capacity required, especially for queries made on smartphones. Users are already transitioning to generative AI-based search because it provides superior answers for many queries.

  The popularity of chat as a search interface is also expected to boost the overall number of queries. As chat technology improves and becomes more capable, smartphones can evolve into true digital assistants. Users will be able to communicate naturally to receive accurate and relevant answers, thanks to the combination of precise on-device personas and large language models (LLMs) that understand text, voice, images, video, and other input modalities. Consequently, models that handle natural language processing, image understanding, video understanding, text generation, and more will be in high demand.

- For laptops and PCs, it boosts productivity tools, enabling more sophisticated content creation and management. Generative AI is revolutionizing productivity by swiftly creating high-quality content from simple prompts. A prime example is Microsoft Office 365 on laptops and PCs. With over 400 million Microsoft Office 365 users globally, integrating generative AI into daily workflows is set to make a substantial impact. Tasks that once took hours or days can now be completed in minutes. Microsoft 365 Copilot leverages the power of large language models (LLMs) combined with user data from the Microsoft Graph and Microsoft 365 apps to transform prompts into powerful productivity tools.

  Office workers can utilize an LLM in the background to read or compose emails in Outlook, draft documents in Word, create presentations in PowerPoint, analyze data in Excel, or collaborate in Teams meetings. Generative AI models, including natural language processing, text-to-text generation, image generation, video generation, and programming, demand significant processing power for these frequently used productivity tasks. Much of this processing can occur on the PC within a device-centric hybrid AI architecture.

- In the automotive sector, generative AI powers advanced digital assistants and supports autonomous driving features, enhancing safety and user experience. Today's AI-driven cockpits, leveraging data from both inside and outside the vehicle, provide highly personalized experiences. Similar to smartphones and PCs, in-vehicle digital assistants keep drivers and passengers seamlessly connected through a hands-free, natural user interface, while also creating new monetization opportunities for the ecosystem.

  These digital assistants can access a user's personal data, such as apps, services, and payment information, as well as sensor data from the vehicle, including cameras, radar, lidar, and cellular vehicle-to-everything (C-V2X). Enterprise APIs enable third-party service providers to integrate their offerings, extending their customer relationships into the vehicle. For example, navigation experiences are greatly enhanced with proactive assistance, offering traffic and weather updates that impact the driver's usual route, recommendations for recharging the vehicle or purchasing a parking permit, and even ordering the user's favorite meal with a simple request.

  The cockpit media experience is also transformed as the vehicle recognizes each occupant and customizes their experience and content, such as music or podcasts. With the rise of in-vehicle augmented reality (AR), digital assistants can tailor displays according to the preferences of the driver or passengers.

  Vehicle maintenance and servicing become more proactive and seamless. By analyzing data such as sensor input, maintenance history, and driving behavior, a digital assistant can predict when maintenance is needed. Using generative AI, the assistant can provide repair information or advise on finding the right service provider, improving vehicle reliability while reducing time and cost.

  Advanced driver assistance systems and autonomous driving (ADAS/AD) solutions often struggle with unusual or unfamiliar objects, especially in poor lighting or challenging weather conditions, leading to unpredictable and sometimes dangerous outcomes. To address this, corner case data must be captured, labeled, and used to retrain models. Generative AI can create simulated corner-case scenarios, predicting the behavior of various road agents like vehicles, pedestrians, cyclists, and motorcyclists. These scenarios help planners decide the drive policy of a vehicle. Both the drive policy stack and perception stack run locally within the vehicle's AI computing capabilities, as strict latency requirements prevent the cloud from playing a role in decision-making for these AI workloads. As ADAS/AD solutions adopt generative AI models with appropriate post-processing, significant energy-efficient AI computing power in vehicles will be essential.

- Extended Reality (XR) benefits from generative AI through the creation of immersive 3D content. Gen-

erative AI holds immense potential for XR, promising to democratize 3D content creation and bring virtual avatars to life. The next generation of AI rendering tools will allow content creators to use various prompts—such as text, speech, images, or video—to generate 3D objects, scenes, and even entire virtual worlds. Additionally, text-to-text large language models (LLMs) will enable the creation of human-like conversations for avatars that are fully voiced and emotive. These advancements will revolutionize how we create and experience immersive content on XR devices.

While the promise of generative AI for XR is undoubtedly exciting, predicting widespread adoption is challenging. However, given the rapid pace of innovation, significant progress is expected in the coming years.

For immersive worlds, text-to-image models like Stable Diffusion will soon enable content creators to generate realistic textures for 3D objects. These capabilities are anticipated to be available on smartphones, and by extension XR devices, within a year. Deployment in XR will require "distributed processing," where the headset handles the perception and rendering stack, while the paired smartphone or cloud runs the generative AI model. In a couple of years, the first text-to-3D and image-to-3D models will likely reach the edge, generating high-quality point clouds of 3D objects. A few years later, these models will improve further, generating high-quality 3D textured objects from scratch. In about a decade, further advancements will enable the generation of entire 3D rooms and scenes from text or images. Eventually, text-to-3D and video-to-3D models might allow us to step into 3D virtual worlds created from scratch, limited only by the user's imagination.

Virtual avatars will follow a similar progression. Text-to-text models, such as the 13 billion parameter LLaMA, will be available on edge devices, generating natural and intuitive conversations for avatars. Text-to-image models will create new textures and outfits for these avatars. In the following years, image-to-3D and encoder/decoder models will generate head and full-body avatars for telecommunication. Eventually, voice prompts, images, or video will be used to create photorealistic, fully animated, intelligent, and mass-producible virtual avatars.

- The Internet of Things (IoT) sees improvements in operational efficiency and customer support through smarter, more responsive systems. AI is already widely used across various IoT verticals, including retail, security, energy and utilities, supply chain, and asset management. It enhances decision-making by analyzing data in near real-time, optimizes operational efficiency, and fosters innovation for competitive differentiation. IoT segments can further benefit from generative AI.

  For instance, in retail, generative AI can enhance both customer and employee experiences. A grocery shopping agent at an onsite kiosk or smart cart can create menus with recipes based on weekly sales specials, budget constraints, and family preferences. Store managers can anticipate off-cycle sales opportunities by preparing for upcoming events. For example, if a sports team is coming to town, a store manager can use generative AI to identify popular branded items and adjust inventory accordingly. Generative AI can also help create new store layouts based on best practices and successful results from other stores in similar communities. It can assist store managers in reorganizing shelves to expand space for the most profitable brands or minimize the prominence of out-of-stock items using data from nearby chain stores.

  In the energy and utilities sector, generative AI can help operation teams create corner-case load scenarios and predict electricity demand along with potential grid failures under unusual circumstances, such as a hot summer with strong winds and localized fires in rural areas. This helps manage resources better and avoid outages. Generative AI can also improve customer service by answering questions about outages or billing.

The hybrid AI architecture, combining on-device and cloud-based processing, is key to unlocking these innovative applications, providing enhanced performance, privacy, and personalization.

### 12.2.2 Benefit of GenAI on device

Hybrid AI architecture, which combines on-device and cloud processing, offers significant benefits in cost, energy efficiency, performance, privacy, security, and personalization. By shifting some processing to edge devices, it reduces the strain on cloud infrastructure and lowers costs, especially as generative AI models grow in complexity. Edge devices also consume less energy and provide reliable performance with lower latency, even during high demand. On-device AI enhances privacy and security by keeping data local, which is crucial for both consumer and enterprise applications. Additionally, hybrid AI enables highly personalized experiences by continuously learning and adapting to user behaviors and preferences, making it a powerful tool for both individual and organizational use [492, 493].

## 12.3 Market drivers and barriers

The introduction of LLMs created a number of applications whether it is for new use cases or to replace traditional approaches in the network management. However, mobile network operators are considering the new revenue possible from the application of LLMs such as the application of LLMs on devices. One way to increase revenue by using LLM, is to reduce labor costs by using finetuned LLMs and enhance customer interactions by using user-friendly chatbots for clients service ordering and deployment.

In this essence, automation is the key for applying LLM in networks. LLMs are also considered in the design of 6G networks. For instance, authors of [209] proposes a split learning system for LLM agents in 6G networks, emphasizing collaboration between mobile devices and edge servers. In this system, multiple LLMs with distinct roles are distributed across these devices and servers to collaboratively perform user-agent interactive tasks. The LLM agents are divided into three modules: perception, grounding, and alignment, which facilitate communication between modules to address extended user requirements related to 6G functions, such as integrated sensing and communication, digital twins, and task-oriented communications. Additionally, the article introduces a novel model caching algorithm for LLMs to enhance model utilization in context, thereby reducing the network costs associated with the collaboration of mobile and edge LLM agents. However, the application of LLM has a number of challenges and limits that will be addressed in more details in Chapter 13. In this chapter, we introduce some of this limits:

- Generalizability: A significant challenge in deploying LLMs for anomaly detection is their generalizability across different domains. Models trained on specific datasets may struggle to perform effectively in new contexts. Strategies such as domain adaptation, multi-task learning, and meta-learning can enhance their adaptability.

- Hallucination and Robustness: LLMs are prone to generating false or misleading information (also called hallucinations), which can undermine their reliability. Implementing rigorous validation mechanisms and adversarial training can mitigate these risks, ensuring more accurate outputs.

- Knowledge Boundary: The knowledge boundary of LLMs limits their ability to address novel events or trends. Continuous learning, transfer learning, and the integration of external knowledge bases can help extend this boundary, improving predictive capabilities.

- Computational Efficiency: The computational demands of LLMs pose challenges for real-time applications. Techniques such as model optimization, hardware acceleration, and cloud-based solutions can enhance efficiency, making LLMs more accessible for organizations.

- Sustainability: If we consider the sustainability of LLMs, many studies bench-marked the energy consumption and Co2 emission of LLM training and inference. If we consider training large models, the energy consumption is related to multiple factors:

  – *Model Size*: larger models typically require more computational resources.

  – *Training Duration*: the length of time spent training the model affects energy consumption.

  – *Data Center Efficiency*: The energy efficiency of the data centers where the training occurs plays a crucial role.

  – *Energy Source*: the carbon footprint depends on whether the energy used is from renewable sources or fossil fuels. Note that according to world nuclear association the coal Co2 emission is around 820g/kwh compared to 12g/kwh for nuclear.

  – *The type of CPU/GPU used*: the choice of CPU or GPU for training and inference significantly impacts energy consumption. Generally, GPUs consume more energy than CPUs. Therefore, in certain situations, training smaller models, like basic BERT, on CPUs can result in lower energy usage.

  Mobile network operators lack control over the energy consumption associated with pre-trained models, which falls under Scope 3 emissions. This makes it challenging for them to adhere to their sustainability goals. However, they can implement strategies to mitigate this issue, such as optimizing prompts, reducing the size of large models, or utilizing small large models (SLMs) when a simpler model with fewer parameters is adequate.

To conclude, the integration of LLMs into 5G+/6G networks presents a transformative opportunity to improve anomaly detection, predictive maintenance, and overall security. By addressing the challenges of generalizability, hallucination, and computational efficiency, LLMs can significantly improve the resilience and reliability of next-generation communication systems. As the field evolves, continued research and innovation will be essential to fully harness the potential of LLMs to secure the future of connectivity.

# PRACTICAL CHALLENGES, OPPORTUNITIES AND FUTURE ROADMAP

# 13 Practical challenges, Opportunities and Future Roadmap

## 13.1 Practical Challenges

### 13.1.1 LTM Inference Speed

The challenge in reducing inference time for Large Telecom Models (LTMs) is vital to achieving real-time performance, particularly in telecommunications networks where low-latency is crucial for dynamic spectrum management, traffic routing, and quality of service. Strategies to optimize inference speed include leveraging hardware accelerators (e.g., FPGAs or telecom-specific ASICs), distributed computing frameworks, or network function virtualization (NFV) optimized for LTM deployment. The use of pruning and quantization techniques can also help to trim unnecessary computations, but balancing accuracy and speed remains a challenge in live networks. The model inferencing latency and tolerance threshold varies across the stack. For instance in AI-Native RAN architecture a LTM model operating at L1 would require far stringent inference time bounds then a LTM operating in SMO for managing orchestration.

### 13.1.2 LTM Model Size

Reducing the size of LTMs while maintaining sufficient accuracy is essential for real-world deployment, especially in edge or network core environments where resources may be constrained. Techniques such as model distillation, weight sharing, and sparsity-aware optimizations help minimize model size without degrading the model's ability to manage large-scale telecom tasks such as traffic prediction and fault detection. Additionally, telecom-specific adaptations of LTMs are emerging, with slimmed-down architectures designed for specific functions within a network, reducing the overall model footprint.

### 13.1.3 LTM Interoperability

In a telecommunications landscape utilizing extensive AI models, such as LLMs, ensuring seamless communication and interactions among diverse entities is essential [494, 495, 496]. The configuration of these AI models and their interoperability must establish a shared, cohesive framework, allowing models to function as a synchronized network, especially in advanced environments like 6G. This involves adopting a standard approach to model configuration that encompasses all necessary attributes, from versioning to architecture specifics, ensuring adaptability across various systems [497].

For example, data serialization plays a foundational role in maintaining a uniform configuration across heterogeneous network components and AI models [498]. Serialization methods ensure that data structures, including model parameters and configurations, can be easily converted, transmitted, and consistently reconstructed. This is particularly crucial for networks with varied hardware and software ecosystems. Common serialization formats, such as JSON or YAML, are instrumental here—each offering unique benefits and trade-offs in terms of human readability, complexity, and error sensitivity [499]. Furthermore, metadata and training context become equally important here, enabling practical deployment across similar use cases and supporting transfer learning (TL) for tasks with semantic similarity. This alignment of model details and training context ensures consistency and also improves the adaptability of AI models within complex network scenarios.

To facilitate interoperability, it's essential to standardize the configuration details for large-scale AI models. This includes defining the model's architecture, training specifics, and system requirements [90]. For example, critical information like model type, version, architecture (e.g., Transformer details), training data summaries, and hyperparameters (learning rate, batch size) must be specified. Additionally, compliance measures, such as security protocols, encryption standards, and privacy policies, should be documented and standardized, enabling secure interactions across networked components.

In practical terms, for example, when configuring AI models for specialized telecom applications like the Network Data Analytics Function (NWDAF), especially when supported by different LLMs, it becomes essential to ensure seamless interoperability. With each LLM potentially varying in architecture, capabilities, and configuration requirements, a unified approach is necessary to allow them to cooperate efficiently. This entails standardizing data formats, aligning model parameters, and synchronizing operational protocols across the LLMs. Such measures ensure that NWDAF can effectively carry out tasks like network analytics, traffic prediction, and anomaly detection without conflicts or disruptions caused by inconsistencies in model interpretation or data handling.

Sophisticated AI-native frameworks such as the *Interconnect* [500] can provide an organized framework for managing data flows and controlling traffic among AI models across nodes. This component enables nodes to cooperate seamlessly, promoting unified decision-making in scenarios like resource allocation and traffic forecasting. Through shared formats, protocols, and APIs, AI Interconnect facilitates a consistent interaction process—essential for a dynamically evolving network environment, where nodes must adapt and renegotiate in real-time.

### 13.1.4 LTM as optimization methods

There is potential for using LTMs as optimization tools within the telecom domain. LTMs can be employed to optimize network parameters, such as resource allocation, load balancing, or spectrum efficiency. They can even be used in conjunction with reinforcement learning models to fine-tune network configurations in real-time, driving improvements in throughput, latency, and network efficiency. Ongoing research is exploring how LTMs can also assist in optimizing traffic management and energy consumption across cellular and IoT networks.

### 13.1.5 Trustworthy AI and LTMs

Despite the growing capabilities of large AI models, particularly in handling multimodal and contextual data, adapting them to the telecom domain poses unique challenges. Communication networks are dynamic and diverse, with data generated distributedly across various nodes and increasingly relying on edge architectures, necessitating partial distributed training of models. Additionally, the telecom sector's cornucopia of applications and services, and its foundational role in our hyper-connected digital society make trustworthiness in LTMs of cardinal importance. Trustworthiness is a multifaceted concept that encompasses several critical dimensions, including robustness, transparency, controllability, explainability, accountability, fairness, and privacy.

Trustworthy AI refers to systems designed, developed, and deployed with a focus on ethical, safe, and transparent practices. Such systems must be reliable, ensuring consistent performance; fair, avoiding biases that could discriminate against specific individuals or groups; accountable, with explainable and justifiable decisions; and transparent, making processes and decisions understandable to stakeholders.

According to ALTAI (Assessment List for Trustworthy AI), trustworthy AI rests on three key pillars maintained throughout the system's lifecycle: (i) lawfulness: adherence to all relevant laws and regulations; (ii) ethical standards: embodiment of ethical principles, human rights, fairness, and societal well-being, and (iii) robustness: resilience and reliability, both technically and socially, to mitigate unintended harm.

Trustworthiness is crucial for LTMs as they operate at scale, handling sensitive data, global communication, and network optimization across highly interconnected nodes. Ensuring privacy, security, and compliance with regulations like GDPR is paramount. Beyond efficient functioning, LTMs must be transparent and auditable to build confidence among users, businesses, and regulators. Key challenges include mitigation of hallucinations (false outputs), reduction of harmful outcomes, and explainability enhancement to make model decisions understandable. Robust training datasets and ethical fine-tuning are essential for improving fairness and mitigating biases.

However, ensuring trustworthy LTMs in dynamic, resource-constrained telecom environments presents challenges such as bias detection, resilience to adversarial attacks, privacy preservation, and operational transparency. These challenges are amplified by the unique demands of telecom networks, where LTMs are integral to tasks like traffic management, fault prediction, and routing sensitive communications.

Key challenges to ensuring trustworthy AI include bias and fairness, where uncontrolled datasets can reinforce biases, leading to unfair outcomes such as imbalanced resource allocation. Hallucinations remain a concern as LTMs may generate plausible but false outputs, undermining reliability. Lack of transparency reduces trust and accountability due to limited interpretability of model decisions. LTMs are also vulnerable to adversarial attacks, where inputs exploit network or data weaknesses, challenging resilience in dynamic environments. Balancing data privacy and security with operational transparency is critical, especially when handling sensitive user data. Ethical content generation requires continuous safeguards to prevent harmful or unsafe outputs, while sustainability poses concerns due to the high energy demands of training and deploying models. Additionally, ensuring human intent alignment is vital to avoid outputs that are technically correct but ethically inappropriate. Finally, regulatory compliance with evolving frameworks, such as GDPR or the EU AI Act, is essential for maintaining legal and ethical standards.

Addressing these challenges demands interdisciplinary collaboration, integrating AI research, ethical frameworks, and regulatory strategies to develop systems that are both powerful and trustworthy. Trustworthy AI is a collaborative endeavor, requiring alignment among developers, users, policymakers, and society to ensure AI serves humanity effectively, equitably, and evolves harmoniously alongside human needs. The dynamic nature of telecom environments—with fluctuating traffic, emerging threats, and evolving user demands—further complicates this effort, requiring LTMs to adapt seamlessly while maintaining trust and reliability.

### 13.1.6 LTM Design Evolution

The design of LTMs is evolving to meet the changing needs of modern telecom networks. Early LTMs focused on static tasks like fault detection, but newer models are increasingly adaptive, supporting dynamic reconfiguration of networks and real-time decision-making. As telecom networks grow more complex, LTM designs must evolve to address multi-layered challenges, such as hybrid cloud-edge deployments, diverse hardware ecosystems, and fluctuating traffic demands.

### 13.1.7 Energy Efficiency

Running LTMs, especially across large, distributed telecom networks, can demand significant energy resources. As telecom infrastructure scales, energy consumption becomes a major concern. Efficient power usage is critical, especially in edge deployments where energy resources might be limited. Managing this requires low-power models, optimized hardware, and intelligent energy management strategies to ensure sustainability in a 24/7 operational environment.

### 13.1.8 Latency in Distributed Networks

LTMs deployed in a distributed telecom network, particularly over large geographical regions, face challenges with latency. Even with local edge processing, the communication between (edge) nodes and central data centers may introduce delays, affecting the timeliness of decision-making processes. Optimizing LTMs for low-latency operations in such complex, multi-tiered environments is critical, especially for applications such as network optimization and traffic management that demand real-time response.

#### 13.1.9 Security and Privacy

With LTMs operating across the telecom infrastructure, they handle sensitive data such as call logs, messaging patterns, and user activity. This makes them a target for potential security breaches. Ensuring that LTMs are equipped with strong encryption, data protection, and authentication methods is essential to prevent unauthorized access and preserve the privacy of user data.

#### 13.1.10 Scalability Across Heterogeneous Networks

Telecom networks are highly heterogeneous, combining 5G, LTE, fiber optics, and legacy systems like 3G in some regions. Deploying LTMs across such diverse infrastructures poses a significant challenge. LTMs must be adaptable and scalable, capable of working efficiently across different network types and architectures. This challenge becomes more pronounced when the network is scaled to accommodate millions of devices.

#### 13.1.11 Dynamic Network Environments

Telecom networks experience constant fluctuations in traffic patterns, user mobility, and varying service demands. LTMs must be dynamic and able to adapt in real-time to these conditions. Models need to be flexible enough to adjust network configurations based on real-time analysis and evolving scenarios, which adds complexity to their design and deployment. To overcome these challenges, several advanced methods have been developed to address the variability and uncertainty in telecom environments, including: Transfer Learning, Meta-Learning and Reinforcement Learning. Next we will briefly introduce them and highlight how they can be utilized to combat the dynamicity in Telecom networks.

**Transfer Learning** involves training a model on a large dataset (source task) and then adapting it to a different but related task (target task) with, ultimately, a smaller dataset. This allows the model to use the knowledge gained from the source task to perform better on the target task. Due to a dynamic environment, transfer learning can be challenging in a time-evolving target domain. Therefore, continuous transfer learning helps by exposing the model to a wider range of data during the initial training phase, making it more adaptable to new and unseen data patterns. Adapting a pre-trained LLM to a specific domain is typically preferable to training an LLM from scratch on domain-specific data. The trained LLMs can be robust and generalise well to handle different telecom scenarios.

In general, transfer learning can significantly benefit LTMs in several practical ways. On the one hand, training LTMs on massive datasets (data such as network performance, customer behavior, and device usage collected from telecom networks) is computationally expensive and time consuming. On the other hand, telecom data can be sensitive and difficult to obtain in large quantities. Transfer learning can help accelerate model development and reduce training costs while overcoming the lack of data by transferring previously held knowledge (e.g., general features and patterns related to network congestion or user behavior) [154].

The different use case where LLMs can exploit Transfer Learning for can be listed as:

- Network management and monitoring: LLMs can be used to process and interpret network logs, error messages or performance metrics and then generate recommended solutions [333]. In [319], transfer learning and non-task-specific telecom data have been incorporated to improve generalisation capabilities when dealing with unseen fault reports. By transferring knowledge from pre-trained models (on general technical texts or IT-related documents) to telecom-specific problems, LTMs can help predict network failures, optimise configurations or provide real-time recommendations [501]. Fine-tuning these models on network-specific data (e.g. fault logs, SNMP data) helps them understand nuanced issues and make predictions based on historical patterns.

- Fraud Detection and Security: Large-scale fraud detection can be improved by using LTMs for anomaly detection through the use of transfer learning [501]. For example, a pre-trained model may have learnt to detect unusual patterns in general transaction data, and this can be transferred to the telecommunications sector. By fine-tuning on telecom-specific datasets (e.g. unusual call patterns, subscription fraud), LTMs can help identify suspicious activity faster.

- Text Mining for Regulation: Telecom operators need to comply with industry regulations, which often involves scanning and analysing large volumes of documents. Pre-trained LTMs can be fine-tuned to identify and extract relevant information from contracts, policies and legal documents [502]. This process of transfer learning allows models to understand regulatory language more effectively than if they were trained from scratch.

- Customer Support: Due to their strong ability to classify customer comments and extract useful feedback, LTMs can be used together with transfer learning to accelerate sentiment analysis based on customer feedback [503].

**Meta-Learning**, commonly known as "learning to learn", is an emerging paradigm in ML that aims to design models that are capable of quickly adapting to new and unseen tasks with minimal additional training rather than being designed to perform well on just one specific task. Unlike traditional ML approaches, which require a significant amount of retraining when exposed to new data, meta-learning utilizes previous learning experiences to adapt for new but related tasks with minimal training, leveraging common patterns and prior knowledge to generalize across different tasks, therefore, reducing the time and computational resources needed for adaptation. Such feature is critical for enabling LTMs given that wireless networks operate in complex and dynamic environments, where factors such as network topology, user mobility, traffic patterns, and interference can change drastically in short periods and space.

Meta-learning involves generally involves two stages: meta-training and meta-adaptation. The meta-training phase can be characterized as a bi-level optimization problem, where two optimization tasks represent the two "learning" processes of "learning to learn". The inner optimization concentrates on base learning, which employs specified hyper-parameters to generate a policy for a given job . Meanwhile, the outer optimization seeks to learn the hyper-parameters by leveraging data sampled from related tasks. The meta-adaptation step allows the model to effectively adjust to new tasks after meta-training is finished. Using the hyper-parameters learned during meta-training, the model can quickly adjust with a few samples from the same distribution as the test data, making it highly efficient for new task adaptation. One prominent algorithm that follows this two-stage framework is Model-Agnostic Meta-Learning (MAML) [504]. MAML optimizes for an initial set of model parameters that can be rapidly fine-tuned for a new task using just a few gradient updates, with limited task-specific data. During the meta-training phase, MAML learns to create parameters that are highly adaptable, and in the meta-adaptation phase, the model uses these learned parameters to efficiently fine-tune and perform well on new tasks. From practical point of view, Meta-Learning LTMs can be applied but not limited to

- Adaptive traffic management: During events such as sports games or concerts, where user density spikes rapidly near the venue location, meta-learning enables LTMs to adapt swiftly to traffic surges. Having the LTMs pre-trained on diverse traffic patterns, these models can use minimal or even no additional data to optimize the performance, reducing latency and preventing QoS interruptions.

- High-speed rail connectivity: Wireless networks in high-speed rail systems face challenges in maintaining stable connections for passengers moving at over 300 km/h. Meta-learning LTMs, trained on varied mobility patterns, can adapt efficiently to optimize handovers, ensuring seamless connectivity for applications like video conferencing.

- IoT Device Management in Smart Cities: Managing thousands of IoT devices in dynamic environments like smart cities requires efficient task scheduling and resource allocation. Meta-learning LTMs, pre-trained on IoT usage patterns across diverse environments, can adapt to new devices and tasks, ensuring efficient and reliable operations with minimal reconfiguration.

By leveraging meta-learning, LTM can become more intelligent and efficient, significantly reducing the requirement for human interventions and enabling faster, automated responses to dynamic changes in network conditions. As telecom models scale to handle increasingly complex networks with higher densities of connected devices and more demanding applications, meta-learning will emerge as a key enabler, driving the efficient management and optimization of these large-scale systems. Its adaptability will allow telecom operators to meet evolving requirements, maintain network stability, and ensure consistent quality of service in highly variable and resource-intensive environments.

**Reinforcement Learning (RL)** is a machine learning paradigm designed to enable intelligent agents to make sequential decisions by interacting with an environment and learning from feedback. In the domain of telecommunications, RL has found significant applications in LTMs, which manage the complexities of dynamic and variable networks. Telecom systems operate in environments characterized by fluctuating traffic loads, interference, user mobility, and resource constraints, creating a need for adaptive and robust solutions. RL empowers LTMs to address these challenges by learning optimal policies that maximize cumulative rewards, such as enhanced QoS, efficient resource utilization, and energy savings.

The relevance of RL in telecommunications stems from its ability to adapt in real time to changing network conditions. RL enables LTMs to dynamically allocate resources such as bandwidth and spectrum, balance network loads, and manage energy consumption. By continuously learning from feedback, RL-based LTMs can adjust their policies to meet network demands effectively. For instance, RL models can optimize the distribution of traffic across base stations, reducing congestion and improving throughput. Similarly, in energy-intensive telecom infrastructures, RL agents can learn power-saving strategies while maintaining service quality, contributing to sustainable network operations.

The core mechanism of RL relies on three key components: the agent, the environment, and the reward system. In telecom applications, the LTM acts as the agent, making decisions based on the state of the network (the environment). The reward system provides feedback on the effectiveness of these decisions, such as reduced latency or increased network reliability. RL's capability to balance exploration and exploitation is particularly critical in telecom networks. Exploration allows the RL agent to test new actions and discover potentially better policies, while exploitation enables it to apply learned strategies to maximize immediate rewards. This balance is essential for managing the trade-offs inherent in dynamic and resource-constrained environments.

Advanced RL techniques such as Deep Reinforcement Learning (DRL) [505] and Multi-Agent Reinforcement Learning (MARL) [506] are particularly suited for telecom applications. DRL integrates RL with deep neural networks, enabling LTMs to handle the high-dimensional state-action spaces typical of telecom networks. For example, DRL can optimize 5G network slicing by efficiently allocating resources to meet the diverse requirements of enhanced mobile broadband (eMBB) and ultra-reliable low-latency communications (URLLC). MARL, on the other hand, involves multiple RL agents operating in a shared environment, making it ideal for distributed systems such as dynamic spectrum sharing and multi-access edge computing.

Despite its advantages, implementing RL in LTMs presents challenges, including scalability, computational demands, and latency sensitivity. Telecom networks involve billions of devices and interconnected systems, requiring RL models to operate efficiently at scale. Real-time decision-making is also critical, as latency delays can compromise service quality. Moreover, training RL models demands extensive data and computational resources, which can be addressed through decentralized approaches like federated learning and edge computing. Ensuring interpretability of RL decisions remains another hurdle, especially in mission-critical applications where transparency and compliance are essential.

Looking ahead, RL is poised to play a pivotal role in the evolution of telecom networks. Hybrid approaches that combine RL with techniques like transfer learning and meta-learning promise greater adaptability and efficiency. Distributed RL and energy-aware algorithms will further address scalability and sustainability challenges. As telecom networks transition to autonomous and AI-driven systems, RL will become a cornerstone of intelligent automation, enabling self-optimizing, self-healing, and self-configuring capabilities. By leveraging RL, LTMs can ensure efficient, resilient, and future-ready telecom networks that meet the growing demands of an interconnected world.

Overall, future LTMs must be designed with agility and flexibility in mind to accommodate varying network conditions and diverse use cases. This design principle is crucial for ensuring efficient network operations. To achieve these goals:

- Adoption of Classical ML/AI Approaches: LTMs should leverage both supervised and unsupervised learning methods for a wide range of dynamic scenarios [507, 508]. Such methods can be especially effective in use cases like dynamic traffic steering [509] and service-aware resource allocation [510, 511].

- Importance of Reinforcement Learning (RL): As highlighted, RL-based techniques are expected to be vital in expanding LTM capabilities. By learning from continuous interactions with the environment, RL can enable more nuanced decision-making and adapt effectively to real-time conditions.

- Value of Data and Testbed Approaches: Data used to train LTMs in these dynamic environments will become increasingly valuable. Testbed deployments [512] are seen as critical enablers for generating diverse and representative datasets [513, 514], thereby enhancing the training and validation of LTMs under realistic network conditions.

By integrating these considerations, LTMs will be better equipped to manage future networks' complexity, delivering robust and intelligent network operations.

#### 13.1.12 Data Governance and Compliance

Given the global reach of telecom operators, LTMs must comply with various regional regulations on data protection and usage, such as GDPR in Europe or HIPAA for healthcare data in the U.S. Ensuring that these models respect regional compliance while maintaining operational efficiency adds another layer of complexity to their deployment and design.

#### 13.1.13 LTM pre-training of a physical-layer foundation model

In contrast to task-oriented AI solutions, foundation models can help Telecom operators to easily adapt to new situations and applications. To realize a generalized understanding of the physical layer using large Telecom models, selecting effective tokenization and pre-training strategies such as next-sample prediction, masking and denoising is essential. However, this is challenging in wireless networks due to the drastic changes in data representation depending on the use case (e.g., the sampling rate or length can vary based on spectrum manager configuration and considered technologies). Additionally, the entropy of different wireless technologies (e.g., LTE, 5G-NR and WiFi) is generally different because of the differences in modulation scheme, channel bandwidth usage, traffic pattern, and load and signal processing techniques. Building foundation models, typically based on the transformer architecture, involves techniques such as patching and tokenization to ensure a predictable input format. Under the dynamic conditions of telecom use cases, it remains an open challenge how to effectively adapt these techniques due to differences in entropy and data representation. There exist works on investigating tokenization strategies for time series data [515, 516] but not for wireless time series data. Additionally, collecting large datasets for new tasks or use cases is often challenging and demands significant resources. Pre-training foundation models in a self-supervised manner aims to alleviate this by requiring only minor model updates during fine-tuning for new use cases with high variety. Nevertheless, designing effective pre-training strategies for multiple wireless downstream tasks with different data representations remains an open research question.

#### 13.1.14 LTM Hardware requirement and evolution

The hardware requirements for LTMs have evolved with advancements in network infrastructure, particularly with the shift toward 5G and beyond. LTMs typically require high computational power, bandwidth, and low-latency hardware, including telecom-specific processors and accelerators. As telecom hardware evolves, models need to be optimised for newer architectures such as edge computing nodes, NFV infrastructure, and cloud-native networks. Balancing resource usage and performance is key, especially when deploying LTMs in environments with varying levels of connectivity and hardware capacity.The hardware requirements for LTMs include processing units, memory and storage, networking and connectivity, and energy efficiency. The following sections provide an in-depth analysis of the architectures of CPUs, GPUs, TPUs, and NPUs, which play a crucial role in supporting LTMs and other AI models.

**CPUs (Central Processing Units)** are general-purpose processors found in virtually every computing system and primarily handle tasks such as data preprocessing, system coordination, and control flow management in AI systems, CPUs typically consist of multiple cores (ranging from 4 to 96 or more in high-end server models) and employ complex instruction sets like x86 (used by Intel and AMD) or ARM (found in Apple's M-series and mobile processors). These instruction sets enable CPUs to execute a wide range of tasks efficiently, although not optimised for large-scale parallel processing, which is crucial for deep learning models. CPUs are designed to handle SIMD (Single Instruction, Multiple Data) operations [517], where a single instruction is applied across multiple data points. For example, Intel's AVX-512 extension accelerates floating-point and integer operations on multiple data points simultaneously, improving the speed of AI tasks such as data preprocessing and feature extraction. Despite these enhancements, CPUs still lack the parallelism of GPUs and TPUs, which limits performance in training large AI models.Notable CPU products for AI workloads include Intel Xeon Scalable Processors (Ice Lake and Sapphire Rapids), featuring high core counts (up to 60 cores) and support for AVX-512 instructions. AMD's EPYC processors (Milan and Genoa) offer up to 96 cores for high-performance multi-threaded tasks. These processors are commonly used in server environments for data preprocessing, model orchestration, and light inference tasks. For edge AI applications, ARM-based processors like Apple's M1 and M2 series, which integrate CPU, GPU, and NPU components, leverage the ARM architecture to handle machine learning tasks efficiently. In contrast, ARM Cortex processors specifically serve as CPU cores but are often part of larger SoCs that include a separate GPU and NPU for AI tasks in mobile devices and edge computing environments.

**GPUs (Graphics Processing Units)** are specialised processors optimised for parallel computation, making them critical for accelerating deep learning tasks. Unlike CPUs, which have a limited number of cores, GPUs can contain thousands of smaller cores designed to handle large-scale matrix multiplications and tensor operations, which are essential for the training and inference of AI models. The architecture of GPUs relies on SIMT (Single Instruction, Multiple Threads), which allows for parallel execution of the same instruction across multiple threads. This architecture is particularly well-suited for AI tasks such as matrix multiplications, convolution operations in neural networks, and large-scale data processing. Modern GPUs are equipped with VRAM (Video RAM), a high-speed memory pool that provides fast access to large datasets and model weights, reducing the bottlenecks associated with slow memory access. For deep learning, GPUs often include specialised hardware units such as Tensor Cores (available in NVIDIA's GPUs), which accelerate deep learning computations, particularly matrix operations, by performing high-efficiency, low-precision calculations (such as FP16 or BF16). GPUs are useful for training large language models (LLMs) or running inference on complex tasks such as image classification, natural language processing (NLP), and reinforcement learning. Leading products in this space include the NVIDIA A100 Tensor Core GPU, built on the Ampere architecture, offering up to 624 teraflops of FP16 performance and equipped with 80GB of HBM2e memory. This GPU is widely used for large-scale training and inference in data centres. Based on the Hopper architecture, the newer NVIDIA H100 Tensor Core GPU provides further performance improvements, particularly in AI-specific tasks. AMD's Instinct MI200, based on the CDNA2 architecture, is another option for AI and high-performance computing (HPC) workloads, offering 128GB of HBM2e memory and over 200 teraflops of FP16 performance. For smaller research projects or individual researchers, NVIDIA's RTX 3090, a consumer-level GPU from the Ampere architecture, is a popular choice, offering 24GB of GDDR6X memory and robust performance for training models with tens of billions of parameters. MALI GPUs are widely used in mobile and embedded systems, designed by ARM. These GPUs, such as the Mali-G78 and Mali-G710, are optimised for AI inference on devices with constrained resources. Supporting OpenCL and Vulkan, they are commonly found in smartphones and IoT devices. MALI GPUs handle AI tasks such as object detection and voice recognition while maintaining high energy efficiency.

**TPUs (Tensor Processing Units)**, developed by Google, are highly specialised processors designed to accelerate tensor operations that are central to AI workloads. Unlike GPUs, which are general-purpose parallel processors, TPUs are specifically designed for matrix multiplications and tensor operations used in deep learning models, particularly within Google's TensorFlow framework. TPUs utilise systolic arrays, specialised hardware units that efficiently handle matrix multiplications by minimising data movement between processing units. This architecture makes TPUs highly efficient for large-scale AI training, allowing them to outperform GPUs in deep learning workloads that rely heavily on matrix calculations. TPUs also feature High Bandwidth Memory (HBM), which is critical for handling large datasets and speeding up the training of massive neural networks. TPUs can be scaled through TPU Pods, which combine multiple TPU chips to enable distributed training across hundreds or even thousands of TPU cores. This makes them ideal for training large language models or multi-modal systems that require immense computational power. Examples of TPUs include the Google TPU v4, which delivers up to 275 teraflops per chip and is designed for maximum energy efficiency and performance-per-watt in AI training tasks. The Google TPU v3 is an earlier generation that offers 123 teraflops of AI performance. Both generations of TPUs are available through Google Cloud, where TPU resources can be rented for AI workloads, providing scalability without needing physical hardware investment.

**NPUs (Neural Processing Units)** are processors specifically designed to accelerate neural network inference, particularly on edge and mobile devices with critical power consumption and efficiency. Unlike GPUs or TPUs, which are designed for both training and inference, NPUs focus on running inference for pre-trained models, often emphasising low-latency, energy-efficient computation. NPUs are typically integrated into System on Chips (SoCs) alongside CPUs and GPUs, allowing mobile and embedded devices to handle AI workloads locally without offloading computation to cloud servers. These processors are optimised for tasks such as convolutional neural networks (CNNs), recurrent neural networks (RNNs), and other neural network architectures that are commonly used in real-time applications like object detection, speech recognition, and language translation. Key features of NPUs include support for quantised models, which reduce the precision of the neural network weights to lower bit-widths (e.g., INT8) without significantly impacting accuracy. This enables NPUs to perform inference faster and with lower power consumption than GPUs or CPUs. NPUs also feature on-chip SRAM (up to 64MB in some models) to minimise latency in accessing data during inference. Leading examples of NPUs include the ARM Ethos series, which provides scalable AI performance for edge devices. The Ethos-U55 is designed for microcontrollers and low-power applications, making it ideal for IoT sensors and wearables. On the other hand, the Ethos-N78 targets higher-end devices like smartphones and delivers up to 2 TOPS, providing efficient solutions for image recognition, speech processing, and other AI tasks. The Ethos NPUs offer high energy efficiency and are optimised for int8 and int16 operations, enhancing their suitability for edge AI applications. Other examples include the Apple Neural Engine (ANE), which is integrated into Apple's M1 and M2 SoCs. The ANE can perform up to 11 trillion operations per second (TOPS), making it ideal for on-device AI applications such as facial recognition and natural language processing. Google's Edge TPU, designed for TensorFlow Lite models, delivers 4 TOPS of AI performance for edge and IoT applications. Huawei's Ascend 910, a high-performance NPU, delivers up to 256 teraflops of FP16 performance, making it suitable for more demanding AI applications.

The architectural distinctions among CPUs, GPUs, TPUs, and NPUs directly influence their efficacy in AI applications. CPUs offer flexibility and are essential for managing diverse computational tasks, while GPUs, TPUs, and NPUs provide the parallelism and specialised processing required for efficient AI model training and inference. The synergy between these hardware components underpins LTM systems' scalability and performance advancements.

**Memory and storage** infrastructure are critical for managing the vast datasets and complex models utilised in LSTM. High-bandwidth memory (HBM) ensures rapid data access and processing speeds. For example, NVIDIA's H100 GPU integrates HBM2e, significantly improving memory bandwidth, which is crucial for training large-scale models. Storage solutions must balance capacity, speed, and reliability. Solid-State Drives (SSDs) offer faster data retrieval compared to traditional Hard Disk Drives (HDDs), thereby reducing bottlenecks in data-intensive AI workflows. Products such as Samsung's PM1733 NVMe SSDs provide high throughput and low latency, facilitating efficient data handling for AI applications. Additionally, advancements in non-volatile memory technologies, such as Intel's Optane DC Persistent Memory, offer a hybrid approach that combines the speed of DRAM with the persistence of traditional storage, enhancing the overall performance of AI systems.

**Networking and Connectivity** are essential for distributed AI systems, facilitating efficient data transfer and communication between computational nodes. High-speed interconnect technologies, such as NVIDIA's NVLink and Mellanox's InfiniBand, provide the necessary bandwidth and low latency critical for large-scale AI training and inference tasks. Data centres that support technologies such as LTM frequently employ advanced networking solutions to enable parallel processing and distributed training across multiple GPUs or TPUs. For example, Google's TensorFlow Research Cloud utilises high-speed networking to interconnect thousands of TPU cores, allowing the training of extensive AI models with remarkable efficiency. Furthermore, advancements in 5G and future wireless technologies are set to enhance connectivity for edge AI applications, facilitating real-time data processing and inference in decentralised environments, thereby expanding the capabilities and reach of AI-driven solutions.

**Energy efficiency** is a critical consideration in the design and deployment of AI hardware, given the substantial power consumption associated with training and operating large models. Energy-efficient processing units, such as NVIDIA's Ampere architecture GPUs, incorporate features like dynamic voltage and frequency scaling (DVFS) to optimise power usage without compromising performance. Data centres are increasingly adopting energy-efficient cooling solutions and renewable energy sources to mitigate the environmental impact of AI workloads. For exam-ple, Google's data centres utilise machine learning algorithms to manage cooling systems dynamically, reducing energy consumption while maintaining optimal operational conditions.

The **Evolution of Hardware** requirements has progressed significantly, driven by the increasing complexity and scale of AI models. Initially, AI systems relied primarily on CPUs, but the advent of deep learning necessitated more specialised processing units capable of handling parallel computations. This evolution led to the emergence of GPUs, followed by TPUs and NPUs, each offering enhanced performance tailored to specific AI workloads. As AI models have grown in size and complexity, exemplified by large-scale models like OpenAI's GPT series and Meta's LLaMA, memory and storage demands have escalated. Training and deploying these models require vast amounts of high-bandwidth memory and advanced storage technologies to manage and process the extensive datasets involved. For instance, GPT-4's intricate architecture demands significant computational resources, driving innovations in GPU and TPU technologies to facilitate efficient training processes. Concurrent with these advancements, the need for robust networking solutions has intensified to support the distributed nature of modern AI training processes. Training models such as GPT-4 or LLaMA involve parallel processing across multiple GPUs or TPUs, relying on high-speed interconnects like NVIDIA's NVLink and Mellanox's InfiniBand to ensure efficient data transfer and synchronisation between computational nodes. This distributed approach not only accelerates training times but also enhances the scalability of AI systems. Energy efficiency has also become a paramount concern, spurring the development of hardware architectures and data centre designs that minimise power consumption while sustaining high computational throughput. Innovations such as energy-efficient GPUs and optimised cooling solutions are essential to support the intensive demands of training large AI models like GPT and LLaMA. The continuous evolution of these hardware components reflects the dynamic interplay between AI advancements and the technological innovations required to support them.

## 13.2 Innovative Opportunities

### 13.2.1 Trustworthy AI and LTMs

Achieving real-time inference and maintaining a small model footprint are critical for deploying LTMs in telecom settings, but these constraints often hinder trust-enhancing features like robust error handling and confidence metrics. Despite these challenges, prediction confidence metrics offer a valuable innovative opportunity by quantifying the reliability of LTM outputs. High-confidence predictions can trigger automated actions, such as fault recovery or traffic rerouting, ensuring faster and more efficient responses. Conversely, low-confidence predictions can prompt further analysis or human intervention, reducing errors and enhancing transparency by adapting to varying levels of uncertainty. This integration enables dynamic, informed decision-making and strengthens trust in LTMs.

Confidence metrics are crucial for detecting anomalies and ensuring system resilience in dynamic environments [518]. A sudden drop in confidence across outputs can signal data drift, model degradation, or adversarial interference, enabling early corrective actions like retraining or threshold recalibration to maintain robust telecom operations. Additionally, confidence metrics balance privacy and transparency: for sensitive user data, high-confidence predictions can allow detailed explanations, while low-confidence predictions limit disclosures to abstract summaries. This approach reduces the risk of exposing sensitive information while preserving interpretability and operational reliability.

Confidence metrics are essential in emerging telecom technologies like Integrated Sensing and Communication (ISAC), where precise coordination between communication and sensing tasks must occur under limited data and high uncertainty. They act as safeguards, ensuring LTM decisions remain reliable and interpretable in complex, overlapping scenarios. For tasks like spectrum allocation or interference management, confidence scores enhance trust in system recommendations, enabling operators to respond effectively to rapidly changing conditions.

To enhance trustworthiness in LTMs, confidence metrics must be embedded as a core architectural component. This requires balancing inference speed, model size, and interoperability with telecom trust demands. Lightweight confidence estimation methods are essential to meet latency requirements, while privacy-preserving techniques like differential privacy ensure explanations don't expose sensitive data. Additionally, confidence metrics improve cross-network synchronization in cascading scenarios, enabling seamless, coordinated responses to threats or anomalies [519].

Embedding confidence-driven strategies into LTM design and deployment enables telecom networks to tackle practical challenges effectively. These approaches enhance resilience, security, and transparency, allowing LTMs to meet the stringent demands of next-generation telecom environments while fostering stakeholder trust.

Conformal prediction is another potential innovation for enhancing trustworthy AI in large generative models. By providing mathematically rigorous uncertainty quantification and calibrated confidence intervals, it ensures the reliability, transparency, and robustness of model outputs across complex, high-dimensional modalities such as text and images. This enables generative models to deliver outputs with statistically valid guarantees—such as confidence scores for text coherence or image fidelity—addressing concerns like overconfidence, bias, and hallucinations while fostering user trust and accountability. Integrating conformal methods also facilitates real-time monitoring and adaptive feedback, ensuring system reliability as models scale in complexity and data size.

### 13.2.2 LTM in O-RAN architecture

The willingness to introduce currently vibrant ideas (such as openness, virtualization, interoperability, modularity, and AI-oriented functioning) to wireless networks has paved the way for the establishment of the O-RAN ALLIANCE. Open RAN, as such, assumes a transformation of the contemporary wireless architecture and functions that will leverage the incorporation of the above-mentioned features while maximizing the usage COTS hardware and minimizing proprietary ones.

The proposed O-RAN architecture [520], as well as designed interfaces and protocols, supports, in general, three control loops, focusing on different functionalities and reflecting three different time scales. The Non-RT (Non-Real Time) control cycle concentrates on 1 second or a longer time-frame, whereas Near-RT (Near-Real Time) operates between 10 milliseconds to 1s period. The Real-Time (RT) loop reflects the changes below 10 milliseconds. By definition, control loops assume the presence of some control and management unit (function, entity) responsible for intelligent reactions to the varying situations in the network. Such a role is entrusted to the RAN Intelligent Controllers, RICs, which operate on a longer (above 1 second) and shorter (between 10 milliseconds and 1 second) scale. These are called, Non- and Near-RT RICs, accordingly. The former RIC is envisaged as a part of the Service Management and Orchestration (SMO) and is connected with the latter RIC via the A1 interface. Following the O-RAN specifications [520], one of the key roles of the Non-RT RIC is to support intelligent optimization of the underlying network. To achieve this goal, a variety of existing machine learning (ML) and artificial intelligence (AI) tools may be utilized. The RIC may benefit from access to rich data analytics and apply AI/ML reasoning. Moreover, it also manages the ML models within SMO and instructs the Near-RT RIC about model changes via the mentioned A1 interface (using ML Model Management Service).

As discussed, AI/ML creates one of the pillars of O-RAN by natively including and embedding artificial intelligence into wireless networks. The variety of envisaged AI/ML applications is huge, starting from traffic management and security applications through energy-efficiency optimization and ending at conflict management or detection. More information about the ML Framework in O-RAN can be found in [521] and in [204].

Modular applications, namely rApps associated with Non-RT RC and xApps with Near-RT RIC, constitute significant components of the overall O-RAN ecosystem. Following the microservice concepts, x/rApps deliver to the mobile network operators complete yet modular functionalities responsible for the execution of specific functions in the wireless network. Examples of such prospective applications are coverage and capacity optimization, energy-saving management or traffic management operating on a long timescale at Non-RT, and traffic steering - optimizing traffic in a shorter time scale at Near-RT RIC. By assumption, these applications may benefit from the utilization of dedicated AI/ML solutions. The AI/ML functionality may be the inherent feature of the application itself, however, the applications may also potentially request access to some existing ML models or tools available through the corresponding RICs. This is because the sole application is not the right place for long-term model training and modification; it could rather rely on a pre-trained model created by the mobile network operator and managed within SMO.

One of the observations that can be made so far is that O-RAN is originally and natively prepared and designed to incorporate various AI/ML, including Large Telecom Models, LTMs. The key aspect is that LTMs are LMs that require large amounts of resources. Analyzing the O-RAN architecture and following the initial agreements on ML framework specified in [204], the SMO may be the entity responsible for managing the whole process related to LTM application in future wireless networks, whereas the tandem of RICs - for its utilization towards better network optimization. One of the key challenges in that respect is the need for a huge amount of computing power and memory to process all the available data effectively in a reasonable time. It is often said that from the implementation and hardware perspective, it is the SMO (and Non-RT RIC) that will have allocated more computing resources and power to operate. Thus, SMO with the overall ML management loop discussed in [521], is naturally designed for incorporation of LTM tools. Moreover, SMO and Non-RT RIC are foreseen to utilize access to rich context information saved in numerous databases and origin from various sources. Thus, Generative AI tools may then be used to empower multi-modal reasoning and optimization of the network. The benefits of multi-modal LTM-based reasoning and inference toward better network optimization are manifolds. First, by processing huge amounts of data collected over a long time from the underlying network, the AI models may reflect various features of network functioning, both from a short and long-term perspective. These models may also reflect the problem of scale, where different decisions may be made for smaller (or fragments of a) networks with a low number of nodes and different for huge networks with multiple base stations and users. Second, the LTM models may benefit from combining various types of information available at the SMO in the form of enrichment information. Such context information may include, for example, the preferences of particular users, radio-environment or radio service maps, bus and tram schedules, or plans of local municipalities in terms of mass events. By intelligent processing of such data, LTMs may lead to better optimization decisions. Third, modular applications—both rApps and xApps—will not be capable of utilizing their own LTM. Thus, incorporating LTM as the native part of the O-RAN architecture will give new opportunities for xApp and rApp designers to better create their own optimization functions.

### 13.2.3 LTM on 6G Digital Twin

Digital Twin (DT) technology, augmented with LTMs, represents a transformative opportunity for advancing the 5G industry into the 6G era. By creating real-time virtual replicas of network infrastructures—including devices, base stations, and communication links—DTs enable precise simulation and optimization of network performance. The integration of LTMs within DTs enhances their intelligence and adaptability, empowering networks to not only mirror physical systems but also analyze, predict, and proactively resolve challenges with unprecedented efficiency. This synergy is essential for managing the complexity of 6G networks, characterized by ultra-dense connectivity, intelligent automation, and dynamic resource allocation.

LTMs play a pivotal role in augmenting DT capabilities, bringing advanced natural language understanding, rea-soning, and decision-making to network management. With their ability to process and interpret vast datasets, LTMs embedded in DTs can extract actionable insights, optimize resource configurations, and facilitate real-time decision-making in a human-like manner. This allows operators to simulate complex scenarios, train AI models on synthetic data, and achieve faster and more accurate predictions of network behavior. The result is a closed-loop system where LTMs enhance the introspective, predictive, and proactive modes of DTs, driving autonomous network operations and continuous self-optimization.

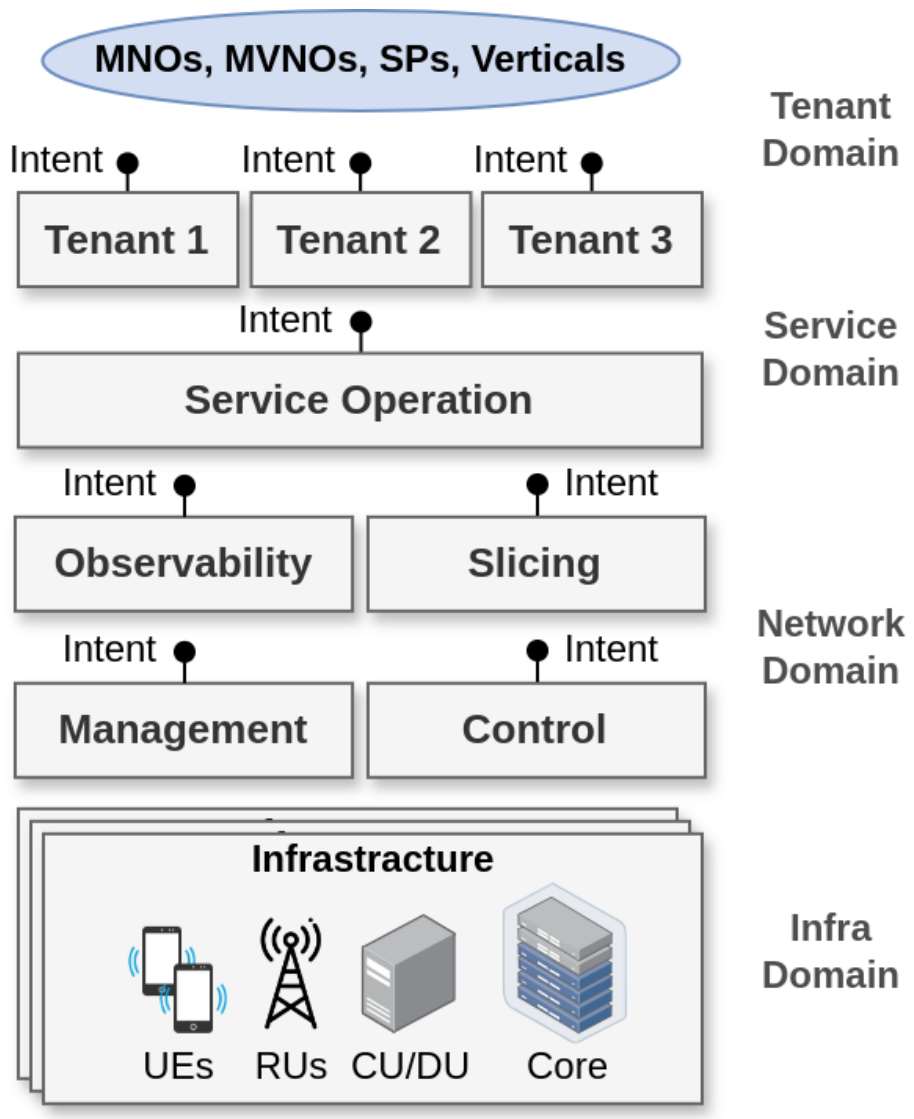

Figure 68: Next-Generation Logical Network Framework

Incorporating LTMs also opens new avenues for automation and innovation in 6G networks. For example, LTMs within DTs can simplify network troubleshooting by interpreting logs and metrics in natural language, enabling seamless collaboration between operators and intelligent systems. They can also streamline policy generation, adapt to shifting network demands, and provide personalized network experiences for end-users. From planning and testing automation to real-time AI training and data-driven decision-making, DTs enriched with LTMs position themselves as an essential innovation for achieving the agility, scalability, and intelligence required in next-generation networks.

## 13.3 Framework towards Next-G

Taking into account all the challenges and innovative enablers of LTM, we consolidate a complete framework. It envisions conceptually the complete architecture, stack, and principles for creating truly autonomous next-generation (Next-G) networks. Fig. 68 shows the complete logical framework consolidated to envision next-generation autonomous networks driven by LTMs. A reference and innovative work explaining and diving into these concepts is MAESTRO [522]. The framework is based on state-of-the-art principles and key enablers for creating truly autonomous networks, which we analyze thoroughly in the next subsections, including cloud-native principles, intent-based networking, multi-agent systems and automatic control systems.

### 13.3.1 Cloud-native Next-G Networks

Cloud-native principles are fundamental to realizing the dynamic network automation required for 6G networks, particularly in integrating and managing LTMs. By leveraging microservices architectures, containerization, and continuous integration/continuous deployment (CI/CD) pipelines, cloud-native designs enable the modular, scalable, and flexible infrastructure needed for 6G's ultra-dynamic environments. These principles allow network functions and LTMs to be deployed, upgraded, and scaled independently, ensuring seamless adaptability to varying traffic loads, user demands, and evolving service requirements. This modularity reduces complexity, shortens development cycles, and enhances the reliability of automated operations.

A key benefit of cloud-native principles is their ability to simplify network configuration and management. Through declarative APIs, automated provisioning tools, and orchestration frameworks like Kubernetes, operators can manage complex network configurations efficiently and uniformly across diverse environments. This streamlines the deployment and optimization of LTMs, which rely on vast computational resources and dynamic dataflows. Moreover, cloud-native automation reduces manual interventions, minimizes configuration errors, and accelerates the rollout of innovative network functions. In 6G, where real-time adaptability and intelligence are paramount, cloud-native principles enable a robust foundation for dynamic orchestration, efficient LTM deployment, and continuous network optimization, making them indispensable for the next generation of network infrastructure.

### 13.3.2 Intent-based Networking (IBN)

Intent-based networking (IBN) is essential for next-generation networks like 6G because it simplifies network management by allowing operators to specify high-level objectives rather than detailed configurations. In highly dynamic and complex environments, traditional manual configuration becomes impractical due to the sheer number of devices, services, and user demands. IBN enables operators to define their desired outcomes—such as latency requirements, security policies, or bandwidth allocations—and the network autonomously adjusts its parameters to fulfill these intents. This level of automation and abstraction not only reduces operational overhead but also enhances agility, allowing the network to adapt in real-time to changing conditions and requirements.

For LTMs, which are advanced AI models tailored for telecom applications, IBN provides a framework that aligns high-level business objectives with network operations. LTMs can interpret the intents specified by operators and translate them into actionable configurations across the network infrastructure. By leveraging IBN, LTMs can more effectively automate network optimization, predict potential issues, and implement proactive measures. This synergy between IBN and LTMs leads to smarter, more responsive networks that can deliver personalized services, improve resource utilization, and enhance overall user experience in next-generation telecom environments.

### 13.3.3 Optimization Methods

Standard optimization methods from control theory and automatic control systems are critical for maintaining stability and ensuring consistent enforcement of network intents in next-generation networks like 6G. These methods provide a mathematically grounded framework for managing the dynamic and variable conditions inherent in telecom networks, such as fluctuating traffic loads, changing channel qualities, and varying user demands. By continuously monitoring network states and applying control algorithms, these methods enable precise adjustments to network configurations, ensuring that the high-level intents—such as latency, throughput, or energy efficiency—are consistently achieved. This stability is crucial for delivering reliable services, even under challenging and unpredictable conditions.

Incorporating optimization techniques into intent enforcement ensures that networks remain resilient and adaptive. For example, methods like model predictive control (MPC) or feedback-based control loops can dynamically adjust resource allocations or optimize routing strategies in response to real-time variations in channel quality or device mobility. These techniques complement the capabilities of LTMs by providing robust mechanisms to implement the decisions derived from AI systems. Together, they form a cohesive system where intents are translated into actionable policies, monitored for compliance, and continually refined to maintain network performance and stability across diverse and evolving conditions. This integration of standard optimization methods ensures that next-generation networks achieve the reliability and efficiency required for the 6G era.

### 13.3.4 Multi-Agent Networking

Multi-agent systems (MAS) play a vital role in next-generation networks, particularly in scenarios where multi-tenant approaches and distributed architectures are prominent as shown in Fig. 68. In such environments, numerous independent entities—such as tenants, operators, and services—must coexist and coordinate within shared network infrastructure. MAS provides a framework where autonomous agents, representing these entities, can collaborate, negotiate, and resolve conflicts in a dynamic and scalable manner. This is crucial for maintaining network efficiency, ensuring fair resource allocation, and mitigating potential disputes in multi-tenant networks. Additionally, MAS enables decentralized decision-making, which is essential for managing the complexity and scale of distributed 6G networks.

LTMs enhance MAS by introducing advanced contextual understanding and natural language capabilities. These capabilities allow agents to interpret nuanced intents, communicate effectively in natural language, and make decisions aligned with overarching network objectives. For instance, LTMs can facilitate seamless collaboration between agents by interpreting and generating policies or resolving conflicts based on contextual knowledge of network conditions and tenant requirements. By combining MAS with LTMs, next-generation networks gain a robust mechanism for coordination, enabling applications such as autonomous resource management, proactive fault handling, and adaptive service provisioning. This synergy ensures that distributed and multi-tenant networks remain cohesive, efficient, and responsive to evolving demands.

### 13.3.5 Small Language Models (SLMs)

Small Language Models (SLMs) are increasingly critical for next-generation networks due to their ability to provide efficient decision-making while significantly reducing overhead and infrastructure demands. Unlike larger models, SLMs require less computational power, memory, and storage, making them ideal for deployment in resource-constrained environments. Their lightweight architecture enables them to fit on a single GPU, ensuring cost-effective scalability and lower energy consumption. Despite their reduced size, SLMs can deliver sufficient accuracy and performance for many telecom applications, including intent interpretation, anomaly detection, and policy generation, making them a practical solution for real-world use cases.

SLMs are particularly well-suited for low-latency and resource-constrained environments, such as those found at the network edge. As edge computing becomes a cornerstone of 6G networks, the ability to deploy intelligent models close to the end-user is essential for minimizing latency and improving service responsiveness. SLMs can efficiently operate in these decentralized settings, enabling localized decision-making and real-time adjustments without relying on centralized infrastructure. This makes them indispensable for edge-based applications like dynamic resource allocation, real-time monitoring, and edge device coordination, ensuring that next-generation networks achieve the agility and efficiency required to meet evolving demands.

GenAINet

IEEE ETI